\documentclass[12pt,  twoside, titlepage]{article}
\pdfoutput=1
\usepackage{german}
\usepackage{amsmath}
\usepackage{phonetic}
\usepackage{graphicx}
\usepackage{rotating}
\usepackage{longtable}
\usepackage{exscale}  % SIEHE HANDBUCH S. 142
\usepackage[nottoc]{tocbibind}  %für Abbildungsverzeichnis in Inhaltsverz
\usepackage{amsthm} % fuer qed Zeichen

\usepackage{graphicx}

\begin{document}

%Gemeinsame Nummerierung
\newtheorem{satz}{Satz}[section]
\newtheorem{bezeichn}[satz]{Bezeichnungen}%[section]
\newtheorem{einebezeichn}[satz]{Bezeichnung}%[section]
\newtheorem{bemerkung}[satz]{Bemerkung}%[section]
\newtheorem{beispiel}[satz]{Beispiel}%[section]
\newtheorem{bemerkungen}[satz]{Bemerkungen}%[section]
\newtheorem{definition}[satz]{Definition}%[section]
\newtheorem{hilfs}[satz]{Hilfssatz}%[section]
\newtheorem{defundbem}[satz]{Definition und Bemerkung}%[section]
\newtheorem{korollar}[satz]{Korollar}%[section]
\newtheorem{allgvor}[satz]{Allgemeine Voraussetzungen}%[section]

\thispagestyle{plain}
%---------
\newfont{\Kall}{cmsy10 at 12pt}

\def\Z{{Z\kern-.3567em Z}}
\def\N{{I\kern-.35em N}}
\def\R{{I\kern-.35em R}}
\def\CC{{C\kern-.65emC}}
\def\Eins{{1\kern-.35em 1}}

\newcommand{\trsp}[1]{{#1}^{\mbox{\sc T}}}
                    % transponiert
\newcommand{\adj}[1]{{#1}^{\dagger}}    % adjungiert
\newcommand{\konjtr}[1]{{#1}^{\mbox{\sc H}}}
                                        % konjugiert transponiert
\newcommand{\cc}[1]{{#1}^{\ast}}        % komplex konjugiert

\newcommand{\reff}[1]{(\ref{#1})}                       % Referenz in Klammern
\newcommand{\mti}[1]{_{\mbox{\footnotesize {#1}}}}   % mathematischer Textindex

\newcommand{\oi}[1]{^{(#1)}}            % oberer Index
\newcommand{\ui}[1]{_{(#1)}}            % unterer Index

\newcommand{\const}{\mbox{const}}

\newcommand{\spur}{\mbox{spur}}

\newcommand{\hquer}{\mbox{\planck}}

\newcommand{\Lzwei}{{\mbox{L}^2}}
\newcommand{\Leins}{{\mbox{L}^1}}
\newcommand{\Leinsloc}{{\mbox{L}^1_{\mbox{loc}}}}
\newcommand{\ACloc}{{\mbox{AC}_{\mbox{loc}}}}

\newcommand{\dy}{{\:\mbox{d}y}}
\newcommand{\dk}{{\:\mbox{d}k}}
\newcommand{\ds}{{\:\mbox{d}s}}
\newcommand{\dr}{{\:\mbox{d}r}}
\newcommand{\dt}{{\:\mbox{d}t}}
\newcommand{\dx}{{\:\mbox{d}x}}
\newcommand{\dkappa}{{\:\mbox{d}\kappa}}
\newcommand{\dlambda}{{\:\mbox{d}\lambda}}

\newcommand{\dnachdk}{{\frac{\mbox{d}}{\mbox{d}k}}}
\newcommand{\dnachdr}{{\frac{\mbox{d}}{\mbox{d}r}}}
\newcommand{\dnachdkappa}{{\frac{\mbox{d}}{\mbox{d}\kappa}}}

\newcommand{\modulus}[1]{\left| {#1} \right|}  % Betrag
\newcommand{\norm}[1]{\left|\left| {#1} \right|\right|}
\newcommand{\normajbj}[1]{\left|\left| {#1} \right|\right|_{[a_j,b_j]}}

\newcommand{\sign}{{\mbox{sign}\,}}

\newcommand{\arccot}{{\mbox{arccot}}}

\newcommand{\supp}{{\mbox{supp}\,}}
\newcommand{\komp}{{\subset \kern-.3567em \subset}}

\newcommand{\kleinoglm}{{\mbox{o}_{\mbox{glm}}}}

\newcommand{\eps}{\varepsilon}
\newcommand{\ro}{\varrho}
\newcommand{\fhi}{\varphi}
\newcommand{\teta}{\vartheta}
\newcommand{\tetas}{\tilde{\vartheta}}
\newcommand{\Rs}{\tilde{R}}

\newcommand{\ros}{\tilde{\varrho}}

\newcommand{\Psvektor}{ \left(
                \begin{array}{c}
                               \Psi_1  \\
                               \Psi_2
                            \end{array}
            \right)
                       }
\newcommand{\Psvektors}{ \begin{array}{c}
                           \tilde{\Psi}_1 \\
                           \tilde{\Psi}_2
                        \end{array}
                       }
\newcommand{\Psis}{\tilde{\Psi}}

\newcommand{\ind}{{_{n;j-1}}}     % st"andig wiederkehrender Index
\newcommand{\indnpluseins}{{_{{n+1};j-1}}}      % st"andig wiederkehrender Index

\newcommand{\wurzel}{{\sqrt{\frac{\lambda-1}{\lambda+1}}}}
                                    %st"andig wiederkehrende Wurzel
\newcommand{\plmin}{{\frac{\lambda+1}{\lambda-1}}}
\newcommand{\minpl}{{\frac{\lambda-1}{\lambda+1}}}

\newcommand{\jinebisn}{{j \in \{1, ..., n\}}}
                                    % st"andig wiederkehrend
\newcommand{\Mj}{{M^{(j)}}}
\newcommand{\Meins}{{M^{(1)}}}
\newcommand{\Mn}{{M^{(n)}}}
\newcommand{\Mjee}{{M^{(j)}_{11}}}
\newcommand{\Mjez}{{M^{(j)}_{12}}}
\newcommand{\Mjze}{{M^{(j)}_{21}}}
\newcommand{\Mjzz}{{M^{(j)}_{22}}}

\newcommand{\Mjs}{{\tilde{M}^{(j)}}}
\newcommand{\Meinss}{{\tilde{M}^{(1)}}}
\newcommand{\Mns}{{\tilde{M}^{(n)}}}
\newcommand{\Mjees}{{\tilde{M}^{(j)}_{11}}}
\newcommand{\Mjeesquad}{{\left(\tilde{M}^{(j)}_{11}\right)^2}}
\newcommand{\Mjezs}{{\tilde{M}^{(j)}_{12}}}
\newcommand{\Mjezsquad}{{\left(\tilde{M}^{(j)}_{12}\right)^2}}
\newcommand{\Mjzes}{{\tilde{M}^{(j)}_{21}}}
\newcommand{\Mjzesquad}{{\left(\tilde{M}^{(j)}_{21}\right)^2}}
\newcommand{\Mjzzs}{{\tilde{M}^{(j)}_{22}}}
\newcommand{\Mjzzsquad}{{\left(\tilde{M}^{(j)}_{22}\right)^2}}

\newcommand{\Aj}{{A_j}}
\newcommand{\Aeins}{{A_1}}
\newcommand{\An}{{A_n}}
\newcommand{\Bj}{{B_j}}
\newcommand{\Beins}{{B_1}}
\newcommand{\Bn}{{B_n}}
\newcommand{\Cj}{{C_j}}
\newcommand{\Ceins}{{C_1}}
\newcommand{\Cn}{{C_n}}

\newcommand{\Ajs}{{\tilde{A}_j}}
\newcommand{\Aeinss}{{\tilde{A}_1}}
\newcommand{\Ans}{{\tilde{A}_n}}
\newcommand{\Bjs}{{\tilde{B}_j}}
\newcommand{\Beinss}{{\tilde{B}_1}}
\newcommand{\Bns}{{\tilde{B}_n}}
\newcommand{\Cjs}{{\tilde{C}_j}}
\newcommand{\Ceinss}{{\tilde{C}_1}}
\newcommand{\Cns}{{\tilde{C}_n}}

\newcommand{\Aajs}{{\tilde{A}_{a_j}}}
\newcommand{\Bajs}{{\tilde{B}_{a_j}}}
\newcommand{\Cajs}{{\tilde{C}_{a_j}}}

\newcommand{\Aaj}{{A_{a_j}}}
\newcommand{\Baj}{{B_{a_j}}}
\newcommand{\Caj}{{C_{a_j}}}
\newcommand{\ffj}{{f_j}}   %ehemals hiess der Befehl \fj  geht nicht mehr seit das phonetic package eingebunden ist
\newcommand{\feins}{{f_1}}
\newcommand{\fn}{{f_n}}
\newcommand{\hj}{{h_j}}
\newcommand{\hjnull}{{h_{j_0}}}
\newcommand{\mj}{{m_j}}
\newcommand{\meins}{{m_1}}
\newcommand{\mn}{{m_n}}
\newcommand{\mjnull}{{m_{j_0}}}
\newcommand{\mi}{{m_i}}

\newcommand{\faj}{{f_{a_j}}}
\newcommand{\maj}{{m_{j}}}

\newcommand{\ffjs}{{\tilde{f}_j}}
\newcommand{\feinss}{{\tilde{f}_1}}
\newcommand{\fns}{{\tilde{f}_n}}
\newcommand{\hjs}{{\tilde{h}_j}}
\newcommand{\hjnulls}{{\tilde{h}_{j_0}}}
\newcommand{\hs}{{\tilde{h}}}
\newcommand{\mjs}{{\tilde{m}_j}}
\newcommand{\ms}{{\tilde{m}}}
\newcommand{\meinss}{{\tilde{m}_1}}
\newcommand{\mns}{{\tilde{m}_n}}
\newcommand{\mjnulls}{{\tilde{m}_{j_0}}}
\newcommand{\mis}{{\tilde{m}_i}}
\newcommand{\Ckkappa}{{C_{k,\kappa}}}
\newcommand{\Ckkappas}{{\tilde{C}_{k,\kappa}}}
\newcommand{\majs}{{\tilde{m}_{j}}}
\newcommand{\fajs}{{\tilde{f}_{j}}}
\newcommand{\hajs}{{\tilde{h}_{j}}}
\newcommand{\hajnulls}{{\tilde{h}_{{j_0}}}}

\newcommand{\ffjsk}{{\tilde{f}_j^{(k)}}}
\newcommand{\mjsk}{{\tilde{m}_j^{(k)}}}
\newcommand{\majsk}{{\tilde{m}_{a_j}^{(k)}}}
\newcommand{\hajsk}{{\tilde{h}_{a_j}^{(k)}}}
\newcommand{\msk}{{\tilde{m}}}
\newcommand{\hjsk}{{\tilde{h}_j^{(k)}}}
\newcommand{\hsk}{{\tilde{h}}}

\newcommand{\aj}{{a_j}}
\newcommand{\ajme}{{a_{j-1}}}
\newcommand{\fsaj}{{\tilde{f}_{a_j}}}
\newcommand{\msaj}{{\tilde{m}_{a_j}}}
\newcommand{\Asaj}{{\tilde{A}_{a_j}}}
\newcommand{\Msaj}{{\tilde{M}_{a_j}}}
\newcommand{\AAA}{{\cal{A}}}

\newcommand{\mass}{m}
\newcommand{\angu}{l}
\newcommand{\Gk}{G_k}
\newcommand{\Fk}{F_k}
\newcommand{\GGj}{{\cal{G}}_j}
\newcommand{\GGjme}{{\cal{G}}_{j-1}}
\newcommand{\GGime}{{\cal{G}}_{i-1}}
\newcommand{\GGnyunten}{{\cal{G}}_\nyunten}
\newcommand{\GGnyukme}{{\cal{G}}_{\nyunten+k-1}}
\newcommand{\GGnyuime}{{\cal{G}}_{\nyunten+i-1}}
\newcommand{\GGnyusme}{{\cal{G}}_{\nyunten+s-1}}
\newcommand{\GGnyuk}{{\cal{G}}_{\nyunten+k}}
\newcommand{\GGnyui}{{\cal{G}}_{\nyunten+i}}

\newcommand{\GGGG}{{{\cal{F}}}} % ,Fehler' bei Ableitung von Spektralfunktion
                          % mit Drehimpuls

\newcommand{\roDj}{{\varrho^{(D_j)}}}
\newcommand{\roDjminus}{{\varrho_{j-1}}}
\newcommand{\roDnminus}{{\varrho_{{n-1}}}}
\newcommand{\roDnb}{{\varrho_{b;n}}}
\newcommand{\roDn}{{\varrho_{n}}}

\newcommand{\roDns}{{\tilde{\varrho}_{n}}}
\newcommand{\roDnbs }{{\tilde{\varrho}_{b;n}}}
\newcommand{\roDnhut}{\widehat{\varrho}_n}

\newcommand{\roDnyobens}{{\tilde{\varrho}_\nyoben}}

\newcommand{\D}{{\cal{D}}} % Diracfactor
\newcommand{\Dmischl}{\tilde{\cal{D}}} % Diracfactor mit Dreh: Def über R bei 1

\newcommand{\C}{{\cal{C}}} % das ist die Konstante mit
                           % \ffj \ge \C ( j\in \N )  (wird bei lemma2B verwendet)
\newcommand{\Cvonk}{{\cal{C}}_{k}} % das ist die Konstante mit
\newcommand{\KC}{{{\cal{K}}}}
\newcommand{\jnull}{{j_0}} % f"ur Lemma2B
\newcommand{\Nnynykappa}{{N(\nyunten, \nyoben, \kappa)}} % f"ur Lemma2B
\newcommand{\Nknynykappa}{{{\cal{N}}_{k}(\nyunten, \nyoben, \kappa)}}
       % f"ur Lemma2B mit Dreh
\newcommand{\Najknynykappa}{{{\cal{N}}_{(a_j\!);k}(\nyunten,\! \nyoben,\! \kappa)}}

\newcommand{\ffjvon}{{f_j\left(\kappa, \teta_{%D_
n;j-1}(\kappa)
                       - \kappa d_j\right)}}

\newcommand{\lambdabn}{{\lambda_{b;\tilde{n}}}}
\newcommand{\phibn}{{\varphi_{b;\tilde{n}}}}
\newcommand{\phibne}{{\varphi_{b;\tilde{n}}^{(1)}}}
\newcommand{\phibnz}{{\varphi_{b;\tilde{n}}^{(2)}}}
\newcommand{\abnquadrat}{{a_{b;\tilde{n}}^2}}
\newcommand{\abnquadrats}{{\tilde{a}_{b;\tilde{n}}^2}}
\newcommand{\RDnnquadrat}{{R_{n}^2}}
\newcommand{\RDjjquadrat}{{R_{j;j}^2}}
\newcommand{\RDnjquadrat}{{R_{n;j}^2}}
\newcommand{\RDnjminusquadrat}{{R_{n;j-1}^2}}

\newcommand{\RDnnquadrats}{{\tilde{R}_{n}^2}}
\newcommand{\RDjjquadrats}{{\tilde{R}_{j;j}^2}}
\newcommand{\RDnjquadrats}{{\tilde{R}_{n;j}^2}}
\newcommand{\RDnjminusquadrats}{{\tilde{R}_{n;j-1}^2}}

\newcommand{\Nblambdalambda}{{N_b (\lambda_1,\lambda_2)}}
\newcommand{\Nblambdalambdas}{{\tilde{N}_b (\lambda_1,\lambda_2)}}

\newcommand{\mynull}{{\mu_L}}
\newcommand{\myDn}{{\mu_{n}}}
\newcommand{\myDnplus}{{\mu_{{n+1}}}}
\newcommand{\myD}{{\mu_{D}}}
\newcommand{\myDj}{{\mu_{j}}}
\newcommand{\myDjsss}{{\tilde{\mu}_{j}}}
\newcommand{\myDjssss}{{\tilde{\mu}_{j}}}
%myDjssss
%\newcommand{\myDjplus}{{\mu_{D_{j+1}}}}
\newcommand{\myDjplus}{{\mu_{{j+1}}}}
\newcommand{\myDsssjplus}{{\tilde{\mu}_{{j+1}}}}
\newcommand{\myDjssspluss}{{\tilde{\mu}_{{j+1}}}}

\newcommand{\myDss }{\tilde{\mu}_D}

\newcommand{\myDjminus}{{\mu_{{j-1}}}}
\newcommand{\myDndTheta}{{\mu_{n;d,\Theta}}}

\newcommand{\myDnschl}{{{\tilde\mu}_{n}}}

\newcommand{\myDjsk}{{\tilde{\mu}_{j;k}}}
\newcommand{\myDjssssk}{{\tilde{\mu}_{j;k}}}
\newcommand{\myDjplussk}{{\tilde{\mu}_{{j+1};k}}}
\newcommand{\myDjsssplussk}{{\tilde{\mu}_{{j+1};k}}}

%^^%\newcommand{\mynulls}{{\tilde{\mu_0}}}
%\newcommand{\mys}{{\tilde{\mu}}}
%\newcommand{\myDns}{{\tilde{\mu}_{D_n}}}
\newcommand{\myDns}{{\tilde{\mu}_{n}}}
\newcommand{\myDnpluss}{{\tilde{\mu}_{{n+1}}}}
\newcommand{\myDs}{{\tilde{\mu}_{D}}}
\newcommand{\myDjs}{{\tilde{\mu}_{j}}}
\newcommand{\myDjpluss}{{\tilde{\mu}_{{j+1}}}}
\newcommand{\myDjminuss}{{\tilde{\mu}_{{j-1}}}}
\newcommand{\myDndThetas}{{\tilde{\mu}_{n;d,\Theta}}}

\newcommand{\mys}{{\tilde{\mu}}}
\newcommand{\myks}{{\tilde{\mu}^{(k)}}}
\newcommand{\mykDns}{{\tilde{\mu}_{D_n}^{(k)}}}
\newcommand{\mykDnpluss}{{\tilde{\mu}_{{n+1;k}}}}
\newcommand{\mykDjs}{{\tilde{\mu_{j;k}}}}
\newcommand{\mykDjpluss}{{\tilde{\mu}_{{j+1};k}}}
\newcommand{\mykDjminuss}{{\tilde{\mu}_{{j-1};k}}}

\newcommand{\nyunten}{{\underline{\nu}}}
\newcommand{\nyoben}{{\overline{\nu}}}
\newcommand{\deltany}{{\Delta\nu}}

\newcommand{\alphabeta}{{[\alpha,\beta]}}
\newcommand{\kappaeizw}{{[\kappa_1,\kappa_2]}}
\newcommand{\doppelInterv}{{[-\beta,-\alpha]\cup[\alpha,\beta]}}

%\newcommand{\K}{{K}}         % Das ist die GEMEINSAME  Konstante aus Lemma4
               % ersetzt durch \KC

\newcommand{\Se}{{\cal{S}}_1}
\newcommand{\Sn}{{\cal{S}}_n}
\newcommand{\Sm}{{\cal{S}}_m}
\newcommand{\Sl}{{\cal{S}}_l}
\newcommand{\Sk}{{\cal{S}}_k}
\newcommand{\Sj}{{\cal{S}}_j}
\newcommand{\Ss}{{\cal{S}}_s}
\newcommand{\Sknull}{{\cal{S}}_{k_0}}
\newcommand{\Pl}{{\cal{P}}_l}
\newcommand{\Plnull}{{\cal{P}}_{l_0}}

\newcommand{\Seschl}{{\tilde{\cal{S}}}_1}
\newcommand{\Snschl}{{\tilde{\cal{S}}}_n}

\newcommand{\W}{\omega}

\newcommand{\RRe}{{{\cal{R}}_1}}
\newcommand{\RRz}{{{\cal{R}}_2}}

\newcommand{\RRn}{{{\cal{R}}}}  % den Index braucht man wohl gar nicht
%\newcommand{\RRn}{{{\cal{R}}_n}}   % f"ur Anhang1 : das ist das geschwungene
          % R, das gleich dem lambda-1 fachen des Pr"uferradiusquadrats ist
\newcommand{\RRjschlange}{{{\tilde{\cal{R}}}_j}}
          % f"ur Anhang1 : das ist das geschwungene
          % R (Schlange, das gleich dem lambda+1 fachen des
          % alternativen Pr"uferradiusquadrats ist

%fuer Anhang 2
\newcommand{\TTnull}{{{\cal{T}}_0}}

%\newcommand{\SS}{{hier mueste ein geschwungenes S kommen}} % und zwar wie folgt:
%\newcommand{\SS}{{\cal{S}}}
% aber das funktioniert seit der umstellung von documentstyle auf dokumentclass nicht mehr

\newcommand{\TS}{{\cal{S}}}
\newcommand{\TSs}{{\tilde{\cal{S}}}}
\newcommand{\OO}{{\cal{O}}}

\newcommand{\dH}{{\partial_H}}
\newcommand{\dHH}{{\partial_H^2}}

\newcommand{\Einheitsmatrix}{{\left(\begin{array}{cc}
                                       1 & 0 \\
                                       0 & 1
                                \end{array}
                          \right)
                         }}
\newcommand{\Nullmatrix}{{\left(\begin{array}{cc}
                                       0 & 0 \\
                                       0 & 0
                                \end{array}
                          \right)
                         }}
\newcommand{\grossO}{{\mbox{O}}}
\newcommand{\grossOglm}{{\mbox{O}_{\mbox{glm}}}}
\newcommand{\grossOfrag}{{\mbox{O}_{\mbox{glm}}}}
   % in Banhang2

\newcommand{\Sschl}{{{S}}}
\newcommand{\sschl}{{{s}}}
\newcommand{\rschl}{{\tilde{r}}}
\newcommand{\drschl}{{\mbox{d}\tilde{r}}}

\newcommand{\gammma}{{\gamma}}

\newcommand{\Konst}{{\cal{K}}}

\newcommand{\Phiee}{{\Phi_{11}}}
\newcommand{\Phiez}{{\Phi_{12}}}
\newcommand{\Phize}{{\Phi_{21}}}
\newcommand{\Phizz}{{\Phi_{22}}}

\newcommand{\Phijs}{{\tilde{\Phi}^{(j)}}}
\newcommand{\Us}{{\tilde{U}}}
\newcommand{\Ujs}{{\tilde{U}^{(j)}}}
\newcommand{\Ujshochminuse}{{\tilde{U}^{(j)-1}}}
\newcommand{\ujrps}{{\tilde{u}^{(j)}_{rp}}}
\newcommand{\ujees}{{\tilde{u}^{(j)}_{11}}}
\newcommand{\ujezs}{{\tilde{u}^{(j)}_{12}}}
\newcommand{\ujzes}{{\tilde{u}^{(j)}_{21}}}
\newcommand{\ujzzs}{{\tilde{u}^{(j)}_{22}}}
\newcommand{\Ujshochmeins}{{\tilde{U}^{(j)-1}}}
\newcommand{\Zj}{{Z^{(j)}}}
\newcommand{\zjp}{{z^{(j)}_{p}}}
\newcommand{\zjep}{{z^{(j)}_{1,p}}}
\newcommand{\zjzp}{{z^{(j)}_{2,p}}}
\newcommand{\zjrp}{{z^{(j)}_{r,p}}}

\newcommand{\zp}{{z_{p}}}
\newcommand{\zep}{{z_{1,p}}}
\newcommand{\zzp}{{z_{2,p}}}
\newcommand{\zrp}{{z_{r,p}}}

\newcommand{\Uee}{{U_{11}}}
\newcommand{\Uez}{{U_{12}}}
\newcommand{\Uze}{{U_{21}}}
\newcommand{\Uzz}{{U_{22}}}

\newcommand{\Xee}{{X_{11}}}
\newcommand{\Xez}{{X_{12}}}
\newcommand{\Xze}{{X_{21}}}
\newcommand{\Xzz}{{X_{22}}}

\newcommand{\Yee}{{Y_{11}}}
\newcommand{\Yez}{{Y_{12}}}
\newcommand{\Yze}{{Y_{21}}}
\newcommand{\Yzz}{{Y_{22}}}

\newcommand{\Uhochmeins}{{U^{-1}}}

\newcommand{\AWP}{{Anfangswertproblem}}

\newcommand{\Nenner}{{N(\underline{\nu}, \overline{\nu};\kappa)}}
       % h"aufig wiederkehrender Nenner bei Lemma 4

\newcommand{\ssigma}{\sigma}

\newcommand{\sigmaess}{{\sigma_{%\mbox{
ess%}
}}}
\newcommand{\sigmas}{{\sigma_{%\mbox{
c%}
}}}
\newcommand{\TqD}{{T}}
\newcommand{\TqDmin}{{T_{q_D}^{\mbox{min}}}}
\newcommand{\TqDmax}{{T^{\mbox{max}}}}
\newcommand{\tauqD}{{\tau}}
\newcommand{\taunull}{{\tau_0}}
\newcommand{\Tnull}{{T_{fr}}}
\newcommand{\Tnulls}{{\tilde{T}_{0}}}
\newcommand{\Tnullmin}{{T_0^{\mbox{min}}}}
\newcommand{\Tnulle}{{\hat{T}_{fr}}}   %eindimensionaler Operator

\newcommand{\Tfr}{{T_{\mbox{fr}}}}

\newcommand{\kleinphi}{{\Psi}}

\newcommand{\kleino}{{\mbox{o}_{\mbox{glm}}}}

\newcommand{\TqDschl}{{\tilde{T}}}
\newcommand{\TqDschlmax}{{\tilde{T}^{\mbox{max}}}}
\newcommand{\TqDschlmin}{{\tilde{T}^{\mbox{min}}}}

\newcommand{\tauschlqD}{{{\tau_q}}}
\newcommand{\Psischl}{{\tilde{\Psi}}}

\newcommand{\cschlange}{{\tilde{c}}}

\newcommand{\OperatorRplusmitDreh}{{T_{\R^+}}}
%----------------------------------------------------------------------------
\newcommand{\behaupt}[1]{{\bf #1}}
\newcommand{\Lemma}[2]{{\bf #1}{\it #2}}
\newcommand{\Hilfslemma}[2]{{\bf #1}{\it #2}}
\newcommand{\Korollar}[2]{{\bf #1}{{\it #2}}}
\newcommand{\Bemerkung}[2]{{\bf #1}{ #2}}
\newcommand{\Behauptung}[2]{{\bf #1}{ #2}}
\newcommand{\AllgVor}[2]{{\bf #1}{\it #2}}
\newcommand{\Satz}[2]{{\bf #1}{\it #2}}
\newcommand{\Bezeichnung}[2]{{\bf #1}{\it #2}}
\newcommand{\Definition}[2]{{\bf #1}{\it #2}}
\newcommand{\Beweis}[2]{{{\it Beweis #1}}{#2}

\hfill$\qed$

$\left.\right. $

}

%----------------------------------------------------------------------------

\newcommand{\REF}[1]{REFERENZ #1}

 \newcommand{\Verweis}[1]{---- Verweis #1 ----}
\newcommand{\Einl}{}
\newcommand{\Begr}{}
 \newcommand{\ODER}{ODER}

\newcommand{\BILD}{BILD
                   \vskip.8cm}

%folgende Befehle evtl noch an allgemeine Notation angleichen
%\newcommand{\Anullll}{{A_0}}
%\newcommand{\Aunendlich}{{A_\infty}}
%\newcommand{\Aeinssl}{{A_1}}
%\newcommand{\Aeinsnull}{{A_{1,0}}}
%\newcommand{\Aohnewas}{{A}}
%\newcommand{\halles}{{h}}
%\newcommand{\hallesnull}{{h_0}}

%\newcommand{\Anullll}{{\tilde{\tau}_{q_D;(0,1]}}}
%\newcommand{\Anullll}{{\tilde{\tau}_{q;(0,1]}}}
\newcommand{\Anullll}{{Z_0}}
\newcommand{\Aunendlich}{{Z_\infty}}
\newcommand{\Aeinssl}{{Q_1}}
\newcommand{\Aeinsnull}{{T_0}}
\newcommand{\Aohnewas}{{Q_0}}
\newcommand{\halles}{{\hat{T}}}
\newcommand{\hallesnull}{{h_0}}
\newcommand{\hkkk}{{h_k}}
\newcommand{\hqdn}{{h_{{n}} }}

% ende von vorspann

%titel
%%%%%%%%%%%%%%%%% Titel

\begin{titlepage}

\end{titlepage}

\begin{center}
\LARGE\bfseries
Singul"arstetiges Spektrum \\
\LARGE\bfseries
kugelsymmetrischer Diracoperatoren
\\[10.5ex]
\large\bfseries
Dissertation an der Fakult"at f"ur Mathematik, Informatik und Statistik
der Ludwig-Maximilians-Universit"at M"unchen
\\
$\left.\right.$

%\\[10.5ex]
\large\bfseries
vorgelegt von
\\
$\left.\right.$
\\
Dipl. Phys. Barbara Janauschek
\\
am 07.05.2014
\\
$\left.\right. $
\\
%\hfill
$\left.\right. $
%\begin{figure}[h]
%%\hscpace{20mm}
%\quad\quad\quad\quad\quad\quad
%\centering
%%  \makebox[\textwidth]{%
%%  \hspace*{-1.8\textwidth}
%\includegraphics[origin=c,width=7cm,clip=true,
%viewport=0.cm 0cm 30cm 30cm]{LMU_siegel}
%%}
%\end{figure}
 \hfill
$\left.\right. $
%\begin{figure}[h]
%\hspace{25mm}
%\centering
%\includegraphics[origin=c,width=7cm,clip=true,natwidth=610,natheight=642,
%viewport=0.cm 0cm 30cm 30cm]{LMU_siegel.eps}
%   \end{figure}

\end{center}

%ende von titel
%\include{deck}
%%%%%\pagebreak 	$\left.\right.$   \pagebreak 
\thispagestyle{plain}
\setcounter{page}{0}
$\left.\right.$
\vfill
%1. Berichterstatter: Prof. Dr. H. Kalf

%2. Berichterstatter: Prof. Dr. 

\pagebreak
$\left.\right.$

1. Berichterstatter: Prof. Hubert Kalf

2. Berichterstatter: Prof. Martin Klaus

Tag der Disputation: 06.06.2014
%\vfill
%abstr
%%%%%%%%%%%%%%%%% Abstract

\thispagestyle{plain}

%\begin{quote}
\pagebreak

{\Large \bf Abstract}

$\left.\right. $

$\left.\right. $

The physical picture of particle behaviour that arises from experimental data is that it belongs to one of the following:
Either the particle always stays near the origin or the particle comes from infinity, is scattered and escapes to infinity.
In the quantum mechanical description these two categories of behaviour are associated with point spectrum and absolutely continuous spectrum respectively. The corresponding spectral measures are point measures or absolutely continuous measures.

According to general results of the measure theory a measure can be decomposed in three parts: a pure point part, an absolutely continuous part and a singular continuous part.

In contrast to the well-known particle behaviour of the other two types of spectrum, the singular continuous spectrum is more difficult to interpret.

For the Schr"odinger operator D.B. Pearson constructed an explicit class of potentials that give rise to purely singular continuous spectrum . This example allows the interpretation of the particle behaviour: 
The particle moves arbitrarily far away from the origin but it feels nevertheless the effect of the potential. Therefore it will recur infinitely often to the vicinity of the origin to run off infinitely often.

The result for the Schr"odinger operator leads to the question whether there can be found similar results in relativistic quantum mechanics.

The aim of this paper is to construct for the first time an explicit potential for the Dirac operator that has purely singular continuous spectrum in $(-\infty, -1]\cup [1,\infty) $.
The characteristic trait of this potential is that it consists of bumps whose distance is growing rapidly.
This allows the particle to depart from the origin arbitrarily far. But the overall effect of the bumps will always lead the particle back to the origin.

\pagebreak
{\Large \bf Zusammenfassung}

$\left.\right. $

$\left.\right. $

In der Experimentalphysik beobachtet man einerseits gebundene Zust"ande und andererseits Streuzust"an\-de. Diesen Zust"anden ordnet die klassische Quantenmechanik folgende Spektren zu: Punktspektrum und absolutstetiges Spektrum. Die zugeh"origen Spektalma"se sind Punktma"se bzw. absolutstetige Ma"se.

In der Ma"stheorie ist statuiert, dass jegliches Ma"s in drei Bestandteile zerlegt werden kann, n"amlich ein Punktma"s, ein absolutstetiges Ma"s und ein singul"arstetiges Ma"s.

Wie l"a"st sich dieses singul"arstetige Spektrum in der Quantenmechanik interpretieren?

F"ur den Schr"odingeroperator wurde durch D. B. Pearson eine explizite Potentialklasse konstruiert, die eine Interpretation des Teilchenverhaltens er\-m"og\-licht: Bei singul"arstetigem Spektrum entfernt sich das Teilchen beliebig weit vom Ursprung, es kehrt aber auch beliebig oft zum Ursprung zur"uck.

Gegenstand der vorliegenden Arbeit ist die Konstruktion einer Potentialklasse, die f"ur Diracsysteme  au"serhalb der zentralen L"ucke $ [-1,1]$ rein singul"arstetiges Spektrum aufweist.

Kennzeichnend f"ur diese Potentiale sind Buckel, deren Abst"ande immer gr"o"ser werden.
Aufgrund der gro"sen Abst"ande zwischen den Buckeln kann sich das Teilchen beliebig weit vom Ursprung entfernen. Die Gesamtheit der Potentialbuckel f"uhrt jedoch auch dazu, dass das Teilchen unendlich oft in Ursprungsn"ahe zur"uckkehren mu"s.

Mit den in der vorliegenden Arbeit konstruierten Potentialen konnte erstmals ein solches Beispiel in der relativistischen Quantenmechanik nachgewiesen werden.

\pagebreak

%ende von abstr

\pagebreak 	
\tableofcontents

\pagebreak
%einl
%%%%%%%%%%%%%%%%% Einleitung

\setcounter{page}{1}
\pagenumbering{arabic}

\hfill
\begin{minipage}[t]{8cm}
\sffamily
Deep in the human unconsciousness is a pervasive need for a logical universe that makes sense. But the real universe is always one step beyond logic.
- Frank Herbert, Dune
\end{minipage}

\section{Einleitung}

Die Ma"stheorie statuiert, dass jedes Ma"s in ein Punktma"s, ein absolutstetiges Ma"s und ein singul"arstetiges Ma"s zerlegt werden kann.
In der mathematischen Physik, und hier insbesondere in der Quantenmechanik, werden die Spektren selbstadjungierter Operatoren untersucht. Nun stellt sich die Frage, ob bei den zugeh"origen Spektralma"sen ebenfalls alle drei Typen von Ma"sen auftreten, und, wenn ja, welche Bedeutung sie haben.

Zum Punktspektrum gibt es zahlreiche Untersuchungen.
Das klassische Beispiel schlechthin ist hier der Schr"odingeroperator mit Coulombpotential, der das Wasserstoffatom im nichtrelativistischen Modell beschreibt (siehe beispielsweise \cite{Messiah}, Kapitel 11).
Das Punktspektrum wird durch gebundene Zust"ande erzeugt. Im Beispiel des Wasserstoffatoms ist dies das Elektron, das sich stets in der N"ahe des Kerns aufh"alt. 

Auch beim absolutstetigen Spektrum besteht der deutliche Bezug zu den experimentellen Beobachtungen. Absolutstetiges Spektrum wird bekanntlich durch Streuzust"ande erzeugt. Man denke beispielsweise an Rutherfords Streuexperiment. Bei diesen Zust"anden kommen die Teilchen aus dem Unendlichen, erfahren eine Wechselwirkung und entfernen sich wieder ins Unendliche (\cite{Messiah2}, Kapitel 19).

In der Experimentalphysik sind das allt"aglich beobachtete Verhalten von Teilchen gebundene Zust"ande und Streuzust"ande.
Deshalb wurde beim Entwickeln von Modellen, die diese Experimente beschreiben, das Augenmerk entsprechend auf zwei Spektraltypen gelegt, n"amlich das Punktspektrum und das absolutstetige Spektrum. Die mathematische Forschung konzentrierte sich somit auf Potentialklassen, die singul"arstetiges Spektrum ausschlie"sen. Als stellvertretende  Beispiele seien \cite{PerrySigalSimon} und \cite{reedSimonIV} genannt. Damit wurde der als Normalfall erachteten Situation des Punktspektums und des absolutstetigen Spektrums Rechnung getragen, ohne das singul"arstetige Spektrum n"aher zu untersuchen. \cite{Kato} stellt in Bemerkung 1.8 des Kapitels X  fest, dass bei Anwendungen der stetige Teil des Spektrums f"ur die meisten selbstadjungierten Operatoren absolutstetig ist.

Nun weckt aber das oben zitierte ma"stheoretische Ergebnis, wenn man sich n"aher damit befa"st, folgende Neugier: Wenn sich zwei Bestandteile eines Ma"ses physikalisch interpretieren lassen, ist das auch f"ur den dritten m"oglich? Was hat es mit singul"arstetigen Ma"sen auf sich?
Gibt es Operatoren mit singul"arstetigem Spektrum und kann man diesem ein wie auch immer exotisch geartetes Teilchenverhalten zuordnen? Gibt es Experimente, bei denen man dieses Verhalten beobachten kann?

Zun"achst konnte zumindest die Existenz von singul"arstetigem Spektrum  mit Methoden der Inversen Streutheorie bewiesen werden (\cite{ARONSHAJN}). Jedoch hat dieser Ansatz das Manko, nur die Existenz nachzuweisen, ohne n"ahere Aussagen zum zugeh"origen Potential zu erm"oglichen. In dieser Situation ohne n"ahere Kenntnis des Potentials bleibt offen, wie singul"arstetiges Spektrum interpretiert werden kann.

F"ur den n"achsten Schritt ist als Ziel gesteckt, ein geeignetes Potential explizit anzugeben, das Anla"s gibt f"ur singul"arstetiges Spektrum. Dies wirft die Frage auf, nach welchen Kriterien man passende Kandidaten f"ur das Potential ausw"ahlen kann.
Es gibt bereits eine ganze Anzahl von S"atzen, die singul"arstetiges Spektrum ausschlie"sen. Beispielsweise hat Lavine eine sogenannte  No-Bump-Bedingung $ \frac{r}{2} \frac{\mbox{d}V}{\mbox{d}r} < \const $ an das Potential $ V $ formuliert, die sicherstellt, dass das singul"arstetige Spektrum leer ist (\cite{LAVINEI}). Wenn also eine derartige Buckelfreiheit ein  absolutstetiges Spektrum garantiert, legt dies die Schlu"sfolgerung nahe, dass die Suche unter den Buckelpotentialen aussichtsreich ist.

D.B. Pearson untersuchte hierzu bereits eine Klassse von eindimensionalen Schr"odingeroperatoren mit Buckelpotentialen und konnte zeigen, dass diese unter gewissen Bedingungen, die im Wesentlichen die Abst"ande zwischen den Buckeln,  jedoch nicht die Form der Buckel betreffen, rein singul"arstetiges Spektrum in $ (0, \infty ) $ aufweisen (\cite{PEARSON}).

Der erste Schwerpunkt von \cite{PEARSON} ist die Formulierung eines abstrakten und allgemein anwendbaren Satzes "uber die Erzeugung singul"arstetiger Ma"se, der anschlie"send bei der Untersuchung der Eigenschaften von Spektralma"sen Anwendung findet.
Ausgangspunkt f"ur den ma"stheoretischen Satz ist die Cantorfunktion  aufgrund ihrer nachgewiesenen Singul"arstetigkeit. 
Definiert ist sie als Grenze 
\begin{displaymath}
\lim_{n \to \infty}  \int_0^r  \prod_{k=1}^n f _k(x) \dx
\end{displaymath}
 mit
 \begin{equation}
 f_k(x) = \frac{3}{2} f (3^{k-1} x)
 \nonumber
 \end{equation}
 wobei die $1$-periodische Funktion $ f $ durch 
\begin{equation}
f(x) = \left\{
  \begin{array}{c c}
    1 & 0 \le x \le \frac{1}{3} \\
    0 & \frac{1}{3} < x < \frac{2}{3} \\
    1 & \frac{2}{3} \le x \le 1
\nonumber
\end{array}
\right.
\nonumber
\end{equation}
gegeben ist.
Die Cantorverteilung ist singul"ar bez"uglich des Lebesguema"ses. 
Entsprechend zeigt Pearson f"ur eine Folge von Ma"sen $ \mu_n $, die durch Produkte oszillierender Funktionen erzeugt werden, dass diese schwach gegen ein singul"arstetiges Ma"s konvergiert.

Der zweite Teil von \cite{PEARSON} ist der Konstruktion eines Potentials, das zu singul"arstetigem Spektrum f"uhrt, gewidmet. Hierzu wird eine Folge von Operatoren mit jeweils endlich vielen Buckeln bestimmt, deren Grenze die gew"unschten Eigenschaften hat.
Dabei ist das Vorgehen wie folgt:
In jedem Schritt werden zum Vorg"angerpotential weitere Buckel hinzugef"ugt. Hierbei ist daf"ur Sorge zu tragen, dass
der Abstand zwischen den Buckeln in einer geeigneten Art und Weise immer gr"o"ser wird. 
Interessant ist dabei, dass die Gestalt des einzelnen Buckels nicht relevant ist. Lediglich die Gesamtheit der Buckel mu"s wie eine g"anzlich reflektierende Barriere wirken.  Aufgrund der gro"sen Abst"ande zwischen den Buckeln sind die Reflektionen des Teilchens an den einzelnen Buckeln so gut wie unabh"angig voneinander, und die Vorgabe an das Gesamtpotential kann dahingehend interpretiert werden, dass die Gesamttransmissionswahrscheinlichkeit 0 ist. In \cite{KiselevLastSimon} werden zu dieser Idee entsprechend Zufallsvariablen eingef"uhrt, die nicht unabh"angig sein m"ussen, solange die Terme erster Ordnung  klein sind im Vergleich zu den Termen zweiter Ordnung. Kiselev, Last und Simon erzielen dabei f"ur den Schr"odingeroperator auf der Halbachse ein allgemeineres Resultat: Bei vorgegebenem abgeschlossenem Intervall $ S \subset (0,\infty ) $ l"a"st sich  rein singul"arstetiges Spektrum in $ (0,\infty) \setminus S $ und rein absolutstetiges Spektrum in $ S $  erzeugen.

Diese "Uberlegungen zu Zufallspfaden f"uhren bereits zur physikalischen Interpretation der Zust"ande des singul"arstetigen Spetrums.
Hier verhalten sich die Teilchen weder wie gebundene Teilchen, noch wie gestreute Teilchen: 
Dass sich die Teilchen sowohl beliebig weit vom Ursprung entfernen, als auch beliebig oft zum Ursprung zur"uckkehren, macht den au"sergew"ohnlich exotischen Charakter des Teilchenverhaltens aus.

Man hat zun"achst den Eindruck, dass dieses merkw"urdige Teilchenverhalten eine seltene Ausnahme darstellt. Es ist aber tats"achlich das v"ollige Gegenteil der Fall: 
Die Existenz einer ganzen Klasse von Operatoren mit singul"arstetigem Spektrum zieht  n"amlich weitreichende Folgen nach sich.
Nach \cite{Simon} folgt n"amlich, dass Potentiale generisch zu singul"arstetigem Spektrum f"uhren. Sie bilden also bei Weitem die Mehrheit.
Das Erstaunliche ist nun, dass wir in der experimentell untersuchten Welt zwar die bekannten gebundenen bzw. gestreuten Teilchen beobachten, dass sich aber die sich seltsam verhaltenden Exoten bislang einer Beobachtung verschlie"sen. Und dies obwohl die zugeh"origen Potentiale, die Anla"s geben f"ur singul"arstetiges Spektrum, bei weitem in der "Uberzahl gegen"uber dem nicht generischen Fall sind.

Die geschilderten Ergebnisse von Pearson und Simon sind im Bereich der nichtrelativistischen Quantenmechanik angesiedelt.
Deshalb stellt sich die Frage, ob auch in einem relativistischen Modell "ahnliche Ergebnisse abgeleitet werden k"onnen.
Statt eines gew"ohnlichen Differentialausdruckes des zu den Sturm-Liouville-Operatoren z"ahlenden Schr"odingeroperators sind im relativistischen Szenario Differentialgleichungssysteme der Untersuchungsgegenstand.
Da die relativistischen Effekte f"ur kleine Geschwindigkeiten verschwinden - sonst w"aren drastische Unterschiede zu beobachten - , sind aus physikalischen Gr"unden "ahnliche Aussagen zu erwarten, auch wenn sich die Eigenschaften des Schr"odingeroperators nur mit Einschr"ankungen auf den Diracoperator "ubertragen.

Um Erkenntnisse zum singul"arstetigen Spektrum im relativistischen Fall zu gewinnen, liegt in der hier vorgelegten Arbeit im ersten Abschnitt das Augenmerk auf eindimensionalen Diracoperatoren auf der Halbachse. Nach dieser Vorbereitung werden die gewonnenen Ergebnisse auf den dreidimensionalen Raum  f"ur Operatoren mit kugelsymmetrischen Potentialen "ubertragen.

Der erste Teil hat f"ur den drehimpulsfreien Fall die Konstruktion einer Klasse eindimensionaler Diracoperatoren zum Gegenstand, welche ein rein singul"arstetiges Spektrum au"serhalb der zentralen L"ucke $ [-1,1] $ aufweisen. Dieses Ergebnis ist die relativistische Entsprechung zu den Schr"odingeroperatoren mit rein singul"arstetigem Spektrum in $ (0,\infty) $ bei \cite{PEARSON}.

Die Konstruktion eines Potentials, das rein singul"arstetiges Spektrum erzeugt, fu"st auf einer Folge approximierender Operatoren mit endlich vielen Buckeln. Von diesen wird gezeigt, dass die zugeh"origen Spektralma"se schwach gegen ein Grenzma"s  konvergieren, und dass dieses Grenzma"s singul"arstetig ist.

Das Vorgehen bei der Bestimmung der approximierenden Operatoren ist durch das geschickte Platzieren von Buckeln gepr"agt. 
Bestimmendes Charakteristikum der Folge von Buckelpotentialen ist der immer gr"o"ser werdende Abstand zwischen den Buckeln, die mit einer gewissen Anwachsrate immer weiter weg vom Ursprung gesetzt werden. Die Gestalt der Buckel ist - "ahnlich wie im nichtrelativistischen Fall - nachrangig.

Aussagen zur Art des Spektrums werden aus dem Verhalten der L"osungen abgeleitet.
Um f"ur gro"se Abst"ande vom Ursprung besser absch"atzen zu k"onnen, wie sich die L"osungen entwickeln, wird nicht die klassische Diracgleichung untersucht, sondern die klassische Gleichung wird in eine "aquivalente Form transformiert, die den Vorteil besitzt, dass in dieser die Drehimpuls\-terme, statt mit $ \frac{1}{r} $ abzuklingen, mit $ \frac{1}{r^2} $ f"ur $ r > 1 $ abgesch"atzt werden k"onnen. Da in diese Transformation die Teilchenmasse eingeht, sind die Ergebnisse %jedoch 
nicht auf masselose Teilchen "ubertragbar.

Ein weiteres wesentliches Hilfsmittel bei der Untersuchung des L"osungsverhaltens stellt die Pr"ufer-Transformation dar. Hierbei werden die beiden Komponenten der L"osungen durch verallgemeinerte Polarkoordinaten ausgedr"uckt. Bei dieser Art der Darstellung k"onnen die Werte der L"osungen jeweils an den Enden der Buckel anhand ihres Radialanteiles direkt verglichen werden.

 Diese Vergleiche der L"osungswerte an den Buckelenden liefern Produkte von stark oszillierenden Funktionen, die an die oben erw"ahnte Produktdarstellung der Cantorfunktion erinnern.  Durch Hinzunahme und geeignete Positionierung weiterer Potentialbuckel kann das  Verhalten der L"osungen so kontrolliert werden, dass das Spektralma"s des Grenzoperators singul"arstetig ist.

Auf den Resultaten der Untersuchung des drehimpulsfreien Falles baut schlie"slich der zweite Teil der vorliegenden Arbeit auf, der der  Situation auf der Halbachse, nunmehr aber unter Hinzunahme von Drehimpulstermen, gewidmet ist. 

Indem der eindimensionale Operator als Radialanteil eines kugelsymmetrischen Operators verwendet wird, k"onnen die mit diesen Methoden gewonnenen Ergebnisse
% im dritten Teil der vorliegenden Arbeit 
nun sogar auf den dreidimensionalen Raum "ubertragen werden. 

Als Endergebnis ist festzuhalten, dass auch mit Drehimpulstermen die konkrete Konstruktion von Potentialen m"oglich ist, deren zugeh"orige Operatoren rein singul"arstetiges Spektrum au"serhalb der zentralen L"ucke $ [-1,1] $ besitzen.

Ob wir in der uns umgebenden Welt solche Zust"ande des singul"arstetigen Spektrums auch beobachten k"onnen, bleibt als Frage offen. Bei der Suche nach geeigneten Experimenten bietet sich der Bereich der Magnetohydrodynamik an. Bei den  aufgewickelten magnetischen Feldlinien wirken nichtperiodische Potentiale innerhalb von Fusionsreaktoren (\cite{SALAT}).

\pagebreak

%ende von einl

%Ateil1_vorSpektralfunktion
%Teil1
%---------

\hfill
\begin{minipage}[t]{8cm}
\sffamily
Simplicity is the most difficult of all concepts.
- Brian Herbert and Kevin Anderson, Dune: House Corrino 
\end{minipage}

\section{Untersuchung des radialen Anteils des Diracoperators ohne Drehimpulsterme}
\label{KapitelohneDreh}

In der Einleitung wurde bereits darauf hingewiesen, dass in der vorliegenden Arbeit Operatoren mit Buckelpotentialen untersucht werden, da diese aussichtsreiche Kandidaten f"ur Operatoren mit rein singul"arstetigem Spektrum sind. 

%Feb2013++++++++++
Das Vorgehen zur Separation des Diracoperators im $ \R^3 $ f"ur kugelsymmetrische Potentiale ist beispielsweise in Abschnitt 20.3 von  \cite{WEIDMANNBAND2} aufgezeigt.
Im Folgenden ist der Untersuchungsgegenstand der radiale Anteil. 

%Feb2013++++++++++++++++++

Zun"achst wird der Begriff Buckelpotential exakt definiert. 

Als Ausgangslage wird hierbei die Situation ohne Drehimpulsterme zugrunde gelegt. Unter dieser Voraussetzung werden Aussagen "uber das L"osungsverhalten abgeleitet.
Als Hilfsmittel hierf"ur wird die sogenannte Pr"ufer-Transfor\-mation eingef"uhrt. Mittels dieser Transformation l"a"st sich das Verhalten der L"osungen auf den potentialfreien Strecken zwischen den Buckeln kompakt beschreiben. Dies erm"oglicht den direkten Vergleich der entsprechenden Werte der L"osungen an den Enden verschiedener Buckel.

\subsection{Allgemeine Bezeichnungen und Definitionen}

Die bereits mehrfach genannten Buckelpotentiale sind wie folgt definiert:
\begin{bezeichn}
$\left.\right.$

Eine Funktion $ 0 %\neq 
< q%_D 
\in 
C(\R^+)
%AUG2013 \Leinsloc (\R_0^+) 
$ hei"se
\label{S_Buckelpotential}
{\bf Buckelpotential}, wenn es eine Folge von Buckelabst"anden
 $ D := \left(d_j\right)_{j\in \N} $, Folgen von Buckelbreiten
\label{S_Buckelbreite}
$ \left(\alpha_j\right)_{j\in \N} \in \left(\R^+\right)^{\N} $ und
Buckelh"ohen
\label{S_Buckelhoehe}
$ \left(H_j\right)_{j\in\N_{0}} \in \left(\R_0^+\right)^{\N} $, sowie
eine Folge von Buckel\-pro\-fi\-len
$ \left(W_j\right)_{j\in \N} $ mit $ 0 < W_j \in 
C[0,\alpha_j]
%AUG2013 \Leins([0,\alpha_j])
 $,
%$ W_j \neq 0 \in \Leins([0,\alpha_j] $
%mit
 $ \int_0^{\alpha_j} W_j =1 $
 ($ j \in \N $) gibt, so dass gilt
\begin{equation}
   q
%_D 
(r) =  \left\{
              \begin{array}{cl}
                0             & 0 \le r \le a_1          \\
                H_j W_j(r-a_j)  & a_j < r < b_j \quad
                                   (j \in \N) \\
                0             & b_j \le r < a_{j+1}
                                   \quad(j \in \N).
                \end{array}
               \right.
\end{equation}
Dabei ist f"ur $ j \in \N $
\label{S_buckelanfang}
\begin{equation}
     a_j := \sum_{i=1}^{j-1} (d_j + \alpha_j) + d_j
\end{equation}
der Anfang, und
\begin{equation}
   b_j := a_j + \alpha_j
\label{S_Buckelende}
\end{equation}
das Ende des $ j $-ten Buckels.
Mit $ H_j $ ist die H"ohe und  mit $ \alpha_j $ die Breite des $ j $-ten
Buckels bezeichnet. Die Form des Buckels ist durch
$ W_j \in 
C[0,\alpha_j] \label{buckelform}
%AUG2013\Leins([0,\alpha_j]) 
$ gegeben.
Der Abstand des ersten Buckels vom Ursprung ist $ d_0 $. F"ur
$ j \ge 2 $ ist $ d_j $ der Abstand zwischen dem $ (j-1) $-ten und
dem $ j $-ten Buckel.
Im Folgenden wird noch die
Definition $ b_0 := 0 $ von Nutzen sein.

Eine Funktion $ q
%_D 
\in 
%AUG2013\Leinsloc(\R_0^+) 
C(\R^+) $ wird {\bf Potential mit
endlich vielen Buckeln} genannt, wenn es $ n \in \N $ gibt mit
$ H_n \neq 0 $ und $ H_j = 0 \quad (j > n) $. Ein solches
$ n $-Buckelpotential wird dann mit $ q_{%D_
n} $\label{endlBuckjelpotential} bezeichnet, wobei
$ D_n := (d_1, \dots, d_n) $ das $ n $-Tupel der Buckelabst"ande ist, von
denen das Potential abh"angt.

Bei $ q
%_{D}
 $ handelt es sich um ein {\bf Potential mit identischen
Buckeln}, falls
es $ H > 0 $, $ \alpha > 0 $ und
$ 0 %\neq 
<W \in  
%AUG2013\Leinsloc[0, \alpha]
C[0, \alpha] $ gibt, so dass
$ H_j = H $, $ \alpha_j = \alpha $ und $ W_j = W $ f"ur jedes \label{idBuckkel}
$ j \in \N $ ist.

Entsprechend weist ein Potential mit einer endlichen Anzahl $ n $ von
Buckeln identische Buckel auf, wenn gilt:

Es gibt $ H > 0 $, $ \alpha > 0 $ und
$ 0 %\neq 
< W \in 
%AUG2013\Leinsloc[0, \alpha] 
C[0, \alpha]$ derart, dass
$ H_1 = \dots = H_n = H $, $ \alpha_1 = \dots = \alpha_n = \alpha $ und
$ W_1 = \dots = W_n = W $ ist und $ H_j = 0 $ f"ur $ j >n $ gilt.
\label{Buckepotentialbezeichnungen}
\end{bezeichn}

\begin{figure}[hp]
\includegraphics[origin=c,width=18.9cm,clip=true,natwidth=610,natheight=642,
viewport=0.5cm 0cm 25cm 20cm, angle=90]{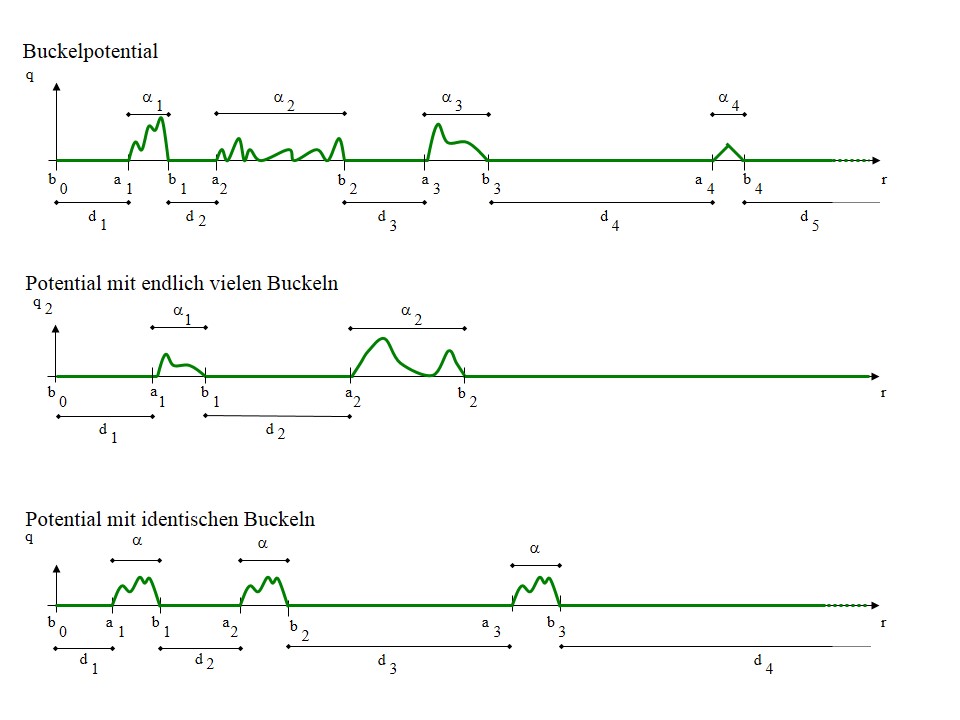}
\caption{\label{BildBuckelpotentiale} Buckelpotentiale}
\end{figure}

Beispiele f"ur solche Buckelpotentiale sind in Abbildung \ref{BildBuckelpotentiale}
gezeigt.

F"ur ein Potential mit vorgegebener Anzahl $ n $ von Buckeln werden im Folgenden Aussagen "uber das L"osungsverhalten abgeleitet.

Sei $ n \in \N $ gegeben.

% Feb13 In Abschnitt \ref{Separation}
% Feb13 wurde der Differentialausdruck (\ref{radialausdruckdirac}) f"ur den Radialanteil des Diracoperators allgemein f"ur % % Feb13 kugelsymmetrische Potentiale $ v $ hergeleitet.

Mit der obigen Definition eines $n$-Buckel-Potentials $ q_{%D_
n} $ ergibt sich somit 
die zentrale Gleichung als das Diracsystem 
%__Auf $ \R_0^+ $ wird das System
%
\begin{equation}
    \Psvektor  '
   =
   \left( \begin{array}{cc}
             0                     & -q_{%D_
             n} + 1 + \lambda \\
             q_{%D_
             n} + 1 - \lambda & 0
          \end{array}
   \right)
    \Psvektor  
 %   \nonumber 
\label{zentraleGleichung}
\end{equation}
 auf $ \R_0^+ $ mit Randbedingung bei $ 0 $
wird
\begin{equation}
   \Psi_1(0) \sin \eta + \Psi_2 (0) \cos \eta = 0
\label{randbedingung}
\end{equation}
mit $ \eta \in [0,\pi) $ gefordert.

%AUG2013 Da $ W_j \in \Leins([0,\alpha_j] )$ $ (j \in \N) $ ist, ist die Ableitung in (\ref{zentraleGleichung}) als die fast "uberall existierende Ableitung lokal absolut stetiger Funktionen zu verstehen.

Da nichttriviale
L"osungen von (\ref{zentraleGleichung})
keine Nullstellen in $ \R^2 $ besitzen, lassen sich f"ur
$ \lambda \in \R, \modulus{\lambda} > 1 $ verm"oge
der folgenden verallgemeinerten
Pr"u\-fer-Trans\-for\-mation neue abh"angige Variablen einf"uhren:

\begin{defundbem}
\label{Ateil1DefPrueferdef}

{\bf Verallgemeinerte Pr"ufer-Transfor\-mation}

Durch
\begin{equation}
  \begin{array}{ccl}
    \Psi_1 & = & R \cos \teta \\
    \Psi_2 & = & R \frac{\kappa}{1 + \lambda} \sin \teta =
                 R \sqrt{\frac{\lambda-1}{\lambda+1}} \sin \teta
  \end{array}
\label{pruefer}
\end{equation}
seien der {\bf (verallgemeinerte) Pr"ufer-Radius} $ R $ \label{S_prueferradius} und der
{\bf (verallgemeinerte) Pr"ufer-Winkel} $ \teta $ \label{S_prueferwinkel} definiert,
wobei
\begin{equation}
  \kappa := \sign{(\lambda)} \sqrt{\lambda^2 -1} \quad(\lambda \in \R,
                                                    \modulus{\lambda} > 1)
%\nonumber                                                     
\end{equation}
\label{S_kappa}ist\footnote{\cite{HughesSchmidt} verwenden eine "ahnliche Definition, die auch in der zentralen L"ucke fortgesetzt wird:
\begin{displaymath}
\omega (\lambda) =\left\{\begin{array}{ll}
-\sqrt{\lambda^2-1}, & \lambda \in (-\infty,-1] \\
i \sqrt{1-\lambda^2}, & \lambda \in (-1,1) \\
\sqrt{\lambda^2-1}, & \lambda \in [1,\infty)
\end{array}
\right.
\end{displaymath}.}.
(Dann ist umgekehrt $ \lambda = \sign (\kappa) \sqrt{\kappa^2 +1} \quad
(\kappa \in \R^* := \R \backslash \{0\}) $.) \label{Rstern}

Der
Pr"ufer-Radius ist eindeutig, der Pr"ufer-Winkel eindeutig bis auf eine
additive Konstante aus $ 2\pi \Z $ bestimmt.
Man beachte, dass wegen $ \modulus{\lambda} > 1 $ gilt:
$ \sign(\lambda) = \sign(\lambda \pm 1) $.
\end{defundbem}

\begin{figure}[t]
\includegraphics[origin=c,width=13cm,clip=true, natwidth=610,natheight=642,
viewport=0.cm 3cm 25.5cm 16cm]{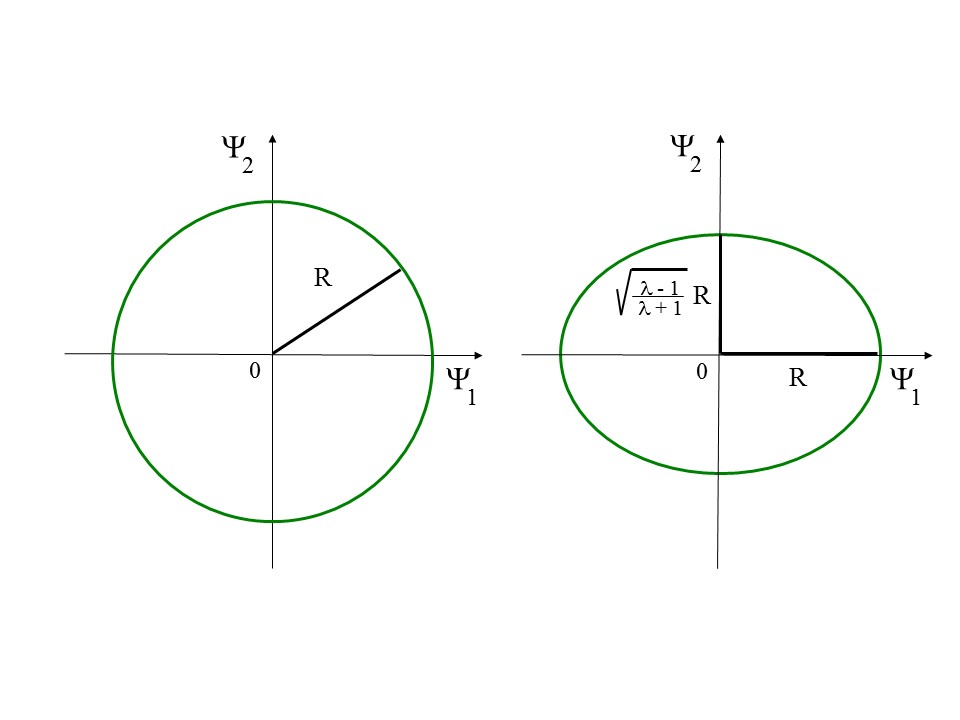}
\caption{\label{PrueferKreisEllipse} Klassische und verallgemeinerte Pr"ufer-Transformation}
\end{figure}

Im Vergleich zur klassischen Pr"ufer-Transformation
\begin{displaymath}
  \begin{array}{ccl}
    \Psi_1 & = & R \cos \teta  \\
    \Psi_2 & = & R  \sin \teta
  \end{array} \nonumber
\end{displaymath}
werden bei der hier verwendeten Verallgemeinerung Polarkoordinaten auf der
durch die Halbmesser $ R $ und
$ R \frac{\kappa}{1+\lambda} = R \sqrt{\frac{\lambda -1}{\lambda +1}} $
bestimmten Ellipse definiert, was in Abbildung \ref{PrueferKreisEllipse} dargestellt
ist.

  Es sei angemerkt, dass die Pr"ufer-Transformation auch bei der Untersuchung diskreter Dirac-Operatoren durch  \cite{CarvalhodeOliveira} eine fundamentale Rolle spielt.

Der Zusammenhang zwischen Potential und dem Verhalten des Pr"ufer-Radius und des Pr"ufer-Winkels wird im folgenden Abschnitt hergestellt.

\subsection{Eigenschaften des Pr"ufer-Radius und des Pr"ufer-Winkels}
\begin{hilfs}
\label{Ateil1lemmaprueferableitungen}

$\left.\right.$

F"ur den Pr"ufer-Radius $ R $ gilt:
\begin{equation}
 (R^2)'  = \frac{2}{\kappa} q_{%D_
 n} \sin 2 \teta R^2
\label{Rquadratableitung}
\end{equation}
bzw.
\begin{equation}
  (\log R)' = \frac{1}{\kappa} q_{%D_
  n} \sin 2 \teta.
\label{Rlogableitung}
\end{equation}

F"ur die Ableitung des Pr"ufer-Winkels $ \teta $ gilt:
\begin{equation}
  \teta '=
    -\kappa + \frac{q_{%D_
    n}}{\kappa} \left(\lambda + \cos 2\teta \right),
\label{prueferwinkelableitung}
\end{equation}
was unabh"angig von $ R $ ist.
\end{hilfs}
%%

%AUG2013Analog zu  (\ref{zentraleGleichung})  ist die Ableitung in (\ref{Rlogableitung}) und (\ref{prueferwinkelableitung}) f"ur lokal absolut stetige Funk\-tionen aufzufassen, da $ q_{D_n} $ lediglich lokal integrierbar ist.

Aus dem Hilfssatz folgt sofort folgende Feststellung:

\begin{bemerkung}
$\left.\right.$

Bei Kenntnis von $ \teta $ erh"alt man nach Hilfssatz \ref{Ateil1lemmaprueferableitungen}  den Pr"ufer-Radius $ R $ durch einfache Integration.
\end{bemerkung}

\Beweis{des Hilfssatzes}
{     %-----------------Hilfslemma---------------------

Die Aussage "uber den Pr"ufer-Winkel wird direkt aus der Definition der verallgemeinerten Pr"ufer-Transformation  abgeleitet. Aus (\ref{pruefer}) liest man 
f"ur den Pr"ufer-Winkel
\begin{displaymath}
 \teta(r) = \left\{
 \begin{array}{l}
   \arctan \left( \sqrt{ \displaystyle
                         \frac{\lambda+1}{\lambda-1}}
                  \displaystyle
                  \frac{\Psi_2(r)}{\Psi_1(r)}
           \right),
  \; \mbox{falls} \; \Psi_1(r) \neq 0
\nonumber
  \\
  \arccot \left( \sqrt{ \displaystyle
                        \frac{\lambda-1}{\lambda+1}}
                 \displaystyle
                 \frac{\Psi_1(r)}{\Psi_2(r)}
          \right),
  \; \mbox{falls} \; \Psi_2(r) \neq 0
 \end{array}
\nonumber
 \right.
 \quad (r\in \R_0^+)
\nonumber
\end{displaymath}
ab.
F"ur seine Ableitung gilt
\begin{displaymath}
  \teta '=
    -\kappa + \frac{q_{%D_
    n}}{\kappa} \left(\lambda + \cos 2\teta \right),
\nonumber
\end{displaymath}
Damit ist der erste Teil der Behauptung des Hilfssatzes \ref{Ateil1lemmaprueferableitungen} gezeigt.

Die zu beweisenden Eigenschaften des Pr"ufer-Radius folgen aus der Tatsache, dass
\begin{equation}
  R^2 = \Psi_1^2 + \frac{\lambda + 1}{\lambda -1} \Psi_2^2
    \qquad \mbox{(verallgemeinerter Pythagoras)}
\label{Rquadrat}
\end{equation}
gilt, was direkt aus der Definition der Pr"ufer-Transformation folgt.
Ableiten von
(\ref{Rquadrat}) liefert
\begin{eqnarray}
\lefteqn{
   (R^2)'  =  2 \Psi_1 \Psi_2
                  \left( - q_{%D_
                  n} + 1 + \lambda
                         + \frac{\lambda+1}{\lambda-1}
                              ( q_{%D_
                              n} + 1 -\lambda )
                  \right)
   =  \frac{4}{\lambda - 1} q_{%D_
   n} \Psi_1 \Psi_2
}
\nonumber \\
  & = & \frac{2}{\lambda - 1} q_{%D_
  n} \sin 2 \teta R^2
            \sqrt{\frac{\lambda-1}{\lambda+1}}
  = \frac{2}{\kappa} q_{%D_
  n} \sin 2 \teta R^2
     \quad \quad \quad \quad \quad \quad \quad \quad \quad \quad
     \quad \quad
\nonumber
%%^\label{Rquadratableitung}
\end{eqnarray}
Wegen $ (\log R)' = \frac{1}{2R^2} (R^2)'  $ erh"alt man somit die zweite Behauptung des Hilfssatzes \ref{Ateil1lemmaprueferableitungen}:
\begin{displaymath}
  (\log R)' = \frac{1}{\kappa} q_{%D_
  n} \sin 2 \teta
\nonumber
%%^\label{Rlogableitung}
\end{displaymath}

}         %-----------------Hilfslemma---------------------
Die Verwendung der verallgemeinerten
Pr"ufervariablen vereinfacht dort, wo das Potential verschwindet, die Situation
erheblich:

Auf den Intervallen $ (b_{j-1}, a_j)$ $ (j \in \{1, \dots, n\}) $
und $ (b_n, \infty) $, auf denen das Potential $ q_{%D_
n} $
den Wert $ 0 $ hat, gilt n"amlich $ \teta '= -\kappa $.
Der Pr"ufer-Winkel zeigt also auf den potentialfreien Strecken lineares Verhalten.
Dadurch erhalten die Werte des Pr"ufer-Winkels an den jeweiligen Buckelenden eine besondere Bedeutung. F"ur diese wird folgende Bezeichnung eingef"uhrt:

\begin{einebezeichn}

$ \left.\right. $

Der  Wert des Pr"ufer-Winkels
einer L"osung von (\ref{zentraleGleichung}) und
(\ref{randbedingung}) am Ende des
$ j $-ten Buckels eines Potentials, dessen Buckelpositionen durch das
Distanztupel $ D_n $ \label{Distanztupel}charakterisiert sind, sei mit 
\begin{equation}
   \teta_{%D_
   n;j} := \teta (b_j; \cdot) \quad (j\in\{0, \dots, n\})
\end{equation}
bezeichnet.\label{S_Thetabuckelende}
Statt $ \teta_{n;n} $ wird die k"urzere Bezeichnung $ \teta_n $ verwendet.
\end{einebezeichn}

Der Zusammenhang zwischen dem Wert des Pr"ufer-Winkels am Ende eines
Buckels und dem Wert am Beginn des n"achsten lautet:
\begin{eqnarray}
  \teta(a_j;\kappa) & = & \teta(b_{j-1};\kappa) - d_j \kappa \nonumber\\
                    & = & \teta_{n;j-1}(\kappa) -d_j \kappa
                           \;\quad(j\in\{1, \dots, n\}).
\end{eqnarray}

Au"serdem gilt
\begin{equation}
    \teta(r;\kappa) = \teta_n (\kappa) -r \kappa
           \quad (r > b_n) .
\label{Ateil1tetalinaer}
\end{equation}

\begin{figure}[h]
\includegraphics[origin=c,width=13.cm,clip=true,natwidth=610,natheight=642,
viewport=0.cm 3cm 23.5cm 19cm]{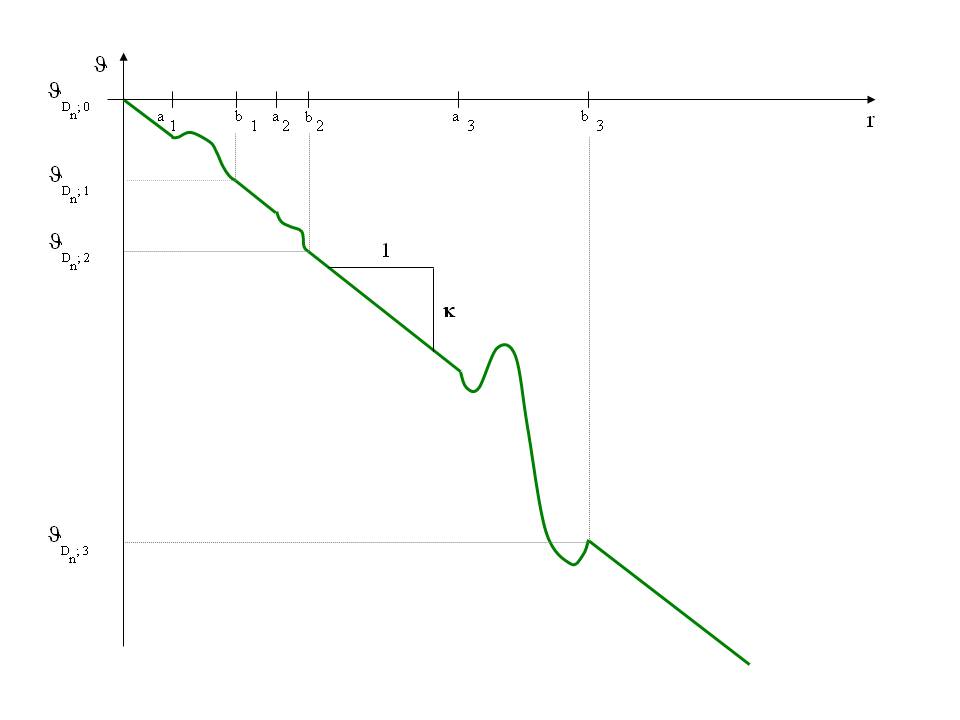}
\caption{\label{BildPrueferwinkel} Schematische Darstellung des Verhaltens
                                   des Pr"ufer-Winkels}
\end{figure}

Der Zusammenhang der Werte des Pr"ufer-Winkels auf den potentialfreien Bereichen
mit den Werten an den jeweiligen Buckelenden ist in
Abbildung \ref{BildPrueferwinkel} veranschaulicht.

Der  Pr"ufer-Winkel hat an der Stelle $ r=0 $ aufgrund der
Randbedingung (\ref{randbedingung}) den Wert
$ \teta_{n;0} = \arctan \left( \sqrt{ \frac{\lambda+1}{\lambda-1}}
                                  \tan \eta \right) $.

Auch der Pr"ufer-Radius verh"alt sich auf den potentialfreien Strecken linear. Wegen  (\ref{Rquadratableitung}) ist er dort sogar konstant. Deshalb sind die Werte des Pr"ufer-Radius an den Buckelenden ebenfalls von besonderer Bedeutung. Auch f"ur den Wert des Pr"ufer-Radius am Ende eines Buckels wird daher eine abk"urzende Bezeichnung definiert:

\begin{einebezeichn}

$ \left.\right. $

Mit
\begin{equation}
  R_{n;j} := R (b_j, \cdot) \quad (j\in \{0, \dots, n\})
  \label{DefRnnnn}
\end{equation}
\label{S_PrueferradiusBuckelende}sei der Wert des Pr"ufer-Radius am Ende des $ j $-ten Buckels
bezeichnet. F"ur $ R_{n;n} $ wird auch die Bezeichnung $ R_n $ genutzt.
\label{Rnabkuerz}
\end{einebezeichn}

Mit dieser Definition l"a"st sich der Zusammenhang des Wertes des Pr"ufer-Radius am Ende eines Buckels und am Beginn des n"achsten Buckels  folgenderma"sen formulieren:
\begin{equation}
   R_{n;j-1} (\kappa) = R(a_j;\kappa) \quad (\jinebisn).
\end{equation}

\begin{figure}[t]
\includegraphics[origin=c,width=13cm,clip=true,natwidth=610,natheight=642,
viewport=0cm 3cm 24.5cm 18cm]{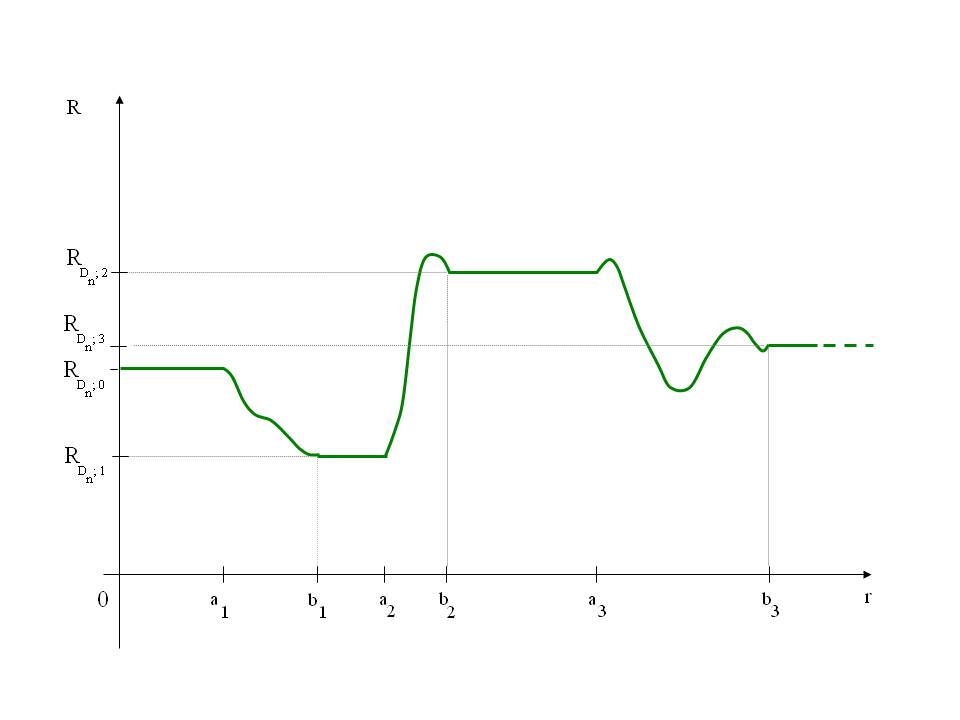}
\caption{\label{BildPrueferradius} Schematische Darstellung des
                                   Pr"ufer-Radius-Verhaltens}
\end{figure}

Das Verhalten von $ R $ in Abh"angigkeit von $ r \in \R_0^+ $ ist in
Abbildung \ref{BildPrueferradius} beispielhaft dargestellt.

\begin{figure}[ht]
\includegraphics[origin=c,width=13.5cm,clip=true,natwidth=610,natheight=642,
viewport=0cm 3cm 22.5cm 19cm]{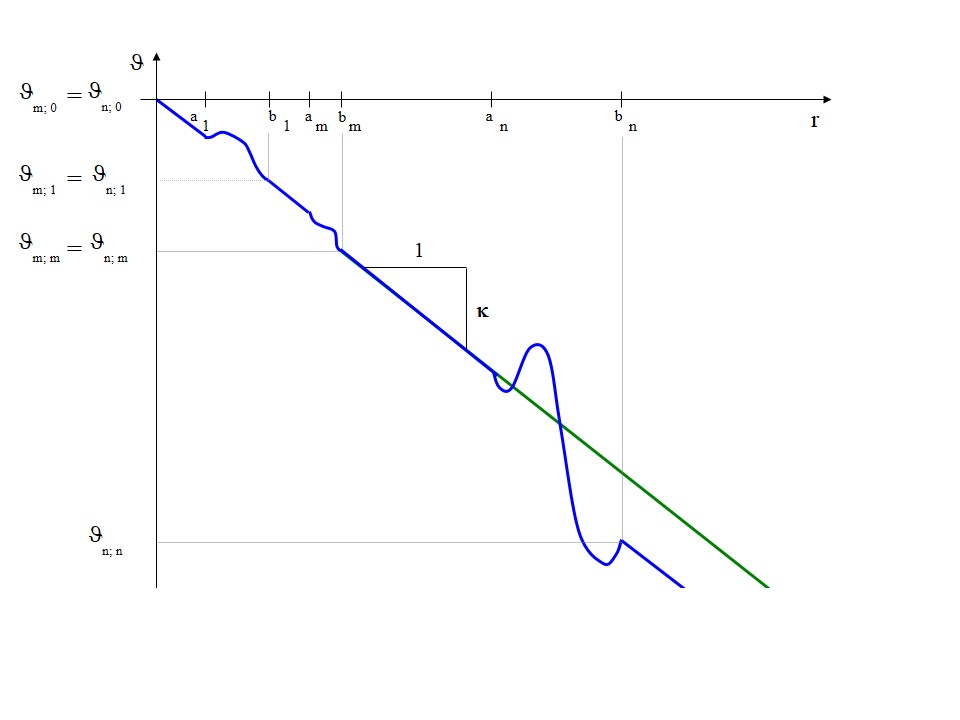}
\caption{\label{Bild_Prueferwinkelvergleich_m_und_n_Buckel} Pr"ufer-Winkel f"ur zwei endliche
Buckelpotentiale mit $ m $ bzw. $ n $ Buckeln, von denen die ersten  $ m $ Buckel "ubereinstimmen}
\end{figure}

\begin{figure}[bht]
\includegraphics[origin=c,width=13.5cm,clip=true,natwidth=610,natheight=642,
viewport=0.5cm 3cm 24.5cm 15cm]{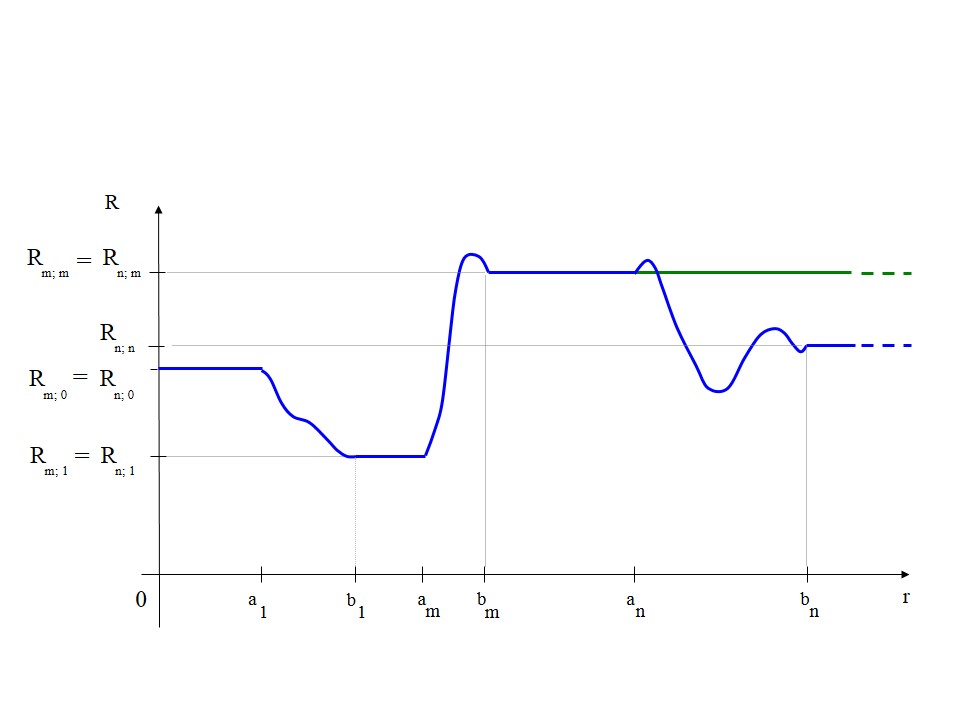}
\caption{\label{Bild_Prueferradiusvergleich_n_m_Buckel} Pr"ufer-Radius f"ur zwei endliche
Buckelpotentiale mit $ m $ bzw. $ n $ Buckeln, von denen die ersten  $ m $ Buckel "ubereinstimmen}
\end{figure}

Man beachte, dass
\begin{equation}
   \begin{array}{rcl}
       R_{n;j}  & = & R_{%D_
       m;j} \\
       \teta_{n;j}  & = & \teta_{%D_
       m;j}
   \end{array}
         \quad (j \in \{0, \dots, \min \{n, m\}\})
         \nonumber 
\end{equation}
gilt. Stimmen n"amlich zwei endliche Buckelpotentiale mit $ m $ bzw. $ n $ Buckeln in ihren ersten $ \min \{n, m\} $ Buckeln bez"uglich deren Form und Lage  "uberein, so sind entsprechend auch die zugeh"origen Werte f"ur die Pr"ufer-Winkel und Pr"ufer-Radien an den ersten $ \min \{n, m\} $ Buckelenden identisch.
\begin{figure}[th]
\includegraphics[origin=c,width=13.9cm,clip=true,natwidth=610,natheight=642,
%viewport=3.5cm 0cm 25cm 20cm, angle=90]{Bild_zweiBuckelpotentiale}
viewport=0cm 9cm 24cm 18.8cm,angle=0]{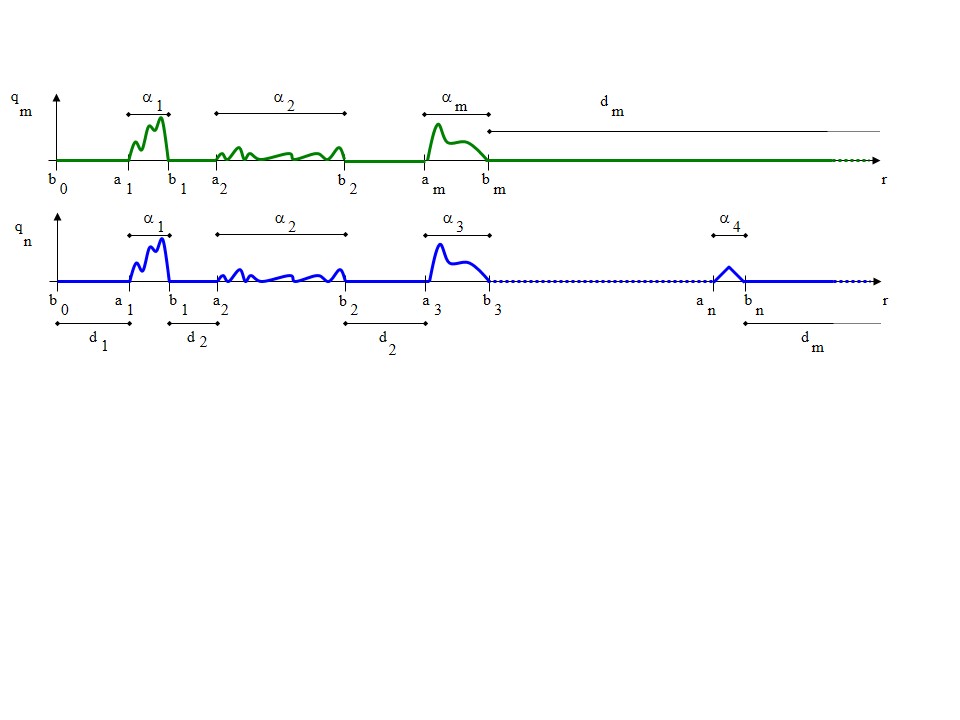}
\caption{\label{Bild m_und_n_Buckel_vergleichen} Zwei endliche Buckelpotentiale mit $ m $ bzw. $ n
$ Buckeln, von denen die ersten  $ m $ Buckel "ubereinstimmen}
\end{figure}
ist. Dies ist f"ur die endlichen Buckelpotentiale aus Abbildung \ref{Bild
m_und_n_Buckel_vergleichen} schematisch in den Abbildungen
\ref{Bild_Prueferwinkelvergleich_m_und_n_Buckel} und \ref{Bild_Prueferradiusvergleich_n_m_Buckel}
dargestellt.

Die Werte der L"osungen von (\ref{zentraleGleichung}) und
(\ref{randbedingung}) k"onnen wegen (\ref{Rlogableitung}) und (\ref{prueferwinkelableitung}) auf den potentialfreien Intervallen
aus ihren Werten, die sie am Ende eines vorangegangenen Buckels
annehmen, ermittelt werden:
\begin{equation}
  \begin{array}{rcl}
     \Psi_1 (r) & = &
       R\ind (\kappa) \cos \left( \teta\ind - \kappa r\right) \\
     \Psi_2 (r) & = &
       R\ind (\kappa) \wurzel \sin \left( \teta\ind - \kappa r\right) \\
  \end{array}
  \; (r \in (b_{j-1};a_j), \jinebisn).
\label{psigleichrcosetc}
\end{equation}
Dabei kann der Ursprung ebenfalls als Buckelende aufgefa"st werden, was in der
Bezeichnung $ b_0 := 0 $ zum Ausdruck gebracht wird.

Wegen (\ref{psigleichrcosetc})  beschreibt der Vektor
$ \left( \begin{array}{c}
           \Psi_1   \\
          \Psi_2
         \end{array}
  \right)
$ der L"osung auf den potentialfreien Strecken
eine gleich\-f"ormige Bewegung auf einer Ellipse, deren Halbmesser im Verh"altnis
$ \sqrt{\frac{\lambda +1}{\lambda -1}} $ zueinander stehen.

Nun fehlt noch f"ur die L"osungen der Zusammenhang der Werte am Buckelanfang und am Buckelende.
Dieser wird durch Transfermatrizen hergestellt:

$\left.\right.$

\begin{bezeichn}

$ \left.\right. $

$ \Mj $ bezeichne die Transfermatrix des $ j$-ten Buckels.\label{transfermatrix}

\end{bezeichn}

$\left.\right.$

$ \Mj $ kann folgenderma"sen bestimmt werden:

Sei $ [0, \alpha_j] $ gegeben. Auf diesem Intervall wird die Gleichung
\begin{displaymath}
  \left( \begin{array}{c}
           u_1 \\
           u_2
         \end{array}
  \right) '
   =
   \left( \begin{array}{cc}
             0                     & -H_j W_j + 1 + \lambda \\
             H_j W_j + 1 - \lambda & 0
          \end{array}
   \right)
  \left( \begin{array}{c}
           u_1 \\
           u_2
         \end{array}
  \right)
\nonumber
\end{displaymath}
betrachtet. Das zugeh"orige Fundamentalsystem mit Anfangswert
$ \left( \begin{array}{cc} 1 & 0 \\
                           0 & 1
         \end{array}
  \right)
$ 
liefert an der Stelle $ \alpha_j $ die Transfermatrix $ \Mj $.

Zum einen ist das Verhalten der L"osungen auf den potentialfreien Strecken zwischen den Buckeln gem"a"s Ergebnis (\ref{psigleichrcosetc}) bekannt. Zum anderen ist das Verhalten der L"osungen bei den Buckeln durch Transfermatrizen gegeben. Fa"st man diese beiden Informationen zusammen, erh"alt man folgenden Hilfssatz, der Auskunft "uber das Wachstumsverhalten der L"osungen insgesamt gibt:

\begin{hilfs}   % Hilfslemma Pr"ufer-Radienverh"altnis ----------
\label{Ateil1lemmaprueferradienverhaeltnis}

$\left.\right.$

F"ur das Verh"altnis der Quadrate der
Pr"ufer-Radien am Ende des $ j $-ten und des $ (j-1) $-ten Buckels gilt:
\begin{eqnarray}
\lefteqn{
   \frac{R_{n;j}^2}{R^2_{n;j-1}}  
  =   \frac{1}{\ffj(\cdot\/;\teta\ind -\cdot d_j)}}
\nonumber \\
 & = & \Aj + \Bj \cos\left( 2 ( \teta\ind - \cdot d_j) \right)
               + \Cj \sin\left( 2 ( \teta\ind - \cdot d_j) \right),
\label{prueferverh}
\end{eqnarray}
wobei
\begin{equation}
   \ffj (\kappa, y)\! := \!
   \frac{1}{\Aj(\!\kappa\!) + \Bj\!(\!\kappa\!) \cos(2y) + \Cj\!(\!\kappa\!) \sin(2y)
   }
  > 0
  \quad (\kappa \in \R^*, y \in \R)
%\label{teil1deffj1}
\label{Ateil1deffj2}
\end{equation}
%???\label{Ateil1deffj2}
%
mit
\begin{eqnarray}
   \Aj & := & \frac{1}{2}
           \left[
              \Mjee^2 + \plmin \Mjze^2 + \minpl \Mjez^2 + \Mjzz^2
           \right]
\label{DefAj}
   \\
   \Bj & := & \frac{1}{2}
             \left[
                  \Mjee^2 + \plmin \Mjze^2 - \minpl \Mjez^2 - \Mjzz^2
             \right]
\label{DefBj}
   \\
   \Cj & := & \sqrt{\minpl}\Mjee\Mjez + \sqrt{\plmin} \Mjze\Mjzz
\label{DefCj}
\end{eqnarray}
definiert ist.
\end{hilfs}   % Hilfslemma Pr"ufer-Radienverh"altnis -----------------------------

\Beweis{}{ % -------Beweis Lemma Pr"ufer-Radienverh"altnis   ---------------

Mit Hilfe der Beziehung
\begin{displaymath}
  \left( \begin{array}{c}
           \Psi_1(b_j;\kappa) \\
           \Psi_2(b_j;\kappa)
         \end{array}
 \right)
 =
 \Mj (\kappa)
  \left( \begin{array}{c}
           \Psi_1(a_j;\kappa) \\
           \Psi_2(a_j;\kappa)
         \end{array}
 \right)
\nonumber
\end{displaymath}
leitet man unter Ber"ucksichtigung des in (\ref{psigleichrcosetc}) beschriebenen Verhaltens der
L"osung $ \Psi $ auf den potentialfreien Strecken 
\begin{eqnarray}
  \Psi_1(b_j;\kappa) & = &
         \Mjee(\kappa) R\ind (\kappa)
               \cos \left( \teta\ind - \kappa d_j\right)
         \nonumber \\
         & & + \Mjez(\kappa) R\ind (\kappa) \wurzel
             \sin \left( \teta\ind - \kappa d_j\right)
\label{Ateil1Psi1}
\end{eqnarray}
und
\begin{eqnarray}
  \Psi_2(b_j;\kappa) & = &
         \Mjze(\kappa) R\ind (\kappa)
                \cos \left( \teta\ind - \kappa d_j \right)
         \nonumber \\
         & & + \Mjzz(\kappa) R\ind (\kappa) \wurzel
             \sin \left( \teta\ind - \kappa d_j \right)
\label{Ateil1Psi2}
\end{eqnarray}
ab.

Das Verh"altnis der Pr"ufer-Radien am Ende des $j$-ten bzw. $(j-1)$-ten Buckels $\frac{R_{n;j}^2}{R^2_{%D_
n;j-1}} $ ist echt positiv. Andernfalls w"are mit  (\ref{Rquadrat})  
$ \Psi $ wegen der Eindeutigkeit der L"osung die Nulll"osung.

Aufgrund des verallgemeinerten Satzes von Pythagoras (\ref{Rquadrat}) ergibt
sich f"ur das Pr"ufer-Radienverh"altnis,
wenn man die soeben gewonnenen Beziehungen (\ref{Ateil1Psi1}) und
(\ref{Ateil1Psi2}) verwendet:

\begin{eqnarray*}
\lefteqn{
   \frac{R_{n;j}^2}{R^2_{%D_
   n;j-1}}  =
     \frac{1}{R^2_{%D_
     n;j-1}} \left( \Psi_1^2(b_j;\cdot) +
                           \frac{\lambda+1}{\lambda-1} \Psi_2^2(b_j;\cdot)
                     \right)
     \nonumber }\\
     & = & \cos^2 (\teta\ind - \cdot d_j) \left[ \Mjee^2 +
               \frac{\lambda+1}{\lambda -1} \Mjze^2 \right]
            \nonumber \\
     &   & + \sin^2  (\teta\ind -\cdot d_j)
               \left[ \frac{\lambda-1}{\lambda+1} \Mjez^2 + \Mjzz^2 \right]
            \nonumber \\
     &   & + 2 \sin  (\teta\ind -\cdot d_j) \cos  (\teta\ind -\cdot
d_j) \cdot
           \nonumber \\
     &   &  \quad\quad\quad\quad\quad\quad\quad\quad
         \cdot \left[ \sqrt{\frac{\lambda-1}{\lambda+1}} \Mjee\Mjez
                  + \sqrt{\frac{\lambda+1}{\lambda-1}} \Mjze\Mjzz
            \right]
     \nonumber   \\
     & = & \frac{1}{2}
           \left[
              \Mjee^2 + \plmin \Mjze^2 + \minpl \Mjez^2 + \Mjzz^2
           \right]
           \nonumber \\
     &   & + \frac{1}{2}
             \left[
                  \Mjee^2 + \plmin \Mjze^2 - \minpl \Mjez^2 - \Mjzz^2
             \right] \cos\left( 2 ( \teta\ind - \cdot d_j) \right)
           \nonumber \\
     &   & + \left[
                  \sqrt{\minpl}\Mjee\Mjez + \sqrt{\plmin} \Mjze\Mjzz
             \right] \sin\left( 2 ( \teta\ind - \cdot d_j) \right)
     \nonumber \\
     & = & \Aj + \Bj \cos\left( 2 ( \teta\ind - \cdot d_j) \right)
               + \Cj \sin\left( 2 ( \teta\ind - \cdot d_j) \right)
     \nonumber \\
     & = & \frac{1}{\ffj(\cdot\/;\teta\ind -d_j\cdot)}.
     \nonumber \\
\end{eqnarray*}

} % -------Beweis Lemma Pr"ufer-Radienverh"altnis   ---------------

\begin{bemerkung}

$\left.\right.$

Die Funktionen $ \Aj $, $ \Bj $ und $ \Cj $
sind f"ur $ \kappa \in \R^* $
reell analytisch,
da die Komponenten der Transfermatrizen $ \Mj $ holomorph im
Spektralparameter $ \lambda $ und dieser wiederum analytisch in
$ \kappa \in \R^* $ ist.
\end{bemerkung}

Bevor die Eigenschaften von $ \ffj $ n"aher untersucht werden, wird die Spektralfunktion des Anfangswertproblems (\ref{zentraleGleichung}), (\ref{randbedingung}) ermittelt.

%ende von Ateil1_vorSpektralfunktion}

%Ateil2_vorgezogen
%Teil2

%-----
\subsection{Die Spektralfunktion \label{AbschnittSpektralfkt}}

\begin{satz}
\label{Ateil2satzspektralfunktionableitung}

$\left.\right.$

F"ur die Ableitung der
zu (\ref{zentraleGleichung}) und (\ref{randbedingung}) zugeh"origen
Spektralfunktion gilt:
\begin{equation}
  \frac{\mbox{d} \roDn}{\dkappa} (\kappa)
  =
  \frac{1}{\pi}
   \left( \prod_{j=1}^n \ffj (\kappa, \teta\ind(\kappa) - \kappa d_j \right)
   \D (\kappa)
   \quad (\kappa \in \R^*),
\label{teil2drodkappa}
\end{equation}
wobei
\begin{equation}
  \D (\kappa) := \frac{{\rm \sign}(\kappa) \sqrt{\kappa^2+1} +1}
                      {{\rm \sign}(\kappa) \sqrt{\kappa^2+1}}
  R(0, \kappa )
  \quad (\kappa \in \R^*)
\label{Ateil2DefD}
\end{equation}
definiert ist.
\end{satz}

\begin{bemerkung}

$\left.\right. $

Damit die Spektralfunktion eindeutig ist, ist noch wahlweise Linksstetigkeit oder Rechtsstetigkeit zu fordern
(\cite{LEVITHANSARGSJANII}). Hier wird ohne Beschr"an\-kung der Allgemeinheit der Linksstetigkeit der Vorzug gegeben. Die Spektralfunktion ist dann bis auf eine additive
Konstante bestimmt. Diese sei so gew"ahlt, dass die Spektralfuktion bei $ -\infty $ den Wert $ 0 $ annimmt.
\end{bemerkung}

Um den Satz zu beweisen, wird das Problem  (\ref{zentraleGleichung}) und
(\ref{randbedingung}) entsprechend dem Vorgehen in Abschnitt {8.1.1}
von \cite{LEVITHANSARGSJANII}\footnote{Die Gestalt von (\ref{teil2zentraleGL}) stimmt mit dem dort untersuchten System bis auf das Vorzeichen von $ \lambda $ "uberein.  \cite{LEVITHANSARGSJANII} betrachten allgemeinere stetige Koeffizienten  $ q_1$ und $ q_2 $ f"ur das System $ y_2'(x) - [ \lambda + q_1(x) ]\; y_1(x) \;=\; 0$, 
 $ y_1'(x)  + [\lambda + q_2 (x) ]\; y_2(x) \;= \; 0 $  } durch

regu\-l"are Probleme approximiert:

Statt auf $ \R^+ $ wird die
Gleichung (\ref{zentraleGleichung}) zun"achst auf einem endlichen Intervall
$ [b_0,b] $ betrachtet (mit $ b> b_n $, damit alle $ n $ Buckel im betrachteten Intervall
enthalten sind). Hier wird die etwas allgemeinere Notation $ b_0 := 0 $ f"ur das linke Ende des betrachteten Intervall verwendet, um die Vergleichbarkeit mit den "Uberlegungen aus Abschnitt \ref{SpektralfktmitDreh} zu erh"ohen.
Bei $ b $ wird eine zus"atzliche Randbedingung gestellt:

\begin{eqnarray}
   \left( \begin{array}{c} \Psi_1 (r) \\
                           \Psi_2 (r)
          \end{array}
   \right) '
   & = &
   \left( \begin{array}{cc}
             0                     & -q_{%D_
             n}(r) + 1 + \lambda \\
             q_{%D_
             n}(r) + 1 - \lambda & 0
          \end{array}
   \right)
   \left( \begin{array}{c} \Psi_1 (r) \\
                           \Psi_2 (r)
          \end{array}
   \right)
\label{teil2zentraleGL}
 \\
&&\quad\quad\quad\quad\quad\quad\quad\quad\quad\quad\quad\quad\quad\quad\quad\quad\quad   \quad (r \in (0,b))
\nonumber \\
  & &
%  \begin{array}{rcl}
%   \Psi_1 (0) \sin \eta - \Psi_2 (0) \cos \eta & = & 0
%   \quad \mbox{mit} \quad \eta \in [0,\pi)    \quad\quad \\
%   \Psi_1 (b) \sin \zeta - \Psi_2 (b) \cos \zeta & = & 0
%   \quad \mbox{mit} \quad \zeta \in [0,\pi).\quad\quad\\
%  \end{array}
 \begin{array}{lcl}
\Psi_1 (b_0) = 1 && \Psi_2 (b_0) = 0 \\
\Psi_1 (b) = 1 && \Psi_2 (b) = 0 
 \end{array}
\label{teil2randbed}
\end{eqnarray}
Da das Spektrum von (\ref{teil2zentraleGL}) und (\ref{teil2randbed}) nur aus
Eigenwerten besteht, ist die zugeh"orige Spektralfunktion eine (bis auf eine
additive Konstante eindeutig bestimmte) Stufenfunktion, welche jeweils bei den
Eigenwerten einen Sprung aufweist. Die H"ohe des Sprunges ist durch
das Inverse der $ \mbox{L}_2 $-Norm der zugeh"origen Eigenl"osung gegeben.

Seien die Eigenwerte dieses Problems mit $ \lambdabn $, die zugeh"origen
Eigenfunktionen, die die Anfangsbedingung
%%%%%Randbedingung
%$ \phibne(0) \sin \eta - \phibnz(0) \cos \eta = 0 $
$ \phibne(0) = 1, \;\phibnz(0) = 0 $
erf"ullen, mit
$ \left( \begin{array}{c} \phibne \\
                          \phibnz
         \end{array}\right) $
$ (\tilde{n} \in \N) $ bezeichnet und

\begin{equation}
  \abnquadrat := \int_0^b \modulus{\phibn(r)}^2 \mbox{d}r
\label{aquadrat}
\end{equation}
defniert.
(vgl. \cite{LEVITHANSARGSJANII}, S. 213)

Dann ist
\begin{equation}
   \roDnb (\lambda)= \left\{
                            \begin{array}{ll}
                               {\displaystyle -\sum_{\lambda < \lambdabn \le 0}
                                      \frac{1}{\abnquadrat}  }
                                & (\lambda < 0)
                                \\
                                {\displaystyle  \sum_{0 < \lambdabn \le \lambda}
                                      \frac{1}{\abnquadrat}    }
                                & (\lambda \ge 0)
                            \end{array}
                          \right.
\label{Ateil2Sprungfunktion}
\end{equation}
die Spektralfunktion des regul"aren Problems vgl. \cite{LEVITHANSARGSJAN}, S. 168\footnote{Dort wird das System 
\begin{eqnarray} y_2'(x) &-& [ \lambda + q_1(x) ]\; y_1(x) \quad=\quad 0 \nonumber
\\ y_1'(x) & + &[\lambda + q_2 (x) ]\; y_2(x) \quad= \quad 0  \nonumber
\end{eqnarray} mit stetigen $ q_1$ und $ q_2 $ untersucht. Die Gestalt stimmt mit (\ref{teil2zentraleGL}) bis auf das Vorzeichen von $ \lambda $ "uberein.}).

\begin{figure}[ht]
\includegraphics[origin=c,width=13cm,clip=true,natwidth=610,natheight=642,
viewport=0.0cm 1cm 25cm 18cm]{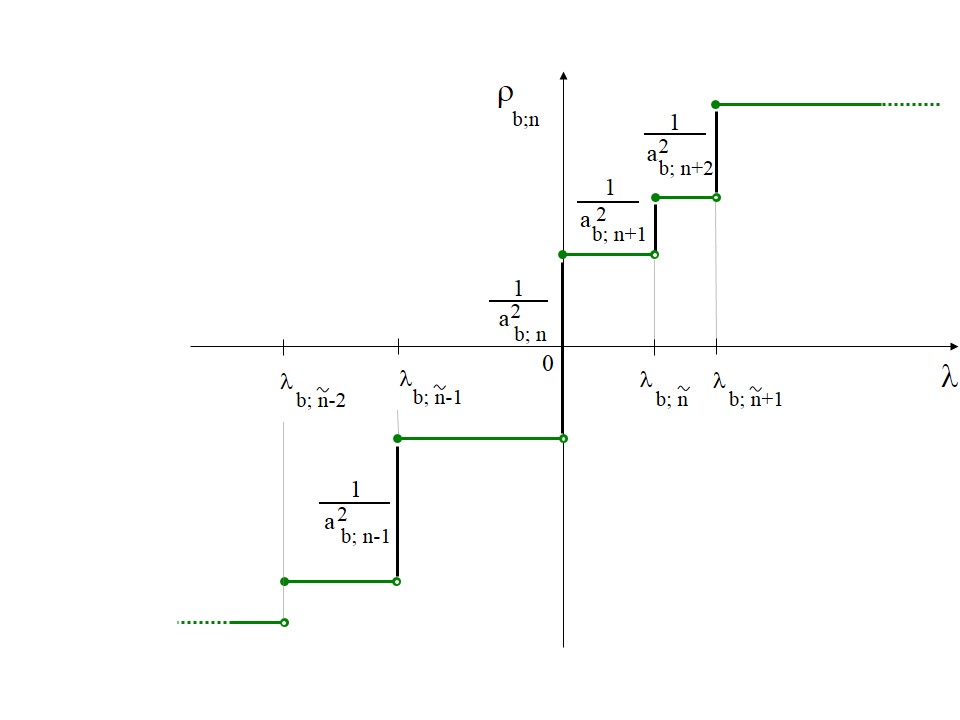}
\caption{\label{Spektralfunktion} Spektralfunktion des regul"aren Problems}
\end{figure}

Das typische Aussehen einer derartigen Spektralfunktion ist in Abbildung
\ref{Spektralfunktion} dargestellt.

Ziel ist es, beim "Ubergang zur Grenze $ b \to \infty $ zum einen die H"ohe der Spr"unge der Spektralfunktion
  und zum anderen die Anzahl der Eigenwerte abzusch"atzen.
Dies ist Inhalt der folgenden beiden Hilfss"atze.
Diese bilden die Grundlage f"ur die Bestimmung des im Satz \ref{Ateil2satzspektralfunktionableitung} angegebenen Differentialquotienten.

Sei $ [\alpha, \beta] \subset R^* $ beliebig.

\begin{hilfs}
\label{aquadrsatz}
$\left.\right. $

Es gilt
\begin{equation}
  \frac{1}{b} \abnquadrat \to \RDnnquadrat (\kappa) \frac{\lambda}{\lambda +1}
  + \kleinoglm (1)
  \quad ( b \to \infty).
  \nonumber 
\end{equation}
Dabei wird durch die Notation $ \kleinoglm (1) $ zum Ausdruck gebracht,
dass die Konvergenz in
%$ \kappa \in [\kappa_1, \kappa_2] $ gleichm"a"sig ist.
$ \kappa \in [\alpha, \beta] $ gleichm"a"sig ist.
\end{hilfs}

\Beweis{}
{

Die Koeffizienten in (\ref{teil2zentraleGL}) werden in einen freien Anteil
\begin{equation}
\Tfr := \left(
\begin{array}{cc}
0 & 1 + \lambda \\
1-\lambda & 0
\end{array}
\right)
\label {deftfrei}
\end{equation}
und den Potentialanteil
\begin{equation}
S(r) :=  \left(
\begin{array}{cc}
0 & s_2(r) \\
s_1 (r)& 0
\end{array}
\right)
\nonumber
\end{equation}
mit $ s_1 (r):=  q_{%D_
n} (r) $, $s_2(r) := -  q_{%D_
n}(r) $ $(r\in [b_0,\infty) $ 
aufgeteilt. Diese allgemein gehaltene Notation wird gew"ahlt, da in Abschnitt \ref{SpektralfktmitDreh} "ahnliche "Uberlegungen angestellt werden.

\begin{equation}
U (r , \kappa)  := 
\left(
\begin{array}{cc}
\cos \kappa r  & - \frac{\kappa}{1-\lambda}\sin \kappa r \\
\frac{\kappa}{1+\lambda}\sin \kappa r & \cos \kappa r
\end{array}
\right)
\quad (r \in [b_0,\infty))
%\label{ungestFS}
\label{UFundamentalsystem}
\end{equation}
ist das Fundamentalsystem zu 
\begin{equation}
u'= \Tfr u,
\nonumber
\end{equation}
das die Randbedingung $ \psi (b_0) = \left(\!\begin{array}{cc}
1 & 0 \\
0 & 1
\end{array}\!
\right)$ 
erf"ullt.
Sei $ u (r) := \left(\!\begin{array}{c}
U_{11}(r) \\
U_{21} (r)
\end{array}\!
\right) $.
Dann gilt f"ur die L"osung $ \kleinphi $ von
\begin{equation}
\kleinphi' = \left(\Tfr + S\right) \kleinphi,
\nonumber
\end{equation}
die die Randbedingung $ \kleinphi(b_0) =  \left(\begin{array}{c}
1 \\
0 
\end{array}
\right) $
erf"ullt (vgl. \cite{EASTHAM}, S. 6):
\begin{equation}
\kleinphi (r, \kappa) = u (r, \kappa) + U (r, \kappa) \int_{b_0}^r U^{-1}(s, \kappa) S(s) \kleinphi(s,\kappa) \ds
\label{integralgl}
\end{equation}

Schreibt man
\begin{eqnarray}
\kleinphi (r, \kappa) &= &u (r, \kappa) + U (r, \kappa) \int_{b_0}^\infty U^{-1}(s, \kappa) S(s) \kleinphi(s,\kappa) \ds
\nonumber \\
& & \quad \quad\quad\quad   -\; U (r, \kappa) \int_{r}^\infty U^{-1}(s, \kappa) S(s) \kleinphi(s,\kappa) \ds
\label{sfdksdf}
\end{eqnarray}
und definiert\footnote{Ein "ahnliches Vorgehen wird in \cite{Coddington}, S. 255, f"ur die Schr"odingergleichung $ Lx=-x'' + qx $, $ x(0) = 0 $ mit $ \int_0^\infty \modulus{q} < \infty $ angewandt, um  die Spektralfunktion als Funktion von $ s $, mit $s^2 := \lambda$ herzuleiten als
$ \frac{\mbox{d} \ro}{\ds} (s) = \frac{2s^2}{\pi A ^2(s)}$.}
\begin{eqnarray}
A(\kappa) \cos (\gamma(\kappa))\! &\!\! :=\!\! & \!
  1 +
  \int_{b_0}^\infty\!\! \cos(\kappa s) s_2(s) \kleinphi_2(s)
    - \frac{1+\lambda}{\kappa} \sin (\kappa s) s_1(s) \kleinphi_1(s) \ds
    \nonumber \\
    & & \label{AA1}\\
    A(\kappa) \sin (\gamma(\kappa))\! &\!\! :=\!\! &\! 
    \int_{b_0}^\infty \!\!\sin(\kappa s) s_2(s) \kleinphi_2(s)
    - \frac{\kappa}{1-\lambda} \cos (\kappa s) s_1(s) \kleinphi(s) \ds
%    \nonumber
\label{AA2}    
\end{eqnarray}
so erh"alt man f"ur (\ref{sfdksdf}) nach Ausmultiplizieren aller Matrizen:
\begin{eqnarray}
\lefteqn{
\kleinphi_1(r,\kappa) =}
\nonumber \\ 
& = &
\cos(\kappa r) A(\kappa) \cos (\gamma (\kappa)) + \sin (\kappa r) A(\kappa) \sin(\gamma(\kappa))
\nonumber \\
& & - \cos(\kappa r) \int_r^\infty \!\left( \cos(\kappa s ) s_2(s) \kleinphi_2(s) 
    - \frac{1+\lambda}{\kappa} \sin( \kappa s ) s_1(s) \kleinphi_1(s) \right)\ds
    \nonumber \\
    & & - \sin(\kappa r) \int_r^\infty \!\left( \sin(\kappa s ) s_2(s) \kleinphi_2(s) 
    - \frac{\kappa}{1-\lambda} \cos( \kappa s ) s_1(s) \kleinphi_1(s) \right)\ds
    \nonumber \\
& = &     
\cos(\kappa r) A(\kappa) \cos (\gamma (\!\kappa\!)) + \sin (\kappa r) A(\kappa) \sin(\gamma(\!\kappa\!))
+ \kleino(1) \quad (r\! \to\! \infty)
\label{psiabsch1} \\
&&
\nonumber\\
\lefteqn{\kleinphi_2(r,\kappa)= }
\nonumber\\
& = &
\frac{-\kappa}{1+\lambda}\sin(\kappa r) A(\kappa) \cos (\gamma (\kappa)) + \frac{-\kappa}{1-\lambda}\cos (\kappa r) A(\kappa) \sin(\gamma(\kappa))
\nonumber \\
& & + \frac{\kappa}{1+\lambda} \sin(\kappa r) \int_r^\infty \!\left(  \cos(\kappa s ) s_2(s) \kleinphi_2(s) 
    - \frac{\lambda+1}{\kappa}  \sin( \kappa s ) s_1(s) \kleinphi_1(s) \right)\ds
    \nonumber \\
    & & -\cos(\kappa r) \int_r^\infty \!\left(\frac{\lambda-1}{\kappa} \sin(\kappa s ) s_2(s) \kleinphi_2(s) 
    + \cos( \kappa s ) s_1(s) \kleinphi_1(s) \right)\ds
    \nonumber \\
& = &     
\cos(\kappa r) A(\kappa) \cos (\gamma (\!\kappa\!)) + \sin (\kappa r) A(\kappa) \sin(\gamma(\!\kappa\!))
+ \kleino(1) \quad (r \to \infty)
\label{psiabsch2} 
%\\
%&&
\end{eqnarray}
Dass die Terme mit den Integralen $ \kleino (1) $ f"ur  $ r\to \infty $ sind, beruht auf den Eigenschaften von $ s_1 $ und $ s_2$, da n"amlich 
$ q_{%D_
n} $ integrierbar ist, und der stetigen Abh"angigkeit von $ \kappa \in[\alpha, \beta] $.
Dies liefert
\begin{eqnarray}
\lefteqn{
\frac{1}{b-b_0} \int_{b_0}^b \modulus{\kleinphi(r,\kappa)}^2 \dr}
\nonumber \\
& = &  \frac{A^2(\kappa)}{b-b_0}\!\int_{b_0}^b\!\!\!\left(
  \cos^2(\kappa r) \cos^2 \gamma\! +\! 2 \cos (\kappa r ) \sin (\kappa r ) \cos \gamma  \sin \gamma\! +\! \sin^2(\kappa r) \sin^2 \gamma
  \right.
  \nonumber \\
  && \quad +\frac{\kappa^2}{(1+\lambda)^2} \sin^2 (\kappa r) \cos^2 \gamma 
     + 2 \frac{\kappa^2}{(-\lambda^2}  \cos (\kappa r ) \sin (\kappa r ) \cos \gamma  \sin \gamma 
     \nonumber \\
   && \left. \quad +  \frac{\kappa^2}{(1-\lambda)^2}\cos^2(\kappa r) \sin^2 \gamma
     + \kleino(1)
  \right) \dr
\nonumber \\
  & = & \frac{A^2(\kappa)}{b-b_0}
  \left[
  b - b_0 -\frac{2}{\lambda+1}\left(\frac{1}{2}(b-b_0) - \frac{1}{4\kappa}(\sin (2\kappa b) -\sin (2\kappa b_0)\right)
  \right.
  \nonumber \\
   & &  
  \quad- \frac{4}{\lambda	^2-1}\sin^2\gamma \left(\frac{1}{2}(b-b_0) -\frac{1}{4\kappa} \left(\sin (2\kappa b) -\sin (2\kappa b_0)\right)
  \right)
  \nonumber \\
   & & \quad\left.+\frac{2}{\lambda^2-1} \sin^2\gamma(b-b_0)    + \kleino(1)
  \right]
  \nonumber \\
  & \to & 
  A^2(\kappa) \frac{\lambda}{\lambda+1}  \quad (b \to \infty)
\label {psiquadratttt}
\end{eqnarray}
Mit Hilfe des verallgemeinerten Satzes von Pythagoras (\ref{Rquadrat}) kann f"ur $ A $ ein Bezug zum Pr"ufer-Radius hergestellt werden: Es ist
\begin{eqnarray}
\lefteqn{
R^2(r, \kappa)=
\kleinphi_1^2(r,\kappa) + \frac{\lambda+1}{\lambda-1} \kleinphi_2^2(r,\kappa)}
\nonumber \\
& = & A^2(\kappa) \left(
\cos^2 (\kappa r) \cos^2 \gamma + 2 \cos(\kappa r ) \sin (\kappa r ) \cos \gamma \sin\gamma
 + \sin^2 (\kappa r) \sin^2 \gamma\right.
 \nonumber \\
 & & +
  \frac{\lambda+1}{\lambda-1}\left(
 \frac{\kappa^2}{(1+\lambda)^2} \sin^2 (\kappa r) \cos^2 \gamma
  + 2 \frac{\kappa^2}{1-\lambda^2} \cos (\kappa r) \sin (\kappa r) \cos \gamma \sin \gamma\right.
  \nonumber \\
  & & \quad \left.\left.
   \frac{\kappa^2}{(1-\lambda)^2} \cos^2(\kappa r ) \sin^2 \gamma
       + \kleino(1)
 \right)
\right)
\nonumber \\
& = & 
 A^2(\kappa) \left(
\cos^2 \gamma + \sin^2 \gamma + \kleino(1) \right)
=  A^2(\kappa) + \kleino(1) \quad (r \to \infty)
\label{rrrrrrrr}
\end{eqnarray}
Da nach Hilfsatz \ref{Ateil1lemmaprueferableitungen} der Pr"ufer-Winkel auf den potentialfreien Strecken konstant ist, ist mit der Bezeichnung aus Definition (\ref{DefRnnnn}) f"ur den Wert des Pr"ufer-Winkels am Ende des letzten Buckels $ R (r, \kappa) = R_{n} $ $ (r \ge b_n)$, so dass $ A(\kappa) = R_{n} $ folgt.

Somit ist
\begin{equation}
  \frac{1}{b} \abnquadrat = \RDnnquadrat (\kappa) \frac{\lambda}{\lambda +1}
  + \kleinoglm (1)
  \quad ( b \to \infty).
  \nonumber
\end{equation}

}  %   Ende Beweis Lemma zu a^2

\begin{hilfs}
\label{anzahlEW}

$\left.\right.$

F"ur die Zahl $ \Nblambdalambda $ der Eigenwerte von
(\ref{teil2zentraleGL}), (\ref{teil2randbed}) in
%im Intervall 
$ (\lambda_1, \lambda_2] $ gilt:
\begin{equation}
   \Nblambdalambda =
   \modulus{\frac{b(\kappa_2-\kappa_1)}{\pi}
           }
   + \grossO (1)
   \quad (b \to \infty).
\label{Ateil2Eigenwertzahlabschaetzung}
\end{equation}
\end{hilfs}

\Beweis{}
{

Dies folgt direkt mit  \cite{WEIDMANN87}, S. 245

\begin{displaymath}
  0 \le \Nblambdalambda =
    \modulus{ \frac{\teta(b;\kappa_2) - \teta(b;\kappa_1)}
                   {\pi}
             + c
            },
\end{displaymath}
wobei $ \modulus{c} \le 1 $ unabh"angig von $ b, \lambda_1, \lambda_2 $ ist.

}   % Beweisende   Zahl der Eigenwerte

\Beweis{von Satz \ref{Ateil2satzspektralfunktionableitung}}{   % Beweis Ableitung Spektralfunktion-----------------

Um den gew"unschten Differentialquotienten
$ \frac{\mbox{d} \roDn}{\dkappa} $ zu bestimmen, werden  
die Vorbereitungen aus Hilfssatz \ref{aquadrsatz} zur $L_2$-Norm der Eigenfunktionen und aus Hilfssatz \ref{anzahlEW} zur Zahl der Eigenwerte f"ur den Differenzenquotienten
$ \frac{ \roDn(\kappa_2) - \roDn(\kappa_1)}{\kappa_2 - \kappa_1} $
zusammengef"uhrt. Es ist damit
\begin{equation}
  \roDnb (\kappa_2) - \roDnb(\kappa_1) 
   = 
   \frac{\frac{b(\kappa_2-\kappa_1)}{\pi} + \grossO(1) }
   {\left(b-b_0\right)\left(R_{%d_n;
   n}^2(\kappa) \frac{\lambda}{\lambda +1}+\kleino(1)\right)}
   \quad (b \to \infty)
   \nonumber
\end{equation}
Somit erh"alt man
\begin{equation}
\frac{  \roDn (\kappa_2) - \roDn(\kappa_1) }{\kappa_2-\kappa_1}
   = 
   \frac{1 }{\pi R_{%n;
   n}^2(\kappa) }
    \frac{\lambda +1}{\lambda}
    \nonumber
\end{equation}

Im Limes $ \kappa_2 \to \kappa_1 $ liefert dies wegen
$\lambda_2 \to \lambda_1 $:
\begin{equation}
  \frac{\mbox{d}\roDn (\kappa)}
       {\dkappa}
  = \frac{1}{\pi \RDnnquadrat(\kappa)}
    \frac{\lambda(\kappa)+1}{\lambda(\kappa)}
%  \quad (\kappa \in \R^*),
  \quad (\kappa \in [ \alpha, \beta ]),
  \label{diffquot}
\end{equation}
beziehungsweise wegen
$ \frac{\dkappa}{\dlambda}
  = \sign (\lambda) \frac{\lambda}{\sqrt{\lambda^2-1}} $
f"ur die Ableitung der Spek\-tral\-funk\-tion nach $ \lambda $:
\begin{eqnarray*}
  \frac{\mbox{d}\roDnhut}{\dlambda} (\lambda)
  & = &
  \frac{\mbox{d}\roDn}{\dkappa} \frac{\dkappa}{\dlambda}(\lambda)
  =
  \frac{1}{\pi \RDnnquadrat (\kappa(\lambda))} \frac{\lambda+1}{\lambda}
   \frac{\lambda}{\sqrt{\lambda^2-1}} \sign(\lambda)
\nonumber \\
  & = &
  \frac{1}{\pi} \frac{1}{\RDnnquadrat(\kappa(\lambda))}
      \sqrt{\frac{\lambda+1}{\lambda-1}}
  \quad (\lambda \in \R, \modulus{\lambda}>1,\, \kappa (\lambda) \in [\alpha, \beta])
\end{eqnarray*}
}   % Beweis Ableitung Spektralfunktion-----------------

Nutzt man die Beziehung (\ref{prueferverh}) bei (\ref{diffquot}) aus, so  erh"alt man die folgende

%----------------------------------------------------------
\begin{definition}

$\left.\right. $

Das zum Anfangswertproblem (\ref{zentraleGleichung}), (\ref{randbedingung}) geh"orige Spektralma"s ist gegeben durch
\begin{equation}
  \myDn (\Sigma) = \int\limits_\Sigma \frac{1}{\pi}
                     \left( \prod_{j=1}^n \ffj (\kappa, \teta_{%D_
                     n;j-1}(\kappa)
                                                       -\kappa d_j
                     \right)
                     \D (\kappa) \dkappa,
\label{teil2defmyDn}
\end{equation}
wobei $ \Sigma \subset R^* = \R \backslash \{0\}$ ein beliebiges
Kompaktum sei.

\end{definition}

%###############

\subsection{Eigenschaften von $ f_j $ und $ m_j $
\label{eigenschaftenfm}}

Das Verh"altnis der Pr"ufer-Radien an aufeinanderfolgenden Buckelenden weist oszillierende Bestandteile auf.
Mittelt man $ \ffj $, so erh"alt man den Wert $ 1 $, was der folgende Hilfssatz besagt: 

\begin{hilfs}
\label{Ateil1lemmafquerm}

$\left.\right.$

Sei $ j \in \{1, \dots, n\} $. Dann gilt f"ur $ \kappa \in \R^* $:
\begin{eqnarray}
  \overline{\ffj}(\kappa) & := &
             \frac{1}{\pi} \int_0^\pi \ffj(\kappa,y)\;\mbox{d}y
  = 1
\label{Ateil1mittelfj}
  \\
  \mj (\kappa)& := & \frac{1}{\pi} \int_0^\pi \log \ffj ( \kappa, y) \dy
  = \log \left(\frac{2}{\Aj(\kappa) + 1}\right)
\label{Ateil1defmj}
\end{eqnarray}
\end{hilfs}

\Beweis{}{ % -------- Beweis Lemma i------------------

Sei $ j \in \{1, \dots, n\} $.
Wegen Definitionen (\ref{DefAj}), (\ref{DefBj}) und
(\ref{DefCj}) gilt
\begin{eqnarray}
\label{Ateil1AquadBquadCquadgleicheins}
   \Aj^2 - \Bj^2 -\Cj^2 & = &
      \Mjee^2 \Mjzz^2 - 2 \Mjee\Mjez\Mjze\Mjzz + \Mjze^2 \Mjez^2
   \nonumber \\
   & = & \left( \mbox{det} \Mj \right)^2 = 1.
%   \nonumber 
\end{eqnarray}
und es gilt $ \Aj > \sqrt{\Bj^2 + \Cj^2}$ . Mit \cite{GroebnerHofteiter}, S. 100, erh"alt man dann damit f"ur das Integral
%Sei $ \kappa \in \R^* $.
%\begin{figure}[t]
%\includegraphics[origin=c,width=11cm,clip=true,
%viewport=1cm 3cm 20cm 16cm]{Bild_Polarkoordinaten}
%\caption{\label{Polarkoord} Polarkoordinaten}
%\end{figure}
%
\begin{eqnarray}
\lefteqn{
  \overline{\ffj}(\kappa)  = 
             \frac{1}{\pi} \int_0^\pi \ffj(\kappa,y)\;\mbox{d}y
 =\frac{1}{\pi} \int_0^\pi
\frac{1}{\Aj(\kappa)+\Bj(\kappa)\cos 2y
                                    +\Cj(\kappa) \sin 2y}
                \;\mbox{d}y}
  \nonumber \\
  & = & \!
             \frac{1}{2\pi} \int_0^{2\pi}\!\!
\frac{1}{\Aj(\kappa)+\Bj(\kappa)\cos x
                                    +\Cj(\kappa) \sin x}
                \mbox{d}x
%  \nonumber \\
%  & = & 
=
  \frac{1}{2\pi}  \frac{2\pi}{\Aj^2 - \Bj^2 -\Cj^2}
  =1,
  \nonumber
\end{eqnarray}
was unabh"angig von $ \kappa $ ist.

Au"serdem gilt f"ur $ \mj $ mit der Substitution $ x = 2y$ und der Parametertransformation $ \Bj =: \ro \cos \xi $, $ \Cj =: - \ro \sin \xi $ und der Ber"ucksichtigung der Periodizit"at und Symmetrie des Cosinus:
\begin{eqnarray}
\lefteqn{   \mj(\kappa)
    =  - \frac{1}{\pi} \int_0^\pi \!\!\log \left( \Aj (\kappa)
               + \Bj(\kappa) \cos 2y + \Cj(\kappa) \sin 2y
\right)
               \;\mbox{d}y }
   \nonumber \\
   & = & - \frac{1}{2\pi} \int_0^{2\pi} \!\!\log \left( \Aj(\kappa) 
                               + \Bj(\kappa) \cos x \!+ \!\Cj(\kappa) \sin x
                                  \right)
\;\mbox{d}x
\nonumber \\
      & = & - \frac{1}{\pi} \!\int_0^{2\pi}\!\! \log \left( \Aj(\kappa) \!+\!
                               \ro \cos (\xi\! +\! x) \right) \dx
%\nonumber \\
%& = &
= - \frac{1}{\pi}\! \int_0^{2\pi} \!\!\log \left( \Aj(\kappa)\! +\!
                               \ro \cos (x) \right) \!\dx
\nonumber \\
& = &- \frac{1}{\pi} \int_0^{\pi}\!\! \log \left( \Aj(\kappa)\! +\!
                               \ro \cos (x) \right)\dx
          -     \frac{1}{\pi} \int_0^{\pi} \!\!\log \left( \Aj(\kappa) \!-\!
                               \ro \cos (x) \right)                     
\dx
\nonumber 
\end{eqnarray}
F"ur die beiden Integrale erh"alt man schlie"slich mit \cite{Bronstein}, S. 69:
\begin{eqnarray*}
   \mj(\kappa)
   & = & -\; 2 \; \frac{1}{2\pi} \pi \log \left(\frac{\Aj(\kappa) + \sqrt{\Aj(\kappa)^2 -
                                                          \ro^2}}{2} \right)
   \nonumber \\
   & = & \log \left( \frac{2}{\Aj(\kappa) + \sqrt{\Aj(\kappa)^2
                             - \Bj(\kappa)^2 - \Cj(\kappa)^2}}\right)
   \nonumber \\
   & = & \log \left(\frac{2}{\Aj(\kappa) + 1}\right)
\label{anhang1mjgleich}
\end{eqnarray*}
Hier wurde zuletzt (\ref{Ateil1AquadBquadCquadgleicheins}) verwendet.

} % -------- Beweis Lemma i------------------

$ \ffj $  kann als Wahrscheinlichkeitsdichte interpretiert werden, da es die  Eigenschaften $ \ffj(\kappa,y) > 0  \: ( \kappa \in \R^*, y \in \R) $
und $ \frac{1}{\pi} \int_0^\pi \ffj(\kappa,y)\dy = 1
                \quad (\kappa \in \R^*) $
aufweist\footnote{
 Zur weiteren Interpretation sei auf Untersuchungen zum Schr"odingeroperator durch  \cite{AVRONundSIMON} verwiesen. Dort wird statt der Hilfsfunktion $ m_j $ in Anlehnung an \cite{Kakutani} die Funktion $ \gamma_n := \int \sqrt{f_n} \mbox{d} \mu $ eingef"uhrt. Das Grenzma"s ist singul"arstetig im Falle $ \prod_{n} \gamma_n = 0 $. Dann konvergiert n"amlich $ \prod_n f_n $ fast "uberall gegen $ 0 $.
\cite{Perry} fa"st das Bild mittels Transmissionswahrscheinlichkeit $ T $ und Reflektionswahrscheinlichkeit $ R $ zusammen: 
$ R \sim g_i^2 a(k) + \grossO (g_i^3) $ und $ T = 1 - R $ liefern  $ R/T \sim  g_i^2  + \grossO(g_i^3) $, wobei $ g_i $ die Buckelh"ohe bezeichnet.
}.

\begin{bemerkung}
\label{Ateil1bmunabhvondj}

$\left.\right.$

Sei $ i \in \N $. In $ \mi $ geht zwar die Gestalt des $ i $-ten Buckels ein, d.h. $ \mi $ h"angt von der Buckelbreite $ \alpha_i $ und vom Buckelprofil $ W_i $ ab.
$ \mi $ ist jedoch unabh"angig
von der Position $ a_i $ des zugeh"origen Buckels und h"angt somit insbesondere nicht von den
 Buckelabst"anden $ d_j $ $(j \in \N) $ ab.
\end{bemerkung}
\begin{hilfs}
\label{Ateil1lemmamnegativ}

$\left.\right.$

 Im Fall eines $ n $-Buckelpotentials mit identischen Buckeln gilt:
\begin{equation}
  m := \mj < 0 \quad (j\in \{1, \dots, n\})
  \nonumber
\end{equation}
\end{hilfs}

F"ur den Beweis dieses Hilfssatzes wird folgende Aussage ben"otigt:

\begin{hilfs}
\label{SpektralfktdurchPotbestimmt}

$\left.\right.$

Die Spektralfunktion bestimmt das Potential eindeutig.
\end{hilfs}

\Beweis{von Hilfssatz
\ref{SpektralfktdurchPotbestimmt}}{

In \cite{GASYMOV} wird das Problem
\begin{eqnarray}
 B  \frac{\mbox{d}y}{\mbox{d}x} + Q(x) y = \lambda y, \quad (x \in \R_0^+) & & \label{gleichuhsf}\\
 y_1(0) = \sin \alpha, \; y_2(0) = - \cos \alpha &&
\end{eqnarray}
betrachtet.
Dabei ist $ B = \left(\begin{array}{cc}0&1\\-1&0\end{array}\right) $, $ \alpha \in \R $ und die Koeffizientenmatrix $ Q(x) = \left(\begin{array}{cc} p(x) & q (x) \\ q(x) & r (x) \end{array} \right) $, $ p, q, r \in L^1_{loc} (\R_0^+) $.

Zun"achst wird dort festgestellt,  dass durch die Spektralfunktion $ \ro ( \lambda ) $ des Problems die Gleichung  (\ref{gleichuhsf}) nicht eindeutig bestimmt ist, da auch die Gleichung
\begin{equation}
B \psi' + \left({\cal A}^{-1} B {\cal  A}' + {\cal A}^{-1} Q{\cal A} \right) \psi = \lambda \psi
\nonumber
\end{equation}
zur selben Spektralfunktion f"uhrt. Dabei ist
\begin{equation}
{\cal A}(x) = \left(
\begin{array}{cc}
\cos \omega(x) & \sin \omega(x)\\
- \sin \omega(x) & \cos\omega(x)
\end{array}
\right)
\nonumber
\end{equation}
mit $ \omega $ absolutstetig und $ \psi = {\cal A }y $.

Die Eindeutigkeit von (\ref{gleichuhsf}) wird jedoch erreicht, wenn man f"ur diese Gleichung eine mit den Worten von \cite{GASYMOV} als kanonisch bezeichnete Form fordert. Diese ist gegeben durch die Bedingung
\begin{eqnarray}
\spur \left(  {\cal A }^{-1} B  {\cal A }  \, + \,  {\cal A }^{-1} Q {\cal A } \right) = 0 &&
\nonumber\\
\psi_1(0) = 0, \; \psi_2(0) = -1&&
\nonumber
\end{eqnarray}
Dann ist n"amlich 
\begin{equation}
\omega(x) = \frac{1}{2} \int_0^x\left(p(s) + r (s) \right)\ds + \alpha
\nonumber
\end{equation}
eindeutig.

Jede Gleichung der Form (\ref{gleichuhsf}) kann also mittels einer orthogonalen Transformation auf die kanonische Form
\begin{equation}
  B y' + Q_1 y = \lambda y
\end{equation}
mit $ \spur \;Q_1 = 0 $ gebracht werden. Die Transformation der Randbedingung liefert $ y_1(0) = 0,\; y_2(0) = -1 $.

Bringt man die Gleichung (\ref{zentraleGleichung}) mit Randbedingung (\ref{randbedingung}) mit der oben definierten Transformation in die kanonische Form, so ist nach \cite{GASYMOV} die zugeh"orige Spektralfunktion eindeutig.
%llllllllllllllll

}

\Beweis{von Hilfssatz \ref{Ateil1lemmamnegativ}}
{

Da im Fall identischer Buckel f"ur die zu den einzelnen Buckeln geh"orenden
Transfermatrizen
$ \Meins = \dots = \Mn =: M $ gilt, ist $ \Aeins = \dots = \An =: A $,
$ \Beins = \dots = \Bn =: B $ und $ \Ceins = \dots = \Cn =: C $, was
$ \feins = \dots = \fn =: f $ und $ \meins = \dots = \mn =: m $
zur Folge hat.

Der Beweis der Behauptung $ m < 0 $
wird in  zwei Schritten gef"uhrt. Zun"achst wird im ersten Schritt die
Annahme $ m \equiv 0 $ zu einem Widerspruch gebracht.

Da $ m $ holomorph ist, mu"s dann im Anschlu"s im zweiten Beweisschritt lediglich noch
ausgeschlossen werden, dass $ m $ an isolierten Punkten $ \kappa \in \R $
verschwindet.

H"atte n"amlich $ m $ auf einem Intervall den Wert $ 0 $, so m"u"ste nach
\cite{KNOPP I}, \S~21,%S. 101
 $\, m $
schon die Nullfunktion sein (Denn es gibt dann eine offene zusammenh"angende Menge in diesem Intervall, die H"aufungspunkt von Nullstellen von $ m $ ist).
%(\REF{KNOPP, Satz auf S. 101 (+ S.98)}).

F"ur den ersten Beweisschritt sei die Widerspruchsannahme
$ m \equiv 0 $.
Dann ist wegen (\ref{Ateil1defmj}) $ A \equiv 1 $, und
aufgrund von $ A^2 - B^2 - C^2 = 1 $ folgt $ B \equiv C \equiv 0 $.

In Definition (\ref{Ateil1deffj2}) eingesetzt liefert dies
f"ur $ j \in \{1, \dots, n\} $:
\begin{equation}
    \frac{R_{%D_
    {n};j-1}^2 (\kappa)}
         {R_{%D_
         {n};j}^2 (\kappa)}
    =  f_j  \equiv 1
    \quad (\kappa \in \R^*).
\label{Ateil1fequiv1}
\end{equation}
Das hei"st, es ist
$ R_{%D_
{n};j-1}^2 = R_{%D_
{n};j}^2 $ $ (j \in \{ 1, \dots, n\}) $.
Somit ist wegen $ R_{%D_
{n};j} > 0 $ $ ( j \in \{ 0, \dots n\} $)
\begin{equation}
  R_{%D_
  {n};j} \equiv R_{%D_
  {n};0} =: R_0
\label{anhang1Rgleich1}
\end{equation}

Wegen (\ref{Ateil1fequiv1}) und (\ref{teil2drodkappa})
gibt es gewisse $ c_{+}, c_{-} \in \R $ mit
\begin{equation}
  \begin{array}{cc}
    \roDn (\lambda(\kappa)) = \roDnminus (\lambda(\kappa)) + c_{+}
       & \kappa > 0 \\
    \roDn (\lambda(\kappa)) = \roDnminus (\lambda(\kappa)) + c_{-}
       & \kappa < 0.
  \end{array}
\label{anhang1refrjglrjminuseins}
\end{equation}
Ziel ist es, f"ur die Spektralfunktion $ \roDn(\lambda) = \roDnminus(\lambda) + c_{-} $ auf ganz
$ \R $ zu zeigen, woraus dann der gew"unschte Widerspruch abgeleitet werden
wird.

Dazu wird folgende Hilfsfunktion $ \RRn $ definiert:
\begin{equation}
    \RRn : \CC \times \R_0^+ \to \CC,
           ( \lambda, r) \mapsto
           (\lambda-1) \Psi_1^2 (r;\lambda) +
           (\lambda+1) \Psi_2^2 (r;\lambda)
\label{anhang1defRRn}
\end{equation}
wobei $ \Psi $ L"osung von (\ref{zentraleGleichung}) und
(\ref{randbedingung}) ist.

$ \RRn( \cdot, b_j) $ ist auf
$ (-\infty,-1) \cup (1,\infty) $
 als holomorphe Funktion der
L"osungskomponenten $ \Psi_1 $ und $ \Psi_2 $ holomorph.
Au"serdem stimmt $ \RRn $  f"ur $ \lambda \in \R, \modulus{\lambda} > 1 $ mit
dem $ (\lambda-1) $-fachen des Quadrats des in (\ref{pruefer}) definierten
Pr"ufer-Radius "uberein, wie der Vergleich mit dem verallgemeinerten Satz von Pythagoras
(\ref{Rquadrat}) zeigt.

Es ist also wegen (\ref{anhang1Rgleich1})
\begin{displaymath}
  \RRn (\lambda, b_n) = (\lambda-1) R_{%D_n;
  n}^2 (\kappa(\lambda)) =
  (\lambda -1) R_0 \quad
  (\lambda \in (-\infty, -1) \cup (1, \infty))
\end{displaymath}
Da $ \RRn (\cdot, b_n ) $ holomorph ist, gilt dann auch auf ganz $ \CC $ die Beziehung 
%
%\begin{displaymath}
$
   \RRn ( \lambda, b_n ) = (\lambda -1) R_0,
   $
%\end{displaymath}
 %
da die holomorphe Fortsetzung eindeutig ist (siehe \cite{KNOPP I}, S. 98 bzw. S.101).
%(\REF{KNOPP S. 98, 101.})

Es ist also
%
%\begin{displaymath}
$
  \RRn(\lambda, b_n) \neq 0 \quad(\lambda\neq1)
  $.
%\end{displaymath}
%

Wegen
%
%\begin{displaymath}
$  \dnachdr \RRn (\lambda, r) = 0 \quad (r \ge b_n)
$
ist dann
\begin{equation}
  \RRn (\lambda, r) = (\lambda -1) R_0
           \quad ( r \in (b_n, \infty), \lambda \in \CC)
\label{anhang1RRJgleichlambda-1}
\end{equation}
Dass  $ \lambda \in [-1, 1] $ kein Eigenwert sein kann, wird mit einer Fallunterscheidung 
$ \lambda \in (-1,1) $ , $ \lambda = -1 $ und $ \lambda = 1 $ mit Hilfe eines Widerspruchbeweises gezeigt: 

Sei angenommen, dass $ \lambda_0 \in (-1,1) $ ein Eigenwert von
(\ref{zentraleGleichung}), (\ref{randbedingung}) w"are. Dann lautete die zugeh"orige Eigenfunktion f"ur $ r \ge b_n $
\begin{displaymath}
   \Psvektor =
   \left( \begin{array}{c}
            \zeta e^{\sigma r } + \xi e ^{-\sigma r} \\
            \frac{\sigma}{1+\lambda_0}
              \left(\zeta e^{\sigma r } + \xi e ^{-\sigma r} \right)
          \end{array}
   \right)
\end{displaymath}
mit gewissen $ \zeta, \xi \in \R $ und $ \sigma = \sqrt{1 - \lambda_0^2} $.

Aus der Forderung nach quadratischer Integrierbarkeit w"urde $ \zeta =0
$
folgen und man h"atte f"ur $ r \ge b_n $ im Widerspruch zu (\ref{anhang1RRJgleichlambda-1}):
\begin{displaymath}
   \RRn (\lambda_0,r) =
       \xi^2 \left((\lambda_0-1) -
                   (\lambda_0 + 1) \frac{\sigma^2}{(1 + \lambda_0)^2} )
             \right) e^{-2\sigma r} = 0
\end{displaymath}

$ \lambda = -1 $ ist kein Eigenwert des
Problems (\ref{zentraleGleichung}),
(\ref{randbedingung}), da die allgemeine L"osung von
(\ref{zentraleGleichung}) f"ur $ r \ge b_j $ die Gestalt
$ \Psi (r) = \left( \begin{array}{c} \zeta \\
                                     -2\zeta r + \xi
             \end{array} \right)
$ mit $ \zeta, \xi \in \R $
hat. Eine solche Funktion ist aber nur f"ur $ \zeta= \xi = 0 $ quadratisch
integrierbar.

Analog zeigt man, dass $ \lambda = 1 $ kein Eigenwert ist.

Somit ist gezeigt, dass
(\ref{zentraleGleichung}), (\ref{randbedingung}) (wohlgemerkt unter der
Widerspruchsannahme $ m \equiv 0 $) keine
Eigenwerte in $ [-1,1] $ besitzt.
Man erh"alt deswegen aus
(\ref{anhang1refrjglrjminuseins}) f"ur $ j \in \{ 1, \dots, n\} $
\begin{displaymath}
    \roDn (\lambda(\kappa)) = \roDnminus (\lambda(\kappa)) + c_{-}.
\end{displaymath}
Da nach Hilfssatz %Bemerkung 
\ref{SpektralfktdurchPotbestimmt}
% \label{verweisaufAnhang_SpektralfktdurchPoteindeutigbestimmt}
die Spektralfunktion das Potential eindeutig bestimmt, 
ist dann aber
$
%\begin{displaymath}
   q_{D_n} = q_{D_{n-1}}
%\end{displaymath}
%
$ 
im Widerspruch zu $ W_n \neq 0 $.

Also ist $ m $ nicht die Nullfunktion.

%--------------

Im zweiten Beweisschritt bleibt nun zu zeigen, dass $ m $ nicht an isolierten Punkten verschwindet.

Falls $ m(\kappa(\lambda_0)) = 0 $ f"ur ein
$ \lambda_0 \in \R, \modulus{\lambda_0} > 1 $ ausf"allt,
geh"ort $ \lambda_0 $ nicht zum Punktspektrum,
 da die allgemeine L"osung von (\ref{zentraleGleichung})
f"ur $ \modulus{\lambda_0} > 1 $ f"ur $ r \ge b_j $ die
Form
\begin{displaymath}
  \begin{array}{ccl}
    \Psi_1 & = & \zeta \cos \sigma r + \xi \sin \sigma r \\
    \Psi_2 & = & \frac{\sigma}{1+\lambda_0}
                 (\zeta \cos \sigma r + \xi \sin \sigma r) \\
  \end{array}
  \quad \zeta, \xi \in \R
\end{displaymath}
mit $ \sigma = \sqrt{\lambda^2-1} $
besitzt. Diese L"osung ist jedoch nur f"ur $ \zeta = \xi = 0 $
quadratisch integrierbar. Es gibt also keine Eigenl"osung.

Es ist also $ m(\kappa) < 0 \quad (\kappa \in \R^*) $.

} % ENDE BEWEIS Lemma m < 0

\vspace{1cm}

Im Fall kleiner werdender Buckel ist das asymptotische Verhalten von
$ \ffj $ bzw. $ \mj $ f"ur gegen $ 0 $ gehende Buckelh"ohe $ H_j $
von Interesse.
Um dieses zu bestimmen, wird der folgende Hilfssatz verwendet, der Aussagen "uber das asymptotische Verhalten der Gr"o"sen $ \Aj $, $ \Bj$ und $ \Cj $ macht.

$\left.\right.$

\begin{hilfs}
\label{Ateil1lemmaABCasymptotik}
\nopagebreak
$\left.\right.$
\nopagebreak
Sei $ j \in \{1, \dots, n\} $.

Es gilt f"ur $ H_j \to 0 $:

\begin{eqnarray}
   \Aj & = & 1 
\nonumber \\ 
& & + 
%
%H_j^2 \frac{4}{\lambda^2}
%                      \left\{ \left[ \int_{a_j}^{b_j} W(s-a_j) \sin 2 \kappa2
%\ds \right] ^2
%    \right.
%    \nonumber \\
%     & &  \left.
%          \quad \quad \quad \quad \quad
%               + \left[ \int_{a_j}^{b_j} W(s-a_j) \cos 2 \kappa2
%\ds \right] ^2
%                      \right\}
%             + \grossO (H_j^3)
%   \nonumber \\
%
%
\frac{1}{2}H_j^2 \left\{
\frac{8}{\kappa^2} \left(\int_{a_j}^{b_j} W(s-a_j) \sin \kappa s \, \cos \kappa s \ds \right)^2
\right.
\nonumber \\
& & \quad \quad\quad 
+  4 \left(\int_{a_j}^{b_j} W(s-a_j) \sin^2 \kappa s \ds \right)  \, \left(\int_{a_j}^{b_j} W(s-a_j) \cos^2 \kappa s \ds \right)
\nonumber \\
    & & 
     \quad \quad\quad+
     \frac{\lambda^2+1}{\lambda^2 -1}
          \left(\int_{a_j}^{b_j} W(s-a_j) \sin^2 \kappa s \ds \right)^2 
\nonumber \\
    & &  \left.
     \quad \quad\quad+
     \frac{\lambda^2+1}{\lambda^2 -1}
        \left( \int_{a_j}^{b_j} W(s-a_j) \cos^2 \kappa s\ds \right)^2 
    \right\}
\nonumber \\
& & + \grossO(H_j^3)
\label{anhang2Ajasymptotik}
   \\
   \Bj & = & - H_j \frac{2}{\kappa} \int_{a_j}^{b_j} W(s-a_j) \sin 2\kappa s
\ds             + \grossO (H_j^2)
\label{anhang2Bjasymptotik}
   \\
   \Cj & = & H_j \frac{2}{\kappa} \int_{a_j}^{b_j} W(s-a_j) \cos 2\kappa s
\ds
             + \grossO(H_j^2)
\label{anhang2Cjasymptotik}
\end{eqnarray}
\end{hilfs}

\Beweis{}
{

%%%%Sei $ j \in \N $.

Sei "ahnlich wie bei (\ref{deftfrei}) mit
\begin{displaymath}
 \Tnull := \left( \begin{array}{cc}
                    0     &  1 + \lambda \\
                  1-\lambda &    0
                  \end{array}
           \right)
\end{displaymath}
wieder der potentialfreie Operator und mit
%
%\begin{displaymath}
% \TS(H) := \left( \begin{array}{cc}
%                    0                  &   - H W(\cdot - a_j)     \\
%                    H W(\cdot - a_j)   &   0
%                  \end{array}
%           \right)
%\end{displaymath}
%
%
\begin{displaymath}
 \TS(H) := \left( \begin{array}{cc}
                    0                  &   - H W(\cdot )     \\
                    H W(\cdot)   &   0
                  \end{array}
           \right)
\end{displaymath}
die St"orung, die als Kopplungsparameter die Buckelh"ohe $ H \in \R $ enth"alt, bezeichnet.

Das Fundamentalsystem 
%%14mit Anfangswert
%%14$ \left( \begin{array}{cc}
%%14                    1 & 0 \\
%%14                    0 & 1
%%14         \end{array}
%%14  \right)
%%14$ von
%%14%
der ungest"orten Gleichung 
\begin{equation}
 \Psi ' = \Tnull \Psi,
\label{anhang2ungestoerteGl}
\end{equation}
betrachtet auf $ [0,\alpha_j] $%%14, lautet:
ist durch (\ref{UFundamentalsystem}) gegeben.
%%14%
%%14\begin{equation}
%%14  U (r) =
%%14  \left( \begin{array}{cc}
%%14          \cos \kappa r             & - \frac{\kappa}{1-\lambda} \sin \kappa
%%14r\\          -\frac{\kappa}{1+\lambda} \sin \kappa r & \cos \kappa r
%%14         \end{array}
%%14  \right)
%%14  \quad
%%14  (r \in [0,\alpha_j]).
%%14\label{UFundamentalsystem}
%%14\end{equation}
%

Das kanonische Fundamentalsystem f"ur die gest"orte Gleichung
\begin{equation}
  \Psi' = [\Tnull + \TS (H)] \Psi
\label{anhang2gestoerteGl}
\end{equation}
mit Anfangswert
$  \left( \begin{array}{cc}
                1 & 0 \\
                0 & 1
          \end{array}
   \right)
$
hei"se $ \Phi (\cdot,  H) $.
Es ist
$ U(r) = \Phi(r, 0) \quad (r \in [0,\alpha_j]) $ und
$ \Mj = \Phi (\alpha_j, H_j) $.

Die Definitionen
%%%%%%(\ref{anhang2DefA}), (\ref{anhang2DefB}) und
%%%%%%(\ref{anhang2DefC})
(\ref{DefAj}), (\ref{DefBj}) und (\ref{DefCj})
werden folgenderma"sen f"ur reelles $ H $
verallgemeinert:
\begin{eqnarray}
   A(H)  := & \frac{1}{2}
           \left[
              \Phiee^2 (b_j,H)+ \plmin \Phize^2(b_j,H) + \minpl \Phiez^2(b_j,H) +
\Phizz^2(b_j,H)
           \right]
           &
\nonumber \\
& &
\label{anhang2DefA}
   \\
   B(H)  := & \frac{1}{2}
             \left[
                  \Phiee^2(b_j,H) + \plmin \Phize^2(b_j,H)
                  - \minpl \Phiez^2(b_j,H) - \Phizz^2(b_j,H)
             \right]
             &
\nonumber \\
& &
\label{anhang2DefB}
   \\
   C(H)  := &\left[
                  \sqrt{\minpl}\Phiee(b_j,H)\Phiez(b_j,H)
                  + \sqrt{\plmin} \Phize(b_j,H)\Phizz(b_j,H)
             \right]
             &
%              \nonumber \\
\label{anhang2DefC}
\end{eqnarray}

(Dann ist $ \Aj = A(H_j) $, $ \Bj = B(H_j) $ und $ \Cj = C(H_j) $ f"ur
$ j \in \{1, \dots, n\} $.)

Verwendet man hier die Integralgleichung (\ref{integralgl}) in der Form
%Es gilt
%
\begin{equation}
\Phi (r) = U(r) + U (r) \int_0^r U^{-1}(s) \TS(s) \Phi(s) \ds,
\label{PhifuerIteraton}
\end{equation}
 so kann man die L"osung iterativ einsetzen, um 
 die Komponenten von $ \Phi $ in Anh"angigkeit von Potenzen von $ H $ zu ermitteln.
 %%14 (vgl. \cite{EASTHAM}, S. 6).

 %%14Setzt man hier $ \Phi(s) $ mit der Gestalt (\ref{PhifuerIteraton}) iterativ ein und verwendet das Fundamentalsystem (\ref{UFundamentalsystem}) der ungest"orten Gleichung, lassen sich

Es ist
\begin{equation}
  \Ujs (x,\kappa) = U (x,\kappa) +  \grossO(H_j) \quad(H_j \to 0 )
\label{BanhangasymptotischesVerhalten_Hj}
\end{equation}

In 0. Ordnung in $ H $ gilt dann:
\begin{eqnarray*}
   A & = & \frac{1}{2}
           \left[
              \Uee^2(b_j) + \plmin \Uze^2(b_j) + \minpl \Uez^2(b_j) +
\Uzz^2(b_j)
           \right]
           + \grossO (H)
   \nonumber \\
     & = & 1 + \grossO (H)
     \nonumber
   \\
   B & = & \frac{1}{2}
             \left[
                  \Uee^2(b_j) + \plmin \Uze^2(b_j) - \minpl \Uez^2(b_j) -
\Uzz^2(b_j)
             \right]
%             \cdot
%          \nonumber \\
%      & &
%        \cdot \cos\left( 2 ( \teta\ind -\cdot d_j) \right)
           + \grossO (H)
   \nonumber \\
     & = & 0 + \grossO (H)
   \\
   C & = & \left[
                  \sqrt{\minpl}\Uee(b_j)\Uez(b_j) + \sqrt{\plmin}
\Uze(b_j)\Uzz(b_j)
             \right]
%\cdot
%       \nonumber \\
%       & &
%           \cdot \sin\left( 2 ( \teta\ind -\cdot d_j) \right)
           + \grossO (H)
   \nonumber \\
   & = & 0 + \grossO (H)
\end{eqnarray*}

Die erste Ordnung von $ A $ in $ H $ verschwindet, w"ahrend man f"ur die erste Ordnung in H von B bzw. C
\begin{equation}
  - \frac{2}{\kappa} \int_{a_j}^{b_j} W(s-a_j) \sin 2 \kappa s \ds
       \nonumber
\end{equation}
%
 %%14erh"alt
 %%14und 
 %%14die erste Ordnung in $ H $ von $ C $ lautet
%
\begin{equation}
   \frac{2}{\kappa} \int_{a_j}^{b_j}  W(s-a_j)  \cos 2 \kappa s \ds   
        \nonumber
\end{equation}
%
%lautet.
erh"alt.
Die zweite Ordnung von  $ A $ in  $ H $ lautet
\begin{eqnarray}
& & \frac{8}{\kappa^2} \left(\int_{a_j}^{b_j} W(s-a_j) \sin \kappa s \, \cos \kappa s \ds \right)^2 
\nonumber \\
& & 
+  4 \left(\int_{a_j}^{b_j} W(s-a_j) \sin^2 \kappa s \ds \right)  \, \left(\int_{a_j}^{b_j} W(s-a_j) \cos^2 \kappa s \ds \right)
\nonumber \\
    & & 
      +
     \frac{\lambda^2+1}{\lambda^2 -1}\left[
          \left(\!\int_{a_j}^{b_j} W(s-a_j) \sin^2 \kappa s \ds \!\right)^2 
%\right.
%\nonumber \\
%&&\quad\quad\quad\quad +     
%\left.
      \! +\! \left(\! \int_{a_j}^{b_j} W(s-a_j) \cos^2 \kappa s\ds \!\right)^2   \right]
\nonumber \\
&& =: {\cal W}_j({b_j}, {a_j}, \kappa)
\label{zweiteOrdAinH}
\end{eqnarray}

} % ENDE BEWEIS Lemma Asymptotik Aj, Bj Cj

Diese Vorbereitungen dienen dazu, das asymptotische Verhalten von $ \ffj $ und $ \mj $ zu beschreiben:

\begin{hilfs}
\label{Ateil1hilfsfjasympt}

$\left.\right.$

Sei $ j \in \{1, \dots, n\}$.
F"ur $ H_j \to 0 $ gilt:
\begin{eqnarray}
%\lefteqn{
  \ffj (\kappa, y)
%}
%\nonumber \\
   & = & \left[
         1 - \frac{2H_j}{\kappa} \int_{a_j}^{b_j} W_j(s-a_j) \sin 2 \kappa s
\ds \cos 2y
         \right.
   \nonumber \\
   & &
         \left.
         + \frac{2H_j}{\kappa} \int_{a_j}^{b_j} W_j(s-a_j) \cos 2 \kappa s \ds
\sin 2y
         + \grossO (H_j^2)
         \right] ^{-1}
              \nonumber
\end{eqnarray}
%
%\ODER
%%
%\begin{equation}
%  \ffj(\kappa, y) ^{-1} =
%         1 - \frac{2H_j}{\kappa} \int_{a_j}^{b_j} W(s-a_j) \sin 2 \kappa s
%\ds \cos 2y
%         + \frac{2H_j}{\kappa} \int_{a_j}^{b_j} W(s-a_j) \cos 2 \kappa s \ds
%\sin 2y
%         + \grossO (H_j^2)
%\end{equation}
%
und
%*****(\ref{anhang2Ajasymptotik}) in (\ref{Ateil1defmj})) eingesetzt liefert
%
\begin{eqnarray}
\lefteqn{
\mj  =
- \frac{H_j^2}{2} \left\{
\frac{8}{\kappa^2} \left(\int_{a_j}^r W_j(s-a_j) \sin \kappa s \, \cos \kappa s \ds \right)^2 
\right.}
\nonumber \\
& & 
+  4 \left(\int_{a_j}^r W_j(s-a_j) \sin^2 \kappa s \ds \right)  \, \left(\int_{a_j}^r W_j(s-a_j) \cos^2 \kappa s \ds \right)
\nonumber \\
    & & 
    +
     \frac{\lambda^2+1}{\lambda^2 -1}\left[\!
          \left(\!\int_{a_j}^r \!W_j(s-a_j) \sin^2 \kappa s \ds \!\right)^2 
%\right.
%\nonumber \\
%&&\quad\quad\quad \quad\quad\quad\quad +     \left.
+
\left.
        \left(\! \int_{a_j}^r W_j(s-a_j) \cos^2 \kappa s\ds \!\right)^2   \right]\!
\right\} 
\nonumber\\
& %Feb 13\le 
< & 0
\label{anhang2asymptotikm}
\end{eqnarray}
%
%wobei 
%\begin {equation}
%\left[\int_{a_j}^{b_j} W(s-a_j) \sin 2 \kappa s \ds
%\right]^2
%                         +\left[\int_{a_j}^{b_j} W(s-a_j) \cos 2 \kappa s \ds
%\right]^2
%> 0
%\end{equation}
%ist.
\end{hilfs}

\Beweis{}
{

Die Asymptotik von $ \ffj $ erh"alt man durch Einsetzen der Ergebnisse von
Hilfssatz \ref{Ateil1hilfsfjasympt}
in (\ref{Ateil1deffj2}).
(\ref{anhang2Ajasymptotik}) in (\ref{Ateil1defmj}) eingesetzt liefert die Behauptung (\ref{anhang2asymptotikm}) f"ur $ m_j $.
 Das Vorzeichen von $ m_j $ liest man bei (\ref{anhang2asymptotikm}) direkt ab.

} % ENDE BEWEIS Lemma Asymptotik fj, mj

%##############

%Teil2

%-----
\subsection{Singul"arstetiges Spektrum \label{singstetSpektrum}}

%??Vergleich mit Pearson f"ur Schr"odinger

Zielsetzung  ist es, bei vorgegebenen Buckelprofilen die Abst"ande zwischen den Buckeln so zu bestimmen, dass das zugeh"orige Randwertproblem rein singul"arstetiges Spektrum besitzt.

Die Folgen f"ur die Buckelh"ohen, Buckelbreiten und
Buckelformen, die die Buckelprofile beschreiben, seien also vorgegeben. Ihre Eigenschaften werden bei der Formulierung des folgenden Satzes
\ref{Ateil2zentralerSatz}
genauer  angegeben.
Zu bestimmen ist
 die Folge $ D := \left(d_j\right)_{j\in \N} $ der Abst"ande zwischen
den Buckeln derartig, dass das im Folgenden formulierte Problem (\ref{Grenzgleichung}) und (\ref{randbedingungGrenzgleichung}) f"ur
$ \lambda \in \R, \modulus{\lambda} > 1 $ rein singul"arstetiges Spektrum
aufweist.

Seien dazu Folgen $ \left(H_j\right)_{j\in \N} \in \R^{\N} $
und $ \left(\alpha_j\right)_{j\in \N} \in (\R^+)^{\N} $
gegeben.
Sei $ W_j \in 
%AUG2013\Leinsloc[0, \alpha_j]
C[0, \alpha_j] \setminus\{0\}\quad (j \in \N ) $ mit
$ \int_0^{\alpha_j} \modulus{W_j} = 1 $.

Betrachtet wird
\begin{equation}
   \Psvektor  '
   =
   \left( \begin{array}{cc}
             0                     & -q%_{D}
 + 1 + \lambda \\
             q%_{D}
 + 1 - \lambda & 0
          \end{array}
   \right)
    \Psvektor 
\label{Grenzgleichung}
\end{equation}
mit der Randbedingung
\begin{equation}
   \Psi_1 (0) = 1, \quad  \Psi_2(0)  = 0.
\label{randbedingungGrenzgleichung}
\end{equation}
Hierbei ist
\begin{equation}
   q%_{D}
 (r) =  \left\{
                    \begin{array}{cl}
                      0               & 0 \le r \le a_1          \\
                      H_i W_i(r-a_i)  & a_i < r < b_i \quad
                                           (i \in \N) \\
                      0               & b_i \le r < a_{i+1}
                                           \quad(i \in \N) \\
                    \end{array}
               \right.
                    \nonumber
\end{equation}
mit
\begin{eqnarray}
   a_i & := & \sum_{j=1}^{i-1} (d_j + \alpha_j) + d_i
                       \quad (i \in \N) 
                            \nonumber\\
   b_i & := & a_i + \alpha_i  \quad (i \in \N).
        \nonumber
\end{eqnarray}

Der Beweis von Satz \ref{Ateil2zentralerSatz} gliedert sich in zwei Teile:
Zun"achst wird gezeigt, dass bei geeigneter Wahl der $ d_j \; (j \in \N) $
zu den in (\ref{teil2defmyDn}) definierten Ma"sen ein
singul"arstetiges Grenzma"s \label{grenzmass} existiert:
\begin{displaymath}
   \lim_{n \to \infty} \myDn (\Sigma) =: \myD (\Sigma)
        \quad (\Sigma \subset \R^* \;\mbox{kompakt}).
\end{displaymath}
Im Anschlu"s wird im Abschnitt \ref{Operatoren} gezeigt, dass dieses Grenzma"s das Ma"s des zu
(\ref{Grenzgleichung}) geh"origen Grenzoperators ist.

%------------------------------------------

\subsubsection{Singul"arstetiges Grenzma"s}

Der zentrale Satz lautet:

%---------
%%
\begin{satz}
\label{Ateil2zentralerSatz}

$\left.\right.$

Seien Folgen $ \left(H_j\right)_{j \in \N} \in R^\N $ und
$ \left(\alpha_j\right)_{j\in \N} \in (\R^+)^\N $ und zu letzteren Funktionen
$ W_j \in 
%AUG2013\Leinsloc[0,\alpha_j] 
C[0,\alpha_j]$ mit
$ \int_0^{\alpha_j} \modulus{W_j} = 1 \quad (j\in \N) $ gegeben mit folgenden
Eigenschaften:

Entweder sind die Buckel mit von Null verschiedener H"ohe
 identisch oder es gilt
$ H_j = 0 $ nur f"ur endlich viele Indizes $ j \in \N $,
$ H_j \to 0 \quad (j \to \infty) $ und
\begin{equation}
\sum_{j=1}^\infty 
%Feb 13-
%Feb 13 \frac{
H_j^2
%Feb 13}{2}
{\cal W}_j 
  = \infty
%\begin{equation}
%  \sum_{j=1}^\infty
%    H_j^2
%    \left( \left[ \int_0^{\alpha_j} W_j(s) \cos 2 \kappa s \ds \right] ^2
%    + \left[ \int_0^{\alpha_j} W_j(s) \sin 2 \kappa s \ds \right] ^2
%    \right)
%{\cal W}_j(0,a_j, \kappa)
%  = \infty,
\label{BedingungBuckelhoehen}
\end{equation}
mit $ {\cal W}_j(b_j
%\alpha_j
,a_j, \kappa) $ gem"a"s Definition (\ref{zweiteOrdAinH}).

Dann gibt es eine Folge $ D := (d_j)_{j\in \N} \in \left(\R^+\right)^\N
$ mit
der Eigenschaft, dass f"ur jedes kompakte Intervall
$ \Sigma \subset \R^* $
\begin{equation}
 \lim_{j \to \infty} \myDj (\Sigma) =: \myD (\Sigma)
      \nonumber
\end{equation}
existiert und ein singul"arstetiges Ma"s auf $ R^* $ definiert.
\end{satz}
\begin{bemerkung}

$\left.\right.$

Die in Abschnitt \ref{eigenschaftenfm} f"ur den Schr"odingeroperator skizzierte Interpretation des Teilchenverhaltens (\cite{Perry}) 
ist hier analog zu sehen: Wenn die Reflektion an den einzelnen Buckeln aufgrund der gro"sen Buckelabst"ande als unabh"angig von den vorangegangenen Reflektionen angesehen werden kann, ist das Verh"altnis der Gesamtreflektionswahrscheinlichkeit zur Gesamttransmissionswahrscheinlichkeit die Summe der Quotienten aus Reflektionswahrscheinlichkeit und Transmissionswahrscheinlichkeit f"ur die einzelnen Buckel. Die Bedingung (\ref{BedingungBuckelhoehen}) bedeutet eine Gesamttransmissionswahrscheinlichkeit von $ 0 $.
\end{bemerkung}

\begin{beispiel}
\label{BSP}
$\left.\right.$

Die Bedingung (\ref{BedingungBuckelhoehen}) wird beispielsweise von den Buckeln der Breite $ \alpha_j = 1 $ mit den Buckelprofilen
 $ W_j \equiv 1 $ $ (j\in\N) $  und den Buckelh"ohen $ H_j = \sqrt{\frac{1}{j}} $  $(j\in \N)$ erf"ullt.
\end{beispiel}

F"ur dieses Beispiel gilt n"amlich
\begin{eqnarray}
\frac{1}{2} H_j^2 \lefteqn{ {\cal W}_j(1, 0, \kappa) = }
\nonumber \\
\!\!& = &\!\!\!\!  \frac{1}{2j} \left[ \frac{8}{\kappa^2} \left[\int_{0}^1 \!\! \sin \kappa s \, \cos \kappa s \ds \right]^2 
\right.
\!
+  4 \left[\int_{0}^1 \!\! \sin^2 \kappa s \ds \right]   \left[\int_{0}^1 \!\! \cos^2 \kappa s \ds \right]+
\nonumber \\
\!\!& & \!\!  \!\! \quad\quad   +\left. 
     \frac{\lambda^2+1}{\lambda^2 -1}\left[
          \left(\int_{0}^1  \sin^2 \kappa s \ds \right)^2 
        \left( \int_{0}^1  \cos^2 \kappa s\ds \right)^2   \right]\right] 
\nonumber \\
\!\!& \ge  & %\!\!\!\! \frac{1}{2j} \!\!\left[
%\! \frac{8}{\kappa^2} \!\left[\!\int_{0}^1 \!\! \sin \!\kappa s \, \cos \!\kappa s \ds \right]^2 \!\!
%
%       +
\frac{2}{j}
    %4
     \left[\!
         \int_{0}^1 \!\! \sin^2 \kappa s \ds \right]   \left[\int_{0}^1 \!\! \cos^2 \kappa s \ds 
         \right]
        % \right]
%%14       \nonumber \\
%%14\!\!&=& %\!\!\!\! \frac{1}{2j}\left( \frac{8}{\kappa^2} \left[\frac{1}{2\kappa} \sin^2\kappa\right]^2
%%14\frac{2}{j}
%%14%+ 4
%%14\left[\frac{1}{2\kappa} - \frac{1}{4\kappa}\sin 2\kappa\right]  \left[\frac{1}{2\kappa} + \frac{1}{4\kappa}\sin 2\kappa\right]
%%14%\right)
%%14       \nonumber \\
%%14\!\!&= & \!\!\!\! \frac{1}{2j}\left( \frac{2\sin^4\kappa}{\kappa^4}  
%%14  + \frac{1}{\kappa^2} - \frac{\sin^2 2\kappa}{(2\kappa)^2}
%%14\right)
%%14\ge \frac{1}{2j}
\label{NRWfuerm}
\end{eqnarray}
%
%%14unabh"angig von $ \kappa \in \R^* $.

$\left.\right. $

Vor dem eigentlichen Beweis von Satz \ref{Ateil2zentralerSatz} sei hier das Vorgehen kurz skizziert:

Ausgehend von einem endlichen Buckelpotential wird in jedem Schritt eine Reihe von zus"atzlichen Buckeln
hinzugenommen, wobei die Abst"ande zwischen neuen Buckeln gewisse Bedingungen erf"ullen, damit Folgendes erreicht
wird:

Zum einen sollen sich die Ma"se, die zu Potentialen geh"oren, die um  zus"atzliche Buckel erweitert wurden,
von den Vorg"angerma"sen kaum unterscheiden (siehe (\ref{satzschritt1})
und (\ref{satzschrittn})).

Zum anderen ist bei der Wahl der neuen Buckelabst"ande darauf zu achten, dass der Tr"ager des zugeh"origen neuen
Ma"ses kleines Lebesguema"s hat (siehe (\ref{satzschritt1S}), (\ref{satzschritt1mynull}) und
(\ref{satzschrittnS}), (\ref{satzschrittnmynull})). Dies sichert beim "Ubergang zur Grenze $ n \to \infty $ die
Singularit"at des Grenzma"ses.

Damit der Grenzoperator kein Punktspektrum in
$ (-\infty, -1) \cup (1, \infty) $ besitzt, werden nur solche Folgen
$ \left(d_j \right)_{j\in \N} $ f"ur die Buckelabst"ande zugelassen, die
eine gewisse Mindestanwachsrate aufweisen. Dadurch wird daf"ur gesorgt,
dass der Pr"uferradius wegen weiter potentialfreier Strecken nicht schnell
genug abf"allt, um quadrat\-integrable L"osungen zu erm"oglichen.

Abgesehen davon fu"st der Beweis dieses Satzes auf folgenden
Kernpunkten: \label{Beweisidee}

Es mu"s sichergestellt werden, dass das Grenzma"s $ \myD $ existiert.
Um dies zu gew"ahrleisten, ist daf"ur Sorge zu tragen, dass die Reihe
\begin{equation}
 \sum_{n=1}^\infty \left( \myDnplus (\Sigma) - \myDn (\Sigma) \right) 
      \nonumber
\end{equation}
f"ur beliebiges Kompaktum $ \Sigma \subset \R^* $ absolut konvergent ist.
Das ist zum Beispiel der Fall, wenn durch geeignete Wahl der
$ d_j \quad (j \in \N) $
\begin{equation}
  \modulus{ \myDnplus (\Sigma) - \myDn (\Sigma) }
  < 2^{-n+2}
  \quad (n \in \N)
\label{teil2DifferenzderMasse}
\end{equation}
erreicht werden kann. (Man k"onnte aber auch jede andere summierbare
Majorante f"ur die Reihe w"ahlen.)

Gem"a"s Konstruktion "ubertr"agt sich die Stetigkeit der
Ma"se $ \myDj \, (j \in \N ) $ auf das Grenzma"s $ \mu_D $.

Die Singularit"at von $ \myD $ wird dadurch erreicht, dass die
approximierenden Ma"se so gew"ahlt werden, dass sie auf Mengen konzentriert
sind, deren Lebesguema"s f"ur gro"se $ n $ beliebig klein wird.

Um diese Eigenschaften des (noch zu bestimmenden) Grenzma"ses $ \myD $ zu
erhalten, wird die Folge $ D $ induktiv bestimmt.
In jedem Schritt m"ussen die neuen Folgenglieder $ d_j $ gen"ugend
gro"s im Vergleich zu den bereits festgelegten Folgengliedern gew"ahlt
werden. (Was \glqq gen"ugend gro"s\grqq \/ bedeutet, wird beim Beweis des
Satzes konkretisiert werden.)

Um die Aussagen f"ur beliebige Intervalle $ \Sigma \subset \R\setminus\{0\} $ zu erhalten, wird w"ahrend des Induktionsbeweises dieser Bereich durch gedoppelte Intervalle $  [-2n,-\frac{1}{2n}] \cup [\frac{1}{2n},2n] \quad (n\in \N) $
ausgesch"opft.

Der Beweis des Satzes st"utzt sich auf die Aussage von mehreren Hilfss"atzen, die im Folgenden bewiesen werden:

Satz \ref{Alemma1} hat die Tatsache zum Gegenstand, dass sich das Integral
"uber das Mittel einer periodischen Funktion nur wenig vom Integral "uber
die Funktion selbst unterscheidet, wenn man diese nur stark genug staucht.
Daran "andert sich nichts, wenn ein stetig differenzierbarer Faktor zu
dieser Funktion hinzutritt.

Nach Hilfssatz \ref{Alemma2} gilt auch im Fall nicht identischer Buckel
\begin{displaymath}
   - \sum_{j=1}^\infty \mj (\kappa) = \infty \quad
      \mbox{gleichm"a"sig in} \quad \kappa \in \alphabeta,
\end{displaymath}
falls die Summe
$ \sum_{j=1}^\infty H_j^2
 %   \left( \left[ \int_0^{\alpha_j} W_j(s) \cos 2 \kappa s \ds \right] ^2
%    + \left[ \int_0^{\alpha_j} W_j(s) \sin 2 \kappa s \ds \right] ^2
%    \right)
{\cal W}_j(0,a_j, \kappa)
$ divergiert.

Satz \ref{Lemma2B} wird f"ur den Beweis von Satz \ref{Alemma4} ben"otigt.

Dass (\ref{teil2DifferenzderMasse}) erreicht werden kann, beruht auf der
Aussage von Hilfssatz \ref{Alemma3}.

Die Aussage von Satz \ref{Alemma3} erm"oglicht den Vergleich des zu einem
$ n $-Buckelpo\-ten\-tial geh"orenden Ma"ses $ \myDn  $ mit dem Ma"s, das entsteht,
wenn man zu den $ n $ Buckeln noch einen weiteren in gen"ugend gro"sem Abstand hinzusetzt.

Hilfssatz \ref{Alemma4} schlie"slich stellt sicher, dass die Ma"se
$ \myDn $ f"ur wachsendes $ n $ auf Mengen mit gegen $ 0 $ gehendem
Lebesguema"s konzentriert sind, wobei gleichzeitig die "ubrigen zuvor gestellten Forderungen
an diese Ma"se erf"ullt sein m"ussen.

%}

%#################

%Lemma1

{\bf Vorbereitungen f"ur den Beweis von Satz \ref{Ateil2zentralerSatz}}

Im folgenden Satz stehen periodische Funktionen im Mittelpunkt. Staucht man eine derartige Funktion stark genug, so unterscheidet sich ihr Integral beliebig wenig von ihrem Mittelwert:

%%\section{,,Lemma 1''}

%------------
%%
\begin{hilfs}      %--------- Satz 1

$ \left.\right.$

$ F: [\alpha, \beta] \to \R $ sei stetig differenzierbar,
$ G:[\alpha, \beta] \times \R \to \R $ sei stetig differenzierbar und
$ \pi $-periodisch im zweiten Argument. Mit
$ \overline{G} := \frac{1}{\pi} \int_0^\pi G( \cdot, y) \dy $ sei die
Mittelung von $ G $ "uber das zweite Argument bezeichnet.
Dann gilt:

 F"ur jedes $ \eps > 0 $ gibt es $ L_0 > 0 $ so, dass f"ur jedes
$ L \ge L_0 $ gilt
\begin{equation}
  \modulus{ \int_\alpha^\beta F(\kappa) \left[ G(\kappa, L\kappa) - \overline{G}(\kappa) \right]
            \dkappa
          }
  < \eps.
       \nonumber
\end{equation}
\label{Alemma1}
\end{hilfs}
%%
%------------

\Beweis{}{  % ------ Beweis Satz 1-----------------------

Da $ G $ stetig differenzierbar und im zweiten Argument $ \pi $-periodisch
ist, gilt
\begin{equation}
  \sup_{ \kappa \in [\alpha,\beta], y \in \R } \modulus{G(\kappa,y)} =
  \sup_{ \kappa \in [\alpha,\beta], y \in [0,\pi] } \modulus{G(\kappa,y)}
  =: B < \infty
\label{supG}
\end{equation}

Zun"achst werden einige Vor"uberlegungen f"ur beliebiges $ L > 1 $ angestellt.

Wegen
$ G(\kappa,L\kappa) - \overline{G}(\kappa) =
  \frac{\mbox{d}}{\mbox{d}\kappa}
      \int_\alpha^\kappa \left(G(\tilde{\kappa},L\tilde{\kappa}) -
                         \overline{G}(\tilde{\kappa})
                    \right)
       \mbox{d}\tilde{\kappa}
$ ist mit partieller Integration:
\begin{eqnarray}
\lefteqn{
  \int_\alpha^\beta F(\kappa) \left[ G(\kappa,L\kappa) - \overline{G}(\kappa) \right] \dkappa
}
\nonumber  \\
   & = & F(\beta) \int_\alpha^\beta
                      \left[ G(\kappa, L\kappa) - \overline{G}(\kappa) \right] \dkappa
     \nonumber \\
   &   &
      - \int_\alpha^\beta \left( F'(\kappa)
                   \int_\alpha^\kappa \left[G(\tilde{\kappa},L\tilde{\kappa}) -
                                                  \overline{G}(\tilde{\kappa})
                                          \right]
       \mbox{d}\tilde{\kappa} \right)
       \dkappa
\label{Lemmaeinsrefeins}
\end{eqnarray}

Die so erhaltenen Terme werden im Folgenden einzeln untersucht.

Bezeichnet $ c_0 $ das kleinste ganzzahlige Vielfache von $ \pi $, das
gr"o"ser ist als $ L\alpha $, und $ c_M $ das gr"o"ste ganzzahlige Vielfache
von $ \pi $, das kleiner als $L \beta $ ist, so erh"alt man eine Zerlegung von
$ [L\alpha,L\beta] $ in  zwei Randst"ucke $ [L\alpha,c_0] $ und
$[c_M,L\beta] $ und ein Mittelst"uck $ [c_0, c_M] $, welches
$ M:= \frac{c_M-c_0}{\pi} $ Perioden der L"ange $ \pi $ enth"alt (siehe Abbildung
\ref{Intervallzerlegung}).

\begin{figure}[ht]
\includegraphics[origin=c,width=12cm,clip=true, natwidth=610,natheight=642,
viewport=0cm 7.5cm 24cm 10.5cm]{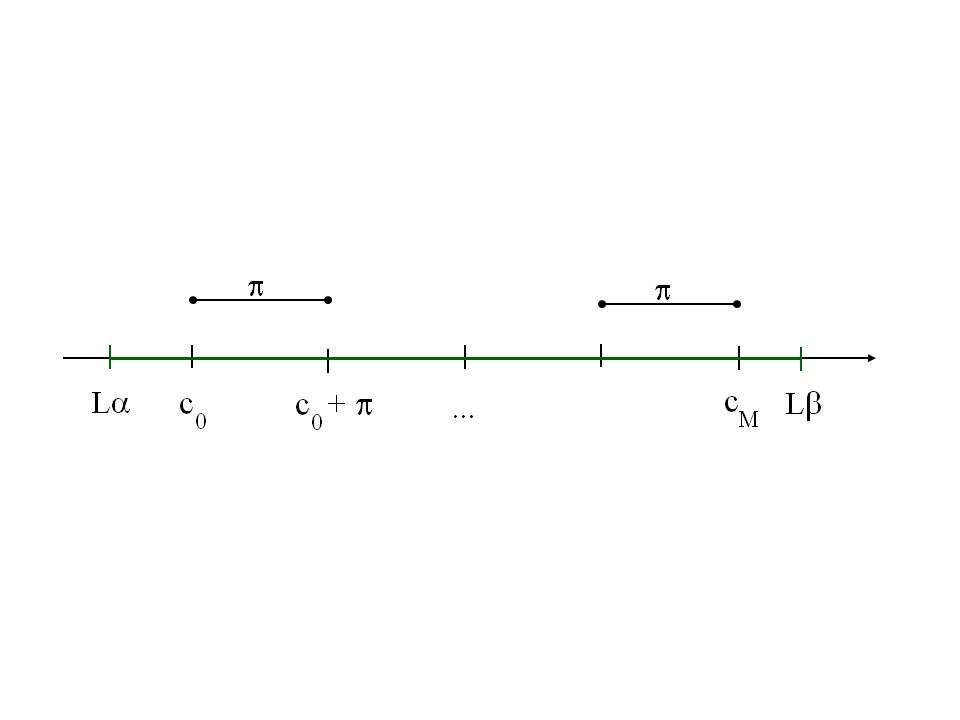}
\caption{\label{Intervallzerlegung} Zerlegung des Intervalls
        $ [L\alpha, L\beta ] $ in $ M +2 $ Intervalle }
\end{figure}

Mit dieser Zerlegung l"a"st sich das Integral des ersten Terms
von (\ref{Lemmaeinsrefeins})
entsprechend als Summe von drei Integralen darstellen:
\begin{eqnarray*}
\lefteqn{ \int\limits_\alpha^\beta
              \left[ G(\kappa, L\kappa) - \overline{G}(\kappa) \right] \dkappa
  \stackrel{y =L\kappa}{=} 
    \frac{1}{L} \int\limits_{L \alpha}^{L \beta}
              \left[ G(\frac{y}{L}, y) - \overline{G}(\frac{y}{L}) \right] \dy}
\nonumber\\
& = & \frac{1}{L} \int\limits_{L \alpha}^{c_0}
              \left[ G(\frac{y}{L}, y) - \overline{G}(\frac{y}{L}) \right] \dy
      + \frac{1}{L} \int\limits_{c_0}^{c_M}
              \left[ G(\frac{y}{L}, y) - \overline{G}(\frac{y}{L}) \right] \dy
\nonumber \\
 &   &
      + \frac{1}{L} \int\limits_{c_M}^{L \beta}
              \left[ G(\frac{y}{L}, y) - \overline{G}(\frac{y}{L}) \right] \dy
\nonumber \\
  & 
 =: & 
 I_1 + I_2 + I_3
\end{eqnarray*}
F"ur $ I_1 $ erh"alt man, wenn man neben (\ref{supG}) noch
%
%\begin{displaymath}
$
%^^^% C:= \sup_{\kappa\in[\alpha,\beta]}\modulus{\overline{G}(\kappa)}
    \sup_{\kappa\in[\alpha,\beta]}\modulus{\overline{G}(\kappa)} \le B
    $
%\end{displaymath}
%
verwendet und ber"ucksichtigt, dass gem"a"s Definition
$ c_0 - L\alpha \le \pi $ ist, folgende Absch"atzung:
\begin{eqnarray}
   \modulus{I_1} & \le &
      \frac{1}{L} \int\limits_{L \alpha}^{c_0}
              \left(
                    \sup_{ \kappa \in [\alpha,\beta], \tilde{y} \in [0,\pi] }
                             \modulus{G(\kappa,\tilde{y})}
                    +
                    \sup_{\kappa\in[\alpha,\beta]}\modulus{\overline{G}(\kappa)}
               \right) \dy
   \nonumber\\
   & = & \frac{1}{L} 2B (c_0-L\alpha)
            \le  \frac{1}{L} 2B \pi
\label{lemma1refIeins}
\end{eqnarray}

Ganz analog sch"atzt man $  \modulus{I_3} $ ab.
%
%%14\begin{displaymath}
%%14   \modulus{I_3}
%%14     =  \frac{1}{L} \int\limits_{c_M}^{L \beta}
%%14              \left[ G(\frac{y}{L}, y) - \overline{G}(\frac{y}{L}) \right] \dy
%%14            \le  \frac{1}{L} 2B \pi
%%14\end{displaymath}
%%14%%14%
%%14ab.

Das Integral $ I_2 $ wird weiter zerlegt in $ M $ Integrale, die jeweils einen 
Integrationsbereich der L"ange $ \pi $ besitzen. 

Stellvertretend f"ur diese wird hier $ [c_0, c_0+\pi] $ n"aher
betrachtet:
\begin{eqnarray*}
\lefteqn{
  \modulus{\int\limits_{c_0}^{c_0+\pi}
              \left[ G(\frac{y}{L}, y) - \overline{G}(\frac{y}{L}) \right] \dy
          } =
} \nonumber \\
%%14  & = & \left|
%%14      \int\limits_{c_0}^{c_0+\pi}
%%14          \left[ G(\frac{c_0}{L}, y) - \overline{G}(\frac{c_0}{L}) \right] \dy
%%14      +
%%14      \int\limits_{c_0}^{c_0+\pi}
%%14           \left[ G(\frac{y}{L}, y) - G(\frac{c_0}{L}, y) \right] \dy
%%14        \right.
%%14  \nonumber \\
%%14  &   & +
%%14        \left.
%%14      \int\limits_{c_0}^{c_0+\pi}
%%14       \left[ \overline{G}(\frac{c_0}{L}) - \overline{G}(\frac{y}{L}) \right] \dy
%%14        \right|
%%14\nonumber \\
  & \le &
      \int\limits_{c_0}^{c_0+\pi} \!\!\!\!\!
         \sup_{x\in[\frac{c_0}{L},\frac{c_0+\pi}{L}],y\in[0,\pi]}\!\!\!\!\!
           \modulus{\! G(\!x , y\!) \!- \!G(\!\frac{c_0}{L}, y\!) \!\dy}
%   \nonumber \\
%  &    & \!\!
    \!\!  + \!\!\int\limits_{c_0}^{c_0+\pi}\!\!\!\!
         \sup_{x\in[\frac{c_0}{L},\frac{c_0+\pi}{L}]}\!
       \modulus{ \overline{G}(\!\frac{c_0}{L}\!) - \overline{G}(\!x\!)} \!\dy
\nonumber \\
  & =   &
      \pi \left[ \sup_{x\in[\frac{c_0}{L},\frac{c_0+\pi}{L}],y\in[0,\pi]}
                 \modulus{ G(x , y) - G(\frac{c_0}{L}, y) }
              +  \sup_{x\in[\frac{c_0}{L},\frac{c_0+\pi}{L}]}
                 \modulus{ \overline{G}(\frac{c_0}{L}) - \overline{G}(x)}
        \right]
\nonumber \\
\end{eqnarray*}
Man beachte dabei, dass gem"a"s der Definition von $ \overline{G} $ die beiden Integrale
%\begin{displaymath}
$
 \int_{c_0}^{c_0+\pi} \overline{G}(\frac{c_0}{L}) \dy =
   \int_{c_0}^{c_0+\pi} G(\frac{c_0}{L},y) \dy 
   $
%\end{displaymath}
 den gleichen Wert aufweisen.

Fa"st man alle Intervalle der L"ange $ \pi $ zusammen, so ergibt sich
\begin{eqnarray*}
  \modulus{I_2}
   & \le &
%%14         \frac{\pi}{L}
%%14%%14         \sum_{j=1}^M
%%14             \left(
%%14             \sup_{x\in[\frac{c_0+(j-1)\pi}{L},\frac{c_0+j\pi}{L}],y\in[0,\pi]}
%%14                 \modulus{ G(x , y) - G(\frac{c_0+(j-1)\pi}{L}, y) }
%%14             \right.
%%14\nonumber \\
%%14   &   & \left.
%%14              + \sup_{x\in[\frac{c_0+(j-1)\pi}{L},\frac{c_0+j\pi}{L}]}
%%14                \modulus{ \overline{G}(\frac{c_0+(j-1)\pi}{L}) - \overline{G}(x)}
%%14         \right)
%%14\nonumber \\
%%14   & = &
         \frac{\pi M}{L}
           \left(
           \sup_{x\in[\frac{c_0+(j_0-1)\pi}{L},\frac{c_0+j_0\pi}{L}],y\in[0,\pi]}
               \modulus{ G(x , y) - G(\frac{c_0+(j_0-1)\pi}{L}, y) }
           \right.
\nonumber \\
   &   & \left.
            + \sup_{x\in[\frac{c_0+(j_0-1)\pi}{L},\frac{c_0+j_0\pi}{L}]}
              \modulus{ \overline{G}(\frac{c_0+(j_0-1)\pi}{L}) - \overline{G}(x)}
         \right)
\nonumber \\
   & =: & \frac{\pi M}{L} H,
\end{eqnarray*}
wobei $ j_0 $ der Index sei, f"ur den das Maximum der Summanden angenommen
wird.

Wegen (\ref{lemma1refIeins}) und der Endlichkeit von  
%
%%14\begin{displaymath}
$
   \sup_{\kappa \in[\alpha, \beta]} \frac{\mbox{d}}{\mbox{d}\kappa} F(\kappa) =: \norm{F'}_{\infty} < \infty
   $
%%14\end{displaymath}
%
und
%
%%14\begin{displaymath}
$
   \modulus{
            \sup_{\tilde{\kappa}\in\alphabeta, \tilde{y}\in[0,\pi]}
               G(\tilde{\kappa},\tilde{y}) +
            \sup_{\tilde{\kappa}\in\alphabeta}
                 \overline{G}(\frac{y}{L})
           }
   =: C < \infty
   $
%%14\end{displaymath}
%
%ausfallen, 
ist f"ur die Absch"atzung des zweiten Terms von (\ref{Lemmaeinsrefeins}) nur
noch folgende Betrachtung, die wieder eine Zerlegung des
%($ \tilde{M} \le M $)
Integrationsintervalls $ [L\alpha,L \kappa] $ in $ \tilde{M} $
($ \tilde{M} \le M $)
Intervalle der L"ange $ \pi $ und zwei Restst"ucke $ [\alpha, c_0] $,
$ [c_0+\tilde{M}\pi, L \kappa] $ mit L"angen kleiner als $ \pi $
ber"ucksichtigt, n"otig:
\begin{eqnarray*}
\lefteqn{ L \modulus{ \int\limits_\alpha^\kappa \left[G(\tilde{\kappa},L\tilde{\kappa}) -
                                          \overline{G}(\tilde{\kappa})
                                    \right]
                      \mbox{d}\tilde{\kappa}
                    }
     =\modulus{ \int\limits_{L\alpha}^{L\kappa}
              \left[ G(\frac{y}{L}, y) - \overline{G}(\frac{y}{L}) \right] \dy
      }
}
\nonumber \\
   & \le &
%%14           \modulus{
%%14                 \int\limits_{L\alpha}^{c_0}
%%14                 \left[ G(\frac{y}{L}, y) - \overline{G}(\frac{y}{L}) \right] \dy
%%14           }
%%14      +
%%14      \sum_{j=1}^{\tilde{M}}
%%14         \modulus{
%%14               \displaystyle{ \int\limits_{c_0+(j-1)\pi}^{c_0+j\pi}    }
%%14               \left[ G(\frac{y}{L}, y) - \overline{G}(\frac{y}{L}) \right] \dy
%%14         }
%%14    \nonumber \\
%%14 &   & +
%%14        \modulus{
%%14              \int\limits_{c_0+\tilde{M}\pi}^{L\kappa}
%%14              \left[ G(\frac{y}{L}, y) - \overline{G}(\frac{y}{L}) \right] \dy
%%14        }
%%14\nonumber\\
%%14  & \le &
      2B \pi +
      \sum_{j=1}^{\tilde{M}}
         \modulus{
               \displaystyle{ \int\limits_{c_0+(j-1)\pi}^{c_0+j\pi}    }
               \left[ G(\frac{y}{L}, y) - \overline{G}(\frac{y}{L}) \right] \dy
         }
    \nonumber \\
 &   & + \int\limits_{c_0+\tilde{M}\pi}^{L\kappa}
             \modulus{
                  \sup_{\tilde{\kappa}\in[\alpha,\beta], \tilde{y}\in[0,\pi]}
              \modulus{G(\tilde{\kappa},\tilde{y})} +
              \sup_{\tilde{\kappa}\in[\alpha,\beta]}
                 \modulus{\overline{G}(\frac{y}{L})} } \dy
\nonumber \\
%%14  & = & 2B \pi +
%%14         \sum_{j=1}^{\tilde{M}}
%%14         \modulus{
%%14               \displaystyle{ \int\limits_{c_0+(j-1)\pi}^{c_0+j\pi}    }
%%14               \left[ G(\frac{y}{L}, y) - \overline{G}(\frac{y}{L}) \right] \dy
%%14         }
%%14        + C (L\kappa -c_0 + \tilde{M}\pi)
%%14\nonumber \\
  & \le & 2B \pi +
         \sum_{j=1}^{M}
         \modulus{
               \displaystyle{ \int\limits_{c_0+(j-1)\pi}^{c_0+j\pi}    }
               \left[ G(\frac{y}{L}, y) - \overline{G}(\frac{y}{L}) \right] \dy
         }
        + C \pi ,
\end{eqnarray*}
was unabh"angig von $ \kappa $ ist. Damit erh"alt man
\begin{eqnarray*}
\lefteqn{
  \modulus{
      - \int_\alpha^\beta \left(  \left(
                        \frac{\mbox{d}}{\mbox{d}\kappa} F(\kappa)
                          \right)
                            \int_\alpha^\kappa \left[G(\tilde{\kappa},L\tilde{\kappa}) -
                                                  \overline{G}(\tilde{\kappa})
                                          \right]
       \mbox{d}\tilde{\kappa} \right)
       \dkappa
  }
   \le
}
\nonumber \\
  & \le & \frac{\norm{F'}_{\infty}}{L} (\beta-\alpha) \pi
    \left[ 2B + C + M H \right]
\end{eqnarray*}

Insgesamt ergibt sich somit:
\begin{eqnarray*}
\lefteqn{
  \modulus{ \int_\alpha^\beta F(\kappa) \left[ G(\kappa, L\kappa) - \overline{G}(\kappa) \right]
            \dkappa
          }
}
\nonumber \\
  & \le & \frac{\pi}{L} \left[ 1 + \norm{F'}_{\infty} (\beta-\alpha) \right] 2B
          + \frac{\pi}{L} \left[2B + \norm{F'}_{\infty}(\beta-\alpha) C \right]
  \nonumber \\
  &     & + \frac{M}{L} \pi \left[ 1 + \norm{F'}_{\infty} (\beta-\alpha)\right] H
\end{eqnarray*}
Die Behauptung des Hilfssatzes folgt wegen $ \frac{M\pi}{L} \to 1 \quad (L \to \infty ) $ und der Tatsache, dass
aufgrund der Stetigkeit von $ G $ und $ \tilde{G} $
\begin{eqnarray*}
  H & = &  \sup_{x\in[\frac{c_0+(j_0-1)\pi}{L},\frac{c_0+j_0\pi}{L}],y\in[0,\pi]}
               \modulus{ G(x , y) - G(\frac{c_0+(j_0-1)\pi}{L}, y) }
\nonumber \\
   &   &
            + \sup_{x\in[\frac{c_0+(j_0-1)\pi}{L},\frac{c_0+j_0\pi}{L}]}
              \modulus{ \overline{G}(\frac{c_0+(j_0-1)\pi}{L}) - \overline{G}(x)}
\nonumber \\
   & \to & 0  \quad (L \to \infty).
\end{eqnarray*}

}  % ------ Beweis Satz 1-----------------------

\begin{korollar}

$\left.\right.$

Sei $ I $ ein kompaktes Intervall. Seien
$ F: [\alpha, \beta] \to \R $ stetig differenzierbar
$ H:[\alpha, \beta] \times \R \times I \to \R $ stetig differenzierbar
und $ \pi $-periodisch im zweiten Argument und
$ z: [\alpha, \beta] \to I $ stetig differenzierbar.
$ \overline{H} := \frac{1}{\pi} \int_0^\pi H( \cdot, y, \cdot) \dy $ sei das
Mittel von $ H $ "uber das zweite Argument.

Dann gilt: F"ur jedes $ \eps > 0 $ gibt es $ L_0 > 0 $ so, dass f"ur jedes
$ L \ge L_0 $ gilt
\begin{equation}
  \modulus{ \int_\alpha^\beta F(\kappa) \left[ H(\kappa, L\kappa, z(\kappa))
         - \overline{H}(\kappa, z(\kappa)) \right]
            \dkappa
          }
  < \eps.
\end{equation}
\label{AKorollarzuLemma1}
\end{korollar}

\Beweis{}{ % ------- Beweis Korollar

\rm
Die Behauptung folgt aus Satz 1 mit
$ G(\kappa, y) = H (\kappa, y, z(\kappa)) $.
} % ------- Beweis Korollar

%Dieses Korollar wird im Fall mit Drehimpuls Verwendung finden.

%_____________________________________________________________________
%nächste Datei reinkopieren

%Lemma2

%\section{,,Lemma 2''}

Die Aussage von Hilfssatz \ref{Ateil1lemmamnegativ}, dass im Fall identischer Buckel $ m < 0 $ ist, kann folgenderma"sen f"ur Buckelpotentiale mit unterschiedlichen Buckeln verallgemeinert werden:

%%%%%%%%%%
%------------
%%
\begin{hilfs}

$\left.\right. $

%Es gilt 
%
Unter den Voraussetzungen des Satzes \ref{Ateil2zentralerSatz} gilt
\begin{equation}
  - \sum_{j=1}^\infty \mj (\kappa) = \infty \quad 
         \mbox{gleichm"a"sig in} \; \kappa \in \alphabeta
         \nonumber
\end{equation}
\label{Alemma2}
\end{hilfs}
\begin{bemerkung}

$\left.\right.$

Dies gilt unabh"angig von der Wahl der Folge 
$ \left( d_j \right)_{j\in \N} $, da $ \mj $ $ \; (j \in \N) $ nach 
Bemerkung \ref{Ateil1bmunabhvondj}
nicht von dieser Folge abh"angt.
\end{bemerkung}
%%

%--------
\Beweis{des Hilfsatzes}{

Im Fall identischer Buckel divergiert 
$ -\sum_{j=1}^\infty \mj(\kappa) = -\sum_{j=1}^\infty m (\kappa) $
gleich\-m"a\-"sig in $ \kappa \in \alphabeta $, da die stetige Funktion
$ m $ nach Hilfssatz \ref{Ateil1lemmamnegativ} echt negativ ist.

Im Fall kleiner werdender Buckel gilt nach Voraussetzung des Satzes
\ref{Ateil2zentralerSatz}
\begin{displaymath}
  \sum_{j=1}^\infty
    H_j^2
%    \left( \left[ \int_0^{\alpha_j} W_j(s) \cos 2 \kappa s \ds \right] ^2
%    + \left[ \int_0^{\alpha_j} W_j(s) \sin 2 \kappa s \ds \right] ^2
%    \right)
{\cal W}_j(0, a_j, \kappa)
  = \infty
\end{displaymath}
Dann gilt aber wegen (\ref{anhang2asymptotikm}):
Zu jedem $ \kappa_0 \in \alphabeta $ und jeder positiven Zahl $ M_0 $ 
gibt es
ein $ n_0(\kappa) \in \N$ so, dass f"ur alle $ n \ge n_0(\kappa) $
\begin{displaymath}
  -\sum_{j=1}^n \mj(\kappa) > M_0
\end{displaymath}
ausf"allt.
%%%%%%%%%
Wegen der Stetigkeit der $ \mj \quad (j \in \{1, \dots, n_0\}) $ gibt es 
$ \delta(\kappa_0) > 0 $ derart, dass f"ur alle
$ \kappa \in (\kappa_0-\delta(\kappa_0),\kappa_0+\delta(\kappa_0)) \cap
  \alphabeta $ gilt
\begin{displaymath}
     -\sum_{j=1}^n \mj(\kappa) > \frac{M_0}{2} =: M .
\end{displaymath}
%
%%%%%%%%%%%
Die offene "Uberdeckung 
\begin{displaymath}
   \bigcup_{\kappa_0\in\alphabeta} \Big( 
             (\kappa_0-\delta(\kappa_0),\kappa_0+\delta(\kappa_0)) \cap
             \alphabeta  
        \Big)
\end{displaymath}
des Kompaktums $ \alphabeta $ enth"alt eine endliche Teil"uberdeckung
\begin{displaymath}
   \bigcup_{i=1}^r \Big( 
             (\kappa_i-\delta(\kappa_i),\kappa_i+\delta(\kappa_i)) \cap
             \alphabeta  
        \Big)
\end{displaymath}
mit einem gewissen $ r \in \N $ und 
$ \kappa_j \in \alphabeta \quad (\{j \in 1, \dots, r \}) $.
%%%%%%%%%
Definiert man
$ n_0 := \max_{i\in\{1, \dots,r\}} n_0(\kappa_i) $, so gilt, dass es 
zu jedem $ M =\frac{M_0}{2} > 0 $ ein nat"urliches $ n_0 $ gibt mit der
Eigenschaft, dass f"ur $ n \ge n_0 $ die Summe $-\sum_{j=1}^n \mj(\kappa) > M $
ausf"allt,
und zwar unabh"angig von $ \kappa $, so dass die Behauptung gezeigt ist.
}

%###############################

%Lemma2aB

%\section{,,Lemma 2B''}
%Bevor Satz \ref{Lemma2B} 
%formuliert wird, werden einige Vor"uberlegungen angestellt.
%%
Eine Vorbereitung f"ur den Beweis von Hilfssatz \ref{Hilfsatzhjquadrat} ist die folgende Feststellung,
dass die Funktionen $ \ffj $ nach unten beschr"ankt sind mit einer gemeinsamen Konstante $ \C $:

\begin{hilfs}

$\left.\right.$

Sei $ \alphabeta \subset \R^* $. 
Dann gibt es ein $ \C > 0 $ mit
\begin{equation}
  \ffj (\kappa, y) \ge \C \quad (j\in \N, \kappa\in\alphabeta, y \in \R). 
\label{lemma2BfgeC}
\end{equation}
\label{Alemma2B}
\end{hilfs}  % ENDE Lemma

\Beweis{}{

Im Fall identischer Buckel ist das trivial, denn 
\begin{displaymath}
   \frac{1}{\ffj(\kappa, y)} =
   A(\kappa) + B(\kappa) \cos 2y + C(\kappa) \sin2y
\end{displaymath}
ist aufgrund der Tatsache, dass die positive Funktion $ f $ stetig 
in $ \kappa $ ist, 
auf $ \alphabeta $ beschr"ankt.

Im Falle kleiner werdender Buckel findet man wegen der in 
Hilfssatz \ref{Ateil1lemmaABCasymptotik}
 angegebenen Asymptotik (\ref{anhang2Ajasymptotik}), 
(\ref{anhang2Bjasymptotik}) und (\ref{anhang2Cjasymptotik})
von $ \Aj $, $ \Bj $ und $ \Cj $ f"ur $ H_j \to 0 $
aufgrund von 
\begin{eqnarray*}
\lefteqn{
   \Aj (\kappa) + \Bj(\kappa) \cos2y + \Cj (\kappa) \sin 2y 
}
\nonumber \\
   & = &
   1 + H_j\frac{2}{\kappa} 
     \left[ \int_0^{\alpha_j} W_j (s) \sin 2\kappa s \ds
            + \int_0^{\alpha_j} W_j (s) \cos 2\kappa s \ds
     \right]
   + \grossOglm(H_j^2)
\nonumber \\
   & \le &
   1 + H_j \frac{4}{\min_{\kappa\in\alphabeta}}
   \underbrace{\int_0^{\alpha_j} \modulus{W_j(s) } \ds}
    _{=1}
   + \grossOglm (H_j^2) 
   \quad(H_j \to 0)
\end{eqnarray*}

ein $ \C > 0 $, das nicht von $ j\in \N $abh"angt, mit der Eigenschaft
\begin{displaymath}
   \Aj (\kappa) + \Bj(\kappa) \cos2y + \Cj (\kappa) \sin 2y
   \le  \frac{1}{\C} \quad (\kappa \in \alphabeta, y \in \R).
\end{displaymath}

}  % Ende Beweis Lemma

Diese Vorbereitung wird im folgenden Hilfssatz verwendet. Neben dem gemittelten Wert $ \mj $ der Gr"o"se $ \log \ffj $ ist  auch die ungemittelte Funktion $ \hj := \log \ffj $ von Interesse. F"ur diese gilt folgende Absch"atzung: 
\begin{hilfs}
\label{Hilfsatzhjquadrat}

$\left.\right.$

Es gibt $ \KC \ge 2 $ mit
\begin{equation}
  \hj^2 \le \KC \left( \ffj - 1 - \hj\right) \quad (j \in \N ),  
\label{lemma2Bhjabschaetzung}
\end{equation}
wobei 
\begin{equation}
   \hj := \log \ffj \quad (j \in \N)
\label{lemma2Bdefhj}
\end{equation}
definiert ist.
\end{hilfs} % Ende Lemma

\Beweis{}
{

Aus
%
%\begin{displaymath}
$   \log t \le t-1 \quad (t \in \R^+)
$ %\end{displaymath}
 folgt 
\begin{displaymath}
   \frac{1}{2} (\log t)^2 
   = \int_1^t \frac{\log s}{s} \ds \le \int_1^t \frac{s-1}{s} \ds
   = t - 1 - \log t \quad (t \ge 1).
\end{displaymath}

Sei nun $ 0 < c < 1 $. Dann gibt es $ \tilde{c} > 0 $ mit
\begin{displaymath}
   \left( \log t \right)^2 
   \le
   \tilde{c} ( t -1 - \log t ) 
   \quad (t \in [c,1]).
\end{displaymath}
(Dies wird beispielsweise von 
$ \tilde{c} := \max_{s \in [c,1]} \frac{(\log s)^2}{s-1-\log s} 
  < \infty 
$
erf"ullt.)

Mit $ {\cal{K}}_c := \max\{2, \tilde{c} \} $ hat man dann:
\begin{equation}
  \left( \log t\right)^2 \le {\cal{K}}_c ( t - 1 -\log t)
  \quad (t \in [c, \infty).
\label{lemma2Bvorueberlegung}
\end{equation}

Die Behauptung (\ref{lemma2Bhjabschaetzung}) folgt aus 
(\ref{lemma2Bvorueberlegung}) mit
$ t = \ffj $, $ c = \C $ von (\ref{lemma2BfgeC})
 und zugeh"origem $ \KC \ge 2 $.
 \label{DefKgeschwungen}
}  %  ENDE Beweis Lemma

\begin{korollar}

$ \left.\right. $

Aus (\ref{lemma2Bhjabschaetzung}) kann man 
\begin{equation}
   0 \le \overline{\hj^2} \le - \KC \mj \quad (j \in \N)
\label{lemma2Babschaetz}
\end{equation}
folgern, indem man "uber das zweite Argument mittelt und 
$ \overline{\ffj} = 1 $ gem"a"s (\ref{Ateil1mittelfj}) und
$ \overline{\hj} = \mj $ f"ur $ j \in \N $ verwendet.

\end{korollar}

Nach diesen Vorbereitungen, die dem Zwecke dienten, sich eine Konstante $
\KC $ zu verschaffen, f"ur die die Absch"atzung (\ref{lemma2Babschaetz})
(unabh"angig von $ j \in \N $) gilt, folgt nun die Formulierung von

%------------
%%
\begin{hilfs}

$\left.\right.$

Sei $ \eps > 0 $ und $ \nyunten \in \N_0 $.
Die nat"urliche Zahl $ \nyoben > \nyunten $ \label{nys} sei so gro"s, dass
\begin{equation}
   \int\limits_\alpha^\beta 
       \frac{1}{- \sum_{i=\nyunten+1}^\nyoben \mi}
   \dkappa
   <
   \frac{2 \eps}{\KC}
            \nonumber
\end{equation}
ausf"allt.
Sei $ \jnull \in \{ \nyunten + 1, \dots, \nyoben\} $, 
$ \Theta: \R^* \to \R $ stetig differenzierbar.
Dann gibt es $ L_{\Theta} > 0 $ so, dass f"ur alle $ L \ge L_\Theta $ gilt:
\begin{equation}
   \int\limits_\alpha^\beta
     \frac{\left[\hjnull (\kappa, \Theta(\kappa) - \kappa L ) \right] ^2}
          {\left[\frac{1}{2} \sum_{i=\nyunten+1}^\nyoben \mi(\kappa)\right]^2}
   < \eps.
            \nonumber
\end{equation}
\label{Lemma2B}
\end{hilfs}
\begin{bemerkungen}

$\left.\right.$

\begin{enumerate}
\item
Dass durch hinreichend gro"ses $ \nyoben $
\begin{displaymath}
   \int\limits_\alpha^\beta 
       \frac{1}{- \sum_{i=\nyunten+1}^\nyoben \mi}
   <
   \frac{2 \eps}{\KC}
\end{displaymath}
gew"ahrleistet werden kann, stellt Hilfsatz \ref{Alemma2} sicher.

\item 
Die stetige Differenzierbarkeit von $ \Theta $ mu"s vorausgesetzt werden, um
Satz \ref{Alemma1} anwenden zu k"onnen. 
In den Situationen, in denen Satz \ref{Lemma2B}
 verwendet wird, wird ohnehin eine sogar 
holomorphe Funktion die Rolle von $ \Theta $ "ubernehmen.
\end{enumerate}
\end{bemerkungen}

\Beweis{von Satz \ref{Lemma2B}}{

Zur Abk"urzung sei 
\begin{displaymath}
  \Nnynykappa := 
% Feb 13       \left[
\frac{1}{2} \sum_{i=\nyunten+1}^\nyoben \mi(\kappa)
%Feb 13\right]^2
\label{Nnyuntenoben}
\end{displaymath}
definiert.

Dass $ \Nnynykappa \neq 0 $ ist, ist im Fall identischer Buckel wegen 
Hilfssatz \ref{Ateil1lemmamnegativ} sofort ersichtlich.

Im Fall kleiner werdender Buckel folgt dies aus 
Hilfssatz \ref{Ateil1hilfsfjasympt}.

Es ist also
\begin{equation}
  \Nnynykappa < 0
\label{Nnynykappa}
\end{equation}

Nach Satz \ref{Alemma1} (angewendet mit 
$ G(\kappa, y) = \hjnull^2 (\kappa, \Theta(\kappa) -y)) $,
$ F = \frac{1}{\Nnynykappa^2} $)
gibt es $ L_\Theta > 0 $ mit
\begin{equation}
  \modulus{\int\limits_\alpha^\beta
             \frac{\hjnull^2 (\kappa, \Theta(\kappa) - \kappa L)
                   - \overline{\hjnull^2} (\kappa)}
                  {\Nnynykappa^2}
           \dkappa
          }
   < \frac{\eps}{2} 
   \quad (L \ge L_\Theta).
\label{lemma2Bref1}
\end{equation}

Au"serdem erh"alt man wegen (\ref{lemma2Babschaetz}) und da 
$ - \mj \ge 0 \quad (j \in \{\nyunten+1, \dots, \nyoben\}) $, 
wenn man die Voraussetzung ausn"utzt:
\begin{eqnarray}
  \int\limits_\alpha^\beta
   \frac{\overline{\hjnull^2}(\kappa)}{\Nnynykappa^2} 
  \dkappa
  & \le & 
  -\KC \int\limits_\alpha^\beta
        \frac{\mjnull}{\Nnynykappa^2} \dkappa
  \le - \KC \int\limits_\alpha^\beta 
       \frac{\sum_{i=\nyunten+1}^\nyoben \mi(\kappa)}{\Nnynykappa^2} \dkappa
\nonumber \\
  &  =  & - \frac{\KC}{4} \int\limits_\alpha^\beta
               \frac{1}
                    {\sum_{i=\nyunten+1}^\nyoben \mi(\kappa)}
           \dkappa
  <  \frac{\KC}{4} \frac{2\eps}{\KC} = \frac{\eps}{2}.
\label{lemma2Bref2}
\end{eqnarray}
Aus (\ref{lemma2Bref1}) und (\ref{lemma2Bref2}) folgt die Behauptung, denn
f"ur $ L \ge L_\Theta $ gilt:
\begin{eqnarray*}
\lefteqn{
  \int\limits_\alpha^\beta
        \frac{\left[ \hjnull^2 (\kappa, \Theta(\kappa) - \kappa L ) \right]^2 }
             {\Nnynykappa^2}
      \dkappa }
\nonumber \\
  & \le &
  \int\limits_\alpha^\beta
   \frac{\overline{\hjnull^2}(\kappa)}{\Nnynykappa^2} 
  \dkappa 
  +
  \modulus{\int\limits_\alpha^\beta
             \frac{\hjnull^2 (\kappa, \Theta(\kappa) - \kappa L)
                   - \overline{\hjnull^2} (\kappa)}
                  {\Nnynykappa^2}
           \dkappa
          }
\nonumber \\
  & < &\eps .
\end{eqnarray*}
}

%###################

%Lemma3

%\section{,,Lemma 3''}

%%

Der folgende Hilfsatz gestattet den Vergleich zwischen dem Ma"s $ \myDndTheta $ und dem $ n $-Buckelma"s $ \myDn $.
Folgende Situationen werden hierbei verglichen: 
Einerseits wird die Situation betrachtet, bei der das $n$-Buckelpotential $
q_{%D_
n} $ durch das  Tupel $ D_n $  von Buckelabst"anden gegeben ist. Diesem wird formal ein Potential gegen"ubergestellt, das in den ersten $ n $ Buckeln mit $ q_{%D_
n} $ "ubereinstimmt, aber einen zus"atzlichen Buckel aufweist. Der $ n\!+\!1$-te Buckel wird dabei durch die zun"achst noch allgemein gehaltenen Gr"o"sen $ \Theta $ und $ d $ beschrieben.
Eine Konkretisierung
erfolgt dann im Beweis von Hilfssatz \ref{Alemma4}.

%%

%------------
%%
\begin{hilfs}

$\left.\right.$

Sei $ n \in \N $, $ d_1, \dots, d_n > 0 $, $ \Theta:\R^* \to \R $ stetig
und $ \eps > 0 $. Dann gibt es $ d_0 > 0 $ mit: Ist $ d \ge d_0 $, so gilt f"ur
jedes Intervall $ \Sigma \subset [\alpha,\beta] $
\begin{equation}
  \modulus{ \myDndTheta (\Sigma) - \myDn (\Sigma) } < \eps,
           \nonumber
\end{equation}
wobei
\begin{eqnarray}
\lefteqn{  \myDndTheta (\Sigma) :=}
\nonumber \\
& := &\!\!\!
   \int\limits_\Sigma \!\left(\!
        \frac{1}{\pi} \left( \prod_{j=1}^n
        \ffj \left(\kappa, \teta\ind (\kappa) \!- \!\kappa d_j\right) \!\right)\!
        f_{n+1} ( \kappa, \Theta(\kappa) \!-\!\kappa d)
        {\cal{D}} (\kappa)\!
        \right)\!\!\!
        \dkappa
\label{mymitdTheteta}
\end{eqnarray}
mit
\begin{equation}
  \D (\kappa) = \frac{{\rm \sign}(\kappa) \sqrt{\kappa^2+1} +1}
                      {{\rm \sign}(\kappa) \sqrt{\kappa^2+1}}
  R(0, \kappa )
  \quad (\kappa \in \alphabeta)
           \nonumber
\end{equation}
wie in (\ref{Ateil2DefD}) definiert ist und $ \myDn $ gem"a"s
(\ref{teil2defmyDn}) zu verstehen ist.
\label{Alemma3}
\end{hilfs}
%%
%------------

\Beweis{von Satz \ref{Alemma3}}{

Sei
\begin{eqnarray}
  \gamma_1 & := &
    \max_{\kappa\in[\alpha,\beta]} \left( \prod_{j=1}^n \ffjvon \right)
            \D (\kappa)
\label{lemma3defgamma1}
  \\
  \gamma_2 & := &
    \max_{\kappa\in[\alpha,\beta]}
       \modulus{\dnachdkappa \left[\prod_{j=1}^n \ffjvon {\D} (\kappa) \right]
               }
\label{lemma3defgamma2}
\end{eqnarray}
$ \gamma_1 $ und $ \gamma_2 $ sind endlich, da die Funktionen
$ \ffj $, $ \teta_{%D_
n;j-1} $ und $ \D $ 
%Feb 13als holomorphe Funktionen 
auf dem
Kompaktum $ \alphabeta $ stetig sind.

Sei $ g_{n+1} := f_{n+1} - 1 $. \label{defgnklein}
Dann ist wegen (\ref{Ateil1mittelfj}) $ \overline{g_{n+1}} = 0 $.

$ d_0 > 0 $ ist so gro"s zu w"ahlen, dass
\begin{equation}
   \frac{\pi}{d_0} \max_{\kappa\in[\alpha,\beta],y\in[0,\pi]}
     \modulus{g_{n+1} (\kappa, y) }
   <
   \frac{\eps}{2} \frac{\pi}{\gamma_1 + (\beta-\alpha)\gamma_2}
   =: \frac{\tilde{\eps}}{2},
\label{lemma3ref1}
\end{equation}
ausf"allt
und f"ur $ \kappa_1, \kappa_2 \in [\alpha,\beta] $
\begin{equation}
    \modulus{\kappa_1-\kappa_2}
    \le \frac{\pi}{d_0}
  \Rightarrow
    \max_{\eta\in\R}
     \modulus{ g_{n+1} (\kappa_1, \Theta(\kappa_1) - \eta)
        -  g_{n+1} (\kappa_2, \Theta(\kappa_2) - \eta)
             }
    < \frac{\tilde{\eps}}{2}\frac{1}{\beta-\alpha}
\label{lemma3ref2}
\end{equation}
gilt.
Letzteres ist m"oglich, da $ \Theta $ stetig und $ g_{n+1} $ auf dem Kompaktum
$ \alphabeta \times [0,\pi] $ gleichm"a"sig stetig ist. (Wegen der Periodizit"at
von $ g_{n+1} $ im 2. Argument gen"ugt es, $ g_{n+1} $ nur
 auf $[\alpha,\beta]\times [0,\pi] $ zu betrachten.)

Sei $ d \ge d_0 $.
Zun"achst werden nun im Folgenden Absch"atzungen f"ur ein beliebiges
Intervall $ \tilde{\Sigma} \subset [\alpha, \beta] $ ermittelt, welche dann im
Anschlu"s beim Beweis der eigentlichen Behauptung angewendet werden.

$ \tilde{\Sigma} $ wird wie in Abbildung \ref{IntervallzerlegungLemma3} skizziert
in $ r $ Teilintervalle $ I_1, \dots, I_r$ mit
$ x_1 := \inf \tilde{\Sigma} $ und
$ x_{i+1} := x_i + \frac{\pi}{d} \quad (i \in \{1, \dots,r-1\}) $, sowie
$ I_i := [x_i, x_{i+1}] \quad (i \in \{1, \dots, r-1\}) $ und
$ I_r := [x_{r-1}, \sup \tilde{\Sigma}] $ zerlegt.
Dabei seien die L"angen
gegeben durch:
\begin{eqnarray*}
  \mynull(I_i) & =  & \frac{\pi}{d} \quad(i \in \{1, \dots, r-1\}) \\
  \mynull(I_r) & \le & \frac{\pi}{d_0},
\end{eqnarray*}
wobei $ \mynull $ das Lebesguema"s\label{Lebesguemass} bezeichnet.
%Dabei ist $ a_1 := \inf \tilde{\Sigma} $ und
%$ a_{i+1} := a_i + \frac{\pi}{d} \quad (i \in \{1, \dots,r-1\}) $, sowie
%$ I_i := [a_i, a_{i+1}] \quad (i \in \{1, \dots, r-1\}) $ und
%$ I_r := [a_{r-1}, \sup \tilde{\Sigma}] $ definiert.

\begin{figure}[ht]
\includegraphics[origin=c,width=12cm,clip=true, natwidth=610,natheight=642,
viewport=0cm 7.0cm 24cm 13.5cm]{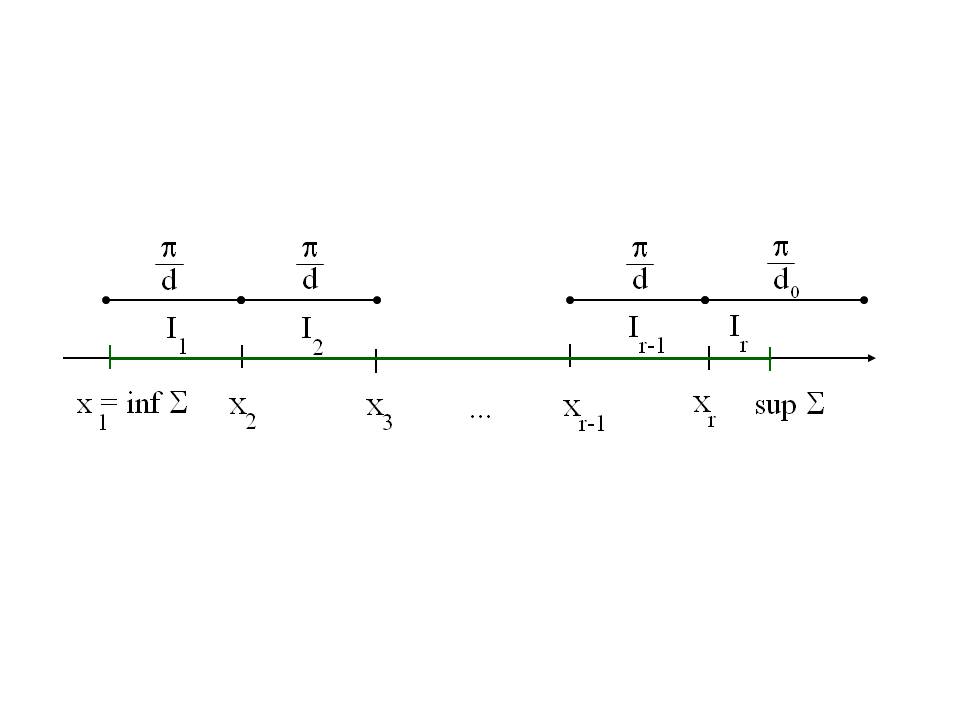}
\caption{\label{IntervallzerlegungLemma3} Zerlegung von $ \tilde{\Sigma} $ in $ r $ Intervalle }
\end{figure}

Wegen (\ref{lemma3ref1}) gilt
\begin{equation}
  \modulus{ \int_{I_r} g_{n+1}(\kappa, \Theta (\kappa) - \kappa d) \dkappa
          }
 \le \frac{\tilde{\eps}}{2}
\label{lemma3ref3}
\end{equation}
F"ur $ i \in \{1, \dots, r-1\} $ und beliebiges $ \kappa_i \in I_i $ erh"alt man,
wenn man beachtet, dass f"ur die
Integrationsvariable $ \kappa $ wegen $ \mu_0(I_i) = \frac{\pi}{d_0} $
stets $ \modulus{\kappa - \kappa_i} \le \frac{\pi}{d_0} $ erf"ullt ist,
und somit (\ref{lemma3ref2}) verwendet werden kann:
\begin{eqnarray}
\lefteqn{
   \modulus{ \int\limits_{I_i} \left[
        g_{n+1} (\kappa, \Theta(\kappa) - \kappa d)
        - g_{n+1} (\kappa_i, \Theta(\kappa_i) - \kappa d) \right]
   \dkappa
   }
} \nonumber \\
 & \le & \int\limits_{I_i}
   \modulus{
        g_{n+1} (\kappa, \Theta(\kappa) - \kappa d)
        - g_{n+1} (\kappa_i, \Theta(\kappa_i) - \kappa d)
   }  \dkappa
\nonumber \\
 & \le &  \mu(I_i)\; \max_{\kappa \in I_i}
      \modulus{
        g_{n+1} (\kappa, \Theta(\kappa) - \kappa d)
        - g_{n+1} (\kappa_i, \Theta(\kappa_i) - \kappa d)
   }
\nonumber \\
 & <  & \frac{\tilde{\eps}}{2}\frac{1}{\beta-\alpha}\frac{\beta-\alpha}{r-1}.
\label{lemma3ref4}
\end{eqnarray}

Au"serdem erh"alt man mit der Substitution
$ y = \Theta(\kappa_i) - \kappa d $ und unter Aus\-n"utzung der Periodizit"at
von $ g_{n+1} $ im zweiten Argument:
\begin{eqnarray}
  \int\limits_{I_i} g_{n+1} (\kappa_i, \Theta(\kappa_i) - \kappa d) \dkappa
  & = & - \frac{1}{d}
        \int\limits_{\Theta(\kappa_i)-d x_i}^{\Theta(\kappa_i)
                                                   -d(x_i+\frac{\pi}{d})}
          g_{n+1}(\kappa_i, y)\dy
\nonumber \\
  & = & - \frac{1}{d} \int\limits_0^\pi g_{n+1}(\kappa_i, y)\dy = 0,
\label{lemma3ref5}
\end{eqnarray}
denn wegen $ \overline{f_{n+1}} = 1 $ gem"a"s (\ref{Ateil1mittelfj})
 ist $ \overline{g_{n+1}} = 0 $.

Fa"st man die Ergebnisse (\ref{lemma3ref3}), (\ref{lemma3ref4}) und
(\ref{lemma3ref5}) f"ur alle
Teilintervalle $ I_i \; (i \in \{1, \dots, r\}) $ zusammen, so erh"alt man,
da
$ \int_{I_i} g_{n+1} (\kappa_i, \Theta (\kappa_i) - \kappa d ) \dkappa = 0 $
ist:
\begin{eqnarray}
\lefteqn{
  \modulus{
    \int_{\tilde{\Sigma}} g_{n+1} (\kappa, \Theta(\kappa) - \kappa d) \dkappa
  }
  \le }
\nonumber \\
 & \le &
    \left(
    \sum_{i=1}^{r-1} \left[
    \left| \int_{I_i} \left(
         g_{n+1} (\kappa, \Theta(\kappa) - \kappa d)
         - g_{n+1} (\kappa_i, \Theta(\kappa_i) - \kappa d) \right) \dkappa
    \right.\right.\right.
    \nonumber \\
    &  &
    \left.\left.\left.
    \quad\quad
       + \int_{I_i} g_{n+1} (\kappa_i, \Theta(\kappa_i) - \kappa d) \dkappa
    \right|
    \right]
    \right)
   + \modulus{ \int_{I_r} g_{n+1} ( \kappa, \Theta(\kappa) - \kappa d)
              \dkappa
     }
\nonumber \\
  & < & \left( \sum_{i=1}^{r-1} \frac{\tilde{\eps}}{2} \frac{1}{r-1}\right)
            + \frac{\tilde{\eps}}{2}
        = \tilde{\eps}.
\label{lemma3ref6}
\end{eqnarray}

Mit diesen Vorbereitungen wird nun die Behauptung des Satzes bewiesen.

Sei nun $ \Sigma \subset \alphabeta $ ein Intervall.

Wenn man
$ g_{n+1} = \frac{\mbox{d}}{\mbox{d}\kappa} {\displaystyle \int_{\inf \Sigma}^\kappa g_{n+1} }$
verwendet und partiell integriert, erh"alt man
\begin{eqnarray*}
\lefteqn{
   \modulus{ \myDndTheta (\Sigma) - \myDn (\Sigma) }
   = }
\nonumber \\
   & = &
   \frac{1}{\pi}
   \modulus{
   \int_\Sigma g_{n+1}(\kappa, \Theta(\kappa) - \kappa d )
        \left( \prod_{j=1}^n
        \ffj \left(\kappa, \teta\ind (\kappa) - \kappa d_j\right) \right)
        \cal{D} (\kappa)
        \dkappa
   }
\nonumber \\
  & \le &
  \frac{1}{\pi}
   \left|
   \int_\Sigma \int_{\inf \Sigma}^\kappa
         g_{n+1}(\tilde{\kappa}, \Theta(\tilde{\kappa}) - \tilde{\kappa} d )
        \; \mbox{d}\tilde{\kappa}
\right. \cdot
\nonumber \\
&&
\left. \quad \cdot  \dnachdkappa\left( \prod_{j=1}^n
        \ffj \left(\kappa, \teta\ind (\kappa) - \kappa d_j\right)
        \cal{D} (\kappa) \right)
        \dkappa
 \right|
  \nonumber \\
  &    &
   + \frac{1}{\pi}
    \modulus{
     \int_{\inf \Sigma}^{\sup \Sigma}
         g_{n+1}(\tilde{\kappa}, \Theta(\tilde{\kappa}) - \tilde{\kappa} d )
        \; \mbox{d}\tilde{\kappa}
          \left( \prod_{j=1}^n
          \ffj \left(\kappa, \teta\ind (\kappa) - \kappa d_j\right) \right)
          \cal{D} (\kappa)
    }
\nonumber \\
  & \le &
    \frac{\gamma_2}{\pi} \int_\Sigma
      \modulus{\int_{inf\Sigma}^{\sup\Sigma}
               g_{n+1}(\tilde{\kappa}, \Theta(\tilde{\kappa}) - \tilde{\kappa} d)
               \; \mbox{d}\tilde{\kappa}
      }
      \dkappa
\nonumber \\
&&
   \quad + \frac{\gamma_1}{\pi}
      \modulus{ \int_{\inf \Sigma}^{\sup \Sigma}
               g_{n+1}(\tilde{\kappa}, \Theta(\tilde{\kappa}) - \tilde{\kappa} d)
              \dkappa
              },
\nonumber
\end{eqnarray*}
wobei die Definitionen (\ref{lemma3defgamma1}) und (\ref{lemma3defgamma2})
verwendet wurden.

Wendet man hierauf (\ref{lemma3ref6}) mit
$ \tilde{\Sigma} = [\inf \Sigma,\kappa] $ bzw.
$ \tilde{\Sigma} = [\inf \Sigma,\sup \Sigma] $ an, so ergibt dies
 schlie"slich
\begin{eqnarray*}
   \modulus{ \myDndTheta (\Sigma) - \myDn (\Sigma) }
   & < &
   \frac{\gamma_2}{\pi} \int_\Sigma \tilde{\eps} \dkappa
   +\frac{\gamma_1}{\pi} \tilde{\eps}
   =
   \frac{\gamma_2}{\pi} \mu_0(\Sigma) \;\tilde{\eps}
   +\frac{\gamma_1}{\pi} \tilde{\eps}
\nonumber \\
   & \le & \tilde{\eps} \left(\frac{\gamma_2(\beta-\alpha)}{\pi}
                              + \frac{\gamma_1}{\pi} \right)
           = \eps,
\nonumber \\
\end{eqnarray*}
da ja wegen $ \Sigma \subset[\alpha,\beta] $ f"ur das Lebesguema"s von $ \Sigma $
gilt:
$ \mynull(\Sigma) \le \beta-\alpha $.

}

Ausgangslage f"ur den folgenden Hilfssatz sind  endliche Buckelpotentiale. Diese stimmen paarweise jeweils bis auf den letzten Buckel "uberein.
Die Aussage des Satzes ist, dass  sich die zugeh"origen Spektralma"se nur geringf"ugig unterscheiden, wenn der jeweils zus"atzliche Buckel weit drau"sen platziert wird.
Au"serdem kann die Position des jeweiligen zus"atzlichen Buckels dabei so gew"ahlt werden, dass das Spektralma"s auf einer Menge mit kleinem Lebesguema"s konzentriert ist.

%##################

%Lemma4.tex

%\section{,,Lemma 4''}

%%
\begin{hilfs}

$\left.\right.$

Sei $ [-\beta,-\alpha] \cup [\alpha,\beta] \subset \R^* $, 
$\nyunten \in \N_0 $, $d_1, \dots, d_\nyunten > 0 $,
$ \eps, \gamma > 0 $. Dann gibt es $ \N \ni \nyoben > \nyunten $ und
$ d_{\nyunten+1} \ge \exp( {\nyunten}^2), \dots, 
  d_\nyoben \ge \exp((\nyoben-1)^2) $ mit:

F"ur jedes Intervall $ \Sigma \subset \doppelInterv $ gilt:
\begin{equation}
   \modulus{\myDjplus (\Sigma) - \myDj(\Sigma) }
   < \gamma 2^{-(j+1)} \quad (j\in\{\nyunten, \dots, \nyoben-1\})
\label{Alemma4Differenzmys}
\end{equation}
und
\begin{eqnarray}
&&    \mynull \left( \left\{ \kappa\in [-\beta,-\alpha] \cup[\alpha,\beta] \Bigg|
\right.\right.
 \nonumber \\
&&
\quad\quad\quad\quad\quad
\left.\left.
      \frac{1}{\pi} 
      \left(\prod_{j=1}^\nyoben
        \ffj (\kappa, \teta_{%D_
        {\nyoben};j-1} (\kappa) - \kappa d_j ) 
      \right) {\cal{D}}(\kappa) 
      > \frac{\eps}{2(\beta-\alpha) }
    \right\} \right)
    \nonumber \\ 
    & & 
 < \eps,  
 \nonumber
\end{eqnarray}
wobei $ \mynull $ das Lebesguema"s bezeichnet.
\label{Alemma4}
\end{hilfs}
\begin{bemerkungen}

$\left.\right.$

\begin{enumerate}

\item 
Die Bedingung 
$ d_{\nyunten+i} \ge \exp \left( \left( \nyunten+i-1\right)^2 \right) $
$ i \in \{1, \dots, \nyoben-\nyunten\}) $ wird lediglich gestellt, um 
in Abschnitt \ref{AAbschnittPunktspektrumohneDreh} 
ausschlie"sen zu k"onnen, dass der Grenzoperator Punktspektrum in
$ (-\infty, -1) \cup (1, \infty) $ besitzt.
%(siehe Abschnitt \ref{AAbschnittPunktspektrum}). 

\item
Statt des in (\ref{Alemma4Differenzmys}) verwendeten Faktors $ 2^{-(j+1)} $ 
k"onnen auch andere positive Zahlen
$ v_j \quad  (j \in \{\nyunten, \dots, \nyoben -1\} $ gew"ahlt werden,
sofern diese einer Folge $ \left( v_j \right)_{j\in \N} $
entstammen, welche die Eigenschaft
$ \sum_{j\in \N} v_j < \infty $ besitzt.

\end{enumerate}

\end{bemerkungen}
%%

%------------

\Beweis{}{

Sei
\begin{equation}
  M:= \frac{1}{\pi} \sup_{\kappa \in \doppelInterv} 
      \left(\prod_{j=1}^\nyunten
        \ffj (\kappa, \teta_{%D_
        {\nyunten};j-1} (\kappa) - \kappa d_j ) 
      \right) {\cal{D}}(\kappa) .
\label{lemma4defM}
\end{equation}

$ M $ ist endlich, da die Funktionen $ \ffj $ $ (j \in \{1, ..., \nyunten \}) $ analytisch sind.

Gem"a"s  Bemerkung \ref{Ateil1bmunabhvondj} h"angt $ \mj \; (j \in \{\nyunten, \dots, \nyoben\}) $  nicht von den noch zu
bestimmenden
$ d_j \; (j \in \{\nyunten, \dots, \nyoben\}) $ ab.

Die nat"urliche Zahl $ \nyoben > \nyunten $ kann aufgrund von 
Hilfssatz \ref{Alemma2} so gro"s
gew"ahlt werden, dass f"ur alle 
$ \kappa \in \doppelInterv $
\begin{equation}
  \Nenner :=
  \frac{1}{2} \sum_{j=\nyunten+1}^\nyoben \mj(\kappa) 
   < \log \left(\frac{1}{M} \frac{\eps}{2(\beta-\alpha)} \right) < 0
\label{lemma4ref1}
\end{equation}
gilt 
und
\begin{eqnarray}
    \displaystyle
    \int\limits_\alpha^\beta 
         \frac{\dkappa}{  \Nenner^2}
    & < \displaystyle\frac{2\eps}{\KC} 
     \nonumber\\
    \displaystyle
    \int\limits_{-\beta}^{-\alpha}
     \frac{\dkappa}{ \Nenner^2}
    & < \displaystyle\frac{2\eps}{\KC} 
     \nonumber
\end{eqnarray}
ausf"allt. 
%***(Es sei daran erinnert, dass 
%***$ \mj  \quad (j \in \{\nyunten, \dots, \nyoben\}) $ nicht von den noch zu 
%***bestimmenden 
%***$ d_j  \quad (j \in \{\nyunten, \dots, \nyoben\}) $ abh"angt.)

Dass dies f"ur gen"ugend gro"ses $ \nyoben $ m"oglich ist, 
gew"ahrleistet Satz \ref{Alemma2}
beziehungsweise ist trivial im Fall konstanter Buckelh"ohen.

Es sei $ \deltany := \nyoben - \nyunten $ definiert.

Im Folgenden werden schrittweise die Abst"ande 
$ d_j $ der Buckel $\nyunten +1 $ bis $ \nyoben $ bestimmt.

{\bf 1. Schritt }

Sei $ d_{\nyunten+1} \ge \exp (\nyunten^2) $.
Um die folgenden Bedingungen (\ref{lemma4schritt10}),
(\ref{lemma4schritt11}) 
und (\ref{lemma4schritt12}) zu erf"ullen, mu"s
$ d_{\nyunten+1} $ gegebenfalls noch vergr"o"sert werden. 
Hilfssatz \ref{Alemma3} (angewendet auf die Intervalle $ [-\beta, -\alpha] $ und
$ \alphabeta $) gew"ahrleistet, dass
\begin{equation}
  \modulus{\mu_{%D
  {\nyunten+1}}(\Sigma)
   - \mu_{%D
   {\nyunten}}(\Sigma) 
  } < \gamma 2^{-(\nyunten+1)}
\label{lemma4schritt10}
\end{equation}
f"ur beliebiges Intervall $ \Sigma \subset \doppelInterv $ 
erreicht werden kann durch Wahl eines gen"ugend gro"sen $ d_{\nyunten+1} $.
(Man beachte, dass (\ref{lemma4schritt10}) bestehen bleibt, wenn man 
$ d_{\nyunten+1} $ gegebenfalls noch weiter vergr"o"sert.)
Bei der Anwendung von Hilfssatz \ref{Alemma3} entsprechen sich jeweils 
$ d $ und $ d_{\nyunten+1} $, $ \Theta $ und $ \teta_{%D_\nyunten;
\nyunten} $,
sowie 
$ \myDndTheta $ und $ \mu_{%D
{\nyunten+1}} $.

Zus"atzlich soll $ d_{\nyunten+1} $ so gro"s sein, dass
\begin{equation}
\displaystyle 
  \int \limits_{-\beta}^{-\alpha} 
  \frac{\left[ h_{\nyunten+1} (\kappa,\teta_{D_\nyunten;\nyunten}(\kappa)
                 - \kappa d_{\nyunten+1} )
        \right]^2
       }
       { \Nenner^2
       } 
  < \frac{1}{8} \frac{1}{\deltany} \eps
\label{lemma4schritt11}
\end{equation}
ist,
was durch Anwendung von Hilfssatz \ref{Lemma2B} ( mit $ i_0 = \nyunten+1 $ , 
$ \Theta(\kappa) = \teta_{%D_\nyunten;
\nyunten} $ und $ L = d_{\nyunten+1} $)
gesichert werden kann. 
Die gleiche Bedingung soll auch bei Integration "uber das Intervall
$ \alphabeta $ erf"ullt sein. 
(Gem"a"s Hilfssatz \ref{Lemma2B} werden diese Forderungen auch noch erf"ullt, wenn 
$ d_{\nyunten+1} $ bei Bedarf weiter vergr"o"sert wird.)

Des weiteren werde $ d_{\nyunten+1} $ unter Umst"anden nochmals vergr"o"sert,
um verm"oge Hilfssatz \ref{Alemma1}
\begin{eqnarray}
  \int\limits_{-\beta}^{-\alpha}
   \left( - 2 m_{\nyunten+1}(\kappa)
   \left( h_{\nyunten+1} (\kappa, \teta_{%D_\nyunten;
   \nyunten} (\kappa) 
                                       - \kappa d_{\nyunten+1} )
          - m_{\nyunten+1} (\kappa)
   \right)
   \right)
         \Nenner^{-2}
  \dkappa
& & 
\nonumber \\
  < \frac{1}{8} \frac{1}{\deltany} \eps
   \quad \quad \quad 
& &
\label{lemma4schritt12}
\end{eqnarray}
und Analoges bei Integration "uber $ \alphabeta $ zu gew"ahrleisten.
($ G(\kappa,y) $ entspricht hier  
$ h_{\nyunten+1} (\kappa, \teta_{%D_\nyunten;
\nyunten} (\kappa) - y) $
und $ F(\kappa) $ der Quotient
$  - 2 m_{\nyunten+1} (\kappa)
         \Nenner^{-2}
$).

Damit ist nun $ d_{\nyunten+1} $ bestimmt (und wird fortan nicht mehr
ver"andert% oder vergr"o"sert werden
).

{\bf \boldmath $  k $. Schritt (mit $ k \in \{2, \dots, \nyoben-\nyunten\} $)}

Sei $ d_{\nyunten+k} \ge \exp ( (\nyunten + k - 1) ^2) $.
Um die folgenden Bedingungen (\ref{lemma4schrittk0}),
(\ref{lemma4schrittk1}), (\ref{lemma4schrittk2}) 
und (\ref{lemma4schrittk3}) zu erf"ullen, mu"s
$ d_{\nyunten+k} $ gegebenfalls noch vergr"o"sert werden. 
Aufgrund von Hilfssatz \ref{Alemma3} (angewendet auf die Intervalle 
$ [-\beta,-\alpha] $ und $ \alphabeta $ mit $d_{\nyunten+k} $ statt
$ d $, $ \teta_{%D_{\nyunten+k-1};
\nyunten+k-1} $ an Stelle von $ \Theta $ und
$ \mu_%D_
{\nyunten+k}%}
 $
f"ur ${\myDndTheta} $) kann durch eine gen"ugend gro"se
Wahl von $ d_{\nyunten+k} $ gew"ahrleistet werden, dass
\begin{equation}
  \modulus{\mu_{%{D
  {\nyunten+k}}(\Sigma)
   - \mu_{%D
   {
   \nyunten+k-1}}(\Sigma) 
  } < \gamma 2^{-(\nyunten+k)}
\label{lemma4schrittk0}
\end{equation}
gilt.

Au"serdem sei $ d_{\nyunten+k} $ so gro"s, dass sowohl
\begin{equation}
\displaystyle 
  \int \limits_{-\beta}^{-\alpha} 
  \left[ h_{\nyunten+k} 
           (\kappa,\teta_{%D_{%\nyunten+k-1};
           \nyunten+k-1}(\kappa)
                 - \kappa d_{\nyunten+k} )
        \right]^2
          \displaystyle 
         \Nenner^{-2}
  < \frac{1}{8} \frac{1}{\deltany} \eps
\label{lemma4schrittk1}
\end{equation}
als auch die analoge Absch"atzung f"ur das Integral "uber $ \alphabeta $
gilt, was aufgrund von  Hilfssatz \ref{Lemma2B} (mit $ i_0 = \nyunten + k $, 
$ \Theta (\kappa) = \teta_{%D_{\nyunten+k-1};
\nyunten+k-1} $ und
$ L = d_{\nyunten+k} $ m"oglich ist. 
 
Zus"atzlich werde $d_{\nyunten+k} $ bei Bedarf noch vergr"o"sert, um verm"oge
Hilfssatz \ref{Alemma1} (mit  $ G(\kappa, y) = 
h_{\nyunten+k} (\kappa, \teta_{%D_{\nyunten+ k-1};
\nyunten+k-1}(\kappa) - y)
$
und
$ F(\kappa) =    
    - \frac{2m_{\nyunten+k}}{\sum_{j=\nyunten+1}^{\nyoben} \mj (\kappa)}
$)
\begin{eqnarray}
& & \int\limits_{-\beta}^{-\alpha}
   \frac{  \left( - 2 m_{\nyunten+k}(\kappa)
       \left( h_{\nyunten+k} 
               (\kappa, \teta_{%D_{\nyunten+k-1};
               \nyunten+k-1} (\kappa) 
                                       - \kappa d_{\nyunten+k} )
          - m_{\nyunten+k} (\kappa)
   \right)\right) }{\Nenner^2}
  \dkappa 
\nonumber \\
& &  < \frac{1}{8} \frac{1}{\deltany} \eps
   \quad \quad \quad \quad 
\label{lemma4schrittk2}
\end{eqnarray}
(und entsprechend auch bei Integration "uber $ \alphabeta $) sicherzustellen.

"Uberdies m"oge $ d_{\nyunten+k} $ so gro"s sein, dass die folgenden 
$ 2 \/(k-2) $ Bedingungen erf"ullt sind: F"ur 
$ s \in \{2, \dots, k-1\} $ soll n"amlich gelten:
\begin{eqnarray}
\lefteqn{
  \int\limits_{-\beta}^{-\alpha}
   \left( h_{\nyunten+k}
                 (\kappa, \teta_{%D_{\nyunten+k-1};
                 \nyunten+k-1} (\kappa) 
                                         - \kappa d_{\nyunten+k} )
     - m_{\nyunten+k}(\kappa) 
   \right) 
  \cdot
}
  \nonumber \\
  & & \quad \cdot
   \left( h_{\nyunten+s} 
               (\kappa, \teta_{%D_{\nyunten+s-1};
               \nyunten+s-1} (\kappa) 
                                       - \kappa d_{\nyunten+s} )
          - m_{\nyunten+s} (\kappa)
   \right)
         \Nenner^{-2}
  \dkappa
  \nonumber \\
  & & 
  \nonumber \\
  & < &
  \frac{1}{8}\frac{1}{k-1} \frac{1}{\deltany} \eps.
     \quad \quad \quad \quad 
\label{lemma4schrittk3}
\end{eqnarray}
und entsprechend bei Integration "uber $ \alphabeta $.
Hierbei wird Hilfssatz \ref{Alemma1} mit
$ G(\kappa,y) = h_{\nyunten+k}(\kappa, \teta_{%D_{\nyunten+k-1},
\nyunten+k-1}
                                - y ) $
und 
\begin{equation}
   F = 
   \left( h_{\nyunten+s} 
               (\cdot, \teta_{%D_{\nyunten+s-1};
               \nyunten+s-1} (\cdot) 
                                       - \cdot\; d_{\nyunten+s} )
          - m_{\nyunten+s} 
   \right)
         \Nenner^{-2}
          \nonumber
\end{equation}
$ 2(k-2) $-mal angewendet.

Aufgrund von (\ref{lemma4schritt10}) und (\ref{lemma4schrittk0}) ist der
erste Teil der Behauptung des Satzes gezeigt.

Als Vorbereitung zum Beweis des zweiten Teils werden folgende
Absch"atzungen durchgef"uhrt:

Es sei $ \kappa \in \doppelInterv $.
Aus 
\begin{equation}
  \frac{1}{\pi} 
   \left( \prod_{j=1}^\nyoben 
     \ffj(\kappa,\teta_{%D_{j-1};
     j-1}(\kappa) - \kappa d_j )
   \right)
   {\cal{D}}(\kappa)
   > 
   \frac{\eps}{2(\beta-\alpha)}
    \nonumber
\end{equation}
folgt gem"a"s Definition (\ref{lemma4defM}) von $ M $
\begin{equation}
   M \left( \prod_{j=\nyunten+1}^\nyoben 
       \ffj(\kappa,\teta_{%D_{j-1};
       j-1}(\kappa) - \kappa d_j )
     \right)
   > 
   \frac{\eps}{2(\beta-\alpha)},
    \nonumber
\end{equation}
woraus durch Logarithmieren
\begin{equation}
   \log \left( \prod_{j=\nyunten+1}^\nyoben 
       \ffj(\kappa,\teta_{%D_{j-1};
       j-1}(\kappa) - \kappa d_j )
     \right)
   > 
   \log \left(\frac{\eps}{2M(\beta-\alpha)}\right)
    \nonumber
\end{equation}
gefolgert werden kann.
Dann gilt auch mit Definition (\ref{lemma2Bdefhj}), wenn man
(\ref{lemma4ref1}) verwendet, das Folgende: 
\begin{equation}
   \sum_{j=\nyunten+1}^\nyoben 
       \hj(\kappa,\teta_{%D_{j-1};
       j-1}(\kappa) - \kappa d_j )
   > 
   \Nenner
    \nonumber
\end{equation}
Dies ist "aquivalent zu (man subtrahiere $ 2 \Nenner $):
\begin{equation}
   \sum_{j=\nyunten+1}^\nyoben 
     \left(
       \hj(\kappa,\teta_{%D_{j-1};
       j-1}(\kappa) - \kappa d_j )
       -\mj(\kappa)
     \right)
   > 
   - \frac{1}{2} \sum_{j=\nyunten+1}^\nyoben \mj (\kappa).
    \nonumber
\end{equation}
Wegen (\ref{lemma4ref1})
 ist die rechte Seite dieser
Ungleichung positiv, so dass Division durch 
$ - \frac{1}{2} \sum_{j=\nyunten+1}^\nyoben \mj (\kappa) = \Nenner $ 
\begin{equation}
  \frac{
   \displaystyle
   \sum_{j=\nyunten+1}^\nyoben 
     \left(
       \hj(\kappa,\teta_{%D_{j-1};
       j-1}(\kappa) - \kappa d_j )
       -\mj(\kappa)
     \right)
  }
  { \displaystyle
     \Nenner }
  > 1
   \nonumber
\end{equation}
liefert und man schlie"slich 
\begin{equation}
  H(\kappa) :=
  \left(
  \frac{
   \displaystyle
   \sum_{j=\nyunten+1}^\nyoben 
     \left(
       \hj(\kappa,\teta_{%D_{j-1};
       j-1}(\kappa) - \kappa d_j )
       -\mj(\kappa)
     \right)
  }
  { \displaystyle
      \frac{1}{2} \sum_{j=\nyunten+1}^\nyoben \mj (\kappa) }
  \right)^2
  > 1
   \nonumber
\end{equation}
hat.
Es gilt also%\footnote{
%$ \mynull $ bezeichnet das Lebesguema"s.}

%
\begin{eqnarray}
 \lefteqn{\quad  \mynull\left(\left\{ \kappa \in \doppelInterv \Bigg|
\right.\right. }
\nonumber \\
&&\quad \quad \quad \quad 
\left.\left.
  \frac{1}{\pi} 
   \left( \prod_{j=1}^\nyoben 
     \ffj(\kappa,\teta_{%D_{j-1};
     j-1}(\kappa) - \kappa d_j )
   \right)
   {\cal{D}}(\kappa)
   > 
   \frac{\eps}{2(\beta-\alpha)}
   \right\} \right)
\nonumber \\
&& \le 
  \mynull \left( \left\{ \kappa \in \doppelInterv \bigg|
  H(\kappa) > 1
  \right\} \right). 
\label{AlemmaGleichung}
\end{eqnarray}
Da $ H(\kappa) \ge 1 \quad (\kappa \in \doppelInterv) $ ist, gilt somit
\begin{eqnarray}
  \mynull(\{ H(\kappa) > 1\} ) 
  & = &
  \int\limits_{\{H(\kappa) > 1\}} 1  \le  
  \int\limits_{\{H(\kappa) > 1\}} H(\kappa) \dkappa 
%\nonumber \\
    \le   
  \int\limits_\doppelInterv H(\kappa) \dkappa.
\nonumber \\
\label{lemma4refH}
\end{eqnarray}
Wegen 
\begin{eqnarray}
\lefteqn{
     -2 \mj (\kappa) 
         \hj (\kappa,\teta_{%D_{j-1};
         j-1} (\kappa) - \kappa d_j)
     + \mj^2 (\kappa)
}
\nonumber \\
   & \le &
     -2 \mj (\kappa) 
      \left(
         \hj (\kappa,\teta_{%D_{j-1};
         j-1} (\kappa) - \kappa d_j)
         - \mj (\kappa)
      \right)
       \nonumber
\end{eqnarray}
ist dann f"ur  $ \kappa \in \doppelInterv $:
\begin{eqnarray}
\lefteqn{
  \sum_{j=\nyunten+1}^\nyoben
  \left( 
   \hj(\kappa,\teta_{%D_{j-1};
   j-1} (\kappa) - \kappa d_j)
   -mj
  \right)^2
  \le
}
\nonumber \\  
& \le & 
  \sum_{j=\nyunten+1}^\nyoben
    \hj^2 (\kappa,\teta_{%D_{j-1};
    j-1} (\kappa) - \kappa d_j)
  \nonumber \\
  & & + 
  \sum_{j=\nyunten+1}^\nyoben
   \left( 
     -2 \mj (\kappa) 
       \left(
         \hj (\kappa,\teta_{%D_{j-1};
         j-1} (\kappa) - \kappa d_j)
         - \mj (\kappa)
       \right)
   \right)
  \nonumber \\
  & & 
  + 
  \sum_{j=\nyunten+1}^\nyoben
  \sum_{i=\nyunten+1}^\nyoben
     \left(
         \hj (\kappa,\teta_{%D_{j-1};
         j-1} (\kappa) - \kappa d_j)
         - \mj (\kappa)
     \right) \cdot
  \nonumber \\
     & & \quad \quad \quad \quad \quad \quad
     \cdot
     \left(
         h_i (\kappa,\teta_{%D_{i-1};
         i-1} (\kappa) - \kappa d_i)
         - m_i (\kappa)
     \right).
\nonumber 
\end{eqnarray}

Diese Absch"atzung f"ur den in (\ref{lemma4refH}) auftretenden Z"ahler
von $ H $
erm"oglicht die Anwendung von (\ref{lemma4schritt11}),
(\ref{lemma4schritt12}),
(\ref{lemma4schrittk1}),
(\ref{lemma4schrittk2}) und
(\ref{lemma4schrittk3}),
so dass man schlie"slich 
\begin{equation}
  \mynull \left( \left\{ \kappa \in \doppelInterv \big|
  H(\kappa) > 1
  \right\} \right)
  < \eps
   \nonumber
\end{equation}
erh"alt, woraus wegen (\ref{AlemmaGleichung}) die Behauptung folgt.
}  %   Ende Beweis

%############################

%Satz

\subsubsection{Beweis von Satz \ref{Ateil2zentralerSatz}\label{BeweiszentralerSatz}}

%---------

Im folgenden Beweis des Satzes \ref{Ateil2zentralerSatz} wird gezeigt,
dass zu vorgegebenen Buckelprofilen sowohl im Falle identischer Buckel
als auch im Fall gegen $ 0 $ gehender Buckelh"ohe - letzteres, wenn Bedingung
(\ref{BedingungBuckelhoehen}) erf"ullt ist - die Abst"ande zwischen
den Buckeln so bestimmt werden k"onnen, dass das zugeh"orige Ma"s
(\ref{teil2defmyDn}) singul"arstetig ist.

Auf Seite \pageref{Beweisidee} wurde die Vorgehensweise des Induktionsbeweises von Satz \ref{Ateil2zentralerSatz} bereits skizziert, der im Folgenden detailliert gef"uhrt wird.

\Beweis{von Satz \ref{Ateil2zentralerSatz}} {

Sei $ \Xi_n := [-2n,-\frac{1}{2n}] \cup [\frac{1}{2n},2n] \quad (n\in \N) $ \label{Xiinterv}
(vgl. Abbildung \ref{Xis}).

\begin{figure}[ht]
\includegraphics[origin=c,width=12cm,clip=true,natwidth=610,natheight=642,
viewport=0cm 6.0cm 24cm 13.5cm]{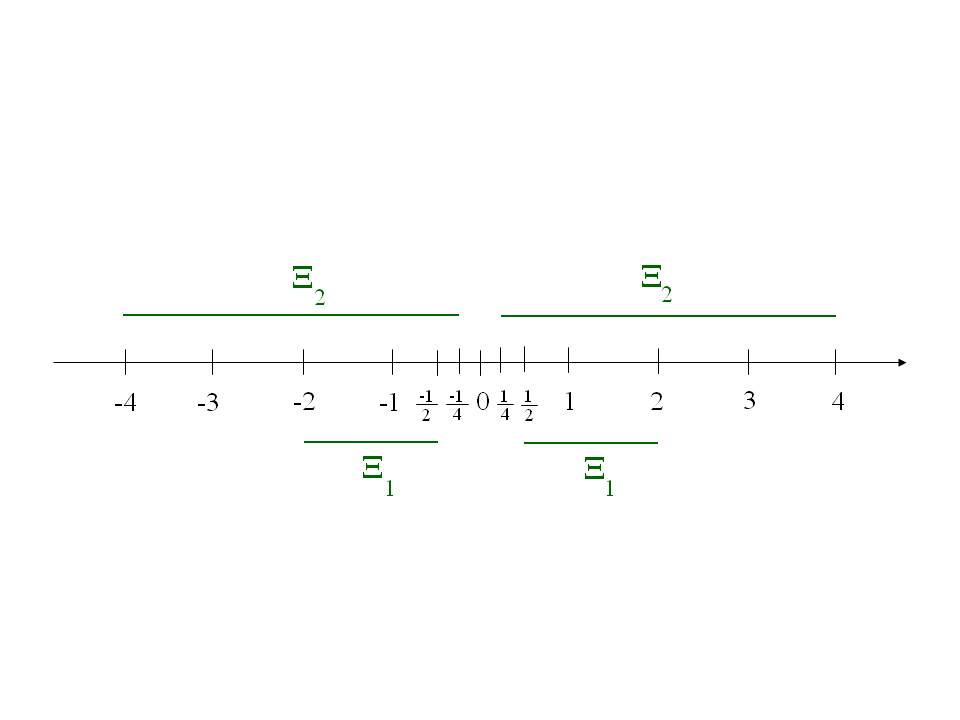}
\caption{\label{Xis} Intervalle $ \Xi_n \, (n\in \N) $ }
\end{figure}

$ (\eps_n)_{n\in\N} $ sei eine streng monotone Nullfolge mit
$ \sum_{n\in\N} \eps_n < \infty $.

Der Beweis wird durch vollst"andige Induktion nach $ \nu_n $, $ (n \in \N) $ gef"uhrt.

1. Schritt:

Nach Hilfssatz \ref{Alemma4}
 gibt es zu $ \nyunten = \nu_0 := 0 $, $ \eps = \eps_1 $,
$ \gamma = \eps_1 $ und $ \alpha=\frac{1}{2} $, $ \beta=2 $ eine
nat"urliche Zahl $ \nu_1 $ und positive Zahlen
\begin{equation}
 d_1 \ge \exp( 0^2), \dots, d_{\nu_1} \ge \exp ((\nu_1-1)^2) 
\label{dschnelleralsexp}
\end{equation}
 so, dass
f"ur jedes kompakte Intervall $ \Sigma \subset \Xi_1 $

\begin{equation}
  \modulus{\mu_{%D
  {j+1}}(\Sigma) - \mu_{%D
  {j}}(\Sigma) }
  < \eps_1 2^{-(j+1)}
  \quad (j\in\{0, \dots, \nu_1-1\})
\label{satzschritt1}
\end{equation}
und
\begin{equation}
    \mynull\left( \Se \right)
    < \eps_2
\label{satzschritt1mynull}
\end{equation}
gilt, wobei
\begin{equation}
   \Se :=
     \left\{ \kappa\in \Xi_1 \Bigg|
      \frac{1}{\pi}
      \left(\prod_{j=1}^{\nu_1}
        \ffj (\kappa, \teta_{%D_{j-1};
        j-1} (\kappa) - \kappa d_j )
      \right) {\cal{D}}(\kappa)
      > \frac{\eps_1}{\mynull(\Xi_1)}
    \right\}
\label{satzschritt1S}
\end{equation}
definiert ist.

$ \Se $ besteht aus endlich vielen Intervallen\footnote{
%holomorph ist\footnote{
      Die Nullstellenmenge der Hilfsfunktion
      $ \frac{1}{\pi}
         \left(\prod_{j=1}^{\nu_1}
           \ffj (\cdot, \teta_{%D_
           {\nyoben};j-1} - \cdot\; d )
         \right) {\cal{D}}
        - \frac{\eps_1}{\mynull(\Xi_1)}
      $
      liegt 
%Feb 13nach \cite{REMMERT}, S. 182 
diskret und abgeschlossen in $ \Xi_1 $.
      Bes"a"se diese Menge der Nullstellen einen H"aufungspunkt, so m"u"ste
      die Hilfsfunktion nach dem Identit"atssatz f"ur analytische Funktionen
      (siehe \cite{KNOPP I}, \S 21% S. 83
      )
      die Nullfunktion sein, was nicht der Fall ist. \label{nullstellenmenge}
%Feb 13      Also besitzt die Hilfsfunktion nur endlich viele Nullstellen in $ \Xi_1 $,
%Feb 13      woraus folgt, dass $ \Se $ aus endlich vielen Intervallen besteht.
}.

Die Anzahl der Intervalle, aus denen $ \Se $ besteht, sei mit $ s_1 $
bezeichnet und $ r_1 := s_1 $ definiert.

$ n $. Schritt:

Hilfssatz \ref{Alemma4} gew"ahrleistet, dass es zu
$ \nyunten = \nu_{n-1} $, $ \eps = \eps_n $,
$ \gamma = \frac{\eps_{n-1}}{r_{n-1}} $ und
$ \alpha=\frac{1}{2n} $, $ \beta=2n $ eine
nat"urliche Zahl $ \nu_n > \nu_{n-1} $ und positive Zahlen
\begin{equation}
d_{\nu_{n-1}+1} \ge \exp( \nu_{n-1}^2), \dots,
  d_{\nu_n} \ge \exp ((\nu_n-1)^2) 
\label{dsschnelleralsexpn}
\end{equation}
 so gibt, dass
f"ur jedes kompakte Intervall $ \Sigma \subset \Xi_n $
\begin{equation}
  \modulus{\mu_{%D
  {j+1}}(\Sigma) - \mu_{%D
  {j}}(\Sigma) }
  < \frac{\eps_{n-1}}{r_{n-1}} 2^{-(j+1)}
  \quad (j\in\{\nu_{n-1}, \dots, \nu_n-1\})
\label{satzschrittn}
\end{equation}
und%\footnote{
%      $ \mynullu_0 $ ist das Lebesguema"s
%}

%
\begin{equation}
    \mynull\left( \Sn \right)
    < \eps_n
\label{satzschrittnmynull}
\end{equation}
ist, wobei
\begin{equation}
   \Sn :=
     \left\{ \kappa\in \Xi_n \Bigg|
      \frac{1}{\pi}
      \left(\prod_{j=1}^{\nu_n}
        \ffj (\kappa, \teta_{%D_{j-1};
        j-1} (\kappa) - \kappa d )
      \right) {\cal{D}}(\kappa)
      > \frac{\eps_n}{\mynull(\Xi_n)}
    \right\}
\label{satzschrittnS}
\end{equation}
definiert ist.

$ \Sn $ besteht wegen einer analogen Argumentation wie f"ur $ \Se $ aus einer endlichen Anzahl $ s_n $ von Intervallen.
%, denn
%die Funktion
%$
%      \frac{1}{\pi}
%      \left(\prod_{j=1}^{\nu_n}
%        \ffj (\cdot, \teta_{D_{\nyoben};j-1} - \cdot\; d )
%      \right) {\cal{D}}
%$
%ist holomorph. 
Sei $ r_n := \max_{j\in\{1, \dots,n\}} s_j $ definiert.

Dass die auf diese Weise induktiv bestimmte Folge $ D := (d_j)_{j\in\N} $
die gew"unschten Eigenschaften besitzt, wird im Folgenden in vier Beweisschritten gezeigt.

Sei dazu $ \Sigma \subset \R^* $ ein kompaktes Intervall.

(Dann gibt es $ \tilde{n} \in \N $ so, dass
$ \Sigma \subset \Xi_n \quad (n \ge \tilde{n}) $
ist.)

%\subsubsection{Existenz des Grenzwertes

%Zun"achst existiert der Grenzwert
{\bf 1. Der Grenzwert der Folge der Ma"se \boldmath $ \left( \myDj \right)_{j\to \infty} $ existiert. }

Es ist n"amlich 
 wegen $ \frac{\eps_{n-1}}{r_{n-1}} \le \frac{\eps_1}{r_1} $,
(\ref{satzschritt1}) und (\ref{satzschrittn})  f"ur
$ m \ge 2 $:
\begin{displaymath}
  \sum_{j=m}^\infty \modulus{\myDjplus (\Sigma)
               - \myDj (\Sigma) }
  \le \sum_{j=m}^\infty \frac{\eps_1}{r_1} 2^{-(j+1)}
  =  \frac{\eps_1}{r_1} 2^{-m} < \frac{\eps_1}{r_1} < \infty
\end{displaymath}
unabh"angig von $ \Sigma $.
Dann ist die Folge $ \left( \myDj (\Sigma) \right)_{j \ge m} $
aufgrund der Konvergenz der obigen Reihe
gleichm"a"sig konvergent.
Das Grenzelement sei mit 

\begin{equation}
  \myD (\Sigma) := \lim_{j \to \infty} \myDj (\Sigma)
\label{satzdefGrenzma}
\end{equation}
bezeichnet.

{\bf 2.{\boldmath $\; \myD $} ist stetig}

%Nun wird die Stetigkeit von $ \myD $ gezeigt.
Sei $ \eps > 0 $.

$ n \in \N $ sei so gro"s, dass f"ur beliebiges kompaktes Intervall
$ \Sigma \subset \R^* $
\begin{displaymath}
  \modulus{ \myDn (\Sigma) - \myD (\Sigma)} < \frac{\eps}{2}
\end{displaymath}
gilt.
Wegen der Stetigkeit von $ \myDn $
(vgl. Gleichung (\ref{teil2drodkappa}))
gibt es dann $ \delta > 0 $ so, dass gilt:
\begin{displaymath}
  \mynull (\Sigma) < \delta
  \Rightarrow
  \myDn (\Sigma) < \frac{\eps}{2}.
\end{displaymath}
Dann ist aber $ \myD $ stetig, denn es ist
\begin{displaymath}
  \mynull(\Sigma) < \delta
    \Rightarrow
  0 \le \myD(\Sigma)
    \le
  \modulus{ \myD(\Sigma) - \myDn(\Sigma) } + \myDn (\Sigma)
    <
  \eps.
\end{displaymath}

{\bf 3. {\boldmath $ \myD $ }ist singul"ar}

%F"ur den Nachweis, dass $ \myD $ singul"ar ist, werden nun einige
%vorbereitende "Uberlegungen angestellt.
Zun"achst werden einige vorbereitende "Uberlegungen angestellt.

Sei $ n \in \N $.

Es ist
\begin{equation}
  \modulus{ \myDn (\Sn) - \myD (\Sn) } < \eps_n.
\label{satzref1}
\end{equation}
Da $ \Sn $ aus $ s_n $ Intervallen besteht, wird zum Beweis dieser Behauptung
zun"achst die Differenz
der Ma"se f"ur ein einzelnes kompaktes Intervall abgesch"atzt.
Sei also $ \Sigma \subset \R^* $ ein kompaktes Intervall.
Dann gibt es eine nat"urliche Zahl
$ \tilde{n} $ so, dass f"ur alle $ n \ge \tilde{n} $ gilt $ \Sigma \subset \Xi_n $.
Mit (\ref{satzschrittn}) (man beachte, dass (\ref{satzschrittn}) f"ur
Dis\-tanzen $ d_j \quad (j\in\{\nu_n +1, \dots, \nu_{n+1}\}) $ auch mit
Faktoren $ \frac{\eps_i}{r_i} \quad (i\in\{1, \dots, n\}) $
statt $ \frac{\eps_n}{r_n} $ gilt, da
$\frac{\eps_i}{r_i} \ge\frac{\eps_n}{r_n} $ $(i \in \{1, \dots, n\})$ ist)
erh"alt man
\begin{eqnarray}
  \modulus{ \myDn (\Sigma) - \myD(\Sigma) } & = &
  \modulus{ \sum_{j=\nu_n+1}^\infty \left(\myDj(\Sigma) -
                                           \myDjminus(\Sigma)
                                    \right)
          }
  \le \frac{\eps_n}{r_n} \sum_{j=\nu_n+1}^\infty 2 ^{-(j+1)}
\nonumber \\
  & = &
  \frac{\eps_n}{r_n} 2 ^{-(\nu_n+1)}
  \le \frac{1}{2}\frac{\eps_n}{r_n}
\label{satzref0}
\end{eqnarray}
F"ur $ \Sn $, das aus $ s_n \le r_n $ Intervallen besteht, liefert dies
die Behauptung (\ref{satzref1}).

Au"serdem gilt f"ur $ \Sn $, welches nach Definition eine Teilmenge von
$ \Xi_n $ ist:
\begin{eqnarray}
  \myDn(\Sn) & = &
   \myDn(\Xi_n) - \myDn(\Xi_n\backslash \Sn)
\nonumber \\
  & = &
   \myDn(\Xi_n) -
   \int\limits_{\Xi_n\backslash \Sn}
      \frac{1}{\pi}
      \left(\prod_{j=1}^{\nu_n}
        \ffj (\kappa, \teta_{%D_{j-1};
        j-1} (\kappa) - \kappa d )
      \right) {\cal{D}}(\kappa)
      \dkappa
\nonumber \\
  & \ge &
   \myDn(\Xi_n) -
   \int\limits_{\Xi_n\backslash \Sn}
      \frac{\eps_n}{\mynull(\Xi_n)}
    \ge
   \myDn(\Xi_n) -
   \int\limits_{\Xi_n}
      \frac{\eps_n}{\mynull(\Xi_n)}
\nonumber \\
  & \ge &
   \myDn(\Xi_n) - \eps_n.
\label{satzref2}
\end{eqnarray}

Dass
\begin{equation}
   \myD(\Sn) \ge \myD(\Xi_n) - 3 \eps_n
\label{satzrefzusingulaer}
\end{equation}
gilt, liegt an folgender Betrachtung, die (\ref{satzref1}) und
(\ref{satzref2}) verwendet:
\begin{eqnarray}
  \myD (\Sn) & \ge &
   \myDn(\Sn) - \modulus{ \myD(\Sn) - \myDn(\Sn) }
   \ge \myDn(\Xi_n) - 2 \eps_n
\nonumber \\
  & \ge & \myD(\Xi_n) - 3 \eps_n,
\label{satzsingulaer}
\end{eqnarray}
wobei Letzteres wegen (\ref{satzref0}), angewendet auf die beiden Intervalle, aus denen $ \Xi_n $ besteht, gilt.

(\ref{satzsingulaer}) einerseits und (\ref{satzschritt1mynull}) und
(\ref{satzschrittnmynull}) andererseits bilden die Grundlage f"ur den
folgenden Nachweis der Singularit"at von $ \myD $.
Sei dazu $ E_l := \sum_{s=l}^\infty \eps_s \quad(l\in \N) $ und
\begin{displaymath}
   {\cal{S}} :=
     \bigcap_{l=1}^\infty \left( \bigcup_{k=l}^\infty {\cal{S}}_k \right)
   =: \bigcap_{l=1}^\infty \Pl
\end{displaymath}
definert.
Dann ist
\begin{displaymath}
  \R^*\backslash {\cal{S}} =
    \bigcup_{l=1}^\infty \left( \bigcap_{s=l}^\infty \R^*\backslash
                                                         \Sl
                         \right).
\end{displaymath}
Da nach (\ref{satzschrittnmynull}) f"ur $ k \in \N $ gilt:
$ \mynull(\Sk) < \eps_k $, ist f"ur $ l \in \N $
\begin{displaymath}
  \mynull( \Pl ) = \mynull \left( \bigcup_{k=l}^\infty \Sk \right)
  \le \sum_{k=l}^\infty \eps_k = E_l < \infty,
\end{displaymath}
da $ \sum_{k=1}^\infty \eps_k $ nach Voraussetzung summierbar ist.
Damit erh"alt man f"ur jedes $ l_0 \in \N $
\begin{displaymath}
  0 \le \mynull\left( \bigcap_{l=1}^\infty \Pl \right)
  \le \mynull\left(\Plnull\right) \le E_{l_0}.
\end{displaymath}
Wegen $ E_{l_0} \rightarrow 0 \quad (l_0 \to \infty) $ gilt somit:
\begin{displaymath}
  \mynull({\cal{S}}) = 0
\end{displaymath}
$ \cal{S} $ ist also eine Lebesguenullmenge.

{ \bf 4. $ \myD$ ist auf $ \cal{S} $ konzentriert.}

%Nun mu"s nur noch gezeigt werden, dass das Ma"s $ \myD $ auf $ \cal{S} $
%konzentriert ist.

Es ist
\begin{eqnarray*}
  \myD\left( \R^* \backslash {\cal{S}} \right)
  & = &
  \myD\left(
       \left( \bigcup_{m=1}^\infty \Xi_m \right) \backslash {\cal{S}}
       \right)
  =
  \myD\left( \bigcup_{m=1}^\infty \Xi_m \backslash {\cal{S}}
      \right)
\nonumber \\
 & \le &
 \sum_{m=1}^\infty \myD\left( \Xi_m \backslash {\cal{S}}
                       \right).
\nonumber
\end{eqnarray*}
F"ur die Summanden gilt dabei                       
\begin{eqnarray*}
  \myD \left( \Xi_m \backslash {\cal{S}} \right)
  & = &
  \myD \left( \Xi_m \cap
                     \left[ \bigcup_{l=1}^\infty
                             \left( \bigcap_{k=l}^\infty
                                      \R^* \backslash \Sk
                             \right)
                     \right]
       \right)
\nonumber \\
& =&
 \myD \left( \bigcup_{l=1}^\infty
              \left( \Xi_m \cap \bigcap_{k=l}^\infty
                                      \R^* \backslash \Sk
              \right)
      \right)
%\nonumber \\
  \le 
 \sum_{l=1}^\infty
   \myD\left( \bigcap_{k=l}^\infty \Xi_m \backslash \Sk
       \right)
 =
 %\sum_{k=l}^\infty 0 =
 0
\end{eqnarray*}
Letzteres folgt aus der folgenden Betrachtung:
%
%%%%%%%%%%%%
%
%f"ur 
F"ur $ k_0 := \max \{ l,m\} $ ist, da ja
$ \Xi_{k_0} \supset \Xi_m $ ist
und (\ref{satzrefzusingulaer}) gilt:
\begin{displaymath}
  \myD\left( \bigcap_{s=l}^\infty \left( %\Xi_m 
  \R^*\backslash \Ss \right)
      \right)
  \le
  \myD\left( \Xi_{k_0} \backslash \Sknull \right)
  \le 3 \eps_{k_0},
\end{displaymath}
Somit folgt%also
\begin{equation}
  \myD\left( \bigcap_{k=l}^\infty \left( \Xi_m \backslash \Sk \right)
      \right)
 = 0,
 \nonumber
 \end{equation}
  was die Behauptung
 \begin{eqnarray*}
  \myD\left( \R^* \backslash {\cal{S}} \right)
  & = & 0
  \end{eqnarray*}
  liefert.
  %\label{satzrefzusingulaerzweite}

}

%#############

%operat(oren)
\subsection{Spektrale Eigenschaften des Grenzoperators $ \TqD$ 
\label{Operatoren}}

Im vorangegangenen Abschnitt wurde eine Folge von Ma"sen $ \left(\myDn\right)_{n\in\N} $ induktiv bestimmt derart, dass das Grenzma"s dieser Folge singul"arstetig ist. Es bleibt zu zeigen, dass dieses Grenzma"s das Spektralma"s eines Differentialoperators $T $ \label{operatorT} ist.

Hierzu werde zun"achst ein Differentialoperator definiert, von dem im Anschlu"s gezeigt wird, dass er die geforderten Eigenschaften aufweist.

Dazu wird das Potential
\begin{equation}
   q%_{D}
 =  \left\{
                \begin{array}{cl}
                  0               & 0 \le r \le a_1          \\
                  H_i W_i(r-a_i)  & a_i < r < b_i \quad (i \in \N) \\
                  0               & b_i \le r < a_{i+1} \quad (i \in \N) \\
                \end{array}
               \right.
\label{satzqD}
\end{equation}
mit
\begin{eqnarray*}
   a_i & := & \sum_{j=1}^{i-1} (d_j + \alpha_j) + d_i
                       \quad (i \in \{ 1, \dots, n\}) \\
   b_i & := & a_i + \alpha_i  \quad (i \in \{ 1, \dots, n\}),
\end{eqnarray*}
betrachtet. Hierbei ist 
 $ D := \left(d_j\right)_{j\in \N} $ die im vorigen
Abschnitt \ref{BeweiszentralerSatz}
definierte Folge von Buckelabst"anden.

Mit diesem Potential 
sei der Differentialausdruck\footnote{
    Dabei sind $ \sigma_2 $ und $ \sigma_3 $ die Pauli-Matrizen
    $
          \sigma_2 = \left(
                         \begin{array}{cc}
                              0 & -i \\
                  i & 0
             \end{array}
             \right)
       $
    und
       $
          \sigma_3 = \left(
                         \begin{array}{cc}
                              1 & 0 \\
                  0 & -1
             \end{array}
             \right)
       $.
\label{S_Pauli-Mmatrizen} 
   }
\begin{equation}
  \tauqD := \sigma_2 p + \sigma_3 + q%_D
\label{OperatorenDifferentialausdruckohneDreh}
\end{equation}
mit $ p := -i \dnachdr $ definiert.

Der zugeh"orige maximale Operator mit Definitionsbereich\label{AClok} 
\begin{displaymath}
  \left\{ %%140310 \Psi 
  u \in \Lzwei\left( \R^+\right)^2 \Big|
             %%140310 \Psi 
             u\in \ACloc\left(\R^+\right)^2,
             \tauqD u
             %%140310 \Psi 
             \in \Lzwei \left(\R^+\right)^2
  \right\}
\end{displaymath}
sei mit $ \TqDmax $ \label{TqDmax}bezeichnet (vgl. \cite{WeidmannmathZ}).
Es gelte die Randbedingung (\ref{randbedingungGrenzgleichung}).

Der freie (maximale) Dirac\-operator $ \Tnulle $\label{eindimfreierOP} (auf $ \R^+ $) mit
Definitionsbereich
\begin{displaymath}
  D(\Tnulle) =
  \left\{ %%140310 \Psi
  u \in \Lzwei(\R^+)^2 \Big| %%140310 \Psi 
  u\in \ACloc (\R^+)^2,
          \taunull\Psi \in \Lzwei (\R^+)^2
  \right\}
\end{displaymath}
(mit $ \taunull := \sigma_2p+ \sigma_3 $)\label{taunull} ist selbstadjungiert %140312\cite{SCHMIDT 95}
\cite{WeidmannmathZ}
und hat wesentliches Spektrum
$ \sigmaess(\Tnulle) = (-\infty, -1] \cup [1, \infty) $. \label{Sigmaess}

%%%%%
\subsubsection{Das wesentliche Spektrum von $\TqD$}
%%%%%%%
%%
\begin{hilfs}
\label{HIlfssatzessspektrum}

$\left.\right. $

Es gilt
\begin{equation}
 \sigmaess(\TqD) \supset \sigmaess (\Tnull)
  \nonumber
\end{equation}
\end{hilfs}

\Beweis{}{

Sei  $ \lambda \in \sigmaess(\Tnull) $. Dann gibt es nach \cite{WEIDMANN76}, Satz 7.24, %%140310 \footnote{
     eine singul"are Folge,
 %%140310 eine Folge
$ \left( \Psi_n\right)_{n\in \N} \subset D(\Tnullmin )
$
%%140310 
zu $ \Tnull $ und $ \lambda $.
%  = \left\{\Psi \in D(\Tnull) \Big| \supp \Psi \komp \R \right\} $ mit
%den Eigenschaften
%$ \norm{\Psi_n}_{\Lzwei(\R)^2} =1 $,
%$ \lim_{n\to\infty} \norm{\left(\Tnullmin -\lambda\right) \Psi_n }
%  _{\Lzwei(\R)^2} =0 $
%und $ \left( \Psi_n, \Phi\right)_{\Lzwei(\R)^2}
%      \stackrel{n \to \infty}{\to} 0 \quad
%     (\Phi \in \Lzwei(\R)^2) $.

Sei nun $ n \in \N $. Da $ \supp \Psi_n \komp \R $ ist, gibt es $ x_n $,
$ y_n \in \R $ mit $ \supp \Psi_n \subset (x_n,y_n) $.
Da die Buckelabst"ande $ \left(d_j\right)_{j \in \N} $ wegen (\ref{dschnelleralsexp}) und (\ref{dsschnelleralsexpn}) schneller als $ \left(\exp (j-1)^2\right)_{j\in\N} $ wachsen, gibt es einen Index $ j(n) \in  \N $ mit $ d_{j(n)} > y_n - x_n $.

%Da f"ur die Folge $ D $ wegen $ d_j > $ exp $ ((j-1)^2)$

%
%\begin{displaymath}
%  \limsup_{j\to \infty} d_j = \infty
%\end{displaymath}
%
%gilt, gibt es $ j(n) \in \N $ mit $ d_{j(n)} > y_n-x_n $.

Die Funktion $  \Psi_n  $ l"a"st sich deshalb so verschieben, dass ihr Tr"ager in einem potentialfreien Bereich zu liegen kommt:
\begin{displaymath}
  \Phi_n := \Psi_n (\cdot + x_n - b_{j(n)} ).
\end{displaymath}

Die Folge $ \left( \Phi_n \right)_{n\in \N} $ ist eine singul"are Folge zu
$ \TqD $ und $ \lambda $ und somit

$ \lambda \in \sigmaess(\TqD) $.
} % ENDE BEWEIS Lemma

\begin{bemerkungen}

$\left.\right.$

\begin{itemize}
\item
Beim Beweis von $ \sigmaess(\TqD) \supset (-\infty,-1]\cup[1,\infty) $ wurde lediglich die Eigenschaft
$ \limsup_{j\to \infty} d_j = \infty $ des Potentials $ q_D $ verwendet. Andere Eigenschaften wie die Buckelh"ohen oder Buckelformen wurden f"ur den Beweis nicht ben"otigt.
\item
Im Fall $ H_j \to 0 \quad (j \to \infty) $ gilt sogar
\begin{displaymath}
   \sigmaess (\TqD) = (-\infty, -1]\cup [1, \infty),
\end{displaymath}
denn einerseits ist nach {\rm \cite{WeidmannmathZ}},
Kor. 6.9.\footnote{
   Es ist hier (mit den Bezeichnungen, die in \cite{WeidmannmathZ} Verwendung finden)
   $ P(r) = \left(\begin{array}{cc}
                    -q%_D 
                    (r) -1 & 0 \\
                    0           & -q%_D
                    (r) + 1
                  \end{array}
            \right)
   \to
    P_0 = \left(\begin{array}{cc}
                    -1 & 0 \\
                    0  & 1
                  \end{array}
            \right)
   $,
   $ P_0 $ hat die Eigenwerte $ -1, 1 $.
}
\begin{displaymath}
   \sigmaess(\TqD) \cap (-1,1) = \emptyset
\end{displaymath}
und andererseits nach {\rm \cite{WeidmannmathZ}}, Satz 6.10
\begin{displaymath}
   \sigmaess (\TqD) \supset \R \setminus  (-1,1),
\end{displaymath}
\end{itemize}
\end{bemerkungen}
%%

%%%%%%%
\subsubsection{Zum Punktspektrum von $ \TqD $}
\label{AAbschnittPunktspektrumohneDreh}

Es gilt der folgende 
\begin{hilfs}

$\left.\right.$

$ \TqD $
besitzt keine Eigenwerte in $ (-\infty,-1)\cup (1, \infty) $.
\end{hilfs}
%%

%5
\begin{bemerkung}

$\left.\right.$

Die Aussage des Hilfssatzes ist bereits unter der alleinigen Voraussetzung
$ d_j \ge \exp\left( (j-1)^2\right) \quad (j \in \N) $ unabh"angig von
der Buckelbeschaffenheit g"ultig.

\end{bemerkung}

\Beweis{}
{

Wegen (\ref{Rlogableitung}) ist n"amlich f"ur $ n \in \N $
(unabh"angig von der Randbedingung, die bei $ 0 $ gestellt wurde):
\begin{displaymath}
  \log ( R(b_n)) =
   \log ( R(a_n) )
    + \int_{a_n}^{b_n} \frac{1}{\kappa} H_n W(r - a_n) \sin 2 \teta
       \dr
\end{displaymath}
und
\begin{displaymath}
  \log (R(r)) = \log ( R(b_{n-1})) \quad(r \in [b_{n-1}, a_n ]).
\end{displaymath}
Folglich gilt, da $ b_n - a_n = \alpha_n $ ist:
\begin{eqnarray*}
  \log ( R(b_n)) & \ge &
   \log ( R(0) )
    - \sum_{i=1}^n
     \frac{1}{\modulus{\kappa}} H_i \int_0^{\alpha_j}
         \modulus{W_i} (r) \dr
\nonumber \\
   & \ge &
      - \sum_{i=1}^n \frac{1}{\modulus{\kappa}}
       \underbrace{
         \sup_{j \in \N} \left\{ H_j \int_0^{\alpha_j} \modulus{W_j} (r) \dr
                         \right\}
       }_{=:\W}
%%14\nonumber \\
%%14   & \ge & 
\ge - n \frac{1}{\modulus{\kappa}} \W,
\end{eqnarray*}

wobei $\W <\infty $ ist, da aufgrund der Voraussetzungen von Satz \ref{Ateil2zentralerSatz} $ \left(H_j\right)_{j\in \N} $ Nullfolge ist und
$  %H_i 
\int_0^{\alpha_j} \modulus{W_i} (r) \dr = 1$
ist.

Dann ist
\begin{displaymath}
  R(r) \ge R(0) \exp \left( - n \frac{1}{\modulus{\kappa}} \W \right)
   \quad (r \in [b_n, a_{n+1}])
\end{displaymath}
Hiermit erh"alt man unter Verwendung der besonderen Gestalt des Potentials
gem"a"s (\ref{satzqD}) und der Tatsache, dass nach Konstruktion
$ d_{j+1} \ge \exp (j^2) \quad(j\in \N) $ gilt:
\begin{eqnarray*}
  \int_{b_n}^\infty \underbrace{R^2(r)}_{>0} \mbox{d}r
  & > &
  \int\limits_{ \bigcup_{j=n}^\infty [b_j, a_{j+1}] }  R^2 (r) \/\mbox{d}r
  \ge
  \int\limits_{ \bigcup_{j=n}^\infty [b_j, a_{j+1}] }
          R^2(0) \exp \left( - j \frac{2}{\modulus{\kappa}} \W \right)
\/
          \mbox{d}r
\nonumber \\
  & = &
  \sum_{j=n}^\infty d_{j+1}
       R^2(0) \exp \left( - \frac{2 j}{\modulus{\kappa}} \W \right)
  \ge
  \sum_{j=n}^\infty R^2(0)
                    \exp \left( j^2 - \frac{2j}{\modulus{\kappa}} \W \right)
\nonumber \\
  & = &
  \sum_{j=n}^\infty
    \exp \left( \left(j - \frac{\W}{\modulus{\kappa}}
                        \right)
          ^2 \right)
                     R^2(0)
                     \exp \left( - \left(\frac{\W}{\modulus{\kappa}}\right)^2
\right)
  = \infty,
\end{eqnarray*}
da $ j - \frac{\W}{\modulus{\kappa}} > 0 \quad (j > \frac{\W}{\modulus{\kappa}})
$ ist.

Die zugeh"orige L"osung ist somit nicht quadratisch integrierbar, da
wegen (\ref{pruefer}) f"ur $ \lambda \in \R, \modulus{\lambda} > 1 $ gilt:
\begin{displaymath}
 \Psi_1^2 + \Psi_2^2
 = R^2 \left( \cos^2\teta + \frac{\lambda-1}{\lambda+1} \sin^2\teta \right)
 \ge R^2 \min \{1, \frac{\lambda-1}{\lambda+1} \}.
\end{displaymath}
Also gibt es kein Punktspektrum in $ (-\infty, -1) \cup (1, \infty) $.

} % ENDE BEWEIS Lemma

Wegen letzterem hat man also f"ur das stetige Spektrum
\begin{equation}
  (-\infty, -1) \cup (1, \infty) \subset \sigmas (\TqD)
\label{refernzspektrum}
\end{equation}
bzw. im Fall $ H_j \to 0 $ sogar:
\begin{equation}
  (-\infty, -1) \cup (1, \infty) = \sigmas (\TqD)
\nonumber
\end{equation}
%

%ende von Ateil2_vorgezogen

\pagebreak

%Bteil0
%Teil0    mit Drehimpuls: Vorbereitung: Lösung auf Intervall (0,1) verschaffen
%---------
\hfill
\begin{minipage}[t]{10cm}
\sffamily
 After a while he says, "<Do you
believe in ghosts?">
 "<No."> I say.
 "<Why not?">
 "<Because they are {\sl un}-sci-en{\sl ti}-fic.">
 The way I say this makes John smile. "<They contain no matter."> I continue,
"<and have no energy and therefore, according to the laws of science, do not exist except in
people's minds.">
 The whiskey, the fatigue and the wind in the trees start mixing in my mind.
"<Of course,"> I add, "<the laws of science contain no matter and have no energy either and
therefore do not exist except in people's minds. It's best to be completely scientific about the
whole thing and refuse to believe in either ghosts or the laws of science. That way you're safe.
That doesn't leave you very much to believe in, but that's scientific too.">
- Robert M. Pirsig,  Zen and the Art of Motorcycle Maintenance
\end{minipage}

\section{Untersuchung des radialen Anteils des po\-ten\-tialfreien Dirac\-operators mit Dreh\-im\-puls\-termen}
\label{KapitelmitDrehLoesungauf01}

Bevor der Radialanteil  des kugelsymmetrischen Operators mit einem Buckelpotential untersucht wird, werden in diesem Abschnitt als Vorbereitung Ergebnisse f"ur den potentialfreien Fall zusammengestellt. Untersuchungsgegenstand ist also der eindimensionale freie Diracoperator mit Drehimpulstermen.

Zun"achst wird in diesem Abschnitt  der Differentialausdruck 
% Feb 13 $ H_k $ aus Hilfssatz~\ref{Reichtradialzuuntersuchen} 
f"ur eine feste Drehimpulsquantenzahl $ k \in \Z \setminus \{0\} $  betrachtet\footnote{
Die vorliegende Arbeit zielt auf Ergebnisse im $ \R^3 $ ab. Im $ \R^2 $ h"atte man f"ur die Dreh\-impulsquantenzahl $ k \in \Z - \frac{1}{2} $.
}.\label{Drehimpulsquantenzahl}

F"ur diese Ausgangssituation ist eine alternative Darstellung der Differen\-tialgleichung g"unstiger, um die Gestalt der L"osungen zu bestimmen. Daher gliedert sich der folgende Abschnitt in zwei Teile. Im ersten wird die Differentialgleichung f"ur den potentialfreien Fall mit Drehimpuls gel"ost. Im zweiten Teil wird diese Differentialgleichung transformiert, so dass sie der aus Abschnitt \ref{KapitelohneDreh}
bekannten Gestalt von (\ref{zentraleGleichung}) entspricht, bei der die Drehimpuls\-terme mit $ \frac{1}{r^2} $ abfallen.
Dieses Abklingverhalten wird bei Absch"atzungen im Abschnitt~\ref{KapitelmitDreh} Verwendung finden.

\subsection{L"osung der potentialfreien Differentialgleichung mit Drehimpulstermen
\label{Abschnittmit GPF}}

Sei $ k \in \Z \setminus \left\{ 0 \right\} $. 

Betrachtet wird auf $ \R^+ $ die Gleichung

\begin{equation}
\hallesnull y := 
 \left[   \sigma_2 p + \sigma_3 + \sigma_1 \frac{k}{r}   \right] y = \lambda y,
\label{klassGleichungmitsigmas}
\end{equation}
mit den Paulimatrizen
\begin{equation}
   \sigma_1 = \left(\begin{array}{cc} 0 & 1 \\
                                                         1 & 0
                              \end{array}
                      \right),
   \,
   \sigma_2 = \left(\begin{array}{cc} 0 & -i \\
                                                         i & 0
                              \end{array}
                      \right),
   \,
   \sigma_3 = \left(\begin{array}{cc} 1 & 0 \\
                                                         0 & -1
                              \end{array}
                      \right)
                       \nonumber
\end{equation}
und dem Impulsterm $
%
%%14\begin{equation}
   p = - i%\nabla
   \frac{\mbox{d}}{\mbox{d}r},
$
%%14.\end{equation}
%

Ausschreiben von (\ref{klassGleichungmitsigmas}) liefert
\begin{equation}
   \left(\begin{array}{c} y_1 \\
            y_2 
           \end{array}
   \right) '
   =
   \left( \begin{array}{cc}
             - \frac{k}{r}                & 1+\lambda  \\
             1-\lambda                    & \frac{k}{r}
          \end{array}
   \right)
   \left(\begin{array}{c} y_1 \\
            y_2 
           \end{array}
    \right) .
\label{klassGleichung}
\end{equation}

Hieraus l"a"st sich folgende Differentialgleichung zweiter Ordnung ableiten:
\begin{equation}
     r^2 y_1''  \, - \, \left[ r^2 (1-\lambda^2)   +  ( k + k^2)  \right] y_1 \, = \, 0
      \nonumber
\end{equation}
Die allgemeine L"osung kann mittels Besselfunktionen $ J_{ \pm \modulus{k+\frac{1}{2}}} $ \label{Besselfunktion}bzw.
Weberfunktionen $  Y_{ \pm \modulus{k+\frac{1}{2}}} $ \label{Weberfunktion}wie folgt dargestellt werden:
%Diese hat die allgemeine L"osung
%
\begin{eqnarray}
   y_1 & = & \sqrt{r} Z_{ \pm \modulus{k+\frac{1}{2}}} \left(-\sqrt{\lambda^2-1}r\right) \nonumber \\
          & = & c_1 \sqrt{r} J_{ \pm \modulus{k+\frac{1}{2}}} \left(-\sqrt{\lambda^2-1}r\right) 
                   + c_2 \sqrt{r} Y_{ \pm \modulus{k+\frac{1}{2}}} \left(-\sqrt{\lambda^2-1}r\right) 
                   \label{allgLsgBessel}
\end{eqnarray}
(siehe \cite{KAMKE}, Abschnitt 132).

Einsetzen zeigt, dass 
\begin{equation}
 \left(\begin{array}{c} v_1 \\
            v_2 
           \end{array}
   \right) 
   =
   \left(\begin{array}{c} 
          \sqrt{r} J_{ \modulus{k+\frac{1}{2}}} \left(-\sqrt{\lambda^2-1}\,r\right) 
          \\
          \sign (k) \, \sqrt{\frac{\lambda-1}{\lambda+1}} \, 
                             \sqrt{r} J_{ \modulus{k-\frac{1}{2}}} \left(-\sqrt{\lambda^2-1}\,r\right) 
           \end{array}
   \right) 
\label{ersterLSG}
\end{equation}
und 
\begin{equation}
 \left(\begin{array}{c} w_1 \\
            w_2 
           \end{array}
   \right) 
   =
   \left(\begin{array}{c} 
          \sqrt{r} Y_{ \modulus{k+\frac{1}{2}}} \left(-\sqrt{\lambda^2-1}\,r\right) 
          \\
          \sign (k) \, \sqrt{\frac{\lambda-1}{\lambda+1} }\, 
                             \sqrt{r} Y_{ \modulus{k-\frac{1}{2}}} \left(-\sqrt{\lambda^2-1}\,r\right) 
           \end{array}
   \right) 
\label{zweiteLsg}
\end{equation}
L"osungen von (\ref{klassGleichung}) sind.

$ v $ und $ w $ sind 
%Feb 13aufgrund der Eigenschaften der Bessel- und Weberfunktionen 
linear unabh"angig.

%Feb 13

Da f"ur alle $ k \in \Z\setminus\{0\} $ gilt $ \modulus{k\pm\frac{1}{2}} \ge + \frac{1}{2} % > -\frac{1}{2}
  $, kann man f"ur die Komponenten von (\ref{ersterLSG}) die Absch"atzung
 $\modulus{ J_{ \modulus{k\pm\frac{1}{2}}} (r)  } \le 
\const ( \modulus{k\pm\frac{1}{2}}) \cdot \modulus{ r} ^{ \modulus{k\pm\frac{1}{2}}}\quad (r \in \R ) $ verwenden (siehe \cite{KOSHLAYKOV}, S. 174).

Die L"osung $ \left(\begin{array}{c} v_1 \\
            v_2 
           \end{array}
   \right) $
ist somit in $ L^2 (0,c) $ f"ur beliebiges $ c > 0 $.

Andererseits ist die zweite L"osung $ \left(\begin{array}{c} w_1 \\
            w_2 
           \end{array}
   \right) $
nicht in in $ L^2 (0,c) $. 
F"ur $n\in \N_o $ kann man n"amlich 
aufgrund der Tatsache, dass $ \modulus{k\pm \frac{1}{2}} \ge \frac{1}{2} $ f"ur 
$ k \in \Z \setminus \{0 \} $ 
ist, 
 f"ur 
$ Y_{n+\frac{1}{2}} (r) = \frac{J_{n+\frac{1}{2}}(r) \cos((n+\frac{1}{2}) \pi )- J_{-n-\frac{1}{2}}(r)}{\sin ( (n+\frac{1}{2})\nu \pi)}$ (siehe \cite{WATSON}, S. 64) die Darstellung
\begin{eqnarray}
\lefteqn{
J_{-n-\frac{1}{2}} (r) = \left(\frac{2}{\pi r}\right)^\frac{1}{2}
   \left[\cos \left( r  + \frac{1}{2}n\pi \right) \sum_{0\le j\le \frac{1}{2}n}
                    \frac{(-1)^j (n+2j)!}{(2j)!(n-2j)! (2r)^{2j}}
  \right.}
\nonumber \\
& &
  \left.  \quad \quad \quad \quad \quad 
        - \sin\left(r + \frac{1}{2}n\pi\right)
           \sum_{0\le j\le \frac{1}{2}(n-1)} \frac{(-1)^j (n+2j+1)!}{(2j+1)! (n-2j-1)!(2r)^{2j+1}}
          \right]
\nonumber
\end{eqnarray}
(siehe \cite{WATSON}, S. 55) verwenden, um 
%f"ur $ n \in \N_0 $ 
zu folgern, dass f"ur 
$ r \to 0 $ gilt
$ Y_{ \modulus{k\pm\frac{1}{2}}} (r)  = \const( \modulus{k\pm\frac{1}{2}})\cdot r ^{-{ \modulus{k\pm\frac{1}{2}}}} $.

Bei $ 0 $ liegt somit der Grenzpunktfall vor.

%%%%%%%%%%%%%%%%%%%%%%%%%%%%%%%%%%%%%%%
\subsection{Transformation der Differentialgleichung
\label{AbschnittTransformation}} %(\ref{klassGleichungmitsigmas})}

Die Drehimpulsterme in Gleichung (\ref{klassGleichungmitsigmas}) fallen mit $ \frac{1}{r} $ ab.
F"ur die weiteren Betrachtungen in Abschnitt \ref{KapitelmitDreh} erweist sich die Transformation der Gleichung 
(\ref{klassGleichungmitsigmas}) als hilfreich, da "ahnlich wie in Abschnitt \ref{KapitelohneDreh} das L"osungsverhalten mit Hilfe der Pr"ufertransformation untersucht wird und Terme mit einem Abklingverhalten $ \frac{1}{r^2} $ abgesch"atzt werden.

\begin{hilfs}
\label{Bteil0Diffglaequivalent}
\nopagebreak
$\left.\right. $

\nopagebreak
Der Differentialoperator
\begin{equation}
\hallesnull =   \sigma_2 p + \sigma_3 + \frac{k}{r} \sigma_1
 \nonumber
\end{equation}
aus  (\ref{klassGleichungmitsigmas}) ist "aquivalent zu folgendem Differentialoperator:
\begin{equation}
   \sigma_2 p + m \sigma_3 + l,
\label{OperatormitgutemAbklinverhalten}
\end{equation}
wobei
\begin{eqnarray}
 \mass(r) & := & \sqrt{ 1 +\frac{k^2}{r^2}}  \quad (r\in \R^+)
%\label{ortsahbMasse}
 \label{ortsabhMasse}
 \\
 \angu(r) & := & \frac{k}{2(r^2+ k^2)} \quad (r\in \R^+)
%\label{angu}
\label{Drehimpulsterm}
\end{eqnarray}
definiert ist\footnote{ Alternativ best"unde die M"oglichkeit, den Term $ m - 1 $ als Gr"o"se zu definieren.}.

In diesem Sinne %Damit 
entspricht damit die Differentialgleichung (\ref{klassGleichung}) der Gleichung
\begin{equation}
   \left( \Psvektors \right) '
   =
   \left( \begin{array}{cc}
             0                             & \mass -1 -\angu + \lambda+1 \\
             \mass -1+ \angu - \lambda+1 & 0
          \end{array}
   \right)
   \left( \Psvektors \right).
\label{BzentraleGleichung}
\end{equation}

\end{hilfs}

\begin{bemerkungen}

$\left.\right.$

$ \mass $ kann als ortsabh"angige Masse aufgefa"st werden. $ \angu $ ist der Dreh\-impuls (vgl.
\mbox{\sc\cite{SCHMIDT 95}}). 

Da in (\ref{ortsabhMasse}) die auf 1 normierte Masse des Teilchens eingeht, ist die Transformation nicht auf den Fall masseloser Teilchen "ubertragbar.
\end{bemerkungen}

\Beweis{von Hilfssatz \ref{Bteil0Diffglaequivalent}}{

Der Beweis folgt \cite{SCHMIDT 95}.

Seien Polarkoordinaten
%
%\begin{equation}
%    \left(\begin{array}{c} m(r)\\
%                                         l(r)
%            \end{array}
%   \right)
%  =:
%  \rho (r)
%  \left(\begin{array}{c} \cos \psi (r)\\
%                                      \sin \psi (r)
%            \end{array}
%   \right)
%\end{equation}
%
\begin{equation}
    \left(\begin{array}{c} 1        \\
                                         \frac{k}{r}
            \end{array}
   \right)
  =:
  \rho (r)
  \left(\begin{array}{c} \cos \psi (r)\\
                                      \sin \psi (r)
            \end{array}
   \right)
    \nonumber
\end{equation}
definiert (vgl. Abbildung \ref{PolarkoordtrGl}). Dabei ist $ \psi $ bis auf eine additive Konstante aus $ 2\pi \Z $ eindeutig bestimmt.
\begin{figure}[t]
\includegraphics[origin=c,width=13.cm,clip=true, natwidth=610,natheight=642,
viewport=0cm 3cm 24.5cm 16cm]{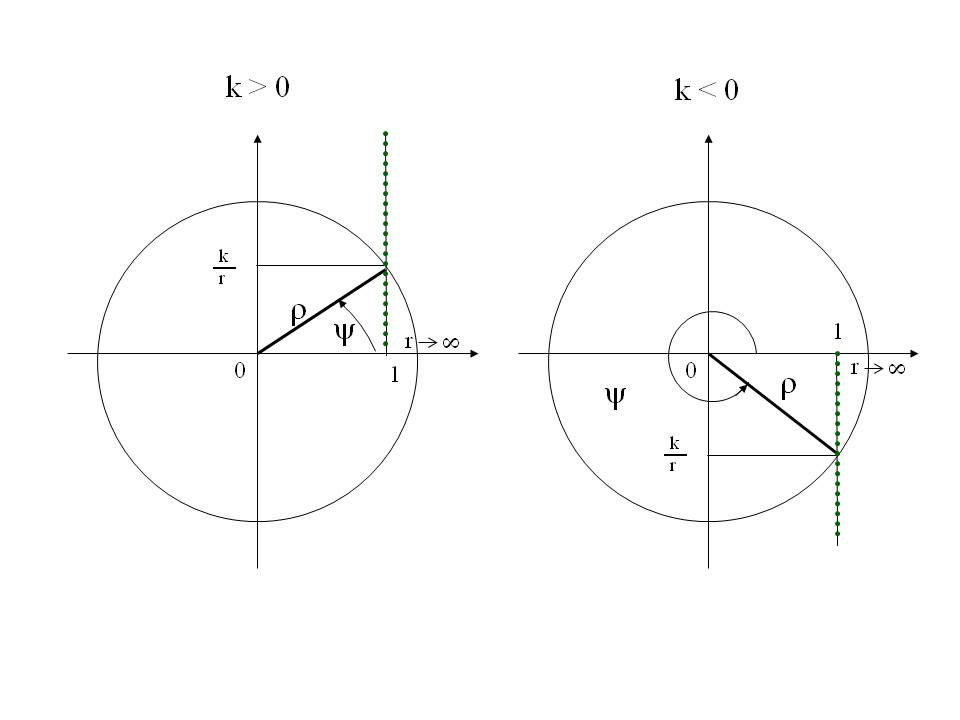}
\caption{\label{PolarkoordtrGl} Polarkoordinaten im Fall $ k \in \N $    bzw. $ k \in - \N $ }
\end{figure}

Dann ist 
%
%%14\begin{equation}
$
   \rho ^2 =  1 +  \frac{k^2}{r^2}
   $
%m^2 +l^2 
%%14\end{equation}
%
und
%
%%14\begin{equation}
$
   \rho = \cos \psi + \frac{k}{r} \sin \psi,
   $
%m \cos \psi + l \sin \psi ,
%%14\end{equation}
%
sowie
%
%%14\begin{equation}
$
  0 = \frac{k}{r} \cos \psi - \sin \psi.
  $
%l \cos \psi -m \sin \psi
%%14\end{equation}
%
F"ur die Ableitung von $ \psi $ gilt:
\begin{equation}
   \psi ' = - \frac{k}{r^2+k^2}
    \nonumber
%\frac{l'm -m'l}{m^2+l^2}
\end{equation}

Dann ist 
%
%%14\begin{equation}
$
   \psi = - \arctan \left(\frac{r}{k}\right) +\const.
   $
%%14\end{equation}
%

Im Fall $ k> 0 $ liest man die Konstante bis auf Vielfache von $ 2\pi \Z $ 
leicht aus der Asymptotik von $ \psi $ f"ur $ r \to \infty $ als $\frac{\pi}{2} $ ab (vgl. Abbildung \ref{Polarkoord_k_pos_Asymptotik}).

\begin{figure}[t]
\includegraphics[origin=c,width=11cm,clip=true, natwidth=610,natheight=642,
viewport=0cm 3cm 23.5cm 16cm]{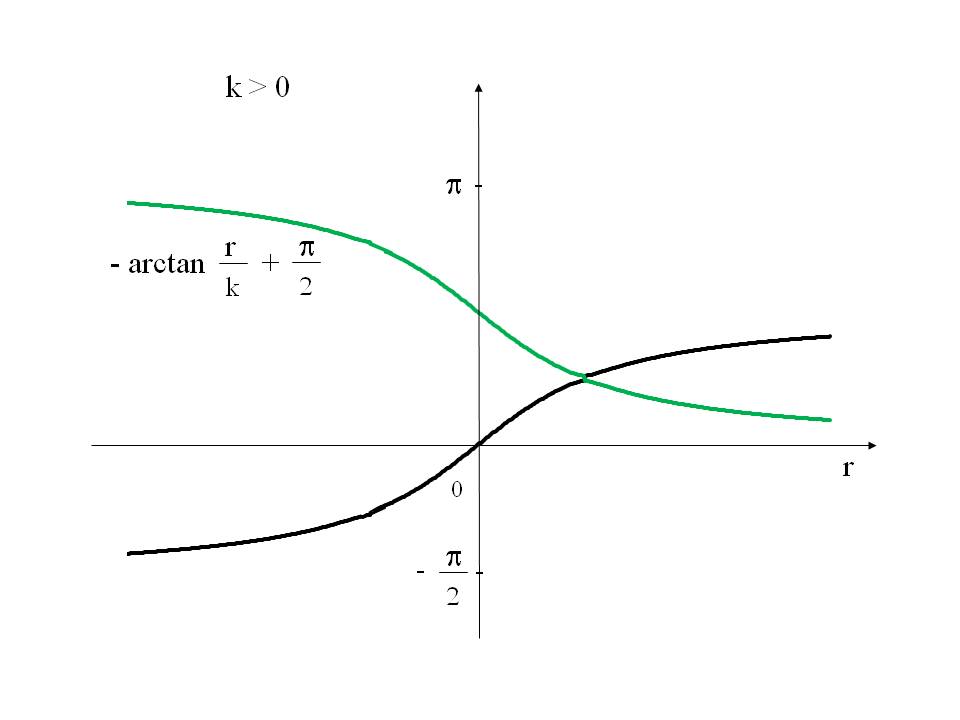}
\caption{\label{Polarkoord_k_pos_Asymptotik} $ k > 0 $: Asymptotik von $ \psi $ f"ur $ r \to \infty $  }
\end{figure}

Es ist also
\begin{equation}
   \psi = - \arctan \left(\frac{r}{k}\right) +\frac{\pi}{2}
\label{psiarctanposk}
\end{equation}

Im Fall $ k < 0 $ erh"alt man (vgl. Abbildung \ref{Polarkoord_k_neg_Asymptotik})
\begin{equation}
   \psi =  \arctan \left(\frac{r}{\modulus{k}}\right) +\frac{3\pi}{2}.
\label{psiarctannegk}
\end{equation}

\begin{figure}[t]
\includegraphics[origin=c,width=11cm,clip=true,natwidth=610,natheight=642,
viewport=0cm 3cm 23.5cm 16cm]{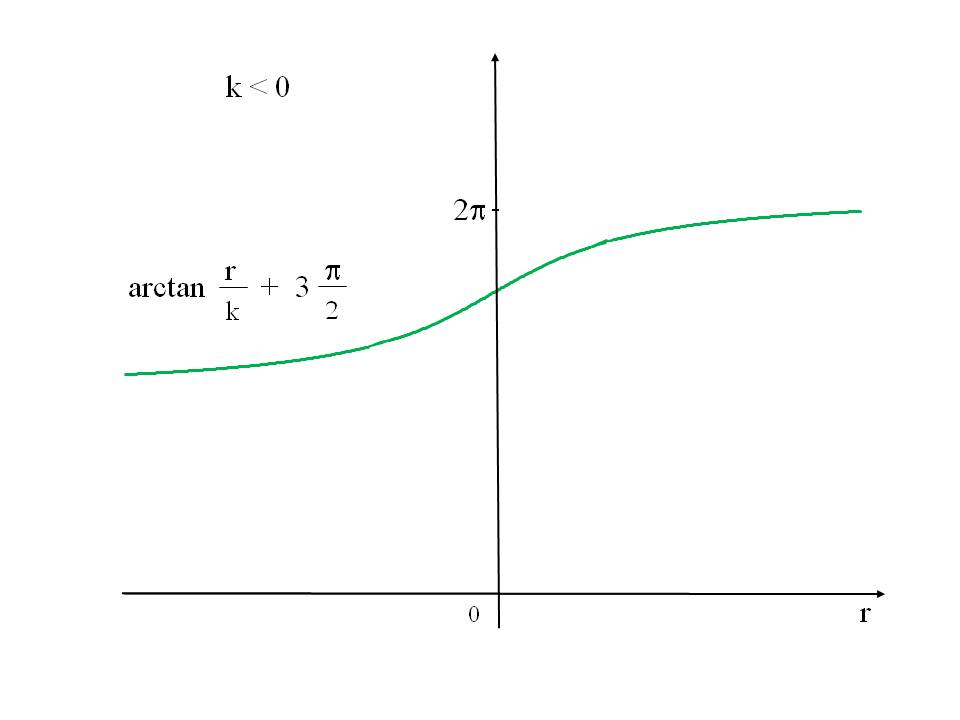}
\caption{\label{Polarkoord_k_neg_Asymptotik}  $ k < 0 $: Asymptotik von $ \psi $ f"ur $ r \to \infty $    }
\end{figure}

Die Matrix
\begin{equation}
  A 
  :=
   \left(\begin{array}{cc}   \cos \frac{\psi}{2} & - \sin \frac{\psi}{2}  \\
                                           \sin \frac{\psi}{2}  & \cos \frac{\psi}{2}
            \end{array}
   \right)
    \nonumber
\end{equation}
ist offensichtlich punktweise orthogonal und hat stetig differenzierbare Komponenten.

Einsetzen von (\ref{psiarctanposk}) bzw. (\ref{psiarctannegk}) liefert nach Vereinfachen
mittels
%
%%14\begin{equation}
$
 \sin (\arctan y) = \sqrt{\frac{y^2}{1+y^2}}
 $
%%14\end{equation}
%
f"ur beliebiges $ k \in \Z \setminus \left\{ 0 \right\} $
\begin{equation}
  A (r)
  = \frac{1}{\sqrt{2}}
   \left(\begin{array}{cc}   \sign{(k)} \sqrt {1 +\sqrt{\frac{r^2}{k^2+r^2}}}  &  \sqrt {1 -\sqrt{\frac{r^2}{k^2+r^2}}}  \\
                                           \sqrt {1 -\sqrt{\frac{r^2}{k^2+r^2}}}                &  \sign{(k)} \sqrt {1 +\sqrt{\frac{r^2}{k^2+r^2}}}
            \end{array}
   \right).
    \nonumber
\end{equation}

Es ist
\begin{equation}
  A^*
  \left( \sigma_2 p + \sigma_3 + \frac{k}{r} \sigma_1
  \right) A
  = 
  \sigma_2 p + \sqrt{1 + \frac{k^2}{r^2}} \sigma_3 + \frac{k}{2\left(r^2 + k^2\right)}.
   \nonumber
\end{equation}

L"ost $ y $ (\ref{klassGleichungmitsigmas}), so ist somit $ A^* y $ eine L"osung von (\ref{BzentraleGleichung}).

Insbesondere kann man f"ur (\ref{BzentraleGleichung}) eine $L^2$-L"osung explizit angeben:

\begin{eqnarray}
 \lefteqn{\left( \Psvektors \right)  = 
       A^* (r) v (r)= }
   \nonumber \\
   & = & 
      \left(\begin{array}{cc}   \cos \frac{\psi(r)}{2} &  \sin \frac{\psi(r)}{2}  \\
                                            - \sin \frac{\psi(r)}{2}  & \cos \frac{\psi(r)}{2}
            \end{array}
   \right)
     \left(\begin{array}{c} 
          \sqrt{r} J_{ \modulus{k+\frac{1}{2}}} \left(-\sqrt{\lambda^2-1}r\right) 
          \\
          \sign (k) \, \frac{\sqrt{\lambda^2-1}}{1+\lambda} \, 
                             \sqrt{r} J_{ \modulus{k-\frac{1}{2}}} \left(-\sqrt{\lambda^2-1}r\right) 
           \end{array}
   \right) 
   \nonumber \\
 & = &
  \frac{1}{\sqrt{2}}
     \left(\begin{array}{l}  
                      \sign{(k)} \sqrt {1 +\sqrt{\frac{r^2}{k^2+r^2}}}  \sqrt{r} J_{ \modulus{k+\frac{1}{2}}} \left(-\sqrt{\lambda^2-1}\, r\right)  \\
    \left.\right. \quad 
         + \sqrt {1 -\sqrt{\frac{r^2}{k^2+r^2}}}       \sign{(k)}      \, \frac{\sqrt{\lambda^2-1}}{1+\lambda} \, 
                          \sqrt{r}    J_{ \modulus{k-\frac{1}{2}}} \left(-\sqrt{\lambda^2-1}\, r\right) 
  \\
 \\
\\
   - \sqrt {1 -\sqrt{\frac{r^2}{k^2+r^2}}}   \sqrt{r}    J_{ \modulus{k+\frac{1}{2}}} \left(-\sqrt{\lambda^2-1}\, r\right) 
 \\
    \left.\right. \quad 
          +   \sqrt {1 -\sqrt{\frac{r^2}{k^2+r^2}}} 
               \, \frac{\sqrt{\lambda^2-1}}{1+\lambda} \, 
                          \sqrt{r}     J_{ \modulus{k-\frac{1}{2}}} \left(-\sqrt{\lambda^2-1}\, r\right) 
            \end{array}
     \right),
  \nonumber \\
\label{LoesungmitDrehohnePotential}
\end{eqnarray}
was den Beweis von Hilfssatz \ref{Bteil0Diffglaequivalent} abschlie"st.   
}

Die Absch"atzung
\begin{displaymath}
0 \le  \sqrt{1+\frac{k^2}{r^2}} -1
 =
 \frac{\left(\sqrt{1+\frac{k^2}{r^2}} -1\right)\left(\sqrt{1+\frac{k^2}{r^2}} +1\right)}{\sqrt{\frac{k^2}{r^2} +1}+1  }
=\frac{1 + \frac{k^2}{r^2} -1 }{\sqrt{1+\frac{k^2}{r^2}} +1}
 \le
 \frac{k^2}{r^2}
 %\quad (r \ge 1)
\end{displaymath}
ist eine Grundlage f"ur die folgende

\begin{bemerkung}

$\left.\right.$

Der Vorteil der Darstellung (\ref{OperatormitgutemAbklinverhalten}) besteht im Abklingverhalten der Masse- und Dreh\-impulsterme f"ur $ r \to \infty $.
Es gilt n"amlich:
\begin{equation}
0 \le 
   \mass (r) - 1 = \sqrt{1 + \frac{k^2}{r^2}} -1 \le \frac{k^2}{r^2}
\label{AbklingverhaltenMasse}
\end{equation}
 und
 \begin{equation}
    \modulus{\angu (r) } = \modulus{\frac{k}{2(r^2+k^2)}} \le \frac{\modulus{k}}{2r^2}.
\label{AbklingverhaltenDrehimpuls}
\end{equation}

\end{bemerkung}
\vfill

%ende von Bteil0

\pagebreak
%Bteil1
%Teil1    mit Drehimpuls
%---------

\begin{figure}[h]
\includegraphics[origin=c,width=8cm,clip=true,natwidth=610,natheight=642,
viewport=0cm .0cm 20.5cm 11.5cm]{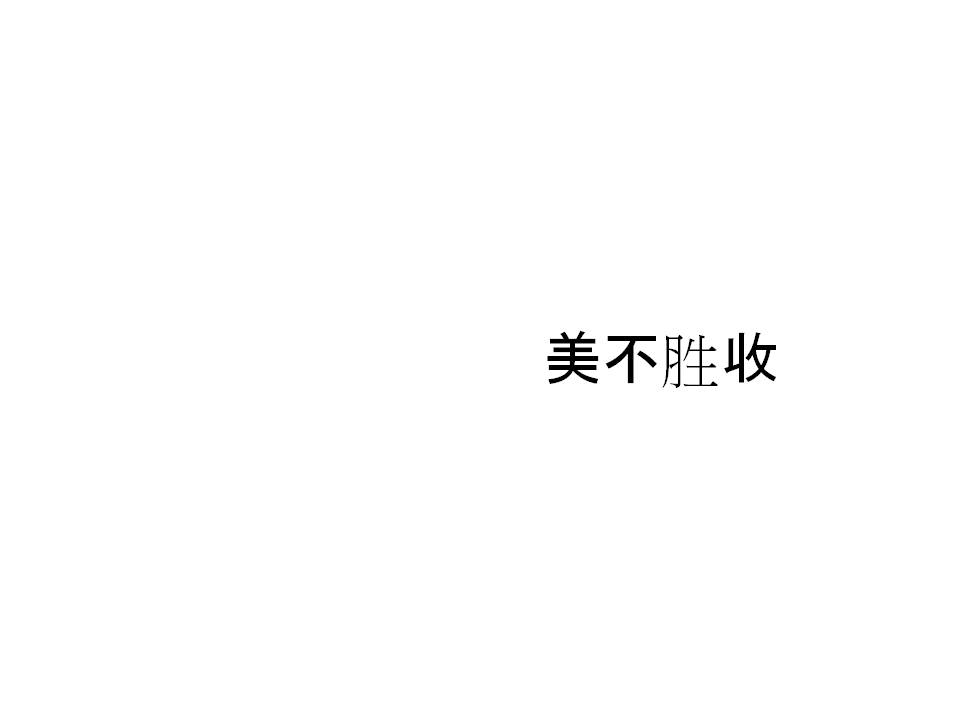}
%\caption{\label{ChinZitat}   }
\end{figure}

\vspace*{-3.8cm}

\hfill
\begin{minipage}[t]{8cm}
\sffamily
Es gibt zu viele sch"one Sachen, als dass man sie alle bewundern k"onnte.
\end{minipage}

\section{Der radiale Anteil des Diracoperators mit Buckelpotential und Drehimpulstermen auf $[1,\infty) $
\label{KapitelmitDreh} \label{intervall1infty}}

Nachdem in Abschnitt \ref{KapitelohneDreh} die vereinfachte Situation des radialen Anteils des Diracoperators unter Vernachl"assigung der Drehimpulsterme untersucht wurde, werden in diesem Abschnitt diese Ergebnisse als Grundlage f"ur die Betrachtung des allgemeinen Falles mit Drehimpulstermen verwendet. Dabei wird die Situation auf das Intervall $ [1,\infty)$ eingeschr"ankt. Im folgenden Abschnitt wird das Intervall $ (0,1] $ untersucht, um schlie"slich die Ergebnisse beider Intervalle f"ur die Halbachse $ \R^+ $ zusammenzufassen.

\subsection{Allgemeine Bezeichnungen und Definitionen}

Sei $ k \in \Z\setminus\{0\} $ eine vorgegebene Drehimpulsquantenzahl.

Die folgenden Betrachtungen werden f"ur ein $n$-Buckelpotential $ q_{%D_
n} $  durchgef"uhrt (Definition wie in Kapitel \ref{KapitelohneDreh} in den Bezeichnungen \ref{Buckepotentialbezeichnungen}).  Sei also $ n \in \N $ und $ q_{%D_
n} $ gegeben.

Ohne Einschr"ankung der Allgemeinheit sei $ d_1 > 1 $.

Ausgangslage ist die auf  $ \R^+ $ betrachtete %die 
Gleichung
\begin{eqnarray}
   \left( \Psvektors \right) '
   & = &
   \left( \begin{array}{cc}
             0                             & \mass-1-q_{%D_
             n}-\angu + \lambda +1\\
             \mass-1+q_{%D_
             n}+\angu - \lambda +1& 0
          \end{array}
   \right)
   \left( \Psvektors \right)
   \nonumber \\
   &&
\label{nocheinVerweis}
%\nonumber \\
% & =:&
%   \tau_{k, q_{D_n}} \left( \Psvektors \right),
\label{BzentraleGleichungmitPotential}
\end{eqnarray}
mit der ortsabh"angigen Masse $ m $ wie in (\ref{ortsabhMasse}) und dem Drehimpuls 
$ l $ wie in (\ref{Drehimpulsterm}).

\begin{bezeichn}

$\left. \right. $

Der zu (\ref{nocheinVerweis})
geh"orende Differentialausdruck sei bezeichnet  mit 
\begin{equation}
\hqdn  := \sigma_2 p + m \sigma_3 + l + q_{%D_
n}.
\label{hqdn}
\end{equation}
\end{bezeichn}

Im Folgenden werden   "Uberlegungen f"ur das Intervall $ [1,\infty) $   angestellt, die anschlie"send mit den Ergebnissen f"ur das Intervall $ (0,1) $ zusammengef"uhrt werden.
Gibt man bei $ r=1 $ eine Randbedingung vor, so k"onnen L"osungen durch die in Abschnitt \ref{KapitelmitDrehLoesungauf01}
gewonnenen $L^2$-L"osungen auf $ (0,1) $ zu L"osungen auf $ \R^+ $ fortgeschrieben werden.

%\subsection{Untersuchung f"ur das Intervalls $ [1,\infty) $
%}

$\left. \right.$

Sei %Feb 13also
 im Folgenden $ r \in [1,\infty) $.

\begin{defundbem}
$\left.\right.$ 

Analog zu (\ref{pruefer}) in Definition \ref{Ateil1DefPrueferdef}
im drehimpulsfreien Fall
 werden hier mit Hilfe folgender verallgemeinerter
Pr"ufertransformation neue abh"angige Variablen eingef"uhrt:
\begin{equation}
  \begin{array}{ccl}
    \Psis_1 & = & \Rs \cos \tetas \\
    \Psis_2 & = & \Rs \sqrt{\frac{\lambda-1}{\lambda+1}} \sin \tetas
  \end{array}
\label{Bpruefer}
\end{equation}

Sei wieder f"ur $ \lambda \in \R, \modulus{\lambda} > 1 $ %definiert.
\begin{equation}
   \kappa := 
   \sign{(\lambda)} \sqrt{\lambda^2 -1}
    \nonumber
\end{equation}
\end{defundbem}

\subsection{Eigenschaften des Pr"ufer-Radius und des Pr"ufer-Winkels}
Es gilt der folgende

\begin{hilfs}
\label{Bteil1lemmaprueferableitungen}

$\left.\right.$

F"ur den Pr"ufer-Radius $ \Rs $ gilt:\label{PrueferradiusmitDreh}
\begin{equation}
  (\log \Rs)' =\sin 2\tetas \left(\frac{1}{\kappa} q_{%D_
  n}  +  \Fk(r;\kappa)\right)
\label{BRlogableitung}
\end{equation}
wobei
\begin{equation}
  \Fk(r;\kappa) := \frac{1}{\kappa} \angu(r)
                  + \frac{\lambda}{\kappa}\left[\mass(r)-1\right]
\label{Bteil1defFk}
\end{equation}
die Eigenschaft
\begin{equation}
  \modulus{ \Fk(r;\kappa)}
  \le \frac{\Ckkappa}{r^2} \quad(r \ge 1, \kappa \in \R^*)
\label{Bteil1Fkabschaetz}
\end{equation}
besitzt mit der Konstanten
\begin{equation}
 \Ckkappa := \frac{\modulus{k}}{2\modulus{\kappa}}
             + \frac{\modulus{\lambda}k^2}{\modulus{\kappa}}.
\label{Bteil1defCkkappa}
\end{equation}

F"ur die Ableitung des Pr"ufer-Winkels $ \tetas $ gilt:\label{PrueferwinkelmitDreh}
\begin{equation}
  \tetas '  =
    -\kappa + \frac{q_{%D_
    n}}{\kappa} \left(\lambda + \cos 2\tetas \right)
    + \Gk(r;\kappa).
\label{Bprueferwinkelableitung}
\end{equation}
Dabei gilt f"ur
\begin{equation}
   \Gk(r;\kappa) :=
        \frac{1}{\kappa}(\lambda + \cos 2\tetas(r)) \angu(r)
        + \frac{1}{\kappa}(1 + \lambda \cos 2\tetas(r))  \left[\mass(r)-1\right]
\label{Bteil1defGk}
\end{equation}
die Absch"atzung
\begin{equation}
  \modulus{\Gk(r;\kappa)}
  \le \frac{\Ckkappas}{r^2} \quad(r \ge 1, \kappa \in \R^*)
\label{BteiliGkabschaetz}
\end{equation}
mit der Konstanten
\begin{equation}
 \Ckkappas := \frac{\left(\modulus{\lambda}+1\right)\left(\modulus{k} + 2 k^2\right)}{2 \modulus{\kappa}}.
\label{Bteil1defCkkappas}
\end{equation}
\end{hilfs}
\begin{bemerkung}

$\left.\right.$

Der Vergleich von (\ref{prueferwinkelableitung}) in Hilfssatz
\ref{Ateil1lemmaprueferableitungen}  und
(\ref{Bprueferwinkelableitung}) zeigt, dass f"ur $ r \to \infty $
die Ableitungen des Pr"ufer-Winkels $ \teta $ im Fall ohne Drehimpuls und die
Ableitungen des Pr"ufer-Winkels $ \tetas $ im Fall mit Drehimpuls asymptotisch
gleiche Struktur aufweisen, da der bei (\ref{Bprueferwinkelableitung}) neu
hinzugekommene Term $ \Gk $ wegen (\ref{BteiliGkabschaetz}) gegen 0 geht.

Stellt man (\ref{Rlogableitung}) und (\ref{BRlogableitung}) einander
gegen"uber, so zeigt sich, dass auch $ \left(\log R\right)' $ und
$ \left(\log \Rs\right)' $ f"ur
$ r \to \infty $ die gleiche Gestalt besitzen, da nach
(\ref{Bteil1Fkabschaetz}) der zus"atzliche Term in (\ref{BRlogableitung})
 f"ur gro"se $ r $ klein wird.

Der Pr"ufer-Winkel verh"alt sich also auf den potentialfreien Strecken fast linear.
Der Pr"ufer-Radius ist dort fast konstant.
\end{bemerkung}

\Beweis{von Hilfssatz \ref{Bteil1lemmaprueferableitungen}}{

Nach Definition (\ref{Bpruefer}) der Pr"ufervariablen erh"alt man  analog wie
im Beweis von Hilfssatz \ref{Ateil1lemmaprueferableitungen} mit entsprechenden
Umformungen
\begin{eqnarray}
    (\log \Rs)' & = & 
   \frac{1}{\kappa} q_{%D_
   n} \sin 2 \tetas
                + \sin 2\tetas
                   \left[ \frac{1}{\kappa} \angu(r)
                         + \frac{\lambda}{\kappa}
                               \left[ \mass(r)-1\right]
                   \right]
   \nonumber \\
   & = &
\frac{1}{\kappa} q_{%D_
n} \sin 2 \tetas + \sin 2\tetas \Fk(r;\kappa),
\label{Bteil1AbleitungRs}
\end{eqnarray}
wobei Definition (\ref{Bteil1defFk}) verwendet wurde.

Mit (\ref{AbklingverhaltenMasse})  und (\ref{AbklingverhaltenDrehimpuls}) kann mit Hilfe der in (\ref{Bteil1defCkkappa}) definierten Konstanten
\begin{displaymath}
 \modulus{\Fk(r;\kappa)}
 =
 \modulus{
   \frac{1}{\kappa }\frac{k}{2(r^2+k^2)} +
     \frac{\lambda}{\kappa} \left(\sqrt{1+\frac{k^2}{r^2}} -1 \right)
 }
 \le
  \frac{\Ckkappa}{r^2} \quad(r \ge 1, \kappa \in \R^*)
\end{displaymath}
abgesch"atzt werden, so dass
der erste Teil der Behauptung 
%des Hilfssatzes 
gezeigt ist.

F"ur die Ableitung des Pr"ufer-Winkels erh"alt man mit 
% unterVerwendung von 
Definition~(\ref{Bteil1defGk})
\begin{eqnarray*}
 \tetas ' &  = &
      -\kappa + \frac{q_{%D_
      n}}{\kappa}(\lambda + \cos 2\tetas(r))
       + \frac{1}{\kappa} (\lambda + \cos2\tetas)\angu(r)
    \nonumber \\
   & &
        + \frac{1}{\kappa}(1 + \lambda \cos 2\tetas(r))  (\mass(r)-1)
   \nonumber \\
    & = &
     -\kappa + \frac{q_{%D_
     n}}{\kappa} \left(\lambda + \cos 2\tetas \right)
    + \Gk(r;\kappa).
\end{eqnarray*}

Es gilt die Absch"atzung
\begin{eqnarray*}
 \modulus{ \Gk(r;\kappa) }
 & = &
 \modulus{ \frac{1}{\kappa}(\lambda + \cos 2\tetas(r)) \angu(r)
        + \frac{1}{\kappa}(1 + \lambda \cos 2\tetas(r))  \left[\mass(r)-1\right]
 }
 \nonumber \\
 & \le &
 \frac{1}{\modulus{\kappa}}(\modulus{\lambda}+1)
 \frac{\modulus{k}}{2r^2}
  + \frac{1}{\modulus{\kappa}}(\modulus{\lambda}+1) \frac{k^2}{r^2}
  \nonumber \\
& = & \frac{1}{2\modulus{\kappa}}(\modulus{\lambda}+1) \left(\modulus{k} + 2 k^2 \right) \frac{1}{r^2}
=
 \frac{\Ckkappas}{r^2}
\end{eqnarray*}
mit der in (\ref{Bteil1defCkkappas}) definierten Konstanten $ \Ckkappas $.
}  % Beweisende

Integriert man (\ref{BRlogableitung}) bzw. (\ref{Bprueferwinkelableitung})
"uber $ [b_{j-1}, a_j] $, so erh"alt man
\begin{equation}
   \Rs(a_j) = \Rs(b_{j-1})
             \exp\left( \int_{b_{j-1}}^{a_j} \sin 2\tetas(s) \Fk(s;\kappa)\ds
                 \right)
\end{equation}
und
\begin{equation}
  \tetas(a_j) =
    \tetas(b_{j-1}) -\kappa d_j + \int_{b_{j-1}}^{a_j} \Gk(s;\kappa )\ds.
\end{equation}

\begin{bezeichn}

$\left.\right. $

F"ur $ j \in \{1, \dots, n\} $ bezeichne bei Vorliegen des $n$-Buckelpotentials $ q_%{D_
n%}
 $
\begin{equation}
   \Rs_{%D_
   n;j}(\kappa) := \Rs(b_j,\kappa).
   \label{Bteil1DefRsj}
\end{equation}
den Wert des Pr"ufer-Radius am Ende des $j$-ten Buckels.
Mit
\begin{equation}
   \tetas_{%D_
   n;j}(\kappa) := \tetas(b_j,\kappa)
\label{tetamitdreh}
\end{equation}
wird der Pr"ufer-Winkel am Ende des $j$-ten Buckels bezeichnet.
Im Fall $ j = n $ wird die k"urzere Bezeichnungs $ \Rs_n $ bzw. $ \tetas_n $ verwendet. \label{Rntetan}
\end{bezeichn}

Mit diesen Bezeichnungen kann man die L"osung $ \Psis $ am Beginn des $ j $-ten Buckels durch den
Wert am Ende des $ (j-1) $-ten Buckels somit
folgenderma"sen ausdr"ucken:
\begin{eqnarray}
 \Psis_1(a_j)
 & = &
 \Rs_{%D_
 n;j-1}(\kappa)
 \exp\left( \int_{b_{j-1}}^{a_j} \sin 2 \tetas (s, \kappa) \Fk(s;\kappa) \ds
     \right)  \cdot
 \nonumber \\
 & & \quad \cdot
 \cos\left( \tetas_{j-1}(\kappa) - \kappa d_j
            + \int_{b_{j-1}}^{a_j}\Gk(s, \kappa) \ds
     \right)
\label{Bteil1psi1aj}
\\
 \Psis_2(a_j)
 & = &
 \sqrt{\frac{\lambda-1}{\lambda+1}}
 \Rs_{%D_
 n;j-1}(\kappa)
 \exp\left( \int_{b_{j-1}}^{a_j} \sin 2 \tetas (s, \kappa) \Fk(s;\kappa) \ds         \right)  \cdot
 \nonumber \\
 & & \quad \cdot
 \sin\left( \tetas_{j-1}(\kappa) - \kappa d_j
            + \int_{b_{j-1}}^{a_j}\Gk(s, \kappa) \ds
     \right)
\label{Bteil1psi2aj}
\end{eqnarray}

\begin{defundbem}

$\left.\right. $

F"ur $ j=0 $ sei
\begin{equation}
   \Rs_{%D_
   n;0} (\kappa):= \Rs(1,\kappa)
   \label{Bteil1DefRbeiEins}
\end{equation}
%
%und
%
\begin{equation}
   \tetas_{%D_
   n;0}(\kappa) := \tetas(1,\kappa)
\end{equation}
definiert. 

Die Stelle $ r=1 $ wird also wie ein Buckelende $ b_0 $ behandelt.
\end{defundbem}

Im folgenden Hilfssatz wird das Verh"altnis
$ \frac{\Rs_{%D_
n;j}^2}{\Rs^2_{%D_
n;j-1}} $ f"ur $ j \in \{ 1, \dots, n\} $
angegeben, das es erm"oglicht, den Wert der L"osung $ \Psis $ an der Stelle
$ a_j $ mit dem Wert der L"osung an der Stelle $ b_0 $ in Beziehung zu
setzen.

\begin{hilfs}
\label{Bteil1lemmaprueferradienverhaeltnis}

$\left.\right.$

F"ur das Verh"altnis der Quadrate der
Pr"ufer-Radien am Ende des $ j $-ten und des $ (j-1) $-ten Buckels gilt f"ur $ j \in \{1, \dots, n\} $:
\begin{eqnarray}
   \frac{\Rs_{%D_
   n;j}^2(\kappa)}{\Rs^2_{%D_
   n;j-1}(\kappa)}
 & = &  \frac{1}{\ffjs(\kappa  ; \tetas\ind -  \: d_j \,\kappa  \:;
                        \int_{b_{j-1}}^{a_j}\Gk (s;\kappa)\ds)},
\label{Bteil1fjDefini}
\end{eqnarray}
wobei
\begin{eqnarray}
  \ffjs (\kappa, y,z)
  &  := &
   \frac{\exp\left(-2\int_{b_{j-1}}^{a_j} \Fk(s;\kappa)\ds\right)}
        {\Ajs(\kappa) + \Bjs(\kappa) \cos(2(y+z)) + \Cjs(\kappa) \sin(2(y+z))
   }
  \nonumber \\
  & &
  \quad\quad\quad\quad\quad\quad\quad\quad\quad\quad\quad
  (\kappa \in \R^*, y \in \R, z \in \R)
\label{Bteil1deffj1}
\end{eqnarray}
mit
\begin{eqnarray}
   \Ajs & := & \frac{1}{2}
           \left[
              {\Mjeesquad} + \plmin \Mjzesquad
              + \minpl \Mjezsquad + \Mjzzsquad
           \right]
\label{BDefAj}
   \\
   \Bjs & := & \frac{1}{2}
             \left[
                  \Mjeesquad + \plmin \Mjzesquad
                  - \minpl \Mjezsquad - \Mjzzsquad
             \right]
\label{BDefBj}
   \\
   \Cjs & := & \sqrt{\minpl}\Mjees\Mjezs + \sqrt{\plmin} \Mjzes\Mjzzs
\label{BDefCj}
\end{eqnarray}
definiert ist.
Dabei ist $ \Mjs $ die zum  $ j $-ten Buckel geh"orige Transfermatrix.
\end{hilfs}

\begin{bemerkung}

$\left.\right.$

Weder der Z"ahler noch der Nenner von $  \ffjs  $ k"onnen verschwinden, da $  \ffjs $ als Quotient von $ \Rs_{%D_
n;j}^2 $ und $ \Rs^2_{%D_
n;j-1} $ definiert ist. W"are einer der Werte des Pr"ufer-Radius $ 0 $, l"age die Nulll"osung vor.
\end{bemerkung}

F"ur den Beweis dieses Hilfssatzes wird noch die folgende Aussage ben"otigt:

\begin{hilfs}

$\left.\right.$

Sei $ \alphabeta \subset \R^* $, $ K \subset \Z\setminus\{0\} $ gegeben.
F"ur das dritte Argument der Funktion $ \ffjs $ in (\ref{Bteil1fjDefini})
gilt:
Es gibt ein kompaktes Intervall $ I_{\alpha, \beta,K} $ mit:
\begin{displaymath}
  \int_{b_{j-1}}^{a_j} \Gk (s, \kappa) \ds \in I_{\alpha, \beta,K}
  \quad (\kappa\in\alphabeta,k\in K)
\end{displaymath}
Dies gilt unabh"angig von $ j \in \N $. Insbesondere kann das Intervall $ I_{\alpha, \beta,K} $ unabh"angig von den Werten $ a_j $ und $ b_{j-1} $ angegeben werden.
\end{hilfs}

\Beweis{dieses Hilfssatzes}{

Da wegen Hilfssatz \ref{Bteil1lemmaprueferableitungen}
\begin{displaymath}
  \modulus{\Gk(s) }  \le \frac{\Ckkappas}{s^2} 
  \quad (s \ge 1)
\end{displaymath}
ist, gilt
\begin{eqnarray}
  \modulus{\int_{b_{j-1}}^{a_j} \Gk(s;\kappa)\ds}
&   \le&
  \Ckkappas\left(-\frac{1}{a_j} + \frac{1}{b_{j-1}}\right)
  \le
  \Ckkappas
\nonumber \\
&  =&
  \frac{\modulus{k}+2k^2}{2\modulus{\kappa}}
  \le
  \frac{\max_{k\in K} (\modulus{k}+2k^2)}
       {\min_{\kappa\in\alphabeta} 2\modulus{\kappa} }.
\nonumber
\end{eqnarray}
Setzt man
\begin{displaymath}
  I_{\alpha, \beta,K} :=
    \left[-\frac{\max_{k\in K} (\modulus{k}+2k^2)}
            {\min_{\kappa\in\alphabeta} 2\modulus{\kappa} },
       \frac{\max_{k\in K} (\modulus{k}+2k^2)}
            {\min_{\kappa\in\alphabeta} 2\modulus{\kappa} }\right],
\end{displaymath}
so ist
\begin{displaymath}
  \int_{b_{j-1}}^{a_j} \Gk(s;\kappa)\ds \in I_{\alpha, \beta,K},
\end{displaymath}
was unabh"angig von $ j $ gilt und insbesondere unabh"angig
davon ist, welche Werte $ b_{j-1} $ und $ a_j $ betrachtet werden.
}

\Beweis{von Hilfssatz \ref{Bteil1lemmaprueferradienverhaeltnis}}{
\nopagebreak

Die Transfermatrix $ \Mjs $ \label{TRansfermatrixmitDreh}f"ur den $ j $-ten Buckel ist gegeben durch den
Wert, den das Fundamentalsystem der auf $ [a_j, a_j + \alpha_j] $
betrachteten Gleichung
\begin{displaymath}
  \left(\begin{array}{c} u_1\\
                         u_2
       \end{array}
  \right)'
 =
 \left( \begin{array}{cc}
           0                             & \mass-H_j W_j-\angu + \lambda \\
           \mass+H_j W_j+\angu - \lambda & 0
        \end{array}
 \right)
  \left(\begin{array}{c} u_1\\
                         u_2
       \end{array}
  \right),
\end{displaymath}
welches den Anfangswert $ \Einheitsmatrix $ hat, an der Stelle
$ a_j+\alpha_j = b_j $
annimmt.
Wegen
%
%\begin{displaymath}
$
   \Psis(b_j) = \Mjs \Psis(a_j)
   $
%\end{displaymath}
%
ist dann mit (\ref{Bteil1psi1aj}) und (\ref{Bteil1psi2aj})
\begin{eqnarray*}
\lefteqn{  \Psis_1(b_j) \!= \!\Mjees \!\Rs_{j-1} \!
                \exp\!\left(\!\! \int_{b_{j-1}}^{a_j} \!\!\!\sin 2\tetas(\!s\!) \Fk(\!s;\kappa\!)\!\ds\!\!
                    \right) 
%%14                    \cdot
%%14         \nonumber \\
%%14         & & \quad \quad \quad \quad
                 \cos\!\left(\!\! \tetas_{j-1} \!-\!
                           \kappa d_j\! +\! \int_{b_{j-1}}^{a_j}\!\!\Gk (\!s;\kappa\!)\!\ds\!
                    \right)}
  \nonumber \\
     & & + \sqrt{\!\frac{\lambda\!-\!1}{\lambda\!+\!1}\! }
\!          \Mjezs\!\exp\!\left(\!\! \int_{b_{j-1}}^{a_j} \!\!\!\sin 2\tetas(\!s\!) \Fk(\!s;\kappa\!)\!\ds
                    \right) 
%%14                      \cdot
%%14         \nonumber \\
%%14         & & \quad \quad \quad \quad
               % \cdot
                \sin\!\left(\! \tetas_{j-1} \!-\!
                           \kappa d_j\! +\! \int_{b_{j-1}}^{a_j}\!\!\!\Gk (\!s;\kappa\!)\!\ds\!
                    \right)
\\
\lefteqn{  \Psis_2(b_j) \!= \!\Mjzes \!\Rs_{j-1}\!
                \exp\!\left(\!\! \int_{b_{j-1}}^{a_j} \!\!\!\sin 2\tetas(\!s\!) \Fk(\!s;\kappa\!)\!\ds \!\!                   \right) 
%         \nonumber \\
%         & & \quad \quad \quad \quad
%                \cdot
                \cos\!\left(\! \tetas_{j-1} \!-\!
                           \kappa d_j \!+ \!\int_{b_{j-1}}^{a_j}\!\!\!\Gk (\!s;\kappa\!)\!\ds
                    \right) }
  \nonumber \\
     & & + \sqrt{\!\frac{\lambda\!-\!1}{\lambda\!+\!1}\! }\!
          \Mjzzs\!\exp\!\left(\! \int_{b_{j-1}}^{a_j}\!\!\! \sin 2\tetas(\!s\!) \Fk(\!s;\kappa\!)\!\ds \!                   \right)  
%         \nonumber \\
    %     & & \quad \quad \quad \quad
        %        \cdot
                \sin\!\left( \!\tetas_{j-1} \!-\!
                           \kappa d_j \!+\! \int_{b_{j-1}}^{a_j}\!\!\!\Gk (\!s;\kappa\!)\!\ds\!
                    \right)
\end{eqnarray*}
Dies liefert
%Hieraus erh"alt man 
analog dem 
%"ahnlich wie im 
Beweis von Hilfssatz
\ref{Ateil1lemmaprueferradienverhaeltnis} nach Verein\-fachen:
\begin{equation}
   \frac{\Rs_{%D_
   n;j}^2}{\Rs^2_{%D_
   n;j-1}}
 =  \frac{1}{\Rs^2_{%D_
 n;j-1}}\!
        \left[\! \Psis_1^2(b_j,\cdot)
               + \frac{\lambda\!+\!1}{\lambda\!-\!1} \Psis_2^2(b_j,\cdot)
       \! \right]
% \nonumber \\
 %& = &
 =
\frac{1}{\ffjs\!(\cdot\/;\tetas\ind \!-\!d_j\cdot;
                        \int_{b_{j-1}}^{a_j}\!\!\Gk (\!s;\!\kappa\!)\!\ds\!)}.
\end{equation}
}  % Ende Beweis

%*****************************************************************

Vor der Untersuchung der Eigenschaften von $ \ffjs $  wird zun"achst die Spektralfunktion des Problems (\ref{BzentraleGleichungmitPotential}) und (\ref
%Feb 13 {randbedingungmitDreh}
{RBbei1}) bestimmt:

%##################

%-----
\subsection{Die Spektralfunktion \label{SpektralfktmitDreh}}

\begin{satz}
\label{datzSPektarlfkt}

$ \left.\right. $

Es gilt f"ur 
%F"ur 
die Ableitung der Spektralfunktion
zu 
\begin{equation}
   \left(\! \Psvektors \!\right) '
   =
   \left( \begin{array}{cc}
             0                             & \mass -1 -q_{%D_
             n}-\angu + \lambda +1\\
             \mass- 1 +q_{%D_
             n}+\angu - \lambda +1 & 0
          \end{array}
   \right)
   \left( \!\Psvektors\! \right) ,
\label{BzentraleGleichungmitPotentialauf1infty}
\end{equation}
%
%(\ref{BzentraleGleichung}) 
%
%und (\ref{Brandbedingung})   !!!! keine Randbedingung da ab 0 bis infty 
mit Randbedingung
\begin{equation}
  \Psis_1 (1) = 1, \quad  \Psis_2(1)  = 0
  \label{RBbei1}
\end{equation}
%2014mit $  \eta \in [0,\pi) $
%Als Randbedingung wird
%
%
 %gilt:
%
\begin{eqnarray}\lefteqn{
  \frac{\mbox{d} \roDns}{\dkappa} (\kappa)
=}
\nonumber \\
  & = &
  \frac{1}{\pi}
   \left( \prod_{j=1}^n \ffjs (\kappa, \tetas\ind(\kappa) - \kappa d_j,
                              \int_{b_{j-1}}^{a_j} \Gk(s, \kappa)\ds)
   \right)
   \Dmischl (\kappa)
%%14    \left(1 + \GGGG_n(\kappa) \right)
% \nonumber \\
%   & & \quad \quad \quad \quad \quad \quad\quad \quad \quad \quad \quad
%       \quad \quad \quad \quad \quad\quad \quad \quad
   \quad (\kappa \in \R^*),
   \nonumber \\
\label{Bteil2drodkappa}
\end{eqnarray}
wobei
\begin{equation}
  \Dmischl (\kappa) := \frac{{\rm \sign}(\kappa) \sqrt{\kappa^2+1} +1}
                      {{\rm \sign}(\kappa) \sqrt{\kappa^2+1}}
  R(1, \kappa )
  \quad (\kappa \in \R^*).
\label{Dgeschwungen}
\end{equation}
\end{satz}

\Beweis{}
{
%%%%%%%%%%%

Um den Satz zu beweisen, wird wie im Beweis von
Satz \ref{Ateil2satzspektralfunktionableitung} vorgegangen, indem
das Problem (\ref{BzentraleGleichungmitPotentialauf1infty}) und (\ref{RBbei1})
%,  (\ref{Brandbedingung})   !!!! keine Randbedingung mehr , da auf 0 intfy berachtet
%
%
zun"achst auf dem endliche Intervall $ (1,b) $ mit zus"atzlicher Randbedingung bei $ b$ untersucht wird:
\begin{eqnarray}
   \left(\! \begin{array}{c} \Psis_1 %(r) 
   \\
                           \Psis_2 %(r)
          \end{array}
  \!  \right) '
   & = & 
   \left( \! \begin{array}{cc}
             0                     & \mass-1-q_{%D_
             n}%(r)
             -\angu + \lambda+1 \\
             \mass -1+ q_{%D_
             n}%(r)
             +\angu - \lambda +1& 0
          \end{array}
          \!
   \right)
   \left( \!\begin{array}{c} \Psis_1 %(r)
    \\
                           \Psis_2 %(r)
          \end{array}
          \!
   \right)
\nonumber \\
& &
   \quad\quad\quad\quad\quad\quad\quad\quad\quad\quad\quad\quad\quad
     \quad   \quad   \quad   \quad   \quad   \quad (r \in (1,b))
%\nonumber \\
\label{Bteil2zentraleGL}
\\
  & &
  \begin{array}{rcl}
   \Psis_1 (b) =1, \quad  \Psis_2 (b)   =  0.&&
%   \quad \mbox{mit} \quad \zeta \in [0,\pi)
\\
  \end{array}
\label{Bteil2randbed}
\end{eqnarray}

Auch hier ist die Spektralfunktion eine Stufenfunktion, die bei den Eigenwerten  $ \lambdabn $ Spr"unge der H"ohe 
$
  \abnquadrats := 
\int_1^b \modulus{\phibn(r)}^2 \mbox{d}r
$
%%14\end{displaymath}
aufweist:\label{Rhonbschlange}
\begin{displaymath}
   \roDnbs (\lambda)= \left\{
                            \begin{array}{ll}
                               {\displaystyle -\sum_{\lambda < \lambdabn \le
0}
                                      \frac{1}{\abnquadrats}  }
                                & (\lambda < 0)
                                \\
                                {\displaystyle  \sum_{0 < \lambdabn \le
\lambda}                                      \frac{1}{\abnquadrats}    }
                                & (\lambda \ge 0)
                            \end{array}
                          \right.
              % } f�r Klammerzuordnung des Editors
\end{displaymath}
(vgl.
\cite{LEVITHANSARGSJANII}, S. 214
Abschnitt 8.1.1\footnote{Die Gestalt von (\ref{BzentraleGleichungmitPotentialauf1infty}) stimmt mit dem von 
 \cite{LEVITHANSARGSJANII} 
 untersuchten System bis auf das Vorzeichen von $ \lambda $ "uberein. Dort gehen stetige Koeffizienten  $ q_1$ und $ q_2 $ in  das System $ y_2'(x) - [ \lambda + q_1(x) ]\; y_1(x) \;=\; 0$, 
 $ y_1'(x)  + [\lambda + q_2 (x) ]\; y_2(x) \;= \; 0 $ ein. }).

Sei $ \alphabeta \subset \R^* $ und $ K \subset \Z\setminus\{0\} $ beliebig.
Dann gilt der folgende

\begin{hilfs}
\label{hilfsnormeigenfkt}

$\left.\right.$

Es gilt
\begin{equation}
  \frac{1}{b-1
} \abnquadrats \to \RDnnquadrats (\kappa) \frac{\lambda}{\lambda
+1}
  + \kleinoglm (1)
  \quad ( b \to \infty). 
  \label{aquadratooo}
\end{equation}
Dabei wird durch die Notation $ \kleinoglm (1) $ zum Ausdruck gebracht,
dass die Konvergenz in $ \kappa \in \alphabeta $
und in $ k \in K $ gleichm"a"sig ist.
\end{hilfs}

\Beweis{}
{
%%%%%%%%%

Der Beweis erfolgt v"ollig analog zum %Vorgehen beim 
Beweis von Hilfssatz \ref{aquadrsatz}. 

Hier ist jedoch das linke Intervallende $ b_0 = 1 $, so dass beispielsweise das Fundamentalsystem (\ref{UFundamentalsystem}) um $ 1 $ verschoben werden muss
(zur Vereinfachung wird  $ \rschl := r - 1 $ verwendet):
\begin{equation}
U (\rschl , \kappa)  := 
\left(
\begin{array}{cc}
\cos \kappa \rschl  & - \frac{\kappa}{1-\lambda}\sin \kappa \rschl \\
\frac{\kappa}{1+\lambda}\sin \kappa \rschl & \cos \kappa \rschl
\end{array}
\right)
\quad (\rschl \in [0,\infty))
 \nonumber
\end{equation}

Die Koeffizienten der Matrix $ S $ lauten hier:
%Bei Letzterem ist f"ur $ r \in [b_0,\infty) $
\begin{eqnarray}
  \sschl_1 (r) & := & \mass(r) -1 +q_%{D_
  n%}
  (r)+\angu (r)
   \nonumber\\
  \sschl_2(r)  & := &  \mass(r)- 1 +q_%{D_
  n%}
  (r)+\angu (r)
   \nonumber
\end{eqnarray}
%definiert.

Diese sind auf $ (1,b) $ (mit $b>1$ beliebig) integrierbar,  denn
 mit (\ref{AbklingverhaltenMasse}) und (\ref{AbklingverhaltenDrehimpuls}) hat man
\begin{eqnarray}
 \modulus{ \mass(r) -1 +\angu (r)}
& \le &
   \frac{k^2}{r^2}
   + 
    \frac{\modulus{k}}{2r^2}
    \le 
\frac{ \const}{r^2}
\label{Abschmundl}
\end{eqnarray}
unabh"angig von $ \kappa $. Das endliche Buckelpotential $ q_{%D_
n} $ ist ohnehin integrierbar.

Deshalb gilt gem"a"s (\ref{psiabsch1}) und (\ref{psiabsch2}):
\begin{eqnarray}
\lefteqn{
\Psis_1(\rschl,\kappa) =}
\nonumber \\ 
& = &
\!\cos(\!\kappa \rschl\!) A(\!\kappa\!) \cos (\!\gamma (\!\kappa\!)\!) + \sin (\!\kappa \rschl\!) A(\!\kappa\!) \sin(\!\gamma(\!\kappa\!)\!)
\nonumber \\
& & - \cos(\!\kappa \rschl\!) \int_\rschl^\infty \!\!\left(\! \cos(\!\kappa s \!) s_2(\!s+1\!) \Psis_2(\!s\!) 
    - \frac{1\!+\!\lambda}{\kappa} \sin( \!\kappa s\! ) s_1(\!s+1\! ) \Psis_1(\!s\!) \!\right)\!\!\ds
    \nonumber \\
    & & - \sin(\!\kappa \rschl\!) \int_\rschl^\infty \!\!\left(\! \sin(\!\kappa s \!) s_2(\!s+1\!) \Psis_2(\!s\!) 
    - \frac{\kappa}{1\!-\!\lambda} \cos( \!\kappa s\! ) s_1(\!s\!) \Psis_1(\!s+1\!) \!\right)\!\!\ds
    \nonumber \\
& = &     
\cos(\kappa \rschl) A(\kappa) \cos (\gamma (\kappa)) + \sin (\kappa \rschl) A(\kappa) \sin(\gamma(\kappa))
+ \kleino(1) \quad (\rschl \to \infty)
\nonumber \\
&&
\nonumber%\label{kkk1b}
\\
\lefteqn{\Psis_2(\rschl,\kappa)= }
\nonumber\\
& = &
\frac{-\kappa}{1\!+\!\lambda}\sin(\!\kappa \rschl\!) A(\!\kappa\!) \cos (\!\gamma (\!\kappa\!)\!) + \frac{-\kappa}{1\!-\!\lambda}\cos (\!\kappa \rschl\!) A(\!\kappa\!) \sin(\!\gamma(\!\kappa\!)\!)
\nonumber \\
& & + \frac{\kappa}{1\!+\!\lambda} \sin(\!\kappa r\!) \int_\rschl^\infty \!\left( \! \cos(\!\kappa s \!) s_2(\!s+1\!) \Psis_2(\!s\!) 
    - \frac{\lambda\!+1\!}{\kappa}  \sin(\! \kappa s \!) s_1(\!s+1\!) \Psis_1(\!s\!) \!\right)\!\!\ds
    \nonumber \\
    & & -\cos(\!\kappa \rschl\!) \int_\rschl^\infty \!\!\left(\!\frac{\lambda\!-\!1}{\kappa} \sin(\!\kappa s\! ) s_2(\!s+1\!) \Psis_2(\!s\!) 
    + \cos(\! \kappa s\! ) s_1(\!s+1\!) \Psis_1(\!s\!)\! \right)\!\!\ds
    \nonumber \\
& = &     
\cos(\kappa \rschl) A(\kappa) \cos (\gamma (\kappa)) + \sin (\kappa \rschl) A(\kappa) \sin(\gamma(\kappa))
+ \kleino(1) \quad (\rschl \to \infty)
\nonumber %\\
%&&
%
%\nonumber%\label{kkk2b}
\end{eqnarray}
%%14Dass die Terme mit den Integralen von $ \rschl $ bis $ \infty $ entsprechend klein werden f"ur gro"se  $ \rschl$, zeigen die folgenden Absch"atzungen:
%
%AAAAAAAAAAAAAAAAAAAAAAAAA
%

Mit (\ref{psiquadratttt}) und (\ref{rrrrrrrr}) erh"alt man f"ur die L"osung $ \Psis $ die Behauptung (\ref{aquadratooo}).

 } % BEWEISENDE Hilfslemma

Nach %der Absch"atzung 
Absch"atzen der $L^2$-Normen der Eigenfunktionen f"ur gro"se $ b $ steht f"ur den Beweis des Satzes \ref{datzSPektarlfkt} noch eine Aussage zur 
Anzahl
%Zahl
 der Eigenwerte aus:
\pagebreak

\begin{hilfs}
\label{anzahlspruenge}
$ \left.\right. $

F"ur die Zahl $  \Nblambdalambdas $ der Eigenwerte von
(\ref{Bteil2zentraleGL}), (\ref{Bteil2randbed})
in $ (\lambda_1, \lambda_2] $ gilt:
\begin{equation}
    \Nblambdalambdas =
   \modulus{\frac{b(\beta-\alpha)}{\pi}
           }
   + \grossO (1)
   \quad (b \to \infty).
   \label{zahlEWe}
\end{equation}
\end{hilfs}

\Beweis{}
{

Die Zahl der Eigenwerte, die man durch Abz"ahlen der Nullstellen des
Pr"ufer-Winkels in Abh"angigkeit von $ \lambda $ erh"alt,
betr"agt nach %\cite{WEIDMANN87}, S.~245, oder 
 \cite{WeidmannmathZ}, Satz~3.1
\begin{eqnarray*}
\lefteqn{
  0 \le  \Nblambdalambdas
   =
    \modulus{ \frac{\tetas(b;\kappa(\lambda_2)) - \tetas(b;\kappa(\lambda_1))}
                   {\pi}
             + c
            }
}
\nonumber \\
 & = &
 \frac{1}{\pi}
 \modulus{
  \tetas_n(\kappa_2) \!-\!\tetas_n(\kappa_1)
 \! -\! (\kappa_2\!-\!\kappa_1)(b\!-\!b_n)
%%14\nonumber \\
%%14& & \quad
  +\! \int_{b_n}^b \!\left( \Gk(r;\kappa_2)
             \!           - \! \Gk(r;\kappa_1) \right)\!\dr\!
   + \!c
}
  \nonumber \\
 & = & 
 %= 
   \modulus{\frac{b(\beta-\alpha)}{\pi}
           }
   + \grossO (1)
   \quad (b \to \infty).
\end{eqnarray*}

} % ENDE BEWEIS Lemma Zahl der Eigenwerte

Nun werden die Aussagen der  Hilfss"atze \ref{hilfsnormeigenfkt} und Hilfssatz \ref{anzahlspruenge} genau wie im Beweis von Satz \ref{Ateil2satzspektralfunktionableitung} verwendet, um die Spektralfunktion zu bestimmen, so dass man ausgehend von 
\begin{equation}
\roDnbs
   (\kappa_2) - \roDnbs(\kappa_1) 
   = 
   \frac{\frac{b(\kappa_2-\kappa_1)}{\pi} + \grossO(1) }
   {\left(b-b_0\right)\left({\RDnnquadrats}(\kappa) \frac{\lambda}{\lambda +1}+\kleino(1)\right)}
   \quad (b \to \infty)
    \nonumber
\end{equation}
Folgendes
\begin{displaymath}
  \frac{\mbox{d}\roDns (\kappa)}
       {\dkappa}
  = \frac{1}{\pi \RDnnquadrats(\kappa)}
    \frac{\lambda(\kappa)+1}{\lambda(\kappa)}
%  \quad (\kappa \in \R^*),
  \quad (\kappa \in [ \alpha, \beta ])
\end{displaymath}
erh"alt.

 }  % BEWEISENDE SATZ

Hiermit 
%%14Schlie"slich 
kann man %%14mithilfe der Spektralfunktion 
das zugeh"orige Ma" s angeben:

\begin{bemerkung}

$\left.\right.$

Das zu (\ref{Bteil2drodkappa}) geh"orige Ma"s ist gegeben durch
\begin{eqnarray}
\lefteqn{
  \myDns (\Sigma) =}
\nonumber \\
& = & \int\limits_\Sigma \frac{1}{\pi}
                        \left( \prod_{j=1}^n \ffjs (\kappa, \tetas\ind(\kappa) - \kappa d_j,
                              \int_{b_{j-1}}^{a_j} \Gk(s, \kappa)\ds)
   \right)
\label{Bteil2defmyDn}
\end{eqnarray}
f"ur beliebige kompakte Teilmengen $ \Sigma \subset \R \backslash \{0\}$.
\end{bemerkung}

%%%Dass es sich bei diesem Ma"s um ein Spektralma"s handelt 

%####################

%Lemma2neu   Teil B

\subsubsection{Eigenschaften von $ \ffjs $ und  $ \mjs $ }

\begin{bemerkung}

$\left.\right.$

Der folgende Hilfssatz liefert "ahnliche Aussagen wie Hilfsssatz \ref{Ateil1lemmafquerm}, denn wenn man $ \ffjs $ "uber das zweite Argument mittelt,  zeigt sich, dass $  \overline{\ffjs}(\kappa;z) $ ungef"ahr  $1 $ ist.
Der Mittelwert von $ \mjs$ ist ungef"ahr $ \log \left(\frac{2}{\Ajs(\kappa) + 1}\right) $. 
\end{bemerkung}

\begin{hilfs}
\label{Bteil1lemmafquerm}
\nopagebreak
$\left.\right. $
\nopagebreak

Sei $ j \in \{1, \dots, n\} $. Dann gilt f"ur $ \kappa \in \R^*, z \in \R$ :
\begin{eqnarray}
  \overline{\ffjs}(\kappa;z) & := &
             \frac{1}{\pi} \int_0^\pi \ffjs(\kappa,y,z)\;\mbox{d}y
 =  \exp\left(-2\int_{b_{j-1}}^{a_j} \!\Fk(s;\kappa)\!\ds\right)
   \\
  \mjs (\kappa;z)& := & \frac{1}{\pi} \int_0^\pi \log \ffjs ( \kappa,y,z) \dy
  = \log \left(\frac{2}{\Ajs(\kappa) + 1}\right)
     -2\int_{b_{j-1}}^{a_j}\! \Fk(s;\kappa)\!\ds
\nonumber \\
& &
\label{Bteil1defmj}
\end{eqnarray}
\end{hilfs}
\begin{bemerkung}

$ \left.\right. $

   $  \overline{\ffjs} $ und $ \mjs $ h"angen nicht vom jeweiligen zweiten
   Argument ab.\footnote{
        Im Fall ohne Drehimpulsterme war in Bemerkung
        \ref{Ateil1bmunabhvondj}  festgehalten worden,
        dass  $ m_j $ nicht von den Buckelabst"anden abh"angt.
        Nach Hilfssatz \ref{Ateil1lemmafquerm} ist im drehimpulsfreien Fall
        sogar $ \overline{\ffj}(\kappa) = 1 $.
}
 Deshalb wird das zweite Argument
   im Folgenden bei der Notation unterdr"uckt werden.
\end{bemerkung}

\Beweis{von Hilfssatz \ref{Bteil1lemmafquerm}}{

Sei $ j \in \{ 1, \dots, n\} $.
Wegen Definitionen (\ref{BDefAj}), (\ref{BDefBj}) und (\ref{BDefCj})
gilt
\begin{equation}
   \Ajs^2 - \Bjs^2 - \Cjs^2 = \left(\Mjs\right)^2 = 1.
\label{Bteil1AqmBqmCq}
\end{equation}

Sei $ \kappa \in \R^* $, $ z \in \R $.

Aufgrund der Periodizit"at wirkt sich die Verschiebung 
 um $ +2z $ bei Integration von $ 0 $ bis $ \pi $ nicht aus. Nach \cite{GroebnerHofteiter}, S. 100, ist mit einer analogen Rechnung wie beim Beweis von Hilfssatz \ref{Ateil1lemmafquerm}, da die Voraussetzung $ \Aj > \sqrt{\Bj^2 + \Cj^2}$ aufgrund von (\ref{Bteil1AqmBqmCq})
 erf"ullt ist: 
\begin{eqnarray*}
  \overline{\ffjs}(\kappa,z) & = &
%%14             \frac{1}{\pi} \int_0^\pi \ffjs(\kappa,y,z)\;\mbox{d}y
%%14  \nonumber \\
%%14  & = &
  \exp\left( - 2 \int_{b_{j-1}}^{a_j} \Fk(s;\kappa)\ds
      \right)
    \cdot
  \nonumber \\
  & & \quad \cdot  \frac{1}{\pi} \int_0^\pi
  \frac{1}{\Ajs(\kappa)+\Bjs(\kappa)\cos 2(y+z)
                                    +\Cjs(\kappa) \sin 2(y+z)}
                \;\mbox{d}y
  \nonumber \\
  & = &   \exp\left( - 2 \int_{b_{j-1}}^{a_j} \Fk(s;\kappa)\ds
      \right)
\end{eqnarray*}
Dies ist unabh"angig von $ z \in \R $.

Entsprechendes Vorgehen wie beim Beweis von Hilfssatz \ref{Ateil1lemmafquerm} liefert
\begin{eqnarray*}
 \mjs (\kappa, z)
& = &
    \log \left(\frac{2}{\Ajs(\kappa) + 1}\right)
     -2\int_{b_{j-1}}^{a_j} \Fk(s;\kappa)\ds,
\end{eqnarray*}
was unabh"angig von $ z \in \R $ ist.

}

%________________________________--

%Hierher Asymptotok der A,B,C

\subsection{ Asymptotisches Verhalten von $ \Ajs $, $ \Bjs $
          und $ \Cjs $ f"ur $ a_j \to \infty $ 
   \label{BanhangasymptotischesVerhalten}	  }

Sei
\begin{equation}
 \Tnulls := \left( \begin{array}{cc}
      0                              & -H_j W_j(\cdot-a_j) +  1 + \lambda \\
      H_j W_j(\cdot-a_j) + 1-\lambda &    0
                  \end{array}
           \right)
            \nonumber
\end{equation}
und
\begin{equation}
 \TSs := \left( \begin{array}{cc}
                    0             &  \mass-1-\angu \\
                    \mass-1+\angu & 0 
                  \end{array}
           \right).
            \nonumber
\end{equation}

Die Terme mit ortsabh"angiger Masse  und Drehimpuls werden im Folgenden als St"orung $ \TSs $ des Operators $ \Tnulls $ aufgefa"st.

Das Fundamentalsystem mit Anfangswert $ \Einheitsmatrix $ 
von $ \Psis' = \Tnulls\Psis $, 
betrachtet auf $ [a_j, b_j ] $ sei mit $ \Ujs $ bezeichnet\footnote{ NB: Die Gleichung $ \Psis' = \Tnulls\Psis $ ist mit (\ref{anhang2gestoerteGl}) identisch}.

Das Fundamentalsystem von $ \Psis'=(\Tnulls+ \TSs)\Psis $, betrachtet auf 
$ [a_j, b_j ] $, welches Anfangswert $ \Einheitsmatrix $ hat, sei mit 
$ \Phijs $ bezeichnet. 

Erinnert sei an die Definition
$ \kappa := \sign(\lambda) \sqrt{\lambda^2-1} \quad (\lambda \in \R, 
            \modulus{\lambda} > 1) $.

\begin{hilfs}
$\left.\right.$

Es ist f"ur $ j \in \N $ %und $ i_1, i_2 \in \{1,2\} $
\begin{equation}
%\modulus{\Phijs_{i_1 i_2}(r) \, - \,  \Ujs_{i_1 i_2} (r) }
\norm{\Phijs
%_{i_1 i_2}(r) 
\, - \,  \Ujs
%_{i_1 i_2} 
(r) }_1
\le
\const \frac{1}{a_j}
 \nonumber
%\label{AsymptotikvonPhi}
\end{equation}
f"ur $ r \in [a_j, b_j] $
unabh"angig von $ \kappa \in[\alpha,\beta] $.
\end{hilfs}

\Beweis{}{

Ausgangslage ist 
erneut 
%wieder 
eine Integralgleichung nach Art von  \cite{EASTHAM}, S.~6.:
%
%Es ist
%
\begin{equation}
\Phijs(r) =
\Ujs (r) + \Ujs(r) \int_{a_j}^r \Ujs\left.\right.^{-1}(s) \TSs (s) \Phijs (s) \ds
 \nonumber
\end{equation}
%(vgl.
Mit der Absch"atzung (\ref{Abschmundl})
kann man mit dem bereits mehrfach angewandten Verfahren
f"ur $ i_1, i_2 \in\{1,2\} $ folgendes ermitteln:
\begin{eqnarray}
%%14\lefteqn{
\norm{\left(\Phijs(r) - \Ujs(r) \right) 
%_{i_1,i_2}  
}  %%14=}
& \le &
%\sum_{i_3,i_4,i_5=1}^2
%%14\norm{\Ujs} \norm{\Ujs \left.\right.^{-1}}\norm{\Phijs}
%%14\left(k^2 + \frac{\modulus{k}}{2}\right) \int_{a_j}^r \frac{1}{s^2} \ds
%%14\nonumber \\
%%14& \le & %8
%
%%14  \norm{\Ujs} \norm{\Ujs \left.\right.^{-1}}\norm{\Phijs}
%%14\left(k^2 + \frac{\modulus{k}}{2}\right) 
%%14\frac{1}{a_j}
%%14\le
\const \frac{1}{a_j}.
 \nonumber
\end{eqnarray}
%
%ermitteln.

}  % ENDE Beweis Lemma ----------------------------

%###########

%\section{,,Einschub''}

Des Weiteren besitzen die Funktionen $ \fajs $ folgende Eigenschaft, welche analog zur Aussage des Hilfssatzes \ref{Alemma2B} des Kapitels \ref{KapitelohneDreh} ist:
\begin{hilfs}
$\left.\right.$
\nopagebreak

Sei die Folge der Buckelh"ohen $ \left(H_j\right)_{j\in\>N} $ gegeben. Sei $ \left( a_j \right)_{n\in\N} $ eine beliebige Folge von Buckelpositionen mit der Eigenschaft, dass 
$    \left(\frac{1}{\aj}\right)_{j\in\N} $
schneller gegen $ 0 $ geht als $ \left(H_j\right)_{j\in\N} $.                       

Sei $ k \in \Z\setminus\{0\} $ und  $ I \subset \R $ ein kompaktes Intervall.  

Dann gilt
\begin{equation}
   \ffjs (\kappa, y, z) \ge \Cvonk
   \quad
   (j\in \N, \kappa \in \alphabeta, y \in \R, z \in I).
%\nonumber \\
%& &  {unabh"angig \,von} \left(a_j\right)_{j \in \N} \in \AAA
 \nonumber
\end{equation}

\end{hilfs}

\Beweis{}{

Es ist 
\begin{displaymath}
  \ffjs (\kappa, y, z)
   =
  \frac{\exp\left(-2\int_{b_{j-1}}^{a_j} \Fk(s;\kappa)\ds\right)}
       {\Ajs(\kappa) + \Bjs(\kappa) \cos(2(y+z)) + \Cjs(\kappa) \sin(2(y+z))}
\end{displaymath}

F"ur den Z"ahler von $ \ffjs $ gilt unter Verwendung von $ b_j \ge 1 \; (j \in \N) $, da wegen (\ref{Bteil1Fkabschaetz})
\begin{displaymath}
  \modulus{\int_{b_{j-1}}^{a_j} \Fk(s;\kappa)\ds }
  \le
  \int_1^\infty \frac{\Ckkappa}{s^2} \ds = \Ckkappa
\end{displaymath}
ist,
\begin{displaymath}
  \exp\left( -2 \int_{b_{j-1}}^{a_j} \Fk(s;\kappa)\ds\right)
  \ge
  \exp\left(-2\Ckkappa\right),
\end{displaymath}
was unabh"angig von $ \left(a_j\right)_{j \in \N} $ ist.

Im Folgenden wird der Nenner abgesch"atzt.
Mit den Definitionen (\ref{BDefAj}), (\ref{BDefBj}) und (\ref{BDefCj}) gilt f"ur das Doppelte des Nenners von $  \ffjs $
\begin{eqnarray}
 \lefteqn{  2\modulus{\Ajs(\kappa) + \Bjs(\kappa) \cos(2(y+z)) + \Cjs(\kappa) \sin(2(y+z))}
 }
 \nonumber \\
 & = &  \left|
           \left[
              {\Mjeesquad} + \plmin \Mjzesquad
              + \minpl \Mjezsquad + \Mjzzsquad
           \right] \right.
           \nonumber \\
           &&
           +  
             \left[
                  \Mjeesquad + \plmin \Mjzesquad
                  - \minpl \Mjezsquad - \Mjzzsquad
             \right]  \cos(2(y+z))
             \nonumber \\
             && 
             \left.+ 2 \left[\sqrt{\minpl}\Mjees\Mjezs + \sqrt{\plmin} \Mjzes\Mjzzs\right] \sin(2(y+z)) \right|
             \nonumber \\
  & \le &
          %     \left|
               % \left[
              \Mjeesquad + \plmin \Mjzesquad + \minpl \Mjezsquad + \Mjzzsquad
%           \right]
              % \right.
%           \nonumber \\
%    &&             
               +% \left[
                  \Mjeesquad + 
                  \nonumber \\
                  && +\plmin \Mjzesquad 
%                  \nonumber \\
    %              &&+ \minpl \Mjezsquad^2 + \Mjzzsquad^2
   %          \right] %\cos(2(y+z)) 
%     \nonumber \\
    % && 
    + 2 \sqrt{\minpl}\modulus{\Mjees\Mjezs} + 2\sqrt{\plmin} \modulus{\Mjzes\Mjzzs }.
    %+ ....
  %  \nonumber \\
%    && + ..... bla
\label {zwischenerg}
\end{eqnarray}

Dabei wurde verwendet, dass mit $ \tilde{\Phi} $  und $ \tilde{ U} $ 
%%14aus (\ref{AsymptotikvonPhi}) 
beispielsweise
\begin{eqnarray}
\lefteqn{  \Mjeesquad = \tilde{\Phi}_{11}(\alpha_j)^2
  = \left( \tilde{\Phi}_{11}(\alpha_j) - \tilde{U}_{11}\right)^2     + 2  \tilde{U}_{11}\left(   \tilde{\Phi}_{11}(\alpha_j) -  \tilde{U}_{11}\right)
   + \tilde{ U}_{11}^2}
   \nonumber \\
  & \le &  
  \const \frac{1}{a_j^2} +  2 \norm{\tilde{U}} \const \frac{1}{a_j} + \Mjee^2
  \nonumber \\
  & = &  \const \frac{1}{a_j^2} +  2 \norm{\tilde{U}} \const \frac{1}{a_j} 
  + \left(\Mjee - U_{11} \right)^2 + 2 \left( \Mjee - U_{11} \right)U_{11} + U_{11}^2
  \nonumber \\
  & \le &\const \frac{1}{a_j^2} +  2 \norm{\tilde{U}} \const \frac{1}{a_j}  
  \const H_j^2 + 
  + 2  \const H_j \cdot 1 + 1^2
  \label{einMquadratabschaetzen}
\end{eqnarray}

mit $ U $ aus (\ref{UFundamentalsystem}) unter Ber"ucksichtigung von  (\ref{BanhangasymptotischesVerhalten_Hj})
abgesch"atzt werden kann.

F"ur (\ref{zwischenerg}) erh"alt man,  da $ \frac{\lambda+1}{\lambda-1} $, 
$ \frac{\lambda-1}{\lambda+1} $ f"ur $ \kappa \in \alphabeta $ beschr"ankt sind, aufgrund der Eigenschaften der Folgen 
$    \left(\frac{1}{\aj}\right)_{j\in\N} $ und $ \left(H_j\right)_{j\in\N} $
\begin{eqnarray}
 \lefteqn{  2\modulus{\Ajs(\kappa) + \Bjs(\kappa) \cos(2(y+z)) + \Cjs(\kappa) \sin(2(y+z))}
 }
 \nonumber \\
 & \le &  \const H_j +  \grossO (H_j^2) \quad\quad\quad\quad\quad\quad\quad\quad\quad\quad\quad
  \nonumber
 \end{eqnarray}
Damit ist der Nenner von $ \ffjs $ gr"o"ser oder gleich $ \frac{1}{ \const H_j } $ plus h"ohere %Terme 
Potenzen in $ H_j $.

Fa"st man die Ergebnisse f"ur Z"ahler und Nenner zusammen, folgt die Behauptung.

}

%######################

%\subsubsection{,,Lemma 2''}
\label{Lemma2mitDreh}

Die Aussage $ - \sum_{j=1}^\infty \mj (\kappa) = \infty  $ von Hilfssatz \ref{Alemma2}  ist auch in der allgemeinen Situation mit Drehimpuls f"ur kleiner werdende Buckel g"ultig:

%------------
%%
\begin{hilfs}
\label{Summemschlange}
$\left.\right.$

Sei $ \alphabeta \subset \R^* $ und $ I \subset \R $ ein endliches Intervall.

Seien Folgen $ \left(H_j\right)_{j \in \N} \in R^\N $ und
$ \left(\alpha_j\right)_{j\in \N} \in (\R^+)^\N $ und zu letzteren Funktionen
$ W_j \in 
C[0,\alpha_j] $ mit $ \int_0^{\alpha_j}\modulus{W_j} =1 $
gegeben mit folgenden Eigenschaften:

$ H_j = 0 $ nur f"ur endlich viele Indizes $ j \in \N $,
$ H_j \to 0 \quad (j \to \infty) $ und
\begin{equation}
%  \sum_{j=1}^\infty
%    H_j^2
%    \left( \left[ \int_0^{\alpha_j} W_j(s) \cos 2 \kappa s \ds \right] ^2
%    + \left[ \int_0^{\alpha_j} W_j(s) \sin 2 \kappa s \ds \right] ^2
%    \right)
%
 \sum_{j=1}^\infty
H_j^2
{\cal W}_j 
  = \infty
   \nonumber
\end{equation}
Dabei ist $ {\cal W}_j({b_j}, a_j, \kappa) $ wie in (\ref{zweiteOrdAinH}) definiert.

Sei  $  \left(\aj\right)_{j\in \N} \in (\R^+)^{\N} $ eine beliebige Folge, f"ur die $ \left(\frac{1}{\aj}\right)_{j\in \N} $  schneller gegen $ 0 $ geht als 
                $  \left(H_{j+1}^3\right)_{j\in \N}$ .

%Sei $ \sup_{j\in\N} \alpha_j < \infty $.
%
%
%

 Dann gibt es zu jedem $ Q \in \R^+ $ eine nat"urliche Zahl $ \nu $ so, dass
\begin{equation}
  - \sum_{j=1}^\nu \mjs(\kappa, z ) > Q \quad (\kappa \in \alphabeta, z \in I).
   \nonumber
\end{equation}
\end{hilfs}

\rm

\Beweis{%von Hilfssatz \ref{Summemschlange}
}{
\nopagebreak

%\Beweis{}{
Auswertung des Fundamentalsystems $\Phijs $ an der Stelle $ b_j $ liefert  f"ur die Komponenten der 
Tranfermatrix $ \Mjs $:
\begin{equation}
 \left[\Mjs\right]_{i_1,i_2} = \Phijs_{r,p}(b_j)
 = \Mj_{i_1,i_2} + \grossOfrag\left(\frac{1}{a_j}\right) 
   \quad (a_j \to \infty)
\label{AsymptotikM}
\end{equation}
%

%%%%%
Damit 
%Es 
gilt f"ur $ \kappa \in \alphabeta, z \in I $
mit der Definition (\ref{BDefAj}) von $ \Ajs $
%wegen des Ergebnisses  (\ref{Banhang2Ajschlange}) 
gem"a"s Definition (\ref{Bteil1defmj}) von $ \mjs $, indem man f"ur die einzelnen Terme von $ \Ajs $ die analogen "Uberlegungen wie bei (\ref{einMquadratabschaetzen})
anstellt:
\begin{eqnarray}
\lefteqn{ \mjs(\kappa, z) 
  = 
 - \log \left( \frac{\Ajs(\kappa) + 1}{2}\right)
  + 2\int_{b_{j-1}}^{\aj} \Fk(s;\kappa)\ds}
  \nonumber \\
 & = & 
 - \log \left(\frac{\frac{1}{2}
\left[
{\Mjeesquad} + \plmin \Mjzesquad
+ \minpl \Mjezsquad + \Mjzzsquad
\right] + 1}{2} \right)
    \nonumber \\
    && \quad  + 2\int_{b_{j-1}}^{\aj} \Fk(s;\kappa)\ds
   \nonumber \\
 & = &
 - \log\left(
 \frac{\frac{1}{2}
\left[
{\Mjee^2} + \plmin \Mjze^2
+ \minpl \Mjez^2 + \Mjzz^2 +  \grossO\left(\frac{1}{\aj}\right)
\right] + 1}{2} \right)
    \nonumber \\
    && \quad  + 2\int_{b_{j-1}}^{\aj} \Fk(s;\kappa)\ds
 \nonumber \\
 &=&
 - \log \left(\frac{\Aj(\kappa) + 1}{2} + \grossO\left(\frac{1}{\aj}\right)
      \right)
  + 2\int_{b_{j-1}}^{\aj} \Fk(s;\kappa)\ds
 \nonumber \\
 & \ge &
 -  \log \left(\frac{\Aj(\kappa) + 1}{2} + \grossO\left(\frac{1}{\aj}\right)
      \right)
 - 2 \Ckkappa \frac{1}{\aj} ,
\label{mabschaetzen}
\end{eqnarray}
da wegen (\ref{Bteil1Fkabschaetz}) 
\begin{equation}
  2\int_{b_{j-1}}^{\aj} \Fk(s;\kappa)\ds
  \ge -2 \Ckkappa \left( \frac{1}{\aj} - \frac{1}{b_{j-1}} \right)
  \ge  - 2 \Ckkappa \frac{1}{\aj} .
\label{Blemma2absch}
\end{equation}

Mit %(\ref{Banhang2Ajschlange}) und 
(\ref{zweiteOrdAinH})
%Im Falle kleiner werdender Buckel 
erh"alt man f"ur (\ref{mabschaetzen}) 
%mit (\ref{Banhang2asymptotik})
%
\begin{eqnarray*}
 \mjs(\kappa, z) 
 &  \ge  & 
  -   \frac{H_j^2}{2} 
             { \cal W}_j(r,a_j,\kappa) 
  +  \grossO (H_j^3)
  - 2 \Ckkappa \frac{1}{\aj} 
 \nonumber \\
 & = &
 - \frac{H_j^2}{2} 
             \frac{1}{\lambda^2 -1} 
               \underbrace{
                  { \cal W}_j(r,a_j,\kappa) 
               }_{ > 0} 
             + \grossO (H_{j+1}^3)
\nonumber	     
\end{eqnarray*}
Somit folgt 
\begin{displaymath}
 - \sum_{j=1}^\nu \mjs(\kappa, z ) > Q \quad (\kappa \in \alphabeta, z \in I)
\end{displaymath}
f"ur jede Folge $ \left(\aj\right)_{j\in\N} $,
da nach Voraussetzung
\begin{displaymath}
  \sum_{j\in\N}  \frac{H_j^2}{2}
           { \cal W}_j(r,a_j,\kappa) 
  = \infty
\end{displaymath}
gilt.

}  % Ende Beweis Lemma2neu

Zur Vorbereitung des Beweises des zentralen Satzes \ref{BzentralerSatzmitDreh}
dient der folgende 
\begin{hilfs}
$\left.\right.$

Sei $  \left(\aj\right)_{j\in\N}  $ eine beliebige Folge von Buckelpositionen. 
Sei $ k \in \Z\setminus\{0\} $, $ \alphabeta \subset \R^* $.
Sei $ \eps > 0 $, $ \nyunten\in\N_0 $ und $ I \subset \R $ ein kompaktes 
Intervall. 

$ \nyoben > \nyunten $ sei so gro"s, dass unabh"angig von $ z \in I $
\begin{equation}
 \int_\alpha^\beta 
  \frac{1}{- \sum_{j=\nyunten+1}^\nyoben \mjs(\kappa) } \dkappa
  < \frac{2\eps}{3\KC} 
\label{Lemma2b1Vor}
\end{equation}
und
\begin{equation}
  \int_\alpha^\beta
  \frac{\exp\left(2 \Ckkappa \exp(\nyunten^2)\right) -1}
    { \left[ \frac{1}{2} \sum_{j=\nyunten+1}^\nyoben \mjs (\kappa, z)\right]^2}
  \dkappa
  < \frac{\eps}{3 \KC}
\label{Lemma2b2Vor}
\end{equation}
ist mit $\Ckkappa $ aus (\ref{Bteil1defCkkappa})  und $ \KC $ gem"a"s (\ref{lemma2Bhjabschaetzung}).

Sei $ j_0 \in\{\nyunten + 1, \dots, \nyoben\} $ und 
$ \Theta : \R^* \to \R $ stetig differenzierbar.
Dann gibt es $ L_\Theta \in \N $ so, dass f"ur alle $ L > L_\Theta $ gilt:
\begin{equation}
 \int_\alpha^\beta 
 \frac{\left[ \hjnulls (\kappa, \Theta(\kappa) - \kappa L, z)\right]^2}
    { \left[ \frac{1}{2} \sum_{j=\nyunten+1}^\nyoben \mjs (\kappa, z)\right]^2}
  \dkappa
 < \eps 
 \quad (z \in I%, \left(a_j\right)_{j\in\N} \in \AAA
 )
  \nonumber
\end{equation}
Dabei ist $  \hjnulls $ folgenderma"sen definiert:
\begin{equation}
 \hjnulls (\kappa,y, z) := \log \mjnulls (\kappa, y, z) .
 \label{defhjnullschl}
\end{equation}
%definiert.
\label{Blemma2B}
\end{hilfs}
 Im drehimpulsfreien Fall ist die entsprechende Behauptung in Hilfssatz \ref{Lemma2B} formuliert.

\begin{bemerkung}

$\left.\right.$

Satz \ref{Summemschlange} 
%aus Abschnitt \ref{Lemma2mitDreh}
gew"ahrleistet, dass es  
$ \nyoben > \nyunten $ gibt, so dass unabh"angig von $ z \in I $
und
unabh"angig von der Folge der Buckelpositionen  $ \left(\aj\right)_{j\in\N} 
%\in  \AAA 
$ 
die Voraussetzungen (\ref{Lemma2b1Vor}) und (\ref{Lemma2b2Vor}) erf"ullt werden.
\end{bemerkung}

\Beweis{}
{

Wegen 
\begin{displaymath}
  \majs = - \log \left( \frac{\Aajs +1}{2}\right) 
                 - 2 \int_{b_{j-1}}^{a_j} \Fk(s;\cdot) \ds 
              = \log\left(\frac{\Aj+1}{2}\right)
                 + \grossOglm\left(\frac{1}{a_{j-1}}\right)
\end{displaymath}
gibt es eine von $ \left(\aj\right)_{j\in\N} $ abh"angende Konstante mit
$
%%14\begin{displaymath}
  \modulus{\majs-\maj} \le \const \frac{1}{a_{j-1}}.
%%14\end{displaymath}
$
Es ist also
\begin{displaymath}
  \modulus{
    \sum_{j=\nyunten+1}^\nyoben (\majs -\maj) 
  }
    \le 
    \const \frac{\nyoben-\nyunten}{a_\nyunten}.
\end{displaymath}

Somit gibt es wegen (\ref{Nnynykappa}) 
\begin{equation}
  \Nknynykappa
  :=
  \inf_{\stackrel{\left(a_j\right) _{i \in \N}
  %1403\in \AAA
  }{z \in I}}
  \left\{
     - \frac{1}{2} \sum_{j=\nyunten+1}^\nyoben \majs (\kappa,z)
  \right\}
  > 0       .
\label{einN}
\end{equation}

Es gilt dann wegen des Korollars \ref{AKorollarzuLemma1} zu Hilfsatz \ref{Alemma1} 
(mit $ \frac{1}{\Nknynykappa^2} $ f"ur 
$ F $ und $ \hajnulls^2 $ f"ur $ H $ eingesetzt)
f"ur  $ L > L_\Theta $, (mit $ \L_\Theta $ aus Korollar \ref{AKorollarzuLemma1}):
\begin{eqnarray}
%%14\lefteqn{
 \modulus{\!
 \int_\alpha^\beta \!\!
  \frac{\hajnulls^2(\!\kappa, \!\Theta(\!\kappa\!) \!\!- \!\!\kappa L ,\! z\!)\! -\!
         \overline{\hajnulls^2} (\!\kappa, \!z\!)}
       {\left[\frac{1}{2} \sum_{j=\nyunten+1}^\nyoben \!\majs (\kappa,z)\right]^2}
 }
%%14}
%%14\nonumber \\
%%14 & \le &
%& \le&
\le
 \modulus{
  \int_\alpha^\beta \!\!
  \frac{\hajnulls^2(\!\kappa, \!\Theta(\!\kappa\!) \!\!-\!\! \kappa L ,\! z\!)\! -\!
         \overline{\hajnulls^2} (\!\kappa,\! z\!)}
       {\Nknynykappa^2}
 } 
% \nonumber \\
%& <&
\!<\!
 \frac{\eps}{3}.
\label{Blemma2BGl1}
\end{eqnarray}

Wegen
$ \hajs^2 \le \KC \left(\fajs - 1 - \hajs\right) $ ist andererseits
\begin{displaymath}
  \overline{\hajs^2}
  \le
  \KC\left( \exp\left(-2\int_{b_{j-1}}^{a_j} \Fk(s;\cdot)\ds \right) -1
            - \majs \right),
\end{displaymath}
wobei $ \overline{\hajs} = \majs $ verwendet wurde.
Nutzt man
\begin{eqnarray*}
 \lefteqn{ \exp\left( -2\int_{b_{j-1}}^{a_j} \Fk(s;\cdot)\ds \right) -1 
   \le 
  \exp\left(2 \Ckkappa \left(\frac{1}{a_j} - \frac{1}{b_{j-1}}\right)\right) -1 }
\nonumber \\
 & \le &
 \exp\left(2 \Ckkappa \frac{1}{a_j} \right) -1 
%%14\nonumber \\
%%14 & \le &
 \exp\left( 2 \Ckkappa \exp (-(j-1)^2) \right) - 1 
\nonumber \\
 & \le &
  \exp\left( 2 \Ckkappa \exp(-\nyunten^2)\right) -1
%%14\nonumber \\
%%14&& \quad\quad
  \quad ( j \in \{ \nyunten+1, \dots, \nyoben\}),
   \nonumber
\end{eqnarray*}
so erh"alt man
\begin{eqnarray}
\lefteqn{  \int_\alpha^\beta
  \frac{ \overline{\hajnulls^2}(\kappa, z)}
       {\left[\frac{1}{2} \sum_{j=\nyunten+1}^\nyoben \majs (\kappa,z)\right]^2}
}
\nonumber \\
 &  \le &
  \KC\left(  
    \int_\alpha^\beta 
      \frac{  \exp\left( 2 \Ckkappa \exp(-\nyunten^2)\right) -1 } 
      {\left[\frac{1}{2} \sum_{j=\nyunten+1}^\nyoben \majs (\kappa,z)\right]^2}
      \dkappa
      -
     \int_\alpha^\beta
     \frac{m_{a_{j_0}}}{\left[\frac{1}{2} \sum_{j=\nyunten+1}^\nyoben \majs (\kappa,z)\right]^2}
%      \dkappa
  \right)
\nonumber \\
& \le &
  \KC\left(  
    \int_\alpha^\beta
      \frac{  \exp\left( 2 \Ckkappa \exp(-\nyunten^2)\right) -1 } 
      {\left[\frac{1}{2} \sum_{j=\nyunten+1}^\nyoben \majs (\kappa,z)\right]^2}
      \dkappa
      -
     \int_\alpha^\beta
     \frac{\sum_{j=\nyunten+1}^\nyoben \majs}{\left[\frac{1}{2} \sum_{j=\nyunten+1}^\nyoben \majs (\kappa,z)\right]^2}
%      \dkappa 
  \right) 
\nonumber \\
& = &
 \KC \!\!    \int_\alpha^\beta\!
      \frac{  \exp\!\left( 2 \Ckkappa \exp(\!-\nyunten^2)\!\right)\! -\!1 }
      {\left[\frac{1}{2} \sum_{j=\nyunten+1}^\nyoben \majs (\kappa,z)\right]^2}
    \!  \dkappa
 - 
  \frac{\KC}{4} \! \int_\alpha^\beta \!
  \frac{1}{\sum_{j=\nyunten+1}^\nyoben \majs (\!\kappa\!,\!z\!)}
 \dkappa
%\nonumber \\
% & \le &
\le
\frac{2\eps}{3}
\label{Blemma2BGl2}
\end{eqnarray}
als Absch"atzung.
Addieren von (\ref{Blemma2BGl1}) und (\ref{Blemma2BGl2}) liefert die Behauptung.
}

%######################

%  TEIL B  Lemma 3

%\section{,,Lemma 3''}

Im folgenden Satz  wird ein Vergleich der Ma"se angestellt, die zu den Potentialen
$ q_{%5D_
n} $ und $ q_{%D_
{n+1}} $ geh"oren, welche sich im letzten Buckel 
unterscheiden.
%, angestellt.
Hierbei ist die Position der $ n $ Buckel von $ q_{%D_
n} $ fixiert, 
w"ahrend  der Mindestabstand $ d_0 $ des neuen $ n+1 $-ten Buckels 
hingegen noch zu bestimmen ist. 
In dieser Hinsicht ist das Vorgehen  vergleichbar mit der Strategie bei Hilfssatz \ref{Alemma3} im drehimpulsfreien Fall des Abschnitts \ref{singstetSpektrum}.

\begin{satz}
\nopagebreak
\label{satzreferenz}

$\left.\right. $
\nopagebreak

Sei $ n \in \N $, $ d_1, \dots, d_n > 0 $, 
und $ \eps > 0 $. 

Dann gibt es %%14 $ \tilde{\alpha}_n \ge 1 $ und 
$ d_0 > 
\alpha_n %%14\tilde{\alpha}_n 
$
  mit: Ist $ d_{n+1} \ge d_0 $, so gilt f"ur
jedes Intervall $ \Sigma \subset [\alpha,\beta] $
\begin{equation}
  \modulus{  \myDnpluss (\Sigma) - \myDns (\Sigma) } < \eps,
   \nonumber
\end{equation}
%
%ausf"allt,
 wobei
\begin{eqnarray}
  \myDns (\Sigma)  &  = &
   \int\limits_\Sigma %%14\left(
        \frac{1}{\pi} \left[ \prod_{j=1}^n
        \ffjs \left(\kappa, \tetas\ind (\kappa) - \kappa d_j, 
                        \int_{b_{j-1}}^{a_j}\Gk (s;\kappa)
              \right) 
                            \right]
        {\cal{D}} (\kappa) %%14\, \cdot \,
%%14        \left(1 + \GGGG_n(\kappa)\right) \,
%%14        \Biggr)
        \dkappa
\nonumber %\\
%& &
\end{eqnarray}
das Ma"s zum $ n$-Buckel-Potential bezeichnet.
\begin{eqnarray}
%\lefteqn{
 \myDnpluss  (\Sigma) 
% =}
%\nonumber \\
& = &
   \int\limits_\Sigma %%14\left(
        \frac{1}{\pi} \left[ \prod_{j=1}^{n+1}
        \ffjs \left(\kappa, \tetas\indnpluseins (\kappa) - \kappa d_j, 
                        \int_{b_{j-1}}^{a_j}\Gk (s;\kappa)
              \right) 
                            \right] %%14\cdot
%%14\right.
%%14\nonumber \\
%%14& &% \left.
%%14     \quad\quad 
%%14      \cdot \,
        {\cal{D}} (\kappa)
%%14        \, \cdot \,
%%14        \left(1 + \GGGG_{n+1}(\kappa)\right)
%%14        \Biggr)
        \dkappa
\nonumber 
%\\
%& &
\end{eqnarray}
ist das Ma"s zu dem Potential mit $n+1 $ Buckeln, das in den ersten $ n $ Buckeln einschlie"slich ihrer Position mit dem 
Potential, das zu $ \myDns $ geh"ort, "ubereinstimmt.
\label{Blemma3}
\end{satz}

\Beweis{}{

Sei $ \Sigma \subset [\alpha,\beta] $.
Dann gilt
\begin{eqnarray}
\lefteqn{
    \myDnpluss (\Sigma) - \myDns (\Sigma) = }   
              \nonumber \\
& = & \frac{1}{\pi}
   \int\limits_\Sigma \left\{
        \left( \prod_{j=1}^n                                                
        \ffjs \left(\kappa, \tetas\ind (\kappa) - \kappa d_j, 
                        \int_{b_{j-1}}^{a_j}\Gk (s;\kappa)\right)
        {\cal{D}} (\kappa) \; \cdot 
\right.\right.
                 \nonumber\\
&   & \quad\quad\quad \quad \cdot
\left.\left.
        \left[\tilde{f}_{n+1}    \left(\kappa, \tetas_{D_{n+1};n} (\kappa) - \kappa d_{n+1}, 
                        \int_{b_{j-1}}^{a_j}\Gk (s;\kappa)\right)          
%%14            \, \cdot
%%14\right.´\right.\right.
%%14\nonumber \\
%%14&&
%%14 \quad\quad\quad \quad
%%14%\left.\left.\left.      
%%14      \cdot \left[1 + \GGGG_{n+1}(\kappa)\right]
%%14        \; -\;\left[1 + \GGGG_n(\kappa)\right]
      \right]
      \right)
       \right\}
        \dkappa.
  \nonumber \\
& & \left. \right.
\label{massvergleich}
\end{eqnarray}
Da $ \tilde{f}_{n+1}  $ sowohl im ersten als auch im dritten Argument von $ \kappa $ abh"angt,
wird das dritte Argument im Folgenden in der Notation unterdr"uckt.

Schiebt man folgenderma"sen Terme ein
\begin{eqnarray}
%\lefteqn{  
%\left[\tilde{f}_{n+1}    \left(\kappa, \tetas_{D_{n+1};n} (\kappa) - \kappa d_{n+1} \right)                
%      \cdot \left[1 + \GGGG_{n+1}(\kappa)\right]
%        \; -\;\left[1 + \GGGG_n(\kappa)\right]
%        \right]
%    }   
%%14\lefteqn{  
%%14\left[
\tilde{f}_{n+1}                    
%%14      \cdot \left[1 + \GGGG_{n+1}\right]
%%14        \; -\;\left[1 + \GGGG_n\right]
%%14        \right]
%%14    }   
%%14              \nonumber \\
& = &
%        \left[\tilde{f}_{n+1}    \left(\kappa, \tetas_{D_{n+1};n} (\kappa) - \kappa d_{n+1} \right)
%               - \overline{\tilde{f} \;}_{n+1}  (\kappa)      +         \overline{\tilde{f} \;}_{n+1}  (\kappa)   
%      \cdot \left[1 + \GGGG_{n+1}(\kappa)\right]
%        \; -\;\left[1 + \GGGG_n(\kappa)\right]
%        \right] Argumente weglassen
%%14   \left[
\tilde{f}_{n+1}    
               - \overline{\tilde{f} \;}_{n+1}    \;  + \;        \overline{\tilde{f} \;}_{n+1} ,
%%14                + \tilde{f}_{n+1}   \GGGG_{n+1}
%%14             -1 -
%%14        \GGGG_n\right]
%%14  \nonumber\\  
%%14& = &
%%14        \left[\tilde{f}_{n+1}    
%%14               - \overline{\tilde{f} \;}_{n+1}    \;  + \;        
%%14            \overline{\tilde{f} \;}_{n+1}  -1 \; + \,
%%14           \left( \tilde{f}_{n+1} -  \overline{\tilde{f} \;}_{n+1} \right)   \GGGG_{n+1}
%%14             \; + \;  
%%14\right.
%%14\nonumber \\
%%14& & 
%%14\quad +\left.
%%14\overline{\tilde{f} \;}_{n+1} \GGGG_{n+1}
%%14           -    \GGGG_n\right]
%%14\nonumber \\
%%14&&
 \nonumber
\end{eqnarray}
so kann man (\ref{massvergleich}) termweise absch"atzen:
Es gilt nach Satz \ref{Alemma1}
\begin{equation}
\modulus{\int\limits_\Sigma  \left( \prod_{j=1}^n                                                
        \ffjs     \right)        {\cal{D}} 
   \left[\tilde{f}_{n+1}          - \overline{\tilde{f} \;}_{n+1}  
   \right]
   }   
< \frac{1}{2}\eps
 \nonumber
\end{equation}
f"ur alle $ d_{n+1} \ge d_I $ f"ur gen"ugend gro"ses  $ d_I > 0 $.

Nach Hilfssatz \ref{Bteil1lemmafquerm} ist 
$   \overline{\tilde{f} \;}_{n+1}(\kappa;z)
 =  \exp\left(-2\int_{b_{n}}^{a_{n+1}} \Fk(s;\kappa)\ds\right)
$. Da nach Hilfssatz \ref{Bteil1lemmaprueferableitungen} 
$  \modulus{ \Fk(r;\kappa)}
  \le \frac{\Ckkappa}{r^2} \quad(r \ge 1, \kappa \in \R^*)
$
abgesch"atzt werden kann, ist
\begin{equation}
\modulus{-2\int_{b_{n}}^{a_{n+1}} \Fk(s;\kappa)\ds}
\le 
2 \Ckkappa \left( - \frac{1}{a_{n+1}} + \frac{1}{b_n} \right)
\le 2 \Ckkappa \frac{1}{b_n}.
 \nonumber
\end{equation}
Somit ist
\begin{eqnarray}
       \modulus{
        \overline{\tilde{f} \;}_{n+1} (\kappa) -1
     }
  & = &
  \modulus{
  \exp\left(-2\int_{b_{n}}^{a_{n+1}} \Fk(s;\kappa)\ds\right)
  \quad - \quad 1
  }
  \nonumber \\
& = &  \grossOglm\left( \frac{1}{b_n} \right) \quad (\kappa \in [\alpha,\beta]).
 \nonumber
\end{eqnarray}
Mit gen"ugend gro"sem $ b_n $ kann also erreicht werden, dass
\begin{equation}
\modulus{
\frac{1}{\pi}
   \int\limits_\Sigma         \left( \prod_{j=1}^n    \ffjs   \right)    {\cal{D}} 
   \left(\overline{\tilde{f} \;}_{n+1} (\kappa) -1 \right)
}
<  \frac{1}{2}\eps.
 \nonumber
\end{equation}

Durch gen"ugend gro"se Wahl %%14der $n$-ten Buckelbreite $ \alpha_n $ und 
des Abstandes $ d_{n+1} $ zum neuen 
$ (n+1) $-ten Buckel erh"alt man also
\begin{equation}
  \modulus{  \myDnpluss (\Sigma) - \myDns (\Sigma) } < \eps.
   \nonumber
\end{equation}
%
%erreicht werden.

}

%###############

%       Teil B          Lemma4.tex
 Im 
 folgenden Satz werden  endliche Buckelpotentiale betrachtet. Diese stimmen paarweise jeweils bis auf den letzten Buckel "uberein.
Die Aussage des Satzes ist, dass  sich die zugeh"origen Spektralma"se nur geringf"ugig unterscheiden, wenn der jeweils zus"atzliche Buckel weit drau"sen platziert wird.
% und erforderlichenfalls die Buckelbreite des Vorg"angerbuckels angepa"st wird.
Au"serdem kann die Position des jeweiligen zus"atzlichen Buckels dabei so gew"ahlt werden, dass das Spektralma"s auf einer Menge mit kleinem Lebesguema"s konzentriert ist.

Bei der Formulierung des Satzes wird bei der Notation der Ma"se zus"atzlich die hier zu ber"ucksichtigende Abh"angigkeit von der Drehimpulsquantenzahl $ k $ angegeben.

%%%%%
%%
\begin{satz}\label{hiermitkabha}
$\left.\right.$

Sei $ [-\beta,-\alpha] \cup [\alpha,\beta] \subset \R^* $, 
$ I \subset \R $ ein kompaktes Intervall,
$\nyunten \in \N_0 $, $d_1, \dots, d_\nyunten > 0 $,
$ K \subset \Z\setminus\{0\} $ endlich,
$ \eps, \gamma > 0 $. Dann gibt es $ \N \ni \nyoben > \nyunten $ und
\begin{equation}
  d_{\nyunten+1} \ge \exp( {\nyunten}^2), \dots, 
  d_\nyoben \ge \exp((\nyoben-1)^2) 
\label{Blemma4dgroesserexp}
\end{equation}
%
%und
%
%\begin{equation}
%  \tilde{\alpha}_{\nyunten+1} \ge 1, \dots, 
%  \tilde{\alpha}_\nyoben \ge 1 
%\label{Blemma4alphaschlangedgroessereins}
%\end{equation}
%%
mit:

F"ur jedes $ k \in K $ ist f"ur jedes 
Intervall $ \Sigma \subset \doppelInterv $:
\begin{equation}
   \modulus{\myDjsssplussk (\Sigma) - \myDjssssk(\Sigma) }
   < \gamma 2^{-(j+1)} \quad (j\in\{\nyunten, \dots, \nyoben-1\})
    \nonumber
\end{equation}
und f"ur jedes $ k \in K $ gilt
\begin{equation}
    \mynull \left( \left\{ \kappa\in [-\beta,-\alpha] \cup[\alpha,\beta] \Bigg|
      \frac{\mbox{d}\ros_{\nyoben;k}}{\mbox{d}\kappa}
      > \frac{\eps}{2(\beta-\alpha) }
    \right\} \right)
    < \eps.
     \nonumber
\end{equation}
\label{Blemma4}
\end{satz}
%%
%%%%%
%%%%%
\begin{bemerkung}
$\left.\right.$

 Die Bedingung 
(\ref{Blemma4dgroesserexp}) wird lediglich gestellt, um bei Satz \ref{keinpunktsp}
ausschlie"sen zu k"onnen, dass der Grenzoperator Punktspektrum in
$ (-\infty, -1) \cup (1, \infty) $ besitzt. 

\end{bemerkung}
%%
%%%%%
%------------
%%%%%
\Beweis{}{

Sei
\begin{equation}
  M:= \frac{1}{\pi} \sup_{\stackrel{\kappa \in \doppelInterv}{z\in I, k \in K}}
      \left(\prod_{j=1}^\nyunten
        \ffjsk \left(\kappa,\tetas_{%D_
        {\nyunten};j-1}(\kappa)-\kappa d_j,z \right) 
      \right) {\cal{D}}(\kappa) 
      %%14(1 + \GGGG_{\nyunten} (\kappa)),
\label{Blemma4defM}
\end{equation}

$ M $ ist endlich wegen %%14(\ref{AbschaetzungFehlerFgeschw}) und 
der Stetigkeit der $ \ffjsk $ $ (j \in \{1, ..., \nyunten\}) $.

Sei  
\begin{displaymath}
   \GGj(\kappa) := \int_{b_{j-1}}^{a_j}\Gk(\kappa, s)\ds 
      \quad(j\in \{1, \dots\nyunten\}).
\end{displaymath}
Die nat"urliche Zahl $ \nyoben > \nyunten $ kann aufgrund von 
Hilfssatz \ref{Summemschlange} so gro"s
gew"ahlt werden, dass f"ur alle 
$ \kappa \in \doppelInterv $
\begin{equation}
  \Najknynykappa :=
  \frac{1}{2} \sum_{j=\nyunten+1}^\nyoben \majs(\kappa, z) 
   < \log \left(\frac{1}{M} \frac{\eps}{2(\beta-\alpha)} \right) < 0
   \quad (z \in I) 
\label{Blemma4ref1}
\end{equation}
f"ur jede Wahl von $ \left(a_j\right)_{j\in \N} $ mit 
$ a_j \ge \exp((j-1)^2) $
gilt 
%%%%%
und
\begin{eqnarray*}
    \displaystyle
    \int\limits_\alpha^\beta 
         \frac{\dkappa}{ \Najknynykappa^2}
%%14    & 
< \displaystyle\frac{2\eps}{\KC} ,
%%14\\
    \displaystyle
    \quad\quad
    \int\limits_{-\beta}^{-\alpha}
     \frac{\dkappa}{ \Najknynykappa^2}  
%%14    & 
< \displaystyle\frac{2\eps}{\KC} 
\end{eqnarray*}
ausf"allt (wende Hilfssatz \ref{Summemschlange} f"ur jedes $ k \in K $ an und w"ahle maximales $ \nyoben $). 
%%%%%
Au"serdem soll mit $ C:= \max_{k\in K, \kappa \in [-\beta,-\alpha] \cup [\alpha, \beta]} \Ckkappa $
\begin{displaymath}
  \int_\alpha^\beta
  \frac{\exp\left(2 C \exp(\nyunten^2)\right) -1}
    { \Najknynykappa^2}
  \dkappa
  < \frac{\eps}{3 \KC}
\end{displaymath}
und entsprechendes bei Integration "uber $ [-\beta, -\alpha] $
gelten.
Dass dies f"ur ge\-n"ugend gro"ses $ \nyoben $ m"oglich ist, 
gew"ahrleistet ebenfalls Hilfssatz \ref{Summemschlange}.

Es sei $ \deltany := \nyoben - \nyunten $ definiert.

Im Folgenden werden schrittweise die Buckelabst"ande
$ d_j \; (j\in\{\nyunten+1, \dots, \nyoben\}) $ bestimmt.

{\bf 1. Schritt}

Sei $ k \in K $ beliebig.

Sei $ d_{\nyunten+1}^{(k)} \ge \exp (\nyunten^2) $.
Um die folgenden Bedingungen (\ref{Blemma4schritt10}),
(\ref{Blemma4schritt11}) 
und (\ref{Blemma4schritt12}) zu erf"ullen, mu"s
$ d_{\nyunten+1}^{(k)} $ gegebenfalls noch vergr"o"sert werden. 
Satz \ref{Blemma3} (angewendet auf die Intervalle $ [-\beta, -\alpha] $ und
$ \alphabeta $) gew"ahrleistet, dass
\begin{equation}
  \modulus{\mys_{%D
  {
  \nyunten+1;k}}(\Sigma)
   - \mys_{{\nyunten;k}}(\Sigma) 
  } < \gamma 2^{-(\nyunten+1)}
\label{Blemma4schritt10}
\end{equation}
f"ur beliebiges Intervall $ \Sigma \subset \doppelInterv $ 
erreicht werden kann durch gen"ugend gro"ses $ d_{\nyunten+1}^{(k)} $ (Abstand zum neuen Buckel).
% Buckelbreite gegebenenfalls verbreitern
%%14und $ \tilde{\alpha}_{\nyunten} $ (Verbreiterungsfaktor des $ \nyunten $-ten Buckels).
(Man beachte, dass (\ref{Blemma4schritt10}) bestehen bleibt, wenn man 
$ d_{\nyunten+1}^{(k)} $ gegebenfalls noch vergr"o"sert.)

Zus"atzlich soll $ d_{\nyunten+1}^{(k)} $ so gro"s sein, dass
\begin{equation}
\displaystyle 
\frac{1}{ \left[\log\left(\frac{\eps}{2M(\beta-\alpha)}\right)\right]^2 } \cdot
  \int \limits_{-\beta}^{-\alpha} \!\!
  {\left[ \hsk_{\nyunten+1} (\kappa,\tetas_{%D_\nyunten;
  \nyunten}(\kappa)
                 \!- \!\kappa d_{\nyunten+1}^{(k)} ,\GGnyunten(\kappa))
        \right]^2 \!\!\dkappa
       }
  < \frac{1}{8} \frac{1}{\deltany} \eps,
\label{Blemma4schritt11}
\end{equation}
%
%%%%%
was durch Anwendung von Hilfssatz \ref{Blemma2B} (mit $ j_0 = \nyunten+1 $ , 
$ \Theta(\kappa) = \tetas_{%D_\nyunten;
\nyunten} $ und $ L = d_{\nyunten+1}^{(k)} $)
gesichert werden kann. Dabei ist $  \hsk_{\nyunten+1} := \log \tilde{m}_{\nyunten+1}%^{(k)} $
$.
Die gleiche Bedingung soll auch bei Integration "uber das Intervall
$ \alphabeta $ erf"ullt sein. 
(Gem"a"s Hilfssatz \ref{Blemma2B} werden diese Forderungen auch noch erf"ullt, 
wenn 
$ d_{\nyunten+1}^{(k)} $ bei Bedarf weiter vergr"o"sert wird.)

Des weiteren werde $ d_{\nyunten+1}^{(k)} $ unter Umst"anden noch vergr"o"sert,
um verm"oge 
%\Verweis{Beinschub 2}
Hilfs\-satz \ref{Alemma1}
\begin{eqnarray}
\lefteqn{\frac{1}{ \left[\log\left(\frac{\eps}{2M(\beta-\alpha)}\right)\right]^2} \;\cdot}
\nonumber  \\
&& \cdot \modulus{
  \int\limits_{-\beta}^{-\alpha}
   { - 2 \msk_{\nyunten+1}(\kappa)
   \left( \hsk_{\nyunten+1} \left(\kappa, \tetas_{%D_\nyunten;
   \nyunten} (\kappa) 
                                       - \kappa d_{\nyunten+1}^{(k)} ,
                                          \GGnyunten(\kappa)\right)
          - \msk_{\nyunten+1} (\kappa)
   \right)
}{
         }
  \dkappa}
  \nonumber \\
& & 
\nonumber \\
&&
\quad \quad
  < \frac{1}{8} \frac{1}{\deltany} \eps
   \quad \quad \quad \quad 
\label{Blemma4schritt12}
\end{eqnarray}
und analog bei Integration "uber $ \alphabeta $ zu gew"ahrleisten.
($ H(\kappa,y,z) $ entspricht hier  
$ \hs_{\nyunten+1} (\kappa, \tetas_{%D_\nyunten;
\nyunten} (\kappa) - y,z) $
und $ F(\kappa) $ der Quotient
$  - 2 \msk_{\nyunten+1} (\kappa)
         \Nenner^{-2}
$).
%%%%%
Damit ist nun $ d_{\nyunten+1}^{(k)} $ bestimmt.

Setze $ d_{\nyunten+1} := \max_{k\in K} d_{\nyunten+1}^{(k)} $
($ d_{\nyunten+1} $ ist jetzt fest gew"ahlt und wird fortan nicht mehr
ver"andert). 
%Feb 13oder vergr"o"sert werden).

{\bf \boldmath
$ i $. Schritt (mit $ i \in \{2, \dots, \nyoben-\nyunten\} $)}

Zun"achst wird wieder ein festes $ k \in K $ betrachtet.

Sei $ d_{\nyunten+i}^{(k)} \ge \exp ( (\nyunten + i - 1) ^2) $.

Um die folgenden Bedingungen (\ref{Blemma4schrittk0}),
(\ref{Blemma4schrittk1}), (\ref{Blemma4schrittk2}) 
und (\ref{Blemma4schrittk3}) zu erf"ullen, mu"s
$ d_{\nyunten+i}^{(k)} $ gegebenfalls noch vergr"o"sert werden. 
Aufgrund von Satz \ref{Blemma3} (angewendet auf die Intervalle 
$ [-\beta,-\alpha] $ und $ \alphabeta $ 
% andere Notation in Lemma 4
%mit $d_{\nyunten+i}^{(k)} $ statt
%$ d $, $ \tetas_{D_{\nyunten+i-1};\nyunten+i-1} $ an Stelle von $ \Theta $ und
%$ \mys_{D_{\nyunten+i}} $
%f"ur $ \mys_{D_n}^{(d,\Theta)} $) 
kann durch eine gen"ugend gro"se
Wahl von $ d_{\nyunten+i}^{(k)} $
%%14 und $ \tilde{\alpha}_{\nyunten+i-1} $ 
gew"ahrleistet werden, dass
\begin{equation}
  \modulus{\mys_{{\nyunten+i;k}}(\Sigma)
   - \mys_{D{\nyunten+i-1;k}}(\Sigma) 
  } < \gamma 2^{-(\nyunten+i)}
\label{Blemma4schrittk0}
\end{equation}
gilt.

Au"serdem sei $ d_{\nyunten+i}^{(k)} $ so gro"s, dass sowohl
\begin{equation}
\displaystyle 
  \int \limits_{-\beta}^{-\alpha} 
 \frac{
  \left[ \hsk_{\nyunten+i} 
          \left(\kappa,\tetas_{%D_{\nyunten+i-1};
          \nyunten+i-1}(\kappa)
                 - \kappa d_{\nyunten+i}^{(k)} ,\GGnyuime(\kappa)
          \right)
        \right]^2
 }{
          \displaystyle 
  \left[ \log\left(\frac{\eps}{2M(\beta-\alpha)}\right)\right]^2
 }
  < \frac{1}{8} \frac{1}{\deltany} \eps
\label{Blemma4schrittk1}
\end{equation}
als auch die analoge Absch"atzung f"ur das Integral "uber $ \alphabeta $
gilt, die Hilfssatz~\ref{Blemma2B} (mit $ j_0 = \nyunten + i $, 
$ \Theta (\kappa) = \tetas_{%D_{\nyunten+i-1};
\nyunten+i-1} $ und
$ L = d_{\nyunten+i}^{(k)} $ erm"oglicht.

Zus"atzlich werde $d_{\nyunten+i}^{(k)} $ bei Bedarf noch vergr"o"sert, um 
verm"oge 
%Korollar \ref{AKorollarzuLemma1} zu 
Hilfssatz \ref{Alemma1}
(mit  $ G(\kappa, y) = 
h_{\nyunten+i} (\kappa, \tetas_{%D_{\nyunten+ i-1};
\nyunten+i-1}(\kappa) - y)
$
und
$ F(\kappa) =    
    - \frac{2\msk_{\nyunten+i}}{\sum_{j=\nyunten+1}^{\nyoben} \mjs (\kappa)}
$)
\begin{eqnarray}
  \int\limits_{-\beta}^{-\alpha}\!
   \frac{ - 2 \msk_{\nyunten+i}(\kappa)\!
   \left(\! \hsk_{\nyunten+i} 
               \left(\kappa, \tetas_{%D_{\nyunten+i-1};
               \nyunten+i-1}\! (\kappa) 
                              \!         -  \! \kappa d_{\nyunten+i}^{(k)} ,
                                         \GGnyuime  \!(\kappa)\!\right)
      \!    -   \!\msk_{\nyunten+i} (\kappa) \!
   \right)
   }
   { \left[ \log\left(\frac{\eps}{2M(\beta-\alpha)}\right)\right]^2
   }\!\!
  \dkappa
%%14& & 
%%14\nonumber \\
  < \!\frac{\eps}{8} \!\frac{1}{\deltany} 
 %%14  \quad \quad \quad \quad 
 \nonumber \\
& &
\label{Blemma4schrittk2}
\end{eqnarray}
(und entsprechend auch bei Integration "uber $ \alphabeta $) sicherzustellen.

"Uberdies m"oge $ d_{\nyunten+i}^{(k)} $ so gro"s sein, dass die folgenden 
$ 2 \/(i-2) $ Bedingungen erf"ullt sind. F"ur 
$ s \in \{2, \dots, i-1\} $ gelte n"amlich:
\begin{eqnarray}
\lefteqn{
  \int\limits_{-\beta}^{-\alpha}
   \left( \hsk_{\nyunten+i}
                 \left(\kappa, \tetas_{%D_{\nyunten+i-1};
                 \nyunten+i-1} (\kappa) 
                                         - \kappa d_{\nyunten+i}^{(k)} ,
                                          \GGnyuime(\kappa)\right)
     - \msk_{\nyunten+i}(\kappa) 
   \right) 
  \cdot
}
  \nonumber \\
  & & \cdot
   \left( \hsk_{\nyunten+s} 
               \left(\kappa, \tetas_{%D_{\nyunten+s-1};
               \nyunten+s-1} (\kappa) 
                                       - \kappa d_{\nyunten+s}^{(k)} ,
                                        \GGnyusme(\kappa\right))
          - \msk_{\nyunten+s} (\kappa)
   \right) 
   %\cdot
%\nonumber \\
%& & \quad \cdot
         \Nenner^{-2}
%  \;
\!
  \dkappa
  \nonumber \\
%  & &   \nonumber \\
  & < &
  \frac{1}{8}\frac{1}{i-1} \frac{1}{\deltany} \eps.
     \quad \quad \quad \quad 
\label{Blemma4schrittk3}
\end{eqnarray}
und entsprechend bei Integration "uber $ \alphabeta $.
Hierbei wird Korollar \ref{AKorollarzuLemma1} $2(i-2) $-mal angewendet mit
$ H(\kappa,y) = \hsk_{\nyunten+i}(\kappa, \tetas_{D_{\nyunten+i-1},\nyunten+i-1}
                                - y ) $
und 
\begin{displaymath}
   F = 
   \left( \hsk_{\nyunten+s} 
               (\cdot, \tetas_{%D_{\nyunten+s-1};
               \nyunten+s-1} (\cdot) 
                                       - \cdot\; d_{\nyunten+s}^{(k)} ,
                                         \GGnyusme(\kappa))
          - \msk_{\nyunten+s} 
   \right) \cdot
         \Nenner^{-2}.
\end{displaymath}
Setze $ d_{\nyunten+i} := \max_{k\in K} d_{\nyunten+i}^{(k)} $.
Sei
\begin{displaymath}
  \GGnyui(\kappa) := \int_{b_{\nyunten+i-1}}^{a_\nyunten+i} \Gk(\kappa,s)\ds
  \quad (\kappa \in \R^*).
\end{displaymath}
Aufgrund von (\ref{Blemma4schritt10}) und (\ref{Blemma4schrittk0}) ist der
erste Teil der Behauptung des Satzes gezeigt.

Als Vorbereitung zum Beweis des zweiten Teils werden folgende
Absch"atzungen durchgef"uhrt:

Es sei $ \kappa \in \doppelInterv $.

\begin{hilfs}
$\left.\right.$

Unter den Voraussetzungen des Satzes gilt f"ur die oben bestimmten 
%%%%% 
$ d_j \quad (j\in\{\nyunten+1, \dots \nyoben\}) $ unabh"angig von $ k \in K $
\begin{equation}
 \mynull \left( \left\{ \frac{\mbox{d}\roDnyobens}{\dkappa} 
                        > \frac{\eps}{2(\beta-\alpha)}\right\} \right)
 < \eps.
  \nonumber
\end{equation}
\end{hilfs}
%%
%%%%
\Beweis{}{

Aus 
\begin{displaymath}
  \frac{1}{\pi} 
   \left( \prod_{j=1}^\nyoben 
     \ffjs(\kappa,\tetas_{%D_{j-1};
     j-1}(\kappa) - \kappa d_j 
              ,\GGjme(\kappa))
   \right)
   {\cal{D}} ( \kappa)
   %%14\GGGG_\nyoben(\kappa)
   > 
   \frac{\eps}{2(\beta-\alpha)}
\end{displaymath}
folgt gem"a"s Definition (\ref{Blemma4defM}) von $ M $
\begin{displaymath}
   M \left( \prod_{j=\nyunten+1}^\nyoben 
       \ffjs(\kappa,\tetas_{%D_{j-1};
       j-1}(\kappa) - \kappa d_j 
                 ,\GGjme(\kappa))
     \right)
   > 
   \frac{\eps}{2(\beta-\alpha)},
\end{displaymath}
woraus durch Logarithmieren
\begin{displaymath}
   \log \left( \prod_{j=\nyunten+1}^\nyoben 
       \ffjs(\kappa,\tetas_{%D_{j-1};
       j-1}(\kappa) - \kappa d_j 
               ,\GGjme(\kappa))
     \right)
   > 
   \log \left(\frac{\eps}{2M(\beta-\alpha)}\right)
\end{displaymath}
gefolgert werden kann.
Dann gilt auch mit (\ref{defhjnullschl}), wenn man
(\ref{Blemma4ref1}) verwendet: 
\begin{displaymath}
   \sum_{j=\nyunten+1}^\nyoben 
       \hjs(\kappa,\tetas_{%D_{j-1};
       j-1}(\kappa) - \kappa d_j 
            ,\GGjme(\kappa))
   > 
   \Nenner
\end{displaymath}
Daraus folgt (man subtrahiere $ 2 \Nenner $):
\begin{equation}
%\lefteqn{
   \sum_{j=\nyunten+1}^\nyoben \!\!
    \! \left(\!
       \hjs(\!\kappa\!,\!\tetas_{%D_{j-1};
       j-1}(\!\kappa\!)\! -\! \kappa d_j 
\!            ,\!\GGjme(\!\kappa\!))
      \! -\!\mjs(\!\kappa,\GGjme(\!\kappa\!))\!
     \right) \!
% }    
% \nonumber \\
% & &
   > \!
   - \frac{1}{2} \!\!\sum_{j=\nyunten+1}^\nyoben \!\!\!\mjs (\!\kappa,\!\GGjme(\!\kappa\!)\!)
%   \quad\quad\quad\quad\quad\quad\quad\quad\quad\quad\quad
\end{equation}
Wegen (\ref{Blemma4ref1})

 ist die rechte Seite dieser
Ungleichung positiv, so dass Division durch 

$ - \frac{1}{2} \sum_{j=\nyunten+1}^\nyoben \mjs (\kappa,\GGjme(\kappa)) 
 =-   \Najknynykappa  > 0 $ 
\begin{displaymath}
  \frac{
   \displaystyle
   \sum_{j=\nyunten+1}^\nyoben 
     \left(
       \hjs(\kappa,\tetas_{%D_{j-1};
       j-1}(\kappa) - \kappa d_j 
            ,\GGjme(\kappa))
       -\mjs(\kappa,\GGjme(\kappa))
     \right)
  }
  { \displaystyle
    -   \Najknynykappa }
  > 1
\end{displaymath}
liefert und man schlie"slich 
\begin{eqnarray}
%%14\lefteqn{ 
 H(\kappa) :=
 %%14}
%%14\nonumber \\
%%14& &
  \left(\!
  \frac{
   \displaystyle
   \sum_{j=\nyunten+1}^\nyoben 
     \left(
       \hjs(\kappa,\tetas_{%D_{j-1};
       j-1}(\kappa) - \kappa d_j 
            ,\GGjme(\kappa))
       -\mjs(\kappa,\GGjme(\kappa))
     \right)
  }
  { \displaystyle
      \frac{1}{2} \sum_{j=\nyunten+1}^\nyoben \mjs (\kappa,\GGjme(\kappa)) }
\!  \right)^2
  \!>\! 1
  \nonumber
\end{eqnarray}
hat.
Es gilt also
\begin{eqnarray*}
\lefteqn{
  \mynull\Bigg(\Bigg\{ \kappa \in \doppelInterv \Bigg|
}
  \nonumber \\
  & &
  \quad\quad\quad \left.\left.
  \frac{1}{\pi} 
   \left( \prod_{j=1}^\nyoben 
     \ffjs(\kappa,\tetas_{%D_{j-1};
     j-1}(\kappa) - \kappa d_j 
          ,\GGjme(\kappa))
   \right)
   {\cal{D}}(\kappa)
   > 
   \frac{\eps}{2(\beta-\alpha)}
   \right\} \right)
%}
\nonumber \\
\end{eqnarray*}
\begin{eqnarray*}
 &  & \le
  \mynull \left( \left\{ \kappa \in \doppelInterv \Bigg|
        \right.\right.
 \nonumber \\
 & & \quad\quad\;
        \left.\left.
  \left(\!
  \frac{
%   \displaystyle
   \sum_{j=\nyunten+1}^\nyoben \!
     \left(
       \hjs(\kappa,\tetas_{%D_{j-1};
       j-1}(\kappa)\! - \!\kappa d_j
           ,\GGjme(\kappa) )
      \! -\! \mjs(\kappa,\GGjme(\kappa))
     \right)
  }
  {
%    \displaystyle
      \frac{1}{2} \sum_{j=\nyunten+1}^\nyoben \mjs (\kappa,\GGjme(\kappa)) }\!
  \right)^{\!2}
 \!> \!1
  \right\} \!\right).
\nonumber \\
\end{eqnarray*}
Da $ H(\kappa) \ge 0 \quad (\kappa \in \doppelInterv) $ ist, gilt somit
\begin{eqnarray}
  \mynull(\{ H(\!\kappa\!)\! > \!1\} ) 
%%14  & = &
=
  \int\limits_{\{H(\kappa)  > 1\}}\!\! 1  \le  
  \int\limits_{\{H(\kappa) > 1\}}\!\! H(\kappa) \dkappa 
%%14\nonumber \\
%%14  & \le & 
=
  \int\limits_\doppelInterv \!\!H(\kappa) \dkappa.
\label{Blemma4refH}
\end{eqnarray}
Wegen 
\begin{eqnarray*}
\lefteqn{
     -2 \mjs (\kappa,\GGjme(\kappa)) 
         \hjs (\kappa,\tetas_{%D_{j-1};
         j-1} (\kappa) - \kappa d_j
               ,\GGjme(\kappa))
     + \mjs^2 (\kappa,\GGjme(\kappa))
}
\nonumber \\
   & \le &
     -2 \mjs (\kappa,\GGjme(\kappa)) 
      \left(
         \hjs (\kappa,\tetas_{%D_{j-1};
         j-1} (\kappa) - \kappa d_j
               ,\GGjme(\kappa))
         - \mjs (\kappa,\GGjme(\kappa))
      \right)
\nonumber      
\end{eqnarray*}
ist dann f"ur  $ \kappa \in \doppelInterv $:
\begin{eqnarray*}
\lefteqn{
  \sum_{j=\nyunten+1}^\nyoben
  \left( 
   \hjs(\kappa,\tetas_{%D_{j-1};
   j-1} (\kappa) - \kappa d_j
        ,\GGjme(\kappa))
   -\mjs(\kappa, ,\GGjme(\kappa))
  \right)^2
%  \le&&
}
%\end{eqnarray*}
%\begin{eqnarray*}
\nonumber \\  
& \le & 
  \sum_{j=\nyunten+1}^\nyoben
    \hjs^2 (\kappa,\tetas_{%D_{j-1};
    j-1} (\kappa) - \kappa d_j
            ,\GGjme(\kappa))
  \nonumber \\
 & & +
  \sum_{j=\nyunten+1}^\nyoben
   \Big( 
     -2 \mjs (\kappa,\GGjme(\kappa)) \cdot
  \nonumber \\   
   & &
   \quad \quad \cdot
   \left.
       \left(
         \hjs (\kappa,\tetas_{%D_{j-1};
         j-1} (\kappa) - \kappa d_j
               ,\GGjme(\kappa))
         - \mjs (\kappa,\GGjme(\kappa))
       \right)
   \right)
  \nonumber \\
  & & 
  + 
  \sum_{j=\nyunten+1}^\nyoben
  \sum_{i=\nyunten+1}^\nyoben
     \left(
         \hjs (\kappa,\tetas_{%D_{j-1};
         j-1} (\kappa) - \kappa d_j
               ,\GGjme(\kappa))
         - \mjs (\kappa,\GGjme(\kappa))
     \right)
     \cdot
 \nonumber \\
 & & \quad \quad \cdot
     \left(
         \hsk_i (\kappa,\tetas_{%D_{i-1};
         i-1} (\kappa) - \kappa d_i
                 ,\GGime(\kappa))
         - \msk_i (\kappa,\GGime(\kappa))
     \right).
\nonumber \\
\end{eqnarray*}
%
%%%%
Diese Absch"atzung f"ur den in (\ref{Blemma4refH}) auftretenden Z"ahler
erm"oglicht die Anwendung von (\ref{Blemma4schritt11}),
(\ref{Blemma4schritt12}),
(\ref{Blemma4schrittk1}),
(\ref{Blemma4schrittk2}) und
(\ref{Blemma4schrittk3}),
%f"ur das Ergebnis
so dass man schlie"slich 
\begin{displaymath}
  \mynull \left( \left\{ \kappa \in \doppelInterv \big|
  H(\kappa) > 1
  \right\} \right)
  < \eps
\end{displaymath}
erh"alt.
}
%%%%
Mit dem Beweis des Lemmas ist der Beweis des Satzes abgeschlossen.
%%%%
}  %   Ende Beweis

%Satz

\subsection{Singul"arstetiges Grenzma"s
%,,Beweis des zentralen Satzes im Fall mit Dreh''}
\label{mitdrehsinggrenzmass}}

Dass auch in der Situation mit Drehimpulstermen eine geeignete Wahl der Buckelabst"ande zu einem singul"arstetigen Grenzma"s f"uhrt, ist Inhalt des folgenden Satzes (seine drehimpulsfreie Entsprechung ist Satz \ref{Ateil2zentralerSatz}):

%---------
%%
\begin{satz}
\label{BzentralerSatzmitDreh}

$\left.\right.$

Seien Folgen $ \left(H_j\right)_{j \in \N} \in R^\N $ und
$ \left(\alpha_j\right)_{j\in \N} \in (\R^+)^\N $ und zu letzteren Funktionen
$ W_j \in 
%\Leinsloc[0,\alpha_j]
C[0,\alpha_j] $  gegeben mit folgenden
Eigenschaften:

Es gelte
$ H_j = 0 $ nur f"ur endlich viele Indizes $ j \in \N $,
$ H_j \to 0 \quad (j \to \infty) $ und
\begin{equation}
\sum_{j=1}^\infty  
%Feb 13 \frac{
H_j^2
% Feb 13 }{2}
{\cal W}_j 
  = \infty,
\label{summemjschl}
\end{equation}
wobei $ {\cal W}_j$  wie in (\ref{zweiteOrdAinH})
definiert ist.

Dann gibt es eine Folge $ D := (d_j)_{j\in \N} \in \left(\R^+\right)^\N
$ %????
 mit
der Eigenschaft, dass f"ur jedes kompakte Intervall
$ \Sigma \subset \R^* $ 
%% neu
jedes $ k \in \Z \setminus \{0\} $\label{grenzmassmitschlange}
\begin{equation}
 \lim_{j \to \infty} \myDjsss (\Sigma) =: \myDss (\Sigma)
  \nonumber
\end{equation}
existiert und ein singul"arstetiges Ma"s auf $ R^* $ definiert.
\end{satz}

Im folgenden Beweis des Satzes \ref{BzentralerSatzmitDreh} wird gezeigt,
dass zu vorgegebenen Buckelprofilen, f"ur welche die Bedingung
(\ref{summemjschl}) erf"ullt ist, die Abst"ande $ d_j \; (j \in \N) $ zwischen
den Buckeln so bestimmt werden k"onnen, dass die zugeh"origen Ma"se
(\ref{Bteil2defmyDn}) in der Grenze ein singul"arstetiges Ma"s definieren. 

Vor dem eigentlichen Induktionsbeweis sei hier das Vorgehen skizziert:

In jedem Schritt wird ausgehend von einem endlichen Buckelpotential  eine Reihe von zus"atzlichen Buckeln
hinzugenommen, wobei die Abst"ande zwischen den neuen Buckeln gewisse Bedingungen erf"ullen m"ussen,
 um Folgendes zu erreichen:

Die neu gewonnenen Ma"se der Potentiale mit zus"atzlichen Buckeln unterscheiden sich kaum von den Vorg"angerma"sen (siehe (\ref{Bsatzschritt1})).
Dies wird durch einen gen"ugend gro"sen Abstand zum jeweiligen Vorg"angerbuckel  bewirkt.

Zus"atzlich ist bei der Wahl der neuen Buckelabst"ande darauf zu achten, dass der Tr"ager des zugeh"origen neuen
Ma"ses kleines Lebesguema"s hat (siehe (\ref{Bsatzschritt1S}), (\ref{Bsatzschritt1mynull}),% und
(\ref{BsatzschrittnS}), (\ref{Bsatzschrittnmynull})), was  die Singularit"at des Grenzma"ses gew"ahrleistet.

Gem"a"s Konstruktion "ubertr"agt sich die Stetigkeit der
Ma"se $ \myDjsss \, (n \in \N ) $ auf das Grenzma"s $ \tilde{\mu}_D $.

Eine Mindestanwachsrate der Buckelabst"ande sorgt daf"ur, dass der Grenzoperator kein Punktspektrum besitzt.

\Beweis{von Satz \ref{BzentralerSatzmitDreh}} {

Sei $ \Xi_n := [-2n,-\frac{1}{2n}] \cup [\frac{1}{2n},2n] \quad (n\in \N) $. Diese Intervallschachtelung sch"opft 
% Feb 13 $ (-\infty,-1) \cup (1,\infty) $ 
$ \R \setminus \{0\} $ aus
(vgl. Abbildung \ref{Xis}).

$ (\eps_n)_{n\in\N} $ sei eine streng monotone Nullfolge mit
$ \sum_{n\in\N} \eps_n < \infty $.

Der Beweis wird durch vollst"andige Induktion nach $ \nu_n $ $ (n \in \N) $ gef"uhrt.

1. Schritt:

Nach Satz \ref{Blemma4}
 gibt es zu $ \nyunten = \nu_0 := 0 $, $ \eps = \eps_1 $,
$ \gamma = \eps_1 $ und $ \alpha=\frac{1}{2} $, $ \beta=2 $ eine
nat"urliche Zahl $ \nu_1 $ und positive Zahlen
\begin{equation}
d_1 \ge \exp( 0^2), \dots, d_{\nu_1} \ge \exp ((\nu_1-1)^2) 
\label{growsserexp}
\end{equation}

so, dass
f"ur jedes kompakte Intervall $ \Sigma \subset \Xi_1 $
% neu
 und alle $ k \in K_1 := \{-1\} \cup  \{-1\} $

\begin{equation}
  \modulus{\myDjssspluss(\Sigma) -\myDjssss(\Sigma) }
  < \eps_1 2^{-(j+1)}
  \quad (j\in\{0, \dots, \nu_1-1\})
\label{Bsatzschritt1}
\end{equation}
und
\begin{equation}
    \mynull\left( \Seschl \right)
    < \eps_2
\label{Bsatzschritt1mynull}
\end{equation}
gilt, 
 wobei die Menge $ \Seschl $ wie folgt definiert ist:
\begin{eqnarray}
%%14\lefteqn{ 
  \Seschl :=%%14}
%%14\nonumber \\
%%14&&
     \left\{ \kappa\in \Xi_1 \Bigg|
      \frac{1}{\pi}
      \left(\prod_{j=1}^{\nu_1}
        \tilde{\ffj }(\kappa, \teta_{%D_{j-1};
        j-1} (\kappa) - \kappa d_j )
      \right) {\cal{D}}(\kappa)
%%14\left(1 + \GGGG_{\nu_1}(\kappa) \right)
      > \frac{\eps_1}{\mynull(\Xi_1)}
    \right\}
%%14\nonumber \\
%%14&&
\label{Bsatzschritt1S}
\end{eqnarray}
%
%definiert ist.

$ \Seschl $  besteht aus endlich vielen Intervallen (vgl. die Argumentation im Beweis von Satz \ref{Ateil2zentralerSatz} auf S. \pageref{nullstellenmenge}).
Die Anzahl der Intervalle, aus denen $ \Seschl $ besteht, sei mit $ s_1 $
bezeichnet und $ r_1 := s_1 $ definiert.

$ n $. Schritt:

Satz \ref{Blemma4} gew"ahrleistet, dass es zu
$ \nyunten = \nu_{n-1} $, $ \eps = \eps_n $,
$ \gamma = \frac{\eps_{n-1}}{r_{n-1}} $ und
$ \alpha=\frac{1}{2n} $, $ \beta=2n $ eine
nat"urliche Zahl $ \nu_n > \nu_{n-1} $ und positive Zahlen
\begin{equation}
d_{\nu_{n-1}+1} \ge \exp( \nu_{n-1}^2), \dots,
  d_{\nu_n} \ge \exp ((\nu_n-1)^2) 
\label{grtexp}
\end{equation}
so gibt, dass
f"ur jedes kompakte Intervall $ \Sigma \subset \Xi_n $
% neu
und jedes $ k \in K_n := \{-n, \dots, -1\} \cup \{1, \dots, n\} $ 
\begin{equation}
  \modulus{\myDjssspluss(\Sigma) - \myDjssss(\Sigma) } 
  < \frac{\eps_{n-1}}{r_{n-1}} 2^{-(j+1)}
  \quad (j\in\{\nu_{n-1}, \dots, \nu_n-1\})
\label{Bsatzschrittn}
\end{equation}
und

\begin{equation}
    \mynull\left( \Snschl \right)
    < \eps_n
\label{Bsatzschrittnmynull}
\end{equation}
ist, wobei $ \Snschl $  folgenderma"sen definiert ist:
\begin{eqnarray}
%%14\lefteqn{  
 \Snschl :=
 %%14}
%%14\nonumber \\
%%14&&
     \left\{ \kappa\in \Xi_n \Bigg|
      \frac{1}{\pi}
      \left(\prod_{j=1}^{\nu_n}
       \tilde{ \ffj} (\kappa, \teta_{%D_{j-1};
       j-1} (\kappa) - \kappa d )
      \right) {\cal{D}}(\kappa)
%%14\left(1 + \GGGG_{\nu_n}(\kappa) \right)
      > \frac{\eps_n}{\mynull(\Xi_n)}
    \right\}
%%14\nonumber \\
%%14&&
\label{BsatzschrittnS}
\end{eqnarray}
%
%definiert ist.

"Ahnlich wie $ \Seschl $ besteht $ \Snschl $ ebenfalls aus einer endlichen Anzahl $ s_n $ von Intervallen.
Es sei $ r_n := \max_{j\in\{1, \dots,n\}} s_j $.

Dass die auf diese Weise induktiv 
bestimmte Folge $ D := (d_j)_{j\in\N} $
die gew"unschten Eigenschaften besitzt, wird im Folgenden gezeigt.

%+++++++++++++
%
Sei dazu $ \Sigma \subset \R^* $ ein kompaktes Intervall.

Dann gibt es $ \tilde{n} \in \N $ so, dass
$ \Sigma \subset \Xi_n \quad (n \ge \tilde{n}) $
ist.

%\subsection{
{\bf 1.Existenz des Grenzwertes \boldmath
$ \displaystyle\lim_{j \to \infty} \myDjssss (\Sigma) $}

Der Nachweis erfolgt entsprechend dem Beweis von Satz \ref{Ateil2zentralerSatz} im Abschnitt \ref{BeweiszentralerSatz}.
Der Grenzwert sei bezeichnet mit
\begin{equation}
 \lim_{j \to \infty} \myDjssss  (\Sigma) =: \myDs (\Sigma).
\label{BsatzdefGrenzma}
\end{equation}
%
%%14existiert.

%\subsection{
{\bf 2.Stetigkeit von \boldmath $ \myDs $}

Auch dieser Beweisschritt kann Eins zu Eins  von der entsprechenden Passage beim Beweis von Satz \ref{Ateil2zentralerSatz} "ubernommen werden.

%\subsection{
{\bf 3. Singularit"at von \boldmath $ \myDs $}

Die Singularit"at des Grenzma"ses ergibt sich aufgrund der Konstruktion analog wie im Beweis von Satz  \ref{Ateil2zentralerSatz}.

%\subsection{
{\bf 4. {\boldmath $ \myDs$ }ist auf $ \cal{S} $ konzentriert}

Ebenfalls wie im Beweis des Satzes  \ref{Ateil2zentralerSatz} schlie"st man, dass  $ \myDs$ ist auf $ \cal{S} $ konzentriert ist.

}

%operat(oren)
\subsection{Spektrale Eigenschaften des Grenzoperators $ \TqDschl $ 
\label{BOperatoren}
}

%++++++++++
%  Achtung aus ohne Dreh herauskopiert
%+++++++++++++++++

Dass  das im vorangegangenem Abschnitt ermittelte singul"arstetige Grenzma"s das Spektralma"s eines Differerentialoperators ist, ist Gegenstand dieses Abschnitts.
Hierzu werde zun"achst ein Differentialoperator definiert, von dem im Anschlu"s gezeigt wird, dass er die geforderten Eigenschaften aufweist.

Das Grenzma"s aus Abschnitt \ref{mitdrehsinggrenzmass}
ist gegeben durch die Folge von Buckelprofilen 
$ \left(W_j\right)_{j\in\N} \in 
C[0,\alpha_j]^\N $,   die vorgegebenen Buckelh"ohen  $ \left(H_j\right)_{j \in \N} \in R^\N $ 
%sowie durch  die nach Satz \ref{BzentralerSatzmitDreh} modifizierten Buckelbreiten $   (\tilde{\alpha}_j)_{j\in \N} \in [1,\infty)^\N $ 
und durch die gem"a"s Satz \ref{BzentralerSatzmitDreh} bestimmte  Folge der Buckelabst"ande
 $ D := \left(d_j\right)_{j\in \N} $.

Mit diesen Parametern wird folgendes Potential definiert 
\begin{equation}
   q%_{D}
 =  \left\{
                \begin{array}{cl}
                  0                                                                     & 1 \le r \le a_1          \\
                  %\frac{1}{\tilde{\alpha}_n} 
                  H_n W_n(%\frac{1}{%\tilde{\alpha_n}} 
                  r -a_n)  & a_i < r <b_i
                  %a_i+\alpha_i\tilde{\alpha}_i 
                  \quad (i \in \N) \\
                  0                                                                      & b_i
                  %a_i+\alpha_i\tilde{\alpha}_i  
                  \le r < a_{i+1} \quad (i \in \N) \\
                \end{array}
               \right.
\label{satzqDmitDreh}
\end{equation}
mit
\begin{eqnarray*}
   a_i & := & %\sum_{j=1}^{i-1} (d_j + \alpha_j%\tilde{\alpha}_j) 
%   + d_i
   b_{i-1} + d_i
                       \quad (i \in \{ 1, \dots, n\}) \\
   b_i & := & a_i + \alpha_i %\tilde{\alpha}_i 
   \quad (i \in \{ 1, \dots, n\}).
\end{eqnarray*}
%
%%130310 definiert.

Zu diesem Potential 
sei der Differentialausdruck
\begin{equation}
  \tauschlqD := %\sigma_2 p + \sigma_3 + q_D
  %\hqdn  := 
  \sigma_2 p + m \sigma_3 + l + q%_{D_n}
\label{OperatorenDifferentialausdruck}
\end{equation}
mit $ p := -i \dnachdr $ definiert.
%%140310 
Der zu $\tauschlqD $ geh"orige maximale Operator mit Defini\-tions\-bereich
\begin{displaymath}
  \left\{ %\Psischl 
  u \in \Lzwei\left( [1,\infty)\right)^2 \Big|
             %\Psischl 
             u\in \ACloc\left( [1,\infty)\right)^2,
             \tauschlqD 
             %\Psischl 
             u\in \Lzwei \left( [1,\infty)^2\right)^2
  \right\}
\end{displaymath}
sei mit $ \TqDschlmax $ \label{TqDschlmax}bezeichnet (vgl. \cite{WeidmannmathZ}).

$ \TqDschlmin $ \label{TqDschlmin}sei der entsprechende minimale Operator mit Definitionsbereich

\begin{displaymath}
  D(\TqDschlmin) =
   \left\{ u %\Psischl
    \in D(\TqDschlmax) \Big | \supp u %\Psischl 
    \komp 
\R_0^+
(1,\infty)
   \right\}.
\end{displaymath}

$ \TqDschl $ \label{TqDsch} bezeichne eine selbstadjungierte Fortsetzung von $ \TqDschlmin $.
\label{Tschl}

Die spektralen Eigenschaften von $  \TqDschl  $ werden mit denen des 
freien Dirac\-operators $ \Tnull $
verglichen. \label{freierDiracaud1infty}

$\left.\right. $

%%%%%
\subsubsection{ Das wesentliche Spektrum von $\TqDschl$\label{wesspektrum}}
%%%%%%%
%%
\begin{hilfs}

$\left.\right. $

Es ist
\begin{equation}
 \sigmaess(\TqDschl) = \sigmaess (\Tnull) = (-\infty,-1] \cup [1,\infty).
  \nonumber
\end{equation}
\end{hilfs}

\Beweis{}{

Die Argumentation erfolgt "ahnlich wie beim Beweis von Hilfssatz \ref{HIlfssatzessspektrum} mit Hilfe von  \cite{WEIDMANN76}, Satz 7.24 und liefert

$
   \sigmaess (\TqDschl) \supset \R \setminus  (-1,1).
   $
%%140310 \end{displaymath}
%gezeigt.

Au"serdem ist nach {\rm \cite{WeidmannmathZ}},
Kor. 6.9., $
   \sigmaess(\TqDschl) \cap (-1,1) = \emptyset.
   $%%140310 \footnote{

} % ENDE BEWEIS Lemma

\begin{bemerkung}

$\left.\right.$

Beim Beweis von $ \sigmaess(\TqDschl) \supset (-\infty,-1]\cup[1,\infty) $ wurde lediglich die Eigenschaft
$ \limsup_{j\to \infty} d_j = \infty $ des Potentials $ q%_D 
$ verwendet. Andere Eigenschaften wie die Buckelh"ohen oder Buckelformen wurden f"ur den Beweis nicht ben"otigt.

\end{bemerkung}

$\left.\right.$ 

%%%%%%%
%\subsubsection{
\subsubsection{ Zum Punktspektrum von $ \TqDschl $
\label{AAbschnittPunktspektrummitDreh}}

Es gilt der folgende 
\begin{hilfs}
\label{keinpunktsp}

$\left.\right.$

$ \TqDschl $
besitzt keine Eigenwerte in $ (-\infty,-1)\cup (1, \infty) $.
\end{hilfs}

$\left.\right.$
%5
\begin{bemerkung}

$\left.\right.$

Die Aussage des Hilfssatzes ist bereits unter der alleinigen Voraussetzung
$ d_j \ge \exp\left( (j-1)^2\right) \; (j \in \N) $ unabh"angig von
der Buckelbeschaffenheit g"ultig.

\end{bemerkung}

\Beweis{}
{

Wegen (\ref{BRlogableitung})  und (\ref{Bteil1defFk}) gilt f"ur $ n \in \N $
(unabh"angig von der Randbedingung, die bei $ 0 $ gestellt wurde):
\begin{eqnarray}
  \lefteqn{\log ( \tilde{R}(a_n+\alpha_n%%14\tilde{\alpha}_n)
  ) = \log ( \tilde{R}(a_n) )
   +}
\nonumber
\\
& +&
  \int_{a_n}^{a_n+\alpha_n%%14\tilde{\alpha}_n
  } \left(\frac{1}{\kappa} 
  %%14\frac{1}{\tilde{\alpha}_n}
     H_n W(%%14\frac{1}{\tilde{\alpha}_n}
        r - a_n) \sin 2 \teta
       \dr
 + \sin 2\tetas \Fk(r;\kappa)\right)
\dr
%\nonumber 
%\\
%&&
 \nonumber
\end{eqnarray}
%
%%140310 und
%
\begin{eqnarray}
\lefteqn{  \log (\tilde{R}(r)) = \log ( \tilde{R}(a_{n-1}+\alpha_{n-1}%%14\tilde{\alpha}_{n-1}
)) +}
\nonumber \\
&&
+\int_{a_{n-1}+\alpha_{n-1}%%14\tilde{\alpha}_{n-1}
}^{a_n}\sin 2\tetas F_k(r;\kappa)\dr
\quad(r \in [a_{n-1}+\alpha_{n-1}\tilde{\alpha}_{n-1}, a_n ]).
 \nonumber
\end{eqnarray}
Mit (\ref{Bteil1Fkabschaetz}) %%14und (\ref{intbleibtgleich}) 
erh"alt man
\begin{eqnarray*}
\lefteqn{
  \log ( \tilde{R}(a_n+\alpha_n%%14\tilde{\alpha}_n
  ))  \ge }
  \\
  & \ge &
   \log ( \tilde{R}(1) )
    - \sum_{i=1}^n
     \frac{1}{\modulus{\kappa}} H_i \int_1^{\alpha_j}
         \modulus{W_i} (r) \dr
- \int_1^{a_n+\alpha_n%%14\tilde{\alpha}_n
} \frac{\Ckkappa}{r^2}
\nonumber \\
   & \ge &
      - \sum_{i=1}^n \frac{1}{\modulus{\kappa}}
       \underbrace{
         \sup_{j \in \N} \left\{ H_j \int_0^{\alpha_j} \modulus{W_j} (r) \dr
                         \right\}
       }_{=:\W}
  - \frac{\Ckkappa}{a_n+\alpha_n%%14\tilde{\alpha}_n
  }
\nonumber \\
   & \ge & - n \frac{1}{\modulus{\kappa}} \W
   -  - \frac{\Ckkappa}{a_n+\alpha_n%%14\tilde{\alpha}_n
   },
\end{eqnarray*}

wobei $\W <\infty $ ist, da aufgrund der Voraussetzungen von Satz \ref{Ateil2zentralerSatz} $ \left(H_j\right)_{j\in \N} $ eine Nullfolge ist und
$ % H_i
 \int_0^{\alpha_j} \modulus{W_i} (r) \dr = 1$
ist.
%%140310 
Dann ist
\begin{eqnarray}
  \tilde{R}(r) & \ge&  \tilde{R}(1) \exp \left( - n \frac{1}{\modulus{\kappa}} \W- \frac{\Ckkappa}{a_n+\alpha_n%%14\tilde{\alpha}_n
  } \right)
\nonumber \\
& = & \const  \exp \left( - n \frac{1}{\modulus{\kappa}} \W\right)
   \quad (r \in [b_n, a_{n+1}]).
    \nonumber
\end{eqnarray}
Hiermit erh"alt man unter Verwendung der besonderen Gestalt des Potentials
gem"a"s (\ref{satzqDmitDreh}) und der Tatsache, dass nach Konstruktion
$ d_{j+1} \ge \exp (j^2) \quad(j\in \N) $ gilt und
da $ j - \frac{\W}{\modulus{\kappa}} > 0 \quad (j > \frac{\W}{\modulus{\kappa}}) $
%% 140310 da 
$ j - \frac{\W}{\modulus{\kappa}} > 0 \quad (j > \frac{\W}{\modulus{\kappa}})
$ ist:
\begin{eqnarray*}
  \int_{b_n}^\infty \underbrace{\tilde{R}^2(r)}_{>0} \mbox{d}r
  & > &
  \int\limits_{ \bigcup_{j=n}^\infty [b_j, a_{j+1}] }  \tilde{R}^2 (r) \/\mbox{d}r
  \ge
  \int\limits_{ \bigcup_{j=n}^\infty [b_j, a_{j+1}] }
          \const^2 \exp \left( - j \frac{2}{\modulus{\kappa}} \W \right)
\/
          \mbox{d}r
\nonumber \\
  & = &
  \sum_{j=n}^\infty d_{j+1}
       \const^2 \exp \left( - \frac{2j}{\modulus{\kappa}} \W \right)
  \ge
  \sum_{j=n}^\infty \const^2
                    \exp \left( j^2 - \frac{2j}{\modulus{\kappa}} \W \right)
\nonumber \\
  & = &
  \sum_{j=n}^\infty
    \exp \left( \left(j - \frac{\W}{\modulus{\kappa}}
                        \right)
          ^2 \right)
                     \const^2
                     \exp \left( - \left(\frac{\W}{\modulus{\kappa}}\right)^2
\right)
  = \infty.
\end{eqnarray*}

Die zugeh"orige L"osung ist folglich nicht quadratisch integrierbar, da
aufgrund von (\ref{Bpruefer}) f"ur $ \lambda \in \R, \modulus{\lambda} > 1 $ gilt:
\begin{displaymath}
 \Psischl_1^2 + \Psischl_2^2
 = \tilde{R}^2 \left( \cos^2\teta + \frac{\lambda-1}{\lambda+1} \sin^2\teta \right)
 \ge \tilde{R}^2 \min \{1, \frac{\lambda-1}{\lambda+1} \}.
\end{displaymath}
Somit gibt es kein Punktspektrum in $ (-\infty, -1) \cup (1, \infty) $.

} % ENDE BEWEIS Lemma
%%%%%%%%%

Fa"st man die vorstehenden Ergebnisse zusammen, so erh"alt man
%%140310 Dies flie"st in den Beweis des folgenden Satz ein:

\begin{satz}
\label{singspe}

$\left.\right.$

$ \TqDschl $ besitzt au"serhalb der zentralen L"ucke $\![-1,1]\!$ rein singul"arstetiges Spektrum:
\begin{eqnarray}
\lefteqn{\sigma(\TqDschl)   \cap \bigl((-\infty,-1)\cup(1,\infty)\bigr) =}
\nonumber \\
&= & \sigma_{sc}(\TqDschl)   \cap \bigl( (-\infty,-1)\cup(1,\infty)\bigr) \label{scSPektrum}
\supset  (-\infty,-1)\cup(1,\infty)
 \nonumber
\end{eqnarray}

\end{satz}
\hfill$\qed$

\vfill
%%140310 \Beweis{}{

\pagebreak

\hfill
\begin{minipage}[t]{8cm}
\sffamily
All you really need to know for the moment is that the universe is a lot more complicated than you might think, even if you start from a position of thinking it's pretty complicated in the first place.
- Douglas Adams, Mostly Harmless
\end{minipage}

\section{Der radiale Anteil des Diracoperators mit Drehimpulstermen auf dem Intervall (0,1]}

Nachdem im vorangegangen Abschnitt das Intervall $ [1,\infty ) $ untersucht wurde, wird in diesem Abschnitt das Intervall $ (0,1] $ betrachtet. Ziel ist es, f"ur die Situation auf diesem Intervall zu zeigen, dass rein diskretes Spektrum vorliegt.

Da der erste Buckel aufgrund von $ d_1 > 1$ rechts von 1 liegt, k"onnen die im Abschnitt \ref{KapitelmitDrehLoesungauf01} f"ur den potentialfreien Fall auf der Halbachse $\R^+$ gewonnenen Ergebnisse analog erschlossen werden.

Analog wie in (\ref{klassGleichungmitsigmas}) sei hier f"ur $ (0,1] $ der Operator
\begin{equation}
    \Anullll: = \sigma_2 p + \sigma_3 + \sigma_1 \frac{k}{r} 
\label{Anuuuul}
\end{equation}
definiert, jedoch mit einer Randbedingung bei $ 1 $
\begin{equation}
   y_1(1) \sin \eta +y_2 (1) \cos \eta = 0 \; \mbox{f"ur ein }\;\eta \in [0,\pi) .
\label{randbedingunggg}
\end{equation} 
%mit $ \eta \in [0,\pi) $ 
%definiert.

Bei $ 0 $ liegt der Grenzpunktfall vor, denn die L"osung $ v $ aus (\ref{ersterLSG}) ist auf $ (0,1] $ quadratintegrierbar, w"ahrend $ w $ aus (\ref{zweiteLsg}) bei $ 0 $ nicht quadratintegrierbar ist.

Sei  f"ur ein $ \lambda \in \CC \setminus \R $ mit $ y $ die L"osung von $ \Anullll y = \lambda y $ bezeichnet, die die Randbedingung (\ref{randbedingunggg}) erf"ullt, und mit 
$ W(v,y) $ die Wronskideterminante\label{Wronski} von $ v $ und $ y $, welche nicht von $ r \in (0,1] $ abh"angt.
% (vgl. \cite{LEVITHANSARGSJAN}, Chapter 3.3).

F"ur $ r, s \in (0,1] $ sei die Greensche Funktion definiert:
\begin{equation}
G(r,s) =  \left\{\begin{array}{cc}
\frac{1}{W(v,y)}\, v(r) y^T(s) \;\mbox{f"ur} & r \le s
\\
\frac{1}{W(v,y)} \,y(r) v^T(s) \;\mbox{f"ur} & s<r
\end{array}
\label{Greenschefunktion}\right.
\end{equation}
%definiert.

Da $ W (v,y) $ nicht von $ r $ abh"angt, wird im Folgenden die Absch"atzung f"ur $ W(v,r) G(r,s) $ durchgef"uhrt:
\begin{eqnarray}
\lefteqn{W(v,y)^2 \; \int_0^1 \int_0^1 \modulus{G(r,s)}^2 \ds \;\dr  = }
\nonumber \\
%& = & \int_0^1 \int_0^r \modulus{y(r) v^T(s)}^2 \ds \dr + \int_0^1 \int_r^1 \modulus{v(r) y^T(s)}^2\ds \dr
%\nonumber \\
%& = &
%\int_0^1 \modulus{y(r)}^2 \int_0^r \modulus{v(s)}^2 \ds\dr + \int_0^1\modulus{v(r)}^2 \int_r^1\modulus{y(s)}^2 \ds\dr
%\nonumber \\
&= &\int_0^1  \modulus{y(r)}^2 \int_0^r \modulus{v(s)}^2 \ds\dr + \int_0^1 \modulus{v(r)}^2 \int_0^s %\modulus{v(r)}^2 
\modulus{y(s)}^2 \dr\ds
\nonumber \\
&= & 2 \int_0^1 \modulus{y(r)}^2 \int_0^r\modulus{v(s)}^2 \ds\dr
\label{fuerGreen}
\end{eqnarray}

Dabei wurde die Reihenfolge der Integration "uber  $s $ und $ r $ %wie in der Darstellung \ref{INtbereich} skizziert 
in eine Integration nach $ r $ und $ s $ "uberf"uhrt (Satz von Fubini).

%\begin{figure}[t]
%\includegraphics[origin=c,width=12cm,clip=true,
%viewport=0cm 6cm 21cm 14cm]{Bild_Integrationsbereich}
%\caption{\label{INtbereich} Integrationsbereich}
%\end{figure}

Sei o.E. $ k > 0 $, was die Notation vereinfacht. 

Mit \cite{KOSHLAYKOV}, S. 174, kann man das Integral "uber  $ v $ aufgrund der Gestalt  (\ref{ersterLSG})  mit Hilfe von
$ \modulus{J_\nu (x) } \le \frac{\modulus{\frac{1}{2}x}^\nu}{\Gamma(\nu +1)} $ absch"atzen.
%,  da  hier $ \nu > -\frac{1}{2} $ ist.
%$ \modulus{k+\frac{1}{2}}, \modulus{k-\frac{1}{2}} > -\frac{1}{2} $ ist.
%Sei dabei im folgenden o.E. $ k - \frac{1}{2} > 0 $.

Es ist also
\begin{equation}
\int_0^r \modulus{v(s)}^2 \ds \propto r^{2k+1}.
\label{vzww}
\end{equation}

Die L"osung $ y (r) = \gamma_1 v(r) + \gamma_2 w(r) $ ($ \gamma_1 $, $ \gamma_2 $ geeignet, um die Randbedingung bei 1 zu erf"ullen)
wiederum kann aufgrund der Tatsache, dass  sich f"ur positive $ \nu $ die Funktion $ Y_\nu (r) $ f"ur kleine $ r $ wie $ - \frac{\Gamma(\nu)}{\pi} \left(\frac{^2}{r}\right)^\nu $ verh"alt (\cite{WATSON}, S. 41), absch"atzen:
\begin{equation}
\modulus{y(r)}^2 \propto r^{-2k-1} \; \mbox{f"ur kleine} \; r
 \nonumber
\end{equation}

In der N"ahe von $ 0 $ erh"alt man somit
\begin{equation}
\int_0^\eps \modulus{y(r)}^2 \int_0^r \modulus{v(s)}^2 \ds \dr \propto \int_0^\eps r^{-2k-1} r^{2k+1} \dr = \int_0^\eps 1 \dr.
 \nonumber
\end{equation}

Damit ist gezeigt, dass das Integral aus (\ref{fuerGreen}) trotz des singul"aren Verhaltens von $ y $ bei $ 0 $ endlich ist.

Somit liegt ein Hilbert-Schmidt-Operator vor.

Aus dessen Kompaktheit (\cite{WEIDMANNBAND1}, Kapitel 3.3) folgt, dass rein diskretes Spektrum vorliegt.

Insbesondere  ist also 
\begin{equation}
\sigma_{ac}(\Anullll) = \emptyset.
\label{Anulllacleer}
\end{equation}

Im folgenden Abschnitt werden die Ergebnisse f"ur die beiden Intervalle $ (0,1] $ und $ [1,\infty )$ zusammengef"uhrt.

%ende von Bteil1

\vfill

\pagebreak

%zusammenbaueneindim

\hfill
\begin{minipage}[t]{8cm}
\sffamily
Everything happens to everybody sooner or later if there is time enough. -  George Bernard Shaw, Back to Methuselah
\end{minipage}

\section{Der radiale Anteil des Diracoperators mit Buckelpotential und Drehimpulstermen auf $ \R^+$ }
\label{zusammenbaueneindim}

Zielsetzung dieses Abschnittes ist es, Aussagen "uber das Spektrum des auf ganz $ \R^+$ definierten Operators
\begin{equation}
\halles := \sigma_2 p + \sigma_3 +q%_{D} 
+ \sigma_1 \frac{k}{r}
\label{hallllles}
\end{equation}
zu machen

Da $ q%_{D_n}
 $ auf $ (0,1] $ verschwindet, gilt f"ur den Operator $ \Anullll $ aus (\ref{Anuuuul})
\begin{equation}
    \Anullll = \sigma_2 p + \sigma_3 + q%_{D}
 + \sigma_1 \frac{k}{r}.
  \nonumber
\end{equation}

Definiert man
\begin{equation}
    \Aunendlich: = \sigma_2 p + \sigma_3 +q%_{D} 
+ \sigma_1 \frac{k}{r}
\label{Aunendl}
\end{equation}
auf $ [1,\infty) $ mit der Randbedingung (\ref{randbedingunggg}) bei $ r = 1$, so stimmt
\begin{equation}
\Anullll \oplus \Aunendlich 
 \nonumber
\end{equation}
mit $ \halles $
bis auf eine St"orung vom Rang 1 "uberein\footnote{Hier sei auf \cite{Simon2} verwiesen. Dort wird die allgemeine Situation des gest"orten Operators $A_\alpha := A + \alpha B $ mit $ B := (\varphi, \cdot)\varphi $ betrachtet, wobei $ B $ zwar Rang $ 1 $ hat, aber lediglich bez"uglich der Sequilinearform des Raumes beschr"ankt ist. Der zugrundeliegende Vektor $ \varphi $ muss insbesondere nicht normierbar sein.
}% (vgl. Skizze in Abbildung \ref{Intervallezusammenfuegen}).
.

Nach \cite{ARONSHAJN} bleibt das absolutstetige Spektrum bei endlichdimensionalen St"orungen erhalten.

Deshalb stimmen die absolutstetigen Spektren von $ \Anullll \oplus \Aunendlich  $ und $ \halles $ "uberein:
\begin{equation}
 \sigma_{ac} (\halles) =\sigma_{ac} (\Anullll \oplus \Aunendlich )
\label{zuspekteren}
\end{equation}

Mit (\ref{Anulllacleer}) erh"alt man
\begin{equation}
\sigma_{ac} (\Anullll \oplus \Aunendlich ) = \sigma_{ac} (\Anullll )  \cup \sigma_{ac} ( \Aunendlich ) 
= \sigma_{ac} (\Aunendlich ) .
\label{nocmalspktum}
\end{equation}

Auf das Spektrum von $ \Aunendlich $ wird mittels der Eigenschaften des  ebenfalls auf $ [1,\infty) $ und mit der gleichen Randbedingung (\ref{randbedingunggg}) bei $ r = 1 $ definierten Operators
\begin{equation}
 \Aeinssl := \sigma_2 p + m \sigma_3 + l + q%_{D}
\label{Aeinssl}
\end{equation}
 geschlossen.
 
 Nach Satz \ref{singspe} gilt
\begin{eqnarray}
\sigma_{sc}(\Aeinssl) \cap \bigl( (-\infty,-1)\cup(1,\infty)\bigr)
\supset (-\infty,-1)\cup(1,\infty).
 \nonumber
\end{eqnarray}

Da nach Abschnitt \ref{AbschnittTransformation}  $ \Aeinssl $ und $   \Aunendlich $ unit"ar "aquivalent sind, umfa"st das singul"arstetige Spektrum von $ \Aunendlich $ ebenfalls mindestens die beiden Intervalle  $  \bigl( (-\infty,-1)\cup(1,\infty)\bigr) $.

Erg"anzend sei folgende Schlu"skette dargestellt:
Seien auf $ [1,\infty) $  weitere Operatoren definiert mit jeweils der Randbedingung (\ref{randbedingunggg}):
\begin{equation}
\Aohnewas := \sigma_2 p + \sigma_3 + q%_{D}
\label{Aohnewas}
\end{equation}
und 
\begin{equation}
\Aeinsnull := \sigma_2 p + \sigma_3 
\label{Aeinsnull}
\end{equation}

Die weitere Untersuchung vergleicht $ \Aohnewas $ und $ \Aeinsnull $.

Bezeichne zu $ \lambda \in \CC\setminus\R $ $ \tilde{v} $ die L"osung von $ \Aeinsnull y = \lambda y $, die die Randbedingung 
(\ref{randbedingunggg})  erf"ullt, und $ \tilde{w} $ die L"osung, die bei $ \infty $ quadratisch integrierbar ist. Ohne Einschr"ankung\footnote{ Vgl. die Argumentation in \cite{HughesSchmidt}, die sich darauf st"utzt, da"s verm"oge einer M"obius-Transformation die Titchmarsh-Weylschen $m$-Funktionen f"ur verschiedene Randbedinungen miteinander in Beziehung gesetzt werden k"onnen.}   wird im Folgenden die Randbedingung $ \tilde {v} (1) = \left(\begin{array}{c} 1 \\ 0 \end{array}
\right) 
$ betrachtet.
Weiter sei 
%$ \ssigma := - i \lambda \in \R $.
$\varsigma := \sqrt{1-\lambda^2}$.

Beide L"osungen k"onnen durch das Fundamentalsystem
\begin{equation}
U(r) = \left(
\begin{array}{cc}
e^{ \varsigma \, r}  & - \frac{\varsigma}{1 + \lambda} e^{ - \varsigma \, r}\\
 \frac{\varsigma}{1 - \lambda} e^{  \varsigma\, r} & e^{ - \varsigma \, r}
\end{array}
\right)
 \nonumber
\end{equation}
dargestellt werden:
\begin{equation}
\tilde{v} (r) = \left(
\begin{array}{c}
\frac{1}{2} e^{- \varsigma} \;e^{\varsigma \, r}  + 
 \frac{1}{2}e^{ \varsigma }  e^{ - \varsigma\, r}\\
 \frac{1}{2} e^{- \varsigma}  \frac{\varsigma}{1 - \lambda} e^{  \varsigma\, r} 
 -  \frac{1}{2} e^{- \varsigma}  \frac{1 + \lambda}{\varsigma} e^{ - \varsigma \, r} 
 \end{array}
\right)
 \nonumber
\end{equation}

und 
\begin{equation}
\tilde{w} (r) = \left(
\begin{array}{c}
 - \frac{\varsigma}{1 + \lambda} e^{ - \varsigma \, r}\\
 e^{ - \varsigma\, r}
\end{array}
\right)
 \nonumber
\end{equation}

F"ur diese L"osungen gelten die Absch"atzungen
\begin{equation}
%\modulus{\tilde{v}(r)} \le \const (\ssigma)\, e^{\sqrt{1+\ssigma^2} r}
\modulus{\tilde{v}(r)} \le \const (\lambda)\, e^{\varsigma r}
\label{ewr}
\end{equation}
und 
\begin{equation}
%\modulus{\tilde{w}(r)} \le \const (\ssigma)\, e^{-\sqrt{1+\ssigma^2} r}
\modulus{\tilde{w}(r)} \le \const (\lambda)\, e^{-\varsigma r}.
\label{sdf}
\end{equation}

Sei mit $ \tilde{W}%(\tilde{v}(r, \lambda),\tilde{ w }(r, \lambda)) 
$\label{wronski2}  die zugeh"orige Wronskideterminante, welche nicht von % f"ur 
$ r \in [1,\infty ) $ abh"angt, bezeichnet.
Des weiteren sei die Greensche Funktion
\begin{equation}
\tilde{G} (r,s,\lambda) := 
  \left\{
 \begin{array}{lcl}
\frac{1}{\tilde{W}%(\tilde{v}(r, \lambda), \tilde{w} (r, \lambda)
 } \tilde{ w} (r,\lambda) \tilde{v}^T (s, \lambda) & \mbox{f"ur} & 1 \le s \le r < \infty \\
\frac{1}{\tilde{W}%(\tilde{v}(r, \lambda), \tilde{w} (r, \lambda) 
} \tilde{ v}(r,\lambda) \tilde{w}^T (s,\lambda) & \mbox{f"ur} & 1 \le r < s < \infty
  \end{array}
\right.
\label{green3}
\end{equation}
definiert.

Mit (\ref{ewr}) und (\ref{sdf}) erh"alt man
\begin{equation}
\modulus{\tilde{G} (r,s,\lambda)}
\le
%\const(\ssigma) e^{-\sqrt{1 + \ssigma^2}\modulus{r-s}}
\const(\lambda) e^{-\varsigma \modulus{r-s}}.
 \nonumber
\end{equation}

F"ur $ f \in L^2(1,\infty)^2 $ ist 
\begin{equation}
\left( \left( \Aeinsnull - \lambda\right)^{-1} f \right) (r) = 
  \int_1^\infty \tilde{G} (r,s,\lambda) f(s) \ds \quad (r \ge 1).
   \nonumber
\end{equation}

Dann ist f"ur $ g \in L^2(1,\infty) $ sowohl $ g \left(\Aeinsnull - \lambda\right)^{-1} $ als auch $\left(\Aeinsnull - \lambda\right)^{-1} g $ ein Hilbert-Schmidt-Operator, denn es ist beispielsweise
\begin{equation}
\int_1^\infty\int_1^\infty \modulus{
g(r) \tilde{G} (r,s,\lambda) 
}^2 \ds \dr
\le \int_1^\infty \modulus{g(r)}^2 \int_{-\infty}^\infty e^{-\varsigma\modulus{r-s}}\ds \dr < \infty.
 \nonumber
\end{equation}

Ziel ist es, zu zeigen, da"s 
\begin{equation}
  \left(\Aeinssl - \lambda\right)^{-1} - \left(\Aohnewas-\lambda\right)^{-1}
   \nonumber
\end{equation}
Spurklasse ist\footnote{Spurklasseeigenschaften werden in  \cite{HughesSchmidt} bei einer "ahnlichen Argumentation herangezogen.}.
Dann ist n"amlich mit \cite{Kato}, Kapitel X, Theorem 4.12:
\begin{equation}
\sigma_{ac} (\Aeinssl) = \sigma_{ac} (\Aohnewas).
 \nonumber
\end{equation}

Mit (\ref{refernzspektrum}) aus Kapitel \ref{KapitelohneDreh} erh"alt man\footnote{Die Beweisschritte des Kapitels \ref{KapitelohneDreh} sind v"ollig analog f"ur den auf $ [1,\infty) $ mit Randbedingung bei 1 definierten Operator 
$ \Aohnewas $ durchf"uhrbar.}
\begin{equation}
\sigma_{ac} (\Aeinssl) \cap ((-\infty,-1) \cup (1,\infty)) = \sigma_{ac} (\Aohnewas) \cap ((-\infty,-1) \cup (1,\infty) )
= \emptyset.
\label{mehrspektrum}
\end{equation}

F"ur den Spurklassenachweis berechnet man zun"achst mit Hilfe der zweiten Resolventengleichung (z.B. \cite{WEIDMANNBAND1}, S. 190) %erh"alt man
\begin{eqnarray}
\lefteqn{\left(\Aeinssl - \lambda \right)^{-1}
- \left(\Aohnewas - \lambda \right)^{-1}
= }
\nonumber \\
& = &\left(\Aeinssl - \lambda \right)^{-1}
\left( \Aohnewas - \Aeinssl\right) \left(\Aohnewas - \lambda \right)^{-1}
\nonumber \\
&=&
\left(\Aeinssl - \lambda \right)^{-1}
\left((m-1) \sigma_3 + l\right) \left(\Aohnewas - \lambda \right)^{-1}
\label{ersteResolventeninfo}
\end{eqnarray}
und 
\begin{eqnarray}
\lefteqn{\left(\Aohnewas - \lambda \right)^{-1}
- \left(\Aeinsnull - \lambda \right)^{-1}
= }
\nonumber \\
& = &\left(\Aeinsnull - \lambda \right)^{-1}
\left( \Aeinsnull - \Aohnewas\right) \left(\Aohnewas - \lambda \right)^{-1}
\nonumber \\
&=&
- \left(\Aeinsnull - \lambda \right)^{-1}
\; q%_{D_n} \; 
\left(\Aohnewas - \lambda \right)^{-1},
\label{zweiteResolventeninfo}
\end{eqnarray}
sowie
\begin{eqnarray}
\lefteqn{\left(\Aeinssl - \lambda \right)^{-1}
- \left(\Aeinsnull - \lambda \right)^{-1}
= }
\nonumber \\
& = &\left(\Aeinssl - \lambda \right)^{-1}
\left( \Aeinsnull - \Aeinssl\right) \left(\Aeinsnull - \lambda \right)^{-1}
\nonumber \\
&=&
- \left(\Aeinssl - \lambda \right)^{-1}
\left(q%_{D_n} 
+( m-1) \sigma_3 + l\right) \left(\Aeinsnull - \lambda \right)^{-1}.
\label{dritteResolventeninfo}
\end{eqnarray}
Aus (\ref{dritteResolventeninfo}) leitet man
\begin{eqnarray}
\lefteqn{
 \left(\Aeinssl -\lambda \right)^{-1} = }
\nonumber \\
& & =
\left[1 \; - \;  \left(\Aeinssl -\lambda \right)^{-1} \left( q%_{D_n} 
+ (m-1)\sigma_3 + l \right) 
\right]
\left(\Aeinsnull - \lambda \right)^{-1}
 \nonumber
\end{eqnarray}
ab und aus (\ref{zweiteResolventeninfo})
\begin{equation}
 \left(\Aohnewas -\lambda \right)^{-1} = 
\left(\Aeinsnull - \lambda \right)^{-1}
\left[1 \; - \;  q%_{D_n} 
 \left(\Aohnewas -\lambda \right)^{-1} 
\right].
 \nonumber
\end{equation}
Dies in (\ref{ersteResolventeninfo}) eingesetzt liefert
\begin{eqnarray}
\lefteqn{\left(\Aeinssl - \lambda \right)^{-1}
- \left(\Aohnewas - \lambda \right)^{-1}
= }
\nonumber \\
& = &
\left[1 \; - \;  \left(\Aeinssl -\lambda \right)^{-1} \left( q%_{D_n} 
+ (m-1)\sigma_3 + l \right) 
\right]
\left(\Aeinsnull - \lambda \right)^{-1} \cdot
\nonumber \\
& &\quad\quad  \cdot 
\left((m-1) \sigma_3 + l\right)
\left(\Aeinsnull - \lambda \right)^{-1}
\left[1 \; - \;  q%_{D_n} 
 \left(\Aohnewas -\lambda \right)^{-1} 
\right].
 \nonumber
\end{eqnarray}
Hierbei sind der erste Faktor $ \left[1 \; - \;  \left(\Aeinssl -\lambda \right)^{-1} \left( q%_{D_n} 
+ (m-1)\sigma_3 + l \right) 
\right] $ und der letzte Faktor $ \left[1 \; - \;  q%_{D_n} 
 \left(\Aohnewas -\lambda \right)^{-1} 
\right] $ 
%aufgrund der Eigenschaften  von $ \Aohnewas $ und $ \Aeinssl$ 
beschr"ankt.
Die mittleren Faktoren sind Spurklasse, denn es ist mit den Definitionen f"ur die ortsabh"angige Masse (\ref{ortsabhMasse}) und den Drehimpuls (\ref{Drehimpulsterm})
\begin{eqnarray}
\lefteqn{\left((m-1) \sigma_3 + l\right)
}
\nonumber 
\\
& = &
\left(\left( 1 +\frac{k^2}{r^2}\right)^{\frac{1}{2}} -1 \right)^{\frac{1}{2}}
 \cdot \sigma_3 \cdot
\left(\left( 1 +\frac{k^2}{r^2}\right)^{\frac{1}{2}} -1 \right)^{\frac{1}{2}}
\nonumber \\
& & + 
\sign(k) 
\left(\frac{\sign (k)\; k}{2(r^2+ k^2)}\right)^\frac{1}{2} \cdot \left(\frac{\sign (k)\; k}{2(r^2+ k^2)}\right)^\frac{1}{2}, 
 \nonumber
\end{eqnarray}
so dass man
\begin{eqnarray}
\lefteqn{\left(\Aeinsnull - \lambda \right)^{-1}\left((m-1) \sigma_3 + l\right)
\left(\Aeinsnull - \lambda \right)^{-1}}
\nonumber 
\\
& = &
\left(\Aeinsnull - \lambda \right)^{-1}\sqrt{m-1 } \cdot \sigma_3\sqrt{ m-1 } \left(\Aeinsnull - \lambda \right)^{-1}
\nonumber \\
& & + \left(\Aeinsnull - \lambda \right)^{-1}
\sign(k) \sqrt{\sign (k)\;l} \cdot \sqrt{\sign (k)\;l}  \left(\Aeinsnull - \lambda \right)^{-1} \nonumber
%\\
%& & 
\end{eqnarray}
erh"alt.
Beide Summanden sind jeweils das Produkt zweier Hilbert-Schmidt-Operatoren.

Mit (\ref{zuspekteren}) und (\ref{nocmalspktum}) ist dann wegen (\ref{mehrspektrum})
\begin{eqnarray}
\sigma_{ac}(\halles)   \cap \bigl( (-\infty,-1)\cup(1,\infty) \bigr)= \emptyset
 \nonumber
\end{eqnarray}
und schlie"slich
\begin{eqnarray}
\sigma_{sc}(\halles)   \cap \bigl( (-\infty,-1)\cup(1,\infty) \bigr)=  \bigl( (-\infty,-1)\cup(1,\infty) \bigr).
\label{singspektrum}
\end{eqnarray}
\hfill$\qed$

%ende von zusammenbaueneindim

\vfill
\pagebreak

%Bzusammenbauen}
%Πολλὰ τὰ δεινὰ κ' οὐδὲν ἀνθρώπου δεινότερον πέλει

\hfill
\begin{minipage}[t]{8cm}
\sffamily
$\, \Pi o \lambda\lambda \grave{\alpha} 
\; \tau \grave{\alpha}
\; \delta\varepsilon\iota\nu \grave{\alpha}
\;\kappa '
o\!\!\stackrel{\scriptscriptstyle\supset}{\upsilon}\!\!\delta\grave{\varepsilon}\nu
\;\stackrel{\scriptscriptstyle\supset}{\alpha}\!\!\nu\vartheta\varrho \acute{\omega}\pi o\upsilon
$

$
\; \delta\varepsilon\iota\nu \acute{o}\tau\varepsilon\varrho o \nu
\; \pi\acute{\varepsilon}\lambda\varepsilon\iota. 
$

$\,$- $\;\Sigma o \varphi o \kappa \lambda \tilde{\eta} \varsigma, 
\;\stackrel{\scriptscriptstyle\supset}{\left.\right.}\!\!\!\! A\nu \tau \iota\gamma \acute{o}\nu \eta
$
\end{minipage}

\section{Das Spektrum f"ur den dreidimensionalen kugelsymmetrischen Fall 
%von $ H_{q%_D
%}$} 
}
\label{Bzusammenbauen}

In diesem Kapitel werden die Ergebnisse der vorangehenden Abschnitte zur Untersuchung des radialen Anteils des Diracoperators verwendet, um eine Aussage f"ur den kugelsymmetrischen Fall im $ \R^3 $ zu gewinnen.

Betrachtet wird hier der Differentialoperator
\begin{equation}
H_{q%_D
} := 
%{\boldsymbol \alpha} \cdot {\boldsymbol p} 
%     + \beta m c^2 + V (\cdot)
%=
 {\boldsymbol \alpha} \cdot {\boldsymbol p} 
     + \beta+ q%_D
 (\modulus{\cdot}).
\label{DiracoperatorKugel}
\end{equation}
Dabei bezeichnet $ {\boldsymbol p} $ den mechanischen Impulsvektor\label{impulsvektor} des Teilchens und es ist 
\begin{equation}
 \boldsymbol \alpha = 
\left(\begin{array}{cc} \boldsymbol 0 & \boldsymbol \sigma\\
                                     \boldsymbol \sigma & \boldsymbol 0
       \end{array}
\right)
\nonumber
\end{equation}
definiert, wobei $ \boldsymbol\sigma $ der Vektor der Pauli-Matrizen ist. Das Potential $ q%_D
 (\modulus{\cdot}) $ ist gem"a"s Definition kugelsymmetrisch.
 
 Aufgrund seiner Symmetrieeigenschaften kann der Operator $ H_q $ auf eine Schar gew"ohnlicher Differentialoperatoren $ H_k $ $ (k \in \Z \setminus\{0\}) $ reduziert werden
(siehe \cite{WEIDMANNBAND2}, Abschnitt 20.3).

Es gilt f"ur
$ \left(\begin{array}{c}g_1\\g_2\end{array}\right) \in D(H_k) $:
%Des weiteren  ist
\begin{equation}
H_k\left(\begin{array}{c}g_1(r)\\g_2(r)\end{array}\right)
=
\left(\begin{array}{cc}0 & -1\\
                                    1 & 0
\end{array}
\right)
\left(\begin{array}{c}g'_1(r)\\g'_2(r)\end{array}\right)
+
\left(\begin{array}{cc}
v(r)+1 & \frac{k}{r}\\
\frac{k}{r} & v(r)-1
\end{array}
\right)
\left(\begin{array}{c}g_1(r)\\g_2(r)\end{array}\right)
\label{radialausdruckdirac}
\end{equation}

In der Nomenklatur des Operators $ H_k $ wird die Abh"angigkeit von der Dreh\-impulsquantenzahl zum Ausdruck gebracht.
Diese Abh"angigkeit war im vorbereitenden Abschnitt \ref{KapitelmitDreh} bei der Bezeichnung des %zugeh"origen 
selbstadjungierten Operators $ \TqDschl $ unterdr"uckt worden, da die vorbereitenden Betrachtungen f"ur ein festes $ k \in \Z $ angestellt wurden. 
Die Ergebnisse des Abschnittes \ref{mitdrehsinggrenzmass} sind hingegen bereits f"ur alle $ k \in \Z\setminus \{0\} $ g"ultig.

Ebenso war im Abschnitt \ref{zusammenbaueneindim} auf die explizite Angabe der Abh"angigkeit von $ k $ verzichtet worden.
Im Folgenden wird jedoch die $k$-Abh"angigkeit in der Bezeichnung $  \tilde{T}_{%q_D;
k}$ f"ur die Operatoren auf $  \R^+ $, die sowohl den Drehimpulsanteil als auch den Potentialanteil enthalten,  angegeben.\label{Thaengtvonkab}

Sei mit $ T_{q%_D
;ges} $ \label{dreidT}der durch den Differentialausdruck
$
H_{q%_D
} 
$ %$ H_V $ aus Definition (\ref{DiracoperatorKugel}) 
bestimmte selbstadjungierte dreidimensionale Operator bezeichnet.

Da das Spektrum der orthogonalen Summe von Operatoren der Abschlu"s der Vereinigung der Spektren der Summanden ist (siehe z.B. \cite{WEIDMANNBAND2}, Satz 18.2), gilt mit dem zentralen Satz \ref{singspe} und dem Ergebnis (\ref{singspektrum}) der  "Uber\-legungen aus Abschnitt %Kapitel 
\ref{zusammenbaueneindim}

% S. 202
%
\begin{equation}
\sigma(T_{q%_D
;ges}) = \overline{{{\displaystyle\bigcup }\atop {\scriptscriptstyle k \in \Z\setminus \{0\}}}\sigma(\tilde{T}_{%q_D;
k})}
=\overline{{{\displaystyle\bigcup }\atop {\scriptscriptstyle k \in \Z\setminus \{0\}}}\sigma_{sc}(\tilde{T}_{%q_D;
k})}.
\nonumber
\end{equation}

Damit gilt 
\begin{satz}
%\begin{equation}
%\sigma(T_{q_D;ges})   \cap (-\infty,-1]\cup[1,\infty) = \sigma_{sc}(T_{q_D;ges})  
%
%\end{equation}
\begin{eqnarray}
\lefteqn{\sigma(T_{q%_D
;ges})   \cap \bigl( (-\infty,-1)\cup(1,\infty) \bigr)}
\nonumber \\
&= & \sigma_{sc}(T_{q%_D
;ges})   \cap \bigl((-\infty,-1)\cup(1,\infty)\bigr)
%%140310 \supset  (-\infty,-1)\cup(1,\infty)
\nonumber
\end{eqnarray}

\end{satz}

$ T_{q%_D
;ges} $ besitzt also au"serhalb der zentralen L"ucke rein singul"arstetiges Spektrum (sic!).

Das in Kapitel \ref{KapitelmitDreh} konstruierte Potential $ q%_D
 $ f"uhrt also zu dem bereits in der Einleitung beschriebenen exotischen Teilchenverhalten, bei dem sich die Teilchen sowohl beliebig weit vom Ursprung entfernen, als auch beliebig oft zum Ursprung zur"uckkehren.

%ende von Bzusammenbauen}

%

%symbolverzeichnis
%Symbolverzeichnis

\vfill
\pagebreak

\section{Symbolverzeichnis
\label{Symbolverzeichnis}}

{

\renewcommand{\arraystretch}{1.35}
\begin{longtable}{@{} l @{  } @{\extracolsep{\fill}} p{9.2cm} r @{} r @{} }  %tabular
{\bf Symbol} & {\bf Bedeutung/Definition} & &\\
$ a_j $
   & Buckelanfang
   & S. & \pageref{S_buckelanfang} \\
$  \abnquadrat $
   & $ \int_0^b \modulus{\phibn}^2 \mbox{d}r $
   & S. & \pageref{aquadrat} \\
$   \abnquadrats $
  & $ \int_1^b \modulus{\phibn}^2 \mbox{d}r $
  & S.& \pageref{Rhonbschlange}\\
$ \Aj $ & $
   \frac{1}{2}
           \left[
              \left(\Mjee\right)^2 + \plmin \left(\Mjze\right)^2 + \minpl \left(\Mjez\right)^2 + \left(\Mjzz\right)^2
           \right] $
   &  S.& \pageref{DefAj}
\\
$   \Ajs $
   &  $ \frac{1}{2}
           \left[
              {\Mjeesquad} + \plmin \Mjzesquad
              + \minpl \Mjezsquad + \Mjzzsquad
           \right] $
    & S. &\pageref{BDefAj} \\

% Feb 13 $ B $ 
% Feb 13    & Laplace-Beltrami-Operator
% Feb 13    & S. \pageref{S_LaplaceBeltrami} \\
$ \ACloc\!\left(\!\R^+\!\right)^2 $
  & Raum der lokal absolutstetigen Funktionen auf $ R^+$ 
  & S. &\pageref{AClok} \\

$ b_j $
   & Buckelende
   & S. &\pageref{S_Buckelende} \\
$ \Bj  $
   &  $ \frac{1}{2}
             \left[
                  \left(\Mjee\right)^2 + \plmin \left(\Mjze\right)^2 - \minpl \left(\Mjez\right)^2 - \left(\Mjzz\right)^2
             \right]
     $
   & S. &\pageref{DefBj}\\
$ \Bjs $
   &  $  \frac{1}{2}
             \left[
                  \Mjeesquad + \plmin \Mjzesquad
                  - \minpl \Mjezsquad - \Mjzzsquad
             \right]
   $
    & S. &\pageref{BDefBj} \\
$  \Cj $ 
   & $ \sqrt{\minpl}\Mjee\Mjez + \sqrt{\plmin} \Mjze\Mjzz $
   & S.& \pageref{DefCj} \\
$ \Cjs  $
   & $ \sqrt{\minpl}\Mjees\Mjezs + \sqrt{\plmin} \Mjzes\Mjzzs $
   & S.& \pageref{BDefCj} \\
$ C(\R^+) $
  & Raum der auf $ \R^+ $ stetigen Funktionen
  & S.& \pageref{S_Buckelpotential} \\
$ D_n $
   & $n$-Tupel von Buckelabst"anden
   & S. &\pageref{Distanztupel} \\
$ \D (\kappa) $
   & $ \frac{{\rm \sign}(\kappa) \sqrt{\kappa^2+1} +1}
                      {{\rm \sign}(\kappa) \sqrt{\kappa^2+1}}
  R(0, \kappa )
  \quad (\kappa \in \R^*)  $
   & S.& \pageref{Ateil2DefD} \\
$ \Dmischl(\kappa) $
     & $ \frac{{\rm \sign}(\kappa) \sqrt{\kappa^2+1} +1}
                      {{\rm \sign}(\kappa) \sqrt{\kappa^2+1}}
  \tilde{R}(1, \kappa )
  \quad (\kappa \in \R^*)  $
     & S.& \pageref{Dgeschwungen}\\
$   \ffj (\kappa, y) $
   &   $ \frac{1}{\Aj(\kappa) + \Bj(\kappa) \cos(2y) + \Cj(\kappa) \sin(2y) }$
   & S. &\pageref{Ateil1deffj2}\\
$ \ffjs (\kappa, y,z) $
   & $ \frac{\exp\left(-2\int_{b_{j-1}}^{a_j} \Fk(s;\kappa)\ds\right)}
        {\Ajs(\kappa) + \Bjs(\kappa) \cos(2(y+z)) + \Cjs(\kappa) \sin(2(y+z))} $
   & S. &\pageref{Bteil1deffj1} \\
$  \overline{\ffj}(\kappa) $
   & $  \frac{1}{\pi} \int_0^\pi \ffj(\kappa,y)\;\mbox{d}y $
   &  S. &\pageref{Ateil1mittelfj}\\
$  \overline{\ffjs} (\kappa;z)  $
  &   $ \frac{1}{\pi} \int_0^\pi \ffjs(\kappa,y;z)\;\mbox{d}y $
  &S.& \pageref{Bteil1defmj} \\   
$ \Fk(r;\kappa)  $
   & $  \frac{1}{\kappa} \angu(r)
                  + \frac{\lambda}{\kappa}(\mass(r)-1) $
   & S.& \pageref{Bteil1defFk}\\
%%14$ \GGGG_n $
%%14   & 
%%14   & S. \pageref{AbschaetzungFehlerFgeschw}\\
$ g_{n+1} $
  &  $ f_{n+1} - 1 $
 & S. &\pageref{defgnklein} \\
$ G(r,s;\lambda)$ 
   & Greensche Funktion auf $ (0,1] $
   & S. &\pageref{Greenschefunktion} \\
%$ G_a(r,s;z)$ 
%   & Greensche Funktion auf $ [a,1] $
 % & S. \pageref{Greenschefunktionaaa} \\
 
 $ \Gk(r;\kappa) $ 
   & $         \frac{1}{\kappa}(\lambda + \cos 2\tetas(r)) \angu(r)
        + \frac{1}{\kappa}(1 + \lambda \cos 2\tetas(r))  (\mass(r)-1) $
   & S. &\pageref{Bteil1defGk} \\
 
$ \tilde{G} (r,s,\lambda)  $
   & Greensche Funktion auf $ [1,\infty) $
   & S. &\pageref{green3} \\

%&&\\   

   $ \hallesnull $
   & $ \sigma_2 p + \sigma_3 + \sigma_1 \frac{k}{r}  $
   & S.& \pageref{klassGleichungmitsigmas} \\

$ \hj $ 
   & $ \log \ffj $
   &  S. &\pageref{lemma2Bdefhj}\\

$\hqdn  $ 
   & $  \sigma_2 p + m \sigma_3 + l + q_{D_n} $
   & S. &\pageref{hqdn} \\

   $  \hjnulls  $
   & $\log \tilde{f}_{j_0} $
   & S. &\pageref{defhjnullschl} \\

$ H $
  & Buckelh"ohe bei identischen Buckeln 
  & S. &\pageref{idBuckkel}\\
$ H_j $
   & Buckelh"ohe
   & S. &\pageref{S_Buckelhoehe} \\
% Feb 13 $ H_V  $
% Feb 13    & Hamiltonoperator im $ \R^3 $ mit kugelsymmetrischem Potential $ V $
% Feb 13    & S. \pageref{S_HamiltonianmitkugelsymmPot} \\
 $ H_{q%_D
 }  $
  & Differentialausdruck im $ \R^3 $ mit kugelsymmetrischem Potential $ q%_D 
  $
   & S. &\pageref{DiracoperatorKugel} \\

$ J_{ \pm \modulus{k+\frac{1}{2}}} $
   & Besselfunktion
   & S. &\pageref{Besselfunktion}\\
$ k $ 
   & Drehimpulsquantenzahl 
  & S.& \pageref{Drehimpulsquantenzahl} \\

$ \KC $
   & Konstante $ \ge 2 $
   & S. &\pageref{DefKgeschwungen} \\
$ l $
   & Drehimpuls
   & S. &\pageref{Drehimpulsterm}\\
% Feb 13 $  L_2\left(\R ^3\right)^4$ 
% Feb 13    & $L_2(\R^3) \oplus L_2(\R^3) \oplus L_2(\R^3) \oplus L_2(\R^3) $
% Feb 13        Raum der quadratintegrierbaren vierkomponentigen Funktionen des $ \R^3 $
% Feb 13    & S. \pageref{S_LzweiRdrei}\\
$ m $
   & ortsabh"angige Masse
   & S. &\pageref{ortsabhMasse} \\
$ \mj $
   & $ \frac{1}{\pi} \int_0^\pi \log \ffj ( \kappa, y) \dy $
   & S. &\pageref{Ateil1defmj} \\
   
$     \mjs (\kappa;z)$ 
  & $ \frac{1}{\pi} \int_0^\pi \log \ffjs ( \kappa,y,z) \dy $
  & S. &\pageref{Bteil1defmj} \\

$ \Mj $
   & Transfermatrix des $j$-ten Buckels
   & S. &\pageref{transfermatrix} \\
$ \Mjs $
   & Transfermatrix des $j$-ten Buckels im Fall mit Dreh\-impuls
   & S. &\pageref{TRansfermatrixmitDreh} \\

$  \Nnynykappa $
   & $
% Feb 13 \left[
\frac{1}{2} \sum_{i=\nyunten+1}^\nyoben \mi(\kappa)
% Feb 13\right]^2 
$
   & S. &\pageref{Nnyuntenoben}\\
   
$   \Nblambdalambda $
  & Zahl der Eigenwerte zwischen $ \lambda_1 $ und $ \lambda_2 $ f"ur das endliche Problem
    & S. &\pageref{Ateil2Eigenwertzahlabschaetzung}   \\
    
$ \Nblambdalambdas $
 &   Zahl der Eigenwerte zwischen $ \lambda_1 $ und $ \lambda_2 $ f"ur das endliche Problem mit Drehimpuls
 & S.& \pageref{zahlEWe} \\
  
$   \Nknynykappa $
   & $  \inf_{\stackrel{\left(a_j\right) \in \AAA}{z \in I}}
  \left\{
     - \frac{1}{2} \sum_{j=\nyunten+1}^\nyoben \majs (\kappa,z)
  \right\}  $
   & S.& \pageref{einN}  \\
$ \Najknynykappa  $
   & $ \frac{1}{2} \sum_{j=\nyunten+1}^\nyoben \majs(\kappa, z) $
   & S. &\pageref{Blemma4ref1} \\
${\boldsymbol p} $
   & Impulsvektor
& S.& \pageref{impulsvektor} \\
$ q
%_D 
$
   & Buckelpotential
   & S.& \pageref{S_Buckelpotential}\\
$ q_{%D_
n} $
   & Buckelpotential mit endlich vielen Buckeln
   & S.& \pageref{endlBuckjelpotential} \\
   $ \Aohnewas $
   & $ \sigma_2 p + \sigma_3 + q_{D} $ auf $ [1,\infty) $
   & S. &\pageref{Aohnewas} \\
 $ \Aeinssl$ 
   & $ \sigma_2 p + m \sigma_3 + l + q_{D} $ auf $ [1,\infty ) $
   & S. &\pageref{Aeinssl} \\
$ R $
   & Pr"ufer-Radius
   & S. &\pageref{S_prueferradius} \\
$ R_n $ 
  & Pr"ufer-Radius am Ende des $ n $-ten Buckels
  & S. &\pageref{Rnabkuerz} \\

$  R_{%D_
n;j}  $
   & Pr"ufer-Radius am Ende des $ j $-ten Buckels eines Potentials, dessen Buckelpositionen durch das
       Distanztupel $ D_n $ charakterisiert sind
   & S. &\pageref{S_PrueferradiusBuckelende} \\
$ \Rs $
   & Pr"ufer-Radius im Fall mit Drehimpuls
   & S. &\pageref{PrueferradiusmitDreh} \\
   $ \Rs_n $
& Pr"ufer-Radius am Ende des $ n $-ten Buckels im Fall mit Dreh\-impuls
  & S. &\pageref{Rntetan} \\
$ \Rs_{%D_
n;j} $
   & Wert des Pr"ufer-Radius am Ende des $j$-ten Buckels des $n$-Buckelpotentials $ q_{%D_
   n} $ im Fall mit Drehimpuls
   & S. &\pageref{Bteil1DefRsj} \\
$  \R^* $
   &  $ \R \backslash \{0\} $
   & S.& \pageref{Rstern} \\   

$  s_{1/2} (r)$ 
  & $ \pm q_{%D_
n} (r) $
  & S.& \pageref{deftfrei} \\

$S(r) $
  & $ \left(
\begin{array}{cc}
0 & s_2(r) \\
s_1 (r)& 0
\end{array}
\right)
$ auf $(0,b) $
  & S. &\pageref{deftfrei}\\

$ \Sn $
   & $ \left\{ \kappa\in \Xi_n \bigg|
      \frac{1}{\pi}
      \left(\prod_{j=1}^{\nu_n}\!
        \ffj (\kappa, \teta_{%D_{j-1};
        j-1} (\kappa) - \kappa d )
      \right) {\cal{D}}(\!\kappa\!)
      > \frac{\eps_n}{\mynull(\Xi_n)}
    \right\}
   $ 
   & S. &\pageref{satzschrittnS} \\
$ \Snschl $ 
%   & $   \left\{ \kappa\in \Xi_n \Bigg|
%      \frac{1}{\pi}
%      \left(\prod_{j=1}^{\nu_n}
%       \tilde{ \ffj} (\kappa, \teta_{D_{j-1};j-1} (\kappa) - \kappa d )
%      \right) {\cal{D}}(\kappa)
%\left(1 + \GGGG_{\nu_n}(\kappa) \right)
%      > \frac{\eps_n}{\mynull(\Xi_n)}
%    \right\}
    & $ \left\{
     \kappa\in \Xi_n \bigg|
      \frac{1}{\pi}
      \left(\prod_{j=1}^{\nu_n}
       \tilde{ \ffj} (\kappa, \tetas_{%D_{j-1};
       j-1} (\kappa) - \kappa d )
      \right) {\cal{D}}(\!\kappa\!)
%\left(1 + \GGGG_{\nu_n}(\kappa) \right)
      > \frac{\eps_n}{\mynull(\Xi_n)}
    \right\}
$ %& \\
%&
& S. &\pageref{BsatzschrittnS}
\\

$T $ 
& Grenzoperator
& S. &\pageref{operatorT}  \\

% Feb 13$ T_0^{\mbox{min}} $
% Feb 13   &  minimaler freier Diracoperator im $ \R^3 $ 
% Feb 13   & S. \pageref{S_minfreiDirac} \\
% Feb 13$ \Tnull $
% Feb 13   &   freier Diracoperator im $ \R^3 $   
% Feb 13   & S. \pageref{S_freierDiracRdrei} \\
$\Aeinsnull$
   & $ \sigma_2 p + \sigma_3 $ auf $[1,\infty) $
   & S.& \pageref{Aeinsnull} \\
  
$ \Tfr$
 & $ \left(
\begin{array}{cc}
0 & 1 + \lambda \\
1-\lambda & 0
\end{array}
\right) $
auf $ (0,b) $
& S.& \pageref{deftfrei} \\
   
 $ T_{q%_D
 ;ges} $ 
   & selbstadjungierter Operator mit kugelsymmetrischen Buckelpotential
   & S.& \pageref{dreidT}\\   
   
   $ \TqDmax $
   & maximaler Operator mit Potential $ q%_D 
   $
   & S. &\pageref{TqDmax} \\

  $ \Tnull $
  
   & freier Diracoperator auf $ [1,\infty) $
     & S. &\pageref{freierDiracaud1infty} \\
     
         $  \halles $ 
   & $ \sigma_2 p + \sigma_3 +q%_{D}
     + \sigma_1 \frac{k}{r} $ auf $ \R^+ $
   & S. &\pageref{hallllles} \\

$ \Tnulle $
  &    freier Diracoperator im $ \R^+ $   
  & S. &\pageref{eindimfreierOP}\\
% Jul 13$ \TqD $
% Jul 13  & selbstadjungierte Fortsetzung von $ \TqDmin $
% Jul 13  & S. \pageref{TqD} \\

$ \TqDschl $
   & selbstadjungierte Fortsetzung von $ \TqDschlmin $  auf  $[1,\infty) $
   & S. &\pageref{TqDsch} \\

$ \TqDschlmax $
   & maximaler Operator mit Potential $ q%_D 
   $ auf  $[1,\infty) $
   & S.& \pageref{TqDschlmax} \\

%$ \TqDschl $ 
%   & selbstadjungierte Fortsetzung von $\TqDschlmin $
%   & S. \pageref{TqDsch} \\
% Jul 13$ \TqDmin $
% Jul 13   & minimaler Operator mit Potential $ q_D $
% Jul 13   & S. \pageref{TqDmin} \\
$ \TqDschlmin $
   & minimaler Operator mit Potential $ q%_D 
   $  auf  $[1,\infty) $
   & S. &\pageref{TqDschlmin} \\

$ \tilde{T}_{q%_D
;k}$
   &  selbstadjungierter Operator mit Buckelpotential $ q%_D 
   $ zur Dreh\~impulsquantenzahl $ k $  auf  $[1,\infty) $
   &S.& \pageref{Thaengtvonkab}\\

$ W $
  & Buckelprofil bei identischen Buckeln
    & S.& \pageref{idBuckkel}\\

%$ \tau_{k, q_{D_n}} $
%   & Differentialauasdruck mit Potential und Drehimpulstermen
 %  & S. \pageref{BzentraleGleichungmitPotential} \\
$ W(v,y) $ 
   & Wronskideterminante
   & S. &\pageref{Wronski} \\
$  W_j  $
   & Form des $ j$-ten Buckels 
   & S. &\pageref{buckelform} \\
$ \tilde{W}(\tilde{v}%(r, \lambda)
,\tilde{ w }%(r, \lambda)
) 
$
   & Wronskideterminante
   & S. &\pageref{wronski2} \\
$ {\cal W}_j (a,r,\kappa) $
   & zweite Ordnung von $ A $ in $H$
   & S. &\pageref{anhang2asymptotikm} \\
% Feb 13$ Y_{l,j} $ 
% Feb 13   & Kugelfunktion vom Grad $ l $
% Feb 13   & S. \pageref{Kugelfunktion} \\
$ Y_{ \pm \modulus{k+\frac{1}{2}}} $
   & Weberfunktion
   & S.  &\pageref{Weberfunktion}\\
$     \Anullll $
   & $ \sigma_2 p + \sigma_3 + \sigma_1 \frac{k}{r} $ auf $ (0,1] $
   & S. &\pageref{Anuuuul} \\
$     \Aunendlich $
   & $ \sigma_2 p + \sigma_3 +q_{D}  + \sigma_1 \frac{k}{r} $ auf $ [1,\infty ) $
   & S. &\pageref{Aunendl} \\
&&\\

&&\\

$ \alpha $
  & Buckelbreite bei identischen Buckeln 
  & S. &\pageref{idBuckkel} \\
$ \alpha_j $ 
   & Buckelbreite
   & S. &\pageref{S_Buckelbreite} \\

$ \teta $
   & Pr"ufer-Winkel
   & S. &\pageref{S_prueferwinkel} \\

$ \teta_{%D_
n;j} $
   &  Wert des Pr"ufer-Winkels am Ende des $ j $-ten Buckles eines Potentials, dessen Buckelpositionen durch das
       Distanztupel $ D_n $ charakterisiert sind
   & S. &\pageref{S_Thetabuckelende} \\
 $ \tetas $
   & Pr"ufer-Winkel im Fall mit Drehimpuls
   & S. &\pageref{PrueferwinkelmitDreh} \\
 $ \teta_n  $
   & Pr"ufer-Winkel am Ende des $n$-ten Buckels im Fall mit Dreh\-impuls
   & S. &\pageref{S_Thetabuckelende} \\
$ \tetas_{%D_
n;j} $
   & Pr"ufer-Winkel am Ende des $j$-ten Buckels des $n$-Buckel\-potentials $ q_{%D_
   n} $ im Fall mit Drehimpuls
   & S. &\pageref{tetamitdreh} \\   
$ \tetas_n $ 
   &  Pr"ufer-Winkel am Ende des $n$-ten Buckels im Fall mit Dreh\-impuls
   & S. &\pageref{Rntetan} \\
   
$ \kappa $
   & $ \sign{(\lambda)} \sqrt{\lambda^2 -1} \quad(\lambda \in \R,
                                                    \modulus{\lambda} > 1) $
   & S. &\pageref{S_kappa}\\
$   \myD $
  & $ \lim_{n \to \infty} \myDn $
  &   S. &\pageref{grenzmass} \\

$   \myDss $
  & $ \lim_{n \to \infty} \myDjsss $
  &   S. &\pageref{grenzmassmitschlange} \\

$  \mynull $
   & Lebesguema"s
   & S. &\pageref{Lebesguemass}\\
      
$  \myDn $
   & Spektralma"s des $n$-Buckelproblems
   & S. &\pageref{teil2defmyDn}\\

$ \myDndTheta $
   & $  \int\limits_\Sigma \!\left(\!
        \frac{1}{\pi}\! \left( \!\prod_{j=1}^n\!
        \ffj \!\left(\!\kappa,\! \teta\ind (\!\kappa\!) \!-\! \kappa d_j\!\right)\! \right)\!
        f_{n+1} ( \kappa,\! \Theta(\kappa)\! -\!\kappa d)
        {\cal{D}} (\!\kappa\!)\!
        \right)\!
        \dkappa
$&\\
&
   & S. &\pageref{mymitdTheteta}\\
$ \myDns $
   & $ 
\int\limits_\Sigma \frac{1}{\pi}
                        \left( \prod_{j=1}^n \ffjs (\cdot, \tetas\ind(\cdot) - \cdot d_j,
                              \int_{b_{j-1}}^{a_j} \Gk(s, \cdot)\ds)
   \right)
%   \D (\cdot)
    %\left(1 + \GGGG_n(\cdot)\right)
%
$
%   &
%\\  & 
& 
 S. &\pageref{Bteil2defmyDn}\\

$ \myDjssssk $
  & $ \myDns $ mit expliziter Angabe der $k$-Abh"angigkeit
  &S.& \pageref{hiermitkabha} \\

$ \nyunten, \nyoben $
  & Indizes
  &
   S. &\pageref{nys}  \\

$ \Xi_n  $
   & $ [-2n,-\frac{1}{2n}] \cup [\frac{1}{2n},2n] $
   & S.& \pageref{Xiinterv} \\

$ \roDn $
  & Spektralfunktion zum $n$-Buckelproblem
    & S.& \pageref{teil2drodkappa}  \\

$ \roDnb $
   & Spektralfunktion zum $n$-Buckelproblem auf $ [0,b] $
   & S. &\pageref{Ateil2Sprungfunktion}\\
$ \roDns $ 
   & Spektralfunktion zum $n$-Buckelproblem auf $ [1,\infty) $ mit Dreh\-impuls
   & S. &\pageref{Bteil2drodkappa} \\
$ \roDnbs  $ 
   & Spektralfunktion zum $n$-Buckelproblem auf $ [1, b] $ mit Dreh\-impuls
   & S. &\pageref{Rhonbschlange} \\
%$\ros_{\nyoben;k} $
%   $   
$\sigma_1$, $\sigma_2$, $\sigma_3$  
   & Pauli-Matrizen 
   & S.& \pageref{S_Pauli-Mmatrizen} \\
 $ \sigmas $
   & stetiges Spektrum 
   & S. &\pageref{refernzspektrum}\\
   $ \sigmaess $
  & wesentliches Spektrum
  & S. &\pageref{Sigmaess} \\

$  \sigma_{sc} $
  & singul"arstetiges Spektrum
  & S. &\pageref{scSPektrum} \\
  
$   \tauqD $ 
  &  $ \sigma_2 p + \sigma_3 + q $ auf $ \R^+ $
  & S. &\pageref{OperatorenDifferentialausdruckohneDreh}\\

$ \taunull $
   &  $ \sigma_2 p + \sigma_3 $ auf $ \R^+ $
   & S. &\pageref{taunull} \\

   $  \tauschlqD $
   & $  \sigma_2 p +m  \sigma_3 + l + q%_D
    $ auf $ [1,\infty) $
   & S. &\pageref{OperatorenDifferentialausdruck} \\

\end{longtable}

}
%ende von symbolverzeichnis
\vfill
\pagebreak
\listoffigures

\vfill
\pagebreak
%biblio

\vfill
%ende von biblio
\pagebreak

%\include{lebenslauf}
%%%%%%%%%%%%%%%%%

Bei allen, die mich w"ahrend meines Studiums unterst"utzt haben, m"ochte ich mich vielmals bedanken.

Besonders gilt mein Dank Herrn Prof. Kalf f"ur seine inspirierende und geduldige Betreuung meiner Dissertation.

Es war mir eine Freude, mich mit Prof. Pearson anl"a"slich seines Besuches in M"unchen "uber Buckelpotentiale und singul"arstetige Schr"odingeroperatoren austauschen zu k"onnen.

%dank
%ende von dank

%erklaer
%%%%%%%%%%%%%%%%%

\vfill
\pagebreak

{\large \bf Eidesstattliche Versicherung
}

(Siehe Promotionsordnung vom 12.07.11, \S 8, Abs. 2 Pkt. .5.)

$\left.\right.$

Hiermit erkl"are ich an Eides statt, dass die Dissertation von mir
selbstst"andig, ohne unerlaubte Beihilfe angefertigt ist.

$\left.\right.$

 Barbara Janauschek

M"unchen, den 01.05.2014

\vfill

\textsc{Mathematisches Institut\\
Ludwig-Maximilians-Universit"at M"unchen\\
Theresienstra"se 39, 80333 M"unchen, Germany}

\textsc{E-mail address: janauschekb@web.de}
%ende von erklaer
\end{document}